%% file: thesis.tex
\documentclass[12pt,a4paper] {book}

\usepackage[dvips]{graphicx}
\usepackage{psfrag}
\usepackage{mathrsfs}
\usepackage{fancyhdr}
\usepackage{slashed}
\usepackage{verbatim}
\usepackage{simplewick}
\usepackage{sectsty}
\usepackage{pslatex}
\usepackage{amssymb}
\usepackage{amsmath}
\usepackage{doublespace} 
\fancypagestyle{abstract}{%
\fancyhead{} % get rid of headers on plain pages
}

\fancypagestyle{contents}{%
\fancyhead[L]{\leftmark}
}

\fancypagestyle{chapter}{%
\fancyhead{} % get rid of headers on plain pages
\fancyhead[R]{\rightmark \ \ \ \bfseries \thepage}
}

\fancypagestyle{normal}{%
\fancyhead[L]{\nouppercase\leftmark}
\fancyhead[R]{Hall\quad \thepage}
\fancyfoot{} % clear all footer fields
\addtolength{\headheight}{1.0pt} % make space for the rule
}
\begin{document}
%====================================================
%----------------------------------------------------
%   Bibliography Style
%
\bibliographystyle{alpha}
%
%----------------------------------------------------

\raggedbottom

% NEW COMMANDS
\include{newcommands}

%
%TITLE PAGE
%%%%%%%%%%%%%%%%%%%%%%%%%%%%%%
\pagestyle{empty}
\include{titlepage}

%%%%%%%%%%%%%%%%%%%%%%%%%%%%%%

\cleardoublepage

\newpage

\pagenumbering{roman}
\setcounter{page}{1}

%ABSTRACT + STATEMENT OF ORIGINALITY + ACKNOWLEDGEMENTS
\pagestyle{abstract}
\include{abstract}

\newpage
\thispagestyle{empty}
\cleardoublepage

%%%%%%%%%%%%%%%%%%%%%%%%%%%%%%

\begin{singlespace} 
%SINGLE OR DOUBLE SPACED?

%
%
%%%%%%%%%%%%%%%%%%%%%%%%%%%%%%
%CONTENTS
%
\thispagestyle{contents}
\tableofcontents

%\newpage
%\setcounter{page}{1}
%\listoffigures  

%\newpage
%\setcounter{page}{1}
%\listoftables

%\begin{titlepage}
%\thispagestyle{empty}
%\end{titlepage}
%\cleardoublepage
\newpage
\thispagestyle{empty}
\cleardoublepage

\end{singlespace} 
\newpage
\pagestyle{normal}
\pagenumbering{arabic}
\setcounter{page}{1}

\renewcommand{\topfraction}{0.85}
\renewcommand{\textfraction}{0.1}
\renewcommand{\floatpagefraction}{0.75}
%
%%%%%%%%%%%%%%%%%%%%%%%%%%%%%%%%%%%%
%
% CHAPTERS OF THESIS
%
\include{introduction}

\include{qcdonlattice}

\include{chieft}

\include{method}

\include{nucleonmass}

\include{mesonmass}
\include{nucleonmagmom}
\include{conclusion}

\thispagestyle{plain}

%
% APPENDICES
%
%%%%%%%%%%%%%%%%%%%%%%%%%%%%%%%%%%%%%%%%%%%%%%%%
\appendix

%\include{appendix1}
\include{appendix2}
\include{appendix3}
\include{appendix4}
%
% BIBLIOGRAPHY
%
%%%%%%%%%%%%%%%%%%%%%%%%%%%%%%%%%%%%%%%%%%%%%%%%

\pagestyle{plain}
\addcontentsline{toc}{chapter}{Bibliography}

\bibliography{references}

%\addtocontents{toc}{
%\vspace{4mm}
%\noindent\textbf{Appended Articles 1-6}}

\end{document}

%% file: newcommands.tex
\newcommand{\nc}{\newcommand}
\nc{\lb}{\langle}
\nc{\rk}{\rangle}
\nc{\Blb}{\Big\langle}
\nc{\Brk}{\big\rangle}

\nc{\mi}{\!\!\mid\!\!}
\nc{\ra}{\rightarrow}
\nc{\Ra}{\Rightarrow}
\nc {\cd}{\partial}
\nc {\sla}{\slashed}
\nc{\ro}{\mathrm}
\nc{\ca}{\mathcal}
\nc{\sr}{\mathscr}
\nc{\mb}{\mathbf}

\nc{\Tr}{\ro{Tr}\,}
\nc{\Str}{\ro{Str}}
\nc{\realtrace}{\ro{Re\; Tr}}
\nc{\maxrealtrace}{\ro{max\, Re\; Tr}}
\nc{\ud}{\ro{d}}
\nc{\nn}{\nonumber}

\nc{\pb}{\bar{\psi}}
\nc{\p}{\psi}

\nc{\vp}{\vec{\pi}}
\nc{\vk}{\vec{k}}
\nc{\vq}{\vec{q}}
\nc{\vap}{\varphi}
\nc{\vt}{\vec{\tau}}
\nc{\si}{\sigma}
\nc{\Si}{\Sigma}
\nc{\tSi}{\tilde{\Si}}
\nc{\tsi}{\tilde{\si}}
\nc{\g}{\gamma}
\nc{\G}{\Gamma}
\nc{\la}{\lambda}
\nc{\La}{\Lambda}
\nc{\ep}{\epsilon}
\nc{\de}{\delta}
\nc{\De}{\Delta}
\nc{\om}{\omega}
\nc{\Om}{\Omega}
\nc{\cL}{\ca{L}}
\nc{\cLe}{\ca{L}_{\ro{eff}}}
\nc {\ti}{\tilde}
\nc{\f}{\frac}
\nc{\da}{\dagger}
\nc{\SU}{\ro{SU}}
\nc{\rad}{\langle r^2\rangle}

\nc{\darrow}{\stackrel{\leftrightarrow}{\cd}}
\nc{\darrows}{\stackrel{\leftrightarrow}{\sla{\cd}}}
\nc{\Darrows}{\stackrel{\leftrightarrow}{\sla{D}}}
\nc {\mpisq}{m_{\pi}^2}
\nc{\mc}{\stackrel{\circ}{M}}
\nc{\gc}{\stackrel{\circ}{g}}
\nc{\fc}{\stackrel{\circ}{f}}
\nc{\mr}{\stackrel{\circ}{m}_\rho}
\nc {\eqb}{\begin{equation}}
\nc {\eqe}{\end{equation}}
\nc {\eqab}{\begin{eqnarray}}
\nc {\eqae}{\end{eqnarray}}

%% file: titlepage.tex
\begin{titlepage} 
\begin{center}

\vspace{\stretch{1}}
\includegraphics[scale=0.2]{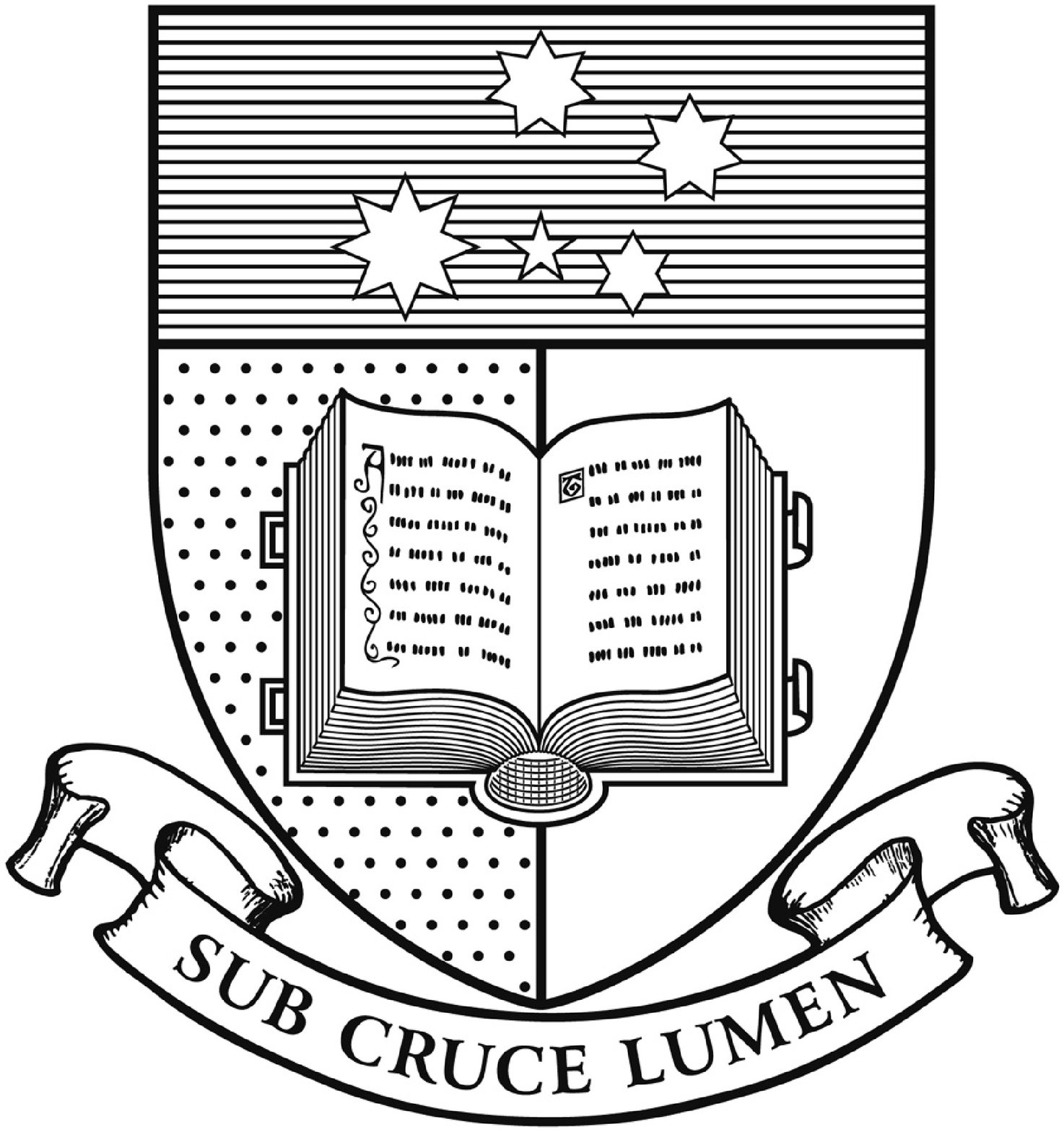} \\
{\Large \scshape The University of Adelaide} \\
{\Large \scshape School of Chemistry and Physics} \\
{\Large \scshape Discipline of Physics} \\

\vspace{\stretch{6}}
%{\LARGE \bf Renormalization Schemes for Chiral 
%Perturabation Theory and the Lattice} \\
%{\LARGE \bf Investigating Intrinsic Scales in the Regularisation
% of Chiral Effective Field Theory} \\

%{\LARGE  \textbf{Chiral Properties of Hadrons in Lattice QCD \\ 
%\vspace{\stretch{1}} using Effective Field Theory}} \\

{\LARGE  \textbf{Chiral Effective Field Theory \\ 
\vspace{\stretch{1}} Beyond the Power-Counting Regime}} \\

\vspace{\stretch{3}}
{\Large Jonathan Michael MacGillivray Hall} \\

\vspace{\stretch{3}}
{\large Supervisors: Prof. Derek B. Leinweber, Dr. Ross D. Young} \\

\vspace{\stretch{3}}
{\large \textit{Special Research Centre for the \\
Subatomic Structure of Matter\\
\emph{and} \\
Department of Physics,\\
University of Adelaide,\\
Australia}

}

\vspace{\stretch{12}}
{\large  June 2011} \\

\pagebreak

\end{center} 
\end{titlepage}

%% file: abstract.tex
%\begin{abstract}

% ~ < 200 words. Aim for 100.

%\end{abstract}

\begin{center}
\subsection*{Abstract}
\end{center}

Chiral effective field theory complements numerical
simulations of quantum chromodynamics on a spacetime lattice.
It provides a model-independent formalism for connecting lattice simulation 
results at finite volume, and at a variety of quark masses, 
to the physical region. 
Knowledge of the power-counting regime of chiral
effective field theory, where higher-order terms of the expansion may be
regarded as negligible, is as important as knowledge of the expansion.
Through the consideration of a variety of renormalization schemes, techniques
 are established to identify the power-counting regime. Within the 
power-counting regime, the results of extrapolation 
are independent of the
renormalization scheme. 

The nucleon mass is considered as a
benchmark for illustrating this %general 
approach. Because the power-counting 
regime is small, the numerical simulation results are also examined to search 
for the possible 
presence of an optimal regularization scale, which may be used to
describe lattice simulation results outside of the power-counting regime. 
%Positive results are reported, which 
%improve on the optimistic application of chiral perturbation theory
%beyond the region of its applicability. 
%
Such an optimal regularization 
scale is found for the nucleon mass. %at chiral order 
%$\ca{O}(m_\pi^3)$. 
The identification of 
%systematic error in the 
an optimal scale, with its associated systematic uncertainty, measures the 
degree to which the lattice QCD simulation results extend beyond the 
power-counting regime, thus quantifying the scheme-dependence of an 
extrapolation. 
%In the case of chiral order $\ca{O}(m_\pi^4\,\ro{log}\,m_\pi)$, the 
%power-counting regime extends further, and an optimal scale is not realized. 
%Consequently, a broader range of acceptable regularization scales leads to 
%statistically consistent extrapolations, and there is no need for special 
%preference towards a particular regularization scale. 

The techniques developed for the nucleon mass
renormalization are applied to the quenched $\rho$ meson mass, which offers 
a unique test case for extrapolation schemes. In the absence of a known 
experimental value, 
 it serves to demonstrate the ability of the extrapolation scheme
 to make predictions without prior phenomenological bias. 
The robustness of the procedure for obtaining an optimal regularization scale 
and performing a reliable chiral extrapolation is %thus 
confirmed.

The procedure developed %in the thesis 
is then applied to the magnetic moment 
and the electric charge radius of the isovector nucleon, to obtain 
a consistent optimal regularization scale. The consistency of the results 
for the 
value of the optimal regularization scale provides strong evidence for the 
existence of 
an intrinsic energy scale for the nucleon-pion interaction. 

%Considering one more example, the quenched $\rho$ meson mass offers 
%a unique testbed for extrapolation schemes. %since the result of the 
%calculation cannot be biased in advance from experiment.
% In the absence of a known experimental value, 
% it serves to demonstrate the ability of the extrapolation scheme
% to make predictions with reduced phenomenological bias. 

\newpage
\thispagestyle{empty}
\cleardoublepage

\newpage

\begin{center}
\subsection*{Statement of Originality}
\end{center}

%This work contains no material which has been accepted for the award
%of any other degree or diploma in any university or other tertiary
%institution and, to the best of my knowledge and belief, contains
%no material previously published or written by another person,
%except where due reference has been made in the text.

%I give consent to this copy of my thesis, when deposited in the University
%Library, being available for loan and photocopying.

This work contains no material which has been accepted for the award
of any other degree or diploma  in any university or other tertiary
institution to Jonathan Hall and, to the best of my knowledge and belief, 
contains no material previously published or written by another person,
 except where due reference has been made in the text. 

I give consent to this copy of my thesis, when deposited in the University
Library, being available for loan and photocopying, subject to the provisions 
of the Copyright Act 1968. 

I also give permission for the digital version of my thesis to be made 
available on the web, via the University's digital research repository, 
the Library catalogue and also through web search engines, 
unless permission has been granted by the University to restrict 
access for a period of time. 

\newpage
 \thispagestyle{empty}
\cleardoublepage

\newpage

\begin{center}
\subsection*{Acknowledgments}
\end{center}

%Acknowledge Rod
Thank you to Professor Derek Leinweber, for his patient and intelligent 
supervision and his sense of humour expressed during our enjoyable 
discussions. Also thank you to Doctor Ross Young for his explanations of 
theory, his patience, and for our numerous, informative conversations. 
Ross' generosity with his time for me has been exemplary, 
and in many cases, an essential component of our successes. 

I also thank Doctor Rod Crewther, who has
 been very informative in the fields of physics 
and education, and Doctor James Zanotti, who has helped marvellously 
with collaboration and support.

Thank you to the staff of the School of Chemistry and Physics for their 
general assistance, and more specifically to Professor Anthony Thomas and 
the staff of the Special Research Centre for the Subatomic Structure of Matter.
 %I also really appreciated the daily companionship of my fellow students.

I give my thanks to my loving family for their support.

It is with hope and faith that we endeavour to extend our learning to reach 
new insights just beyond our present reach.

%% file: introduction.tex
\chapter{Introduction}
\label{chpt:introduction}

\textit{``One measure of the depth of a physical theory is the extent to which it poses serious challenges to aspects of our worldview that had previously seemed immutable.''}
(Greene, B. 1999. \textit{The Elegant Universe} p.386) \cite{Greene} %\cite{Greene})

\section{Prologue}

%Define things such as QED, QCD, ChiPT, LEC's, etc.
%Don't use the acronyms in opening/closing statements
%of chapters and thesis.

%Need to introduce Feynman slash notation?

The theoretical physicist challenges previous theory, 
using original research that enables alternative coherence to emerge, 
as outlined by Bohm \cite{Bohm} (p.223). 
The underlying theory behind the strong force of particle interactions, 
 which is the force 
responsible for the binding of protons and neutrons together in 
atomic nuclei, had been a persistent  
mystery throughout the first half of the Twentieth
 Century. This hitherto unknown force 
% is responsible for binding the protons and neutrons together in atomic 
%nuclei, and this force is 
acts in opposition to the electric Coulomb force 
that repels positively charge 
protons from each other, but is at least two 
orders of magnitude stronger at the distance scale of an atomic nucleus.
%
%at least two orders of magnitude stronger than the electromagnetic force 
%that would act to repel positively charge 
%protons from each other. 
The strong interaction between protons and neutrons, or nucleons,  
is currently most successfully described by the theory of quantum 
chromodynamics (QCD). The advent of the quark model, and the theory of 
the colour force by which the quarks interact, opened a new field 
of research into the internal structure of matter. 

In 1964, Gell-Mann and Zweig independently proposed the existence of 
a new constituent particle, the quark, in order to classify the bewildering 
array of subatomic particles called hadrons \cite{GellMann:1964nj}. 
It was discovered that the 
hadrons  
can be arranged into families that correspond to 
representations of the group $\SU(3)$, and that three 
quark types, or flavours, 
were required to form the fundamental representation of this group. 
It was not until 1968 that the results of deep inelastic scattering 
experiments provided the first evidence of the existence of these 
new elementary particles. 
As more hadrons were discovered, additional quark flavours were proposed.  
It is currently accepted that six 
flavours of quark are required to produce the full range of hadrons observed in 
particle accelerator experiments. Their names, in ascending order of mass, 
are: up, down, strange, charm, bottom and top. Of the six flavours, the 
most recent to be discovered experimentally was the top quark, 
in 1995 at Fermilab, with a mass of 172 GeV \cite{Protopopescu:1995zz}.
 
Each quark has a unit of charge equal to $+2/3$ or $-1/3$ times the 
charge of a proton (units of $+e$). 
An an example, the proton consists of two up quarks 
and a down quark for a total charge of $+1\,e$, whereas a neutron 
consists of two down quarks and an up quark for a total charge of zero. 
However, because quarks have a certain spatial 
distribution inside the nucleon, or indeed any hadron, 
 the internal, high energy 
dynamics as described by the behaviour of quarks %requires 
%a suitable theory, such as QCD, in order to predict 
gives rise to properties such as non-zero magnetic moments for the neutron and 
anisotropic momentum distributions. 
It is clear that in order to describe the internal behaviour of a hadron, 
one cannot assume that a quark behaves as a static source. Instead, the 
dynamics of quarks must be described by a theory, the most successful of 
which is QCD.

QCD connects the quark model of nuclear physics to quantum gauge field theories 
by introducing the quarks as the relevant degrees of freedom inside a hadron. 
The hadrons are formed by confined colour singlets of three quarks called 
baryons, 
or  quark-antiquark pairs, known as mesons.
Quarks are spin-$1/2$ fermions, which also have the properties of colour and 
approximate flavour symmetry. Since fermions, by definition, must 
 satisfy Fermi-Dirac statistics, the fact that each 
%Since
 baryon contains three bound quarks in the same state %already violates 
%the expected statistical behaviour of quarks. 
violates the Pauli Exclusion Principle.  
Therefore, it was necessary to suppose an the existence 
of an additional quantum number, %in a hadronic system, 
%so that the non-integral spin quarks 
%would satisfy Fermi-Dirac statistics. 
known as colour charge, so that each quark may be assigned one of three, 
orthogonal basis states, labelled red, green and blue. 
 Colour is mediated by the related gauge particles of the strong force; 
the gluons, and the 
%associated colour 
%charges are conventionally labelled red, green and blue, and 
%They represent the ``charge'' acted
% on by the strong force. 
 %thus 
 quarks also form a representation of the colour gauge group 
$\SU(3)_{c}$, with eight group generators. 

Mathematically, QCD is a non-Abelian theory. That is, 
the gauge connection of the gluons %, 
%which is a Yang-Mills potential, 
is non-commutative. 
The fact that isolated, unbound quarks are never found in experiment 
is one of the striking consequences of a non-Abelian theory. 
Confinement of the quarks within a hadron is a result of the gluon fields 
exerting a linear potential that increases with distance between quarks 
\cite{Wilson:1974sk}. That is, for a large distance scale, the strong coupling 
 parameter $\alpha_s$ also becomes large. 
 This behaviour %is the opposite to 
contrasts with electromagnetism, %(in greater than two
% dimensions), 
where the electric Coulomb force 
diminishes as two charged particles are separated. %, and increases 
% at short distances, becoming infinite at the Landau Pole. 
 However, quarks experience only a small force from 
the gluon fields as $\alpha_s$ becomes small at short distances 
\cite{Gross:1973id,Politzer:1973fx,Gross:1973ju,Gross:1974cs}. 
This asymptotic freedom is observed when probing the 
internal structure of hadron at high energies, where the small de Broglie 
wavelength of the probe is able to resolve the short distances within the 
composite particle. Near this asymptotically free regime, the methods 
of perturbative quantum field theory are suitable for constructing 
 amplitudes, cross-sections and scattering matrices. However, 
it leads to a difficulty in finding an appropriate method for performing 
a calculation with QCD in the low-energy region. 
Two of the most successful methods 
that will be discussed in this thesis are chiral effective field theory  
($\chi$EFT) and lattice QCD.

%In QCD, flavour symmetry is violated  
%by the finite (and differing) masses of the six quarks; a feature %consequence 
%of a process known as dynamical chiral symmetry breaking, discussed 
%in Chapter \ref{chpt:chieft}. 
%In %a theory corresponding to the low-energy region of QCD, 
%$\chi$PT, approximate 
%$\SU(2)_{f}$ or $\SU(3)_{f}$ symmetry is assumed for the lightest quarks: 
%up, down (and strange). %This is 
%%not an unreasonable supposition, %since the explicit 
%symmetry breaking of the global gauge group, which induces the mesons to have 
%finite (but calculable) mass, can be accommodated in calculations and 
% By making this assumption, the theory can be constructed that incorporates 
%the approximate gauge symmetry, and allows the identification of the 
%symmetry-breaking terms. %This theory, known as chiral perturbation theory 
%($\chi$PT) 
% leads to interesting low-energy physics that agrees with experiment. 
%  A key example of this is the anomalous magnetic moment of the nucleon 
%(proton or neutron), discussed further in Chapter 
%\ref{chpt:nucleonmagmom}.

Using $\chi$EFT, 
one is able to encapsulate the dynamics of a quantum system by writing 
down an `effective' action of low-energy degrees of freedom. 
By imposing symmetries 
 satisfied by QCD, % (in some region of applicability), 
one can expand out 
the formula for an observable property 
 into a series of quantum 
amplitudes that can be arranged 
in order of the importance of their contribution 
by a choice of power-counting scheme: usually in increasing powers of 
mass/energy. These amplitudes can alter, or 
%\emph
{renormalize} the 
 calculation of an observable from its na\"{i}ve value, and landmark success 
has been made in confirming these real and measurable effects by experiment.
For example, the value of the anomalous magnetic dipole moment 
of the electron agrees with experiment to better than twelve significant 
figures. %, and the quantum 
The Casimir effect (1948), which describes the forces arising from the quantum 
vacuum fluctuations, were successfully predicted 
by the gauge field theory of quantum electrodynamics (QED).
In the low-energy, non-perturbative region of QCD, many phenomena can 
be explained by the emergent properties of quark confinement and 
the behaviour of their bound states as hadrons. 
For example, the proton and neutron also have a large anomalous 
component of their magnetic moment. This is due to 
the cloud of interacting fields, which renormalize the core of the observable. 
This `hadron cloud' is one of the unique properties of a quantum 
field theory.
Of the available low-energy effective theories of QCD, 
chiral perturbation theory ($\chi$PT) is the most notable, in its careful 
incorporation of the fundamental symmetries of QCD.
However, the robustness of $\chi$PT 
is confined only to a restrictive region called the power-counting regime. 
Within the power-counting regime, the perturbative expansions that occur 
in $\chi$PT are convergent; the terms of the expansion series are 
ordered such that higher-order terms are sufficiently smaller than 
lower-order terms. The details of the power-counting regime are discussed 
in more detail in Chapter \ref{chpt:chieft}.

Lattice QCD is a discretized version of QCD, where the dynamics are 
evaluated on a finite-sized box with only certain allowed values of 
position (or momentum) separated by a fixed spacing.
Thus, lattice QCD is equivalent to QCD in the limit of infinite 
box size and vanishing lattice spacing. 
Using lattice QCD, one is able to access the non-perturbative, low-energy 
regime of QCD and provide reliable predictions of hadronic behaviour. 
In addition, lattice QCD simulations do not suffer from the common 
 problems of quantum field theory %maladies 
associated with renormalization. 
The discrete lattice spacing and the finite box size of the lattice 
act as an ultraviolet and infrared regulator, respectively. 
Thus, observable quantities evaluated on the lattice are finite 
and calculable. 
Nevertheless, it can be computationally 
expensive to 
evaluate observables at large box sizes, small lattice spacings and 
physical quark masses. To be able to obtain a result using quark masses 
as small as their physical values, 
an extrapolation is a practical alternative to a brute-force 
approach. In addition, the 
corrections to finite-volume effects ought also to be 
calculated for a realistic comparison with experiments. 

\section{Overview and Aims}

The framework of 
%quantum chromodynamics (QCD) 
QCD provides a rich selection of possibilities 
for inquiry. Among these, the low-energy, %\emph
{chiral} dynamics of hadrons 
 provides us with a uniquely successful understanding of many of their 
imporant properties. 

This thesis explores the properties of the aforementioned power-counting regime
  by considering how low-energy constants, which occur in a calculation 
using the methods of $\chi$PT, %of the effective action %(LECs) 
are renormalized, or altered, at different energy scales. 
%by quantum amplitudes 
%constructed from $\chi$PT. 
This knowledge of the power-counting regime, 
in turn, yields insight into the repercussions 
of chiral symmetry breaking in QCD. %, which occurs in the physical universe.

The results of lattice QCD simulations provide an  
%practical 
important application for the %motivation for this 
investigation into $\chi$PT and the power-counting regime. %Few 
%lattice QCD results in the literature are evaluated at quark masses 
%that lie within the power-counting regime, 
Lattice QCD results are typically produced at a variety of 
quark masses larger than the physical quark mass. 
As such, a chiral 
extrapolation to the physical point is required before 
the result can be compared to experiment. 
In addition, experimental results are not constrained by the boundaries 
of a small box only a few fermi in length. It is important to be able to 
quantify how the finite-volume nature of lattice QCD affects calculations. 
Analysis shows that the finite-volume behaviour of QCD on the lattice can 
affect the result of a calculation in non-trivial ways. %A finite-volume 
Being able to perform an extrapolation that takes into account finite-volume 
effects 
is also an important step in understanding the effects 
of a finite-volume box on the dynamics of QCD.

%As of early $2011$, 
The investigation of the power-counting regime has additional importance. 
Few lattice QCD results in the literature are evaluated at quark masses 
that lie within the power-counting regime. 
As such, the powerful tools associated with $\chi$PT may not 
be used legitimately, since 
the chiral power-counting expansion of an observable 
would not be convergent. 
If higher-order terms in the series expansion are not small with 
respect to some power-counting scheme, the result of an extrapolation 
will be scheme-dependent. 
This thesis describes the construction of 
an extended effective field theory %($\chi$EFT) %(E$^2$FT) 
that can be applied outside the power-counting regime by  
%sacrificing scheme-independence, but 
extracting a quantative estimate 
of the extent of the energy scale-dependence, associated with the 
process of regularization in $\chi$PT calculations. It is discovered that %the 
lattice simulation results themselves can 
provide guidance on an optimal choice of regularization scale. 
This optimal scale indicates a possible connection with the 
finite-size of the hadron cloud in the form of an intrinsic scale.

Thus, by analyzing the results from the supercomputer simulations of 
lattice QCD, an %\emph
{intrinsic scale} will be discovered that characterizes
 the finite size of the interaction between the hadron cloud and the 
core of the hadron.

%% file: qcdonlattice.tex
%\chapter{QCD on the Lattice}
\chapter{Lattice QCD}
\label{chpt:qcdonlattice}
%Description

\textit{``While the classical vision of the world is intrinsically limited, nothing restricts the scientific representation. During the conception stage, the method is free to consider all hypotheses, even the most far-fetched, in order to mimic Reality.''}
(Omn\`{e}s, R. 2002. \textit{Quantum Philosophy: Understanding and Interpreting Contemporary Science} p.268) \cite{Omnes}

%INTRO TO QCD

The inception of a discrete, lattice 
approach to quantum chromodynamics (QCD) in 1974 by Wilson marked 
the beginning of a robust, investigative technique 
into the previously inaccessible low-energy region of strong force 
interactions \cite{Wilson:1974sk}. 
By simulating the behaviour of quarks on a lattice, 
bound states of hadrons are formed, exhibiting confinement, and 
the behaviour of particle interactions is correctly predicted: 
a testament to the achievement of QCD as a theory of the strong force. 

Lattice QCD provides non-perturbative techniques for obtaining results 
from the % dynamics of QCD that are applicable  
%in principle, %across a wide range of energy
% regions. 
%to the 
low-energy, chiral dynamics of hadrons. %, and to the result of 
%high energy scattering experiments. 
It involves the construction of a finite-volume box of discrete momenta, 
with calculations performed from first principles.
 The finite box size of the lattice removes any infrared divergences 
that would occur in infinite-volume QCD, and 
 the lattice spacing acts to regulate the ultraviolet behaviour
of observable quantities by limiting the lattice momenta 
to discrete values. 

%Also, pinch stuff from latticeqcd.tex file from honours, if needed.
% A perturbative calculation is often divergent in the low-energy 
%region because of quark confinement. 

%so perturbative schemes 
%make sense in the high energy region of QCD only, as long as there 
%is a suitable renormalization scheme. 
%Although the interaction coupling of QED increases %approaches the %\emph{Landau Pole} 
 %Landau Pole 
%at short distances, 
%\cite{}. LAUNDAU POLE 
%in QCD, quarks become asymptotically free %non-interacting 
%particles for high energy interactions %(%\emph
%{asymptotic freedom}), 
%and can thus be written as a perturbative expansion of its 
%$n$-point Green's Functions.% \cite{}. Scherer? 

In lattice QCD, a Euclidean hypercube is constructed 
with finite length and discrete lattice spacing. 
The quantum field theory can then be 
represented by the functional integrals defined on
 such a box. 
 The momenta can only take the discrete values in the four-box:
\eqb
k_\mu = \frac{2\pi}{a N}n_\mu\,,
\eqe
where $a$ is the lattice spacing, $n_{\mu}$ is an integer array representing 
the lattice sites, and $N$ 
is the number of lattice sites in each direction, such that 
$-N/2 < n_\mu \leq N/2$. \, 
Thus, the maximum value $k_\mu$ can take is $\pi/a$. 
This means that the ultraviolet physics included in our
 lattice is entirely determined by the lattice spacing, 
which thus acts to %remove 
regulate arbitrarily hard momentum contributions 
to quantum field theoretical quantities. 
%The ideal physics 
 The real-world dynamics of QCD are 
%is 
recovered in the limit of vanishing lattice
 spacing (the continuum limit) and the infinite-volume limit. 
%which taken first?

%Since loop integrals are now also discretized on the finite-volume lattice, 
%the three-dimensional integrals encountered in the non-relativistic 
%heavy-baryon limit
%(after the time component has been integrated out) 
%can be replaced by summations over all possible momentum
% values using this procedure \cite{Armour:2005mk}:
%
%\eqb
%\label{eqn:discr}
%\int\! \! \ud^3k \, \,   \approx \frac{1}{L^3} \left( \frac{2 \pi}{a} \right )^%3 \sum_{k_{x},k_{y},k_{z}}\,,
%\eqe
%
%where $L = aN$ is the lattice box length, assumed equal in each spatial 
%direction. 
%It is useful to define the %\emph
%{finite-volume correction} to the loop 
%integral, by convention, subtracting the integral from the sum quantity.
%This technique will be used to correct for finite-volume effects
% encountered in 
%Chapters \ref{chpt:intrinsic} through \ref{chpt:mesonmass}.

%introduce fvcs?

The dynamics of QCD are encoded in the QCD Lagrangian: a quantity 
in quantum field theory that extends the classical notion of the 
difference between the kinetic and potential energy terms to 
a density in spacetime. The generalized kinetic and potential terms 
are constructed from the relevant degrees of freedom: quantum fields 
\cite{WQTF}. 
 %which 
The QCD Lagrangian includes 
a sum of Fermi-Dirac Lagrangians for all quark flavours, 
an interaction term 
and a Yang-Mills term. In tensor form (and summing over repeated indices), 
the Lagrangian reads: 
\begin{align}
\cL_{\ro{QCD}} &= \cL_{\ro{Dirac}} + \cL_{\ro{int}} + \cL_{\ro{YM}} \\
&= \sum_{q}\Big\{\pb_q^i (\g^\mu\darrow_\mu-m_q)\p_q^i 
- \alpha_s\pb_q^i\g^\mu J_a^{ij}\ca{A}^a_\mu\p_q^j\Big\}
- \f{1}{4}G_{\mu\nu}^a G^{\mu\nu}_a\,.
\label{eqn:lattqcdlag}
\end{align}
 The fields $\p_q$ and $\pb_q$ 
are Dirac spinors representing 
different quark flavours and colours, with mass $m_q$. 
(Dirac spinor algebra was introduced in References 
\cite{Dirac:1928hu,Dirac:1928ej}, and 
some of the basic properties of a Dirac spinor can be found in Appendix 
\ref{app:spinors}.) 
The fields $G_{\mu\nu}^a$ are the non-Abelian field strength tensors 
corresponding to the gluon field $\ca{A}_\mu^a$, via the equation:
\eqb
G_{\mu\nu}^a = \cd_\mu \ca{A}_\nu^a - \cd_\nu\ca{A}_\mu^a - 
i\alpha_s f_{abc}\ca{A}_\mu^b \ca{A}_\nu^c\,,
\eqe
where the 
structure constants $f_{abc}$ are defined in Appendix \ref{app:struc}. 
The Yang-Mills term describes the self-interaction 
of the gluon fields, such that the result is invariant with respect to 
a special type of symmetry known as the gauge symmetry. In QCD, the 
gauge symmetry is realized in the Lagrangian by forming representions of a 
mathematical group, in this case, $\SU(3)_c$ (where $c$ stands for 
`colour'). Each term in the Lagrangian must be invariant under transformations 
involving this group. 
The quark spinors form a basis for the fundamental 
representation of the group. The gluon fields, however, are defined 
in the eight-dimensional representation of $\SU(3)$, and the index 
$a$ runs from $1$ through $8$. The matrices $J_a^{ij}$ are the generators 
of the gauge group $\SU(3)$.  A detailed review of the symmetries 
of QCD is included in Chapter \ref{chpt:chieft}. Suffice to say, 
the Lagrangian in Equation (\ref{eqn:lattqcdlag}) will be assumed in 
defining the QCD Action in the following Section.

\section{Functional Methods}
\label{sec:funcmeth}
%talk about c-numbers, grassmann alg, etc, eg. Steven Kerr's talk
%and link to Appendix2: The spinor field identities can be referred to.
%base on Scherer, and DGH.
%Talk about Ward Identities

Lattice QCD relies on a variety of techniques to obtain information 
about the dynamics of QCD. In particular, the path integral 
method of quantization serves as a starting point, where complex 
valued Grassmann fields are used to represent the quark spinors $\p$ and
 their adjoints $\pb$. (For a short summary on the properties of Grassmann 
algebra and Berezin integrals, refer to Appendix \ref{app:spinors}.) 
Before introducing the procedure for calculating
 the expectation values of observables using 
lattice QCD, it is helpful to review the functional methods 
 required for defining the generating functionals and the $n$-point Green's 
Functions. In the following Section, use is made of the functional derivative 
$\f{\de}{\de \ca{J}(x)}$, the properties of which follow analogously 
from the standard derivative of a function \cite{MM}. 

Consider the generating functional technique, choosing a set of fields 
$\Phi = \{\ca{A}_{\mu}^a,\p,\pb\}$, defined by a set of gauge fields 
$\ca{A}_\mu^a$ 
 and Dirac spinors $\p$ \& $\pb$,   
and integrating over all possible paths. 
In general, for a Lagrangian $\cL(\Phi, \cd^{\mu}\Phi)$, 
the corresponding action can be written as follows:
\eqb
\label{eqn:testaction}
S[\Phi] = \int\! \! \ud^4 \!x\,\, \cL(\Phi(x), \cd^{\mu}\Phi(x))\,.
\eqe
The generating functional with source terms $\ca{J}(x_{i})$ 
 takes the form: %explain
\eqb
\label{eqn:genfunc}
\ca{Z}[\ca{J}(x_{i})] = \frac{1}{N} \! \int\! \!
 \ca{D}\Phi\,\, \ro{exp}\left
\{iS[\Phi] - \int\!\!\ud^4 x \,\ca{J}(x_{i})\Phi(x_{i})\right\}\,, 
\quad \int\!\!\ca{D}\Phi \equiv \prod_{i=1}^{\infty}
\int\!\!\ud \Phi_i,
\eqe
with normalization:
\eqb
 N = \int\! \! \mathcal{D}\Phi\,\,\ro{exp}\left\{iS[\Phi]\right\}\,.
\eqe
 The calculation of the $n$-point Green's Functions is performed 
 by taking functional derivatives of 
  the generating functional 
with respect to sources $\ca{J}(x_{i})$, and then setting each source to zero:
\eqb
\tau^{(n)}(x_{1},\cdots,x_{n}) = \frac{1}{N} \! \int\! \! 
\ca{D}\Phi\,\,\Phi_{1}\cdots\Phi_{n}\,\ro{exp}\left\{iS[\Phi]\right\}\,.
\eqe
In order to obtain only the connected diagrams for the generating 
functional, one can define the connected generating functional $\ca{W}$:
\eqb
\ca{W}[\ca{J}] = -i\,\ro{log}\ca{Z}[\ca{J}].
\eqe
The connected (or irreducible)  
$n$-point Green's Functions can then be calculated as 
the time-ordered vacuum expectation values of the fields, with respect to 
the interacting vacuum $|\Omega\rk$:
\eqb
\label{eqn:nptGF}
G^{(n)}(x_{1},\cdots,x_{n}) = \lb \Omega\!\mid\ca{T}[\Phi(x_{1})
\cdots\Phi(x_{n})] \mid\! \Omega\rk  = \f{1}{i^n}\f{\de^{(n)}\ca{W}[\ca{J}]}
{\prod_{i=1}^{n}\de \ca{J}(x_i)}\Bigg|_{\ca{J}=0}\,.
\eqe
%
%up to a normalization constant.
%

%This normalization factor is the partition function 
%for calculating expectation values of observables.
The generating functional of Equation (\ref{eqn:genfunc}) is useful 
%as it is in 
for constructing an expansion of amplitudes. 
This expansion is obtained from the %\emph
{Schwinger-Dyson} equations, 
the set of differential equations satisfied by the generating functional:
\eqb
\f{\de}{\de\Phi(x_i)}S\left[\f{1}{i}\f{\de}{\de \ca{J}}\right]
\ca{Z}[\ca{J}(x_i)] + \ca{J}(x_i)\ca{Z}[\ca{J}] = 0.
\eqe
The Schwinger-Dyson Equations are simply the Euler-Lagrange equations 
of motion for the $n$-point Green's Functions of the gauge field theory. 
They provide a continuum persective to the challenging problems of 
 non-perturbative QCD, as summarized by Roberts and Williams 
\cite{Roberts:1994dr}. 
The investigation of the analytic properties of 
these equations form a crucial component of the study of quark confinement: 
where the strong coupling parameter becomes large. The Schwinger-Dyson Equations
 also shed light onto the process of dynamical chiral symmetry breaking, 
discussed in detail in Chapter \ref{chpt:chieft}. 
%The study of Schwinger-Dyson Equations
%
%However, a suitable approximation and truncation scheme is required to 
%obtain a closed result.
%

%

%From the formalism, 
Physical observables of a system can be obtained conveniently 
using Equation 
(\ref{eqn:nptGF}). 
 To evaluate expectation values $\lb\ca{O}\rk$ numerically,
 it is common practice to remove the difficulties of Minkowski spacetime by
 an analytic continuation to imaginary Euclidean time, or a 
Wick rotation, $t \rightarrow -it$, and $S = iS_{E}$. 
Thus the expectation values become numerically soluble,
 since the highly oscillatory behaviour of the $n$-point Green's Functions
 have been exponentially damped. Thus:
\eqb
\lb \ca{O} \rk = \f{\int\! \! \ca{D}\Phi\,
\ca{O}\,\,\ro{exp}\left\{-S_{E}[\Phi]\right\}} 
{\int\!\! \ca{D}\Phi\,\,\ro{exp}\left\{-S_{E}[\Phi]\right\}} \,,
\eqe
which is of the same form as the correlation function 
in statistical mechanics. Using the Euclidean Action,
 the fermionic part of the partition function can be 
calculated explicitly, leaving an expression in terms of a 
fermion correlation matrix $\ca{M}$:
\eqb
\label{eqn:genfuncwithdet}
\ca{Z} = \int\!\! \ca{D} \ca{A}_{\mu}^a\,\, \ro{det}(\ca{M}[\ca{A}_\mu^a]) 
\,\,\ro{exp}\left\{-S_E[\ca{A}_\mu^a]\right\}\,.
\eqe
%
%
%DISCUSS DIFFERENT ACTIONS, WILSON FERMIONS, DOMAIN WALL, OVERLAP.. ETC.
%WHATEVER IS USED IN THE PROCEEDING CHAPTERS
%Talk about correlation functions (another section/subsection?)

\subsection{Wilson Fermions}

In constructing an action on the lattice,
such as that of Equation (\ref{eqn:testaction}), 
 there is a 
difficulty in implementing the fermion field. This difficulty is known 
as the 
%\emph
{fermion doubling problem}. The problem occurs when solving the
 kinetic part of the Dirac Equation of motion, $(i\slashed{\cd} -m)\psi = 0$, 
on the lattice. %The covariant derivative $D$, as described in 
%Chapter \ref{chpt:chieft}, 
The derivative $\cd$ is taken as an 
average (or a forward-backward average so that the result is Hermitian), 
and the propagator derived is of the form: $\sin(\slashed{p} + m)^{-1}$.  
The correct behaviour of the Green's Function is exhibited as 
$p \rightarrow 0$, but as $p \rightarrow \pi$ the propagator also 
vanishes at the edge of the Brillouin Zone: the fundamental cell 
of a lattice theory with a periodic boundary.  Thus for 
$\sin(\slashed{p}) = 0$ there are $2^{dim}$ degenerate quarks for each flavour,
 which corresponds to sixteen degenerate quarks in four-space. In order 
to amend this, Wilson introduced a five-dimensional operator, which 
increases the mass of the doubler species proportional to lattice spacing
 $a$ \cite{Wilson:1974sk}. Note that as $a \rightarrow 0$ in the continuum 
limit, the Wilson term disappears and does not %unphysically 
alter the dynamics of QCD.
%
%From the Wilson Action, a %\emph
%{Wilson Line} may be defined, 
%which describes the 
%parallel transport for fermions around a closed loop. % which is the
% \emph{plaquette} from which the gauge connection can be derived 
%\cite{Kizilersu}. 
However, %chiral symmetry is violated by the Wilson Action, and large 
%scaling violations occur. 
the Wilson Action violates chiral symmetry. 
This important symmetry ensures the consistent renormalization 
of the low-energy constants of the Lagrangian via the chiral Ward Identities, 
which describe the conservation of a symmetry as applied to 
quantum amplitudes. Chiral symmetry is described in more depth in 
Chapter \ref{chpt:chieft}. 
 %In order to recover chiral symmetry approximately, 
Additionally, a %\emph
 so-called {Clover term} is often added to the 
 Lagrangian, which is proportional to $\bar{\psi}J^a
G_{\mu\nu}^a\psi$. This term is also 
a five-dimensional object, %but as long as extra terms added into the 
% Lagrangian are polynomial in $a$, 
 %they are 
and, like the Wilson Action, is suppressed in the continuum limit. 
In addition, errors of $\mathcal{O}(a)$ 
can be removed, and higher-order errors of $\mathcal{O}(a^2)$ 
can be suppressed by using non-perturbatively improved 
actions \cite{Narayanan:1994gw,Luscher:1996sc,Zanotti:2001yb}.
Lattice QCD simulation results relying on a variety of actions 
are presented in Chapters \ref{chpt:intrinsic} through 
\ref{chpt:nucleonmagmom}, 
and the benefits and %deficiencies 
shortcomings of each one will be addressed as they arise.  
%can be recovered in the continuum limit.

%NEW BIT: HOW TO EXTRACT OBSERVABLES FROM THE LATTICE: THE CORRELATION 
%FUNCTIONS AND THE EFFECTIVE MASS

\subsection{Correlation Functions and the Effective Mass}

Consider the following example regarding the construction of a correlation 
matrix element, and the extraction of the effective mass. 
In applying lattice QCD to the extraction of the mass of the nucleon, 
one defines 
interpolating fields $\chi$ and $\bar{\chi}$,  
which incorporate the structure of a nucleon in terms of its constituent 
quarks. 
%multi-particle state. 
For example, in the case of a proton, 
$\chi = \ep^{abc}(u_a^T\,C\,\g_5\,d_b)u_c$  is a suitable choice,
 since the maximally anti-symmetric 
Levi-Civita symbol $\ep$ ensures a colour-singlet state, 
and the Dirac spin matrix $\g_5$ (defined in Appendix \ref{app:spin}) 
%projects out only the states of positive parity. 
preserves the spinor properties of the interpolating field. 
The fields $u$, $d$ are Dirac spinors 
representing the up and down quarks, respectively, and
 the charge conjugation matrix $C = i\g_0\g_2$ ensures that the 
product of a spinor and its transpose satisfies Lorentz invariance. 

The two-point Green's Function for a proton, or more generally,
 the nucleon, can be expanded by inserting
both a complete set of momentum- and spin-dependent eigenstates 
$|A(q,s)\rk$, 
and a translation operator on the $\chi$ field:
%use of the translation operator: %don't copy!! <-
%
\begin{align}
G^{(2)}(\vec{x},t) &= \lb \Omega|\chi(\vec{x},t)\,\bar{\chi}(0)|\Omega\rk,\\
G^{(2)}(\vec{p},t) &= \sum_{\vec{x},A,\vec{q},s}e^{-i\vec{p}\cdot\vec{x}} 
\lb\Omega|e^{-iq\cdot x}\,\chi(0)\,e^{iq\cdot x}|A(q,s)\rk
\lb A(q,s)|\bar{\chi}(0)|\Omega\rk\\
&= \sum_{\vec{x},A,\vec{q},s}e^{-i(\vec{p}-\vec{q})\cdot\vec{x}} e^{-E_A(\vec{q}) t}\lb\Omega|
\chi(0)|A(q,s)\rk
\lb A(q,s)|\bar{\chi}(0)|\Omega\rk\\
&= \sum_{A,\vec{q},s}\delta(\vec{p}-\vec{q})e^{-E_A(\vec{q}) t}|\lb\Omega|
\chi(0)|A(q,s)\rk|^2\\
&=
\sum_{A,s}|\la_{A,p,s}|^2e^{-E_A(\vec{p}) t}
\psi(\vec{p},s)\,\bar{\psi}(\vec{p},s),
\end{align}
for complex-valued scalar coefficients $\la_{A,p,s}$, $\la^*_{A,p,s}$  and 
spinor fields $\psi$, $\bar{\psi}$ 
defined by the matrix elements:
\eqb
\la_{A,p,s}\psi = \lb\Omega|\chi(0)|A(p,s)\rk;\qquad 
\la^*_{A,p,s}\bar{\psi} = \lb A(p,s)|\chi(0)|\Omega\rk.
\eqe
The mass of the nucleon can then be extracted from the two-point Green's 
Function at zero $3$-momentum, that is, $E_A(\vec{p} = 0) = M_A$. To obtain 
a measure of this quantity from the exponential, one defines the effective mass 
$M_{\ro{eff}}$ by comparing the behaviour of the Green's Function at times 
$t$ and $t+1$:
\eqb
M_{\ro{eff}} = \ro{log}\left(\f{G_2(0,t)}{G_2(0,t+1)}\right).
\eqe
Note that $M_{\ro{eff}}$ is a dimensionless quantity, and the calculation 
of the mass of the nucleon must involve the conversion to physical units 
from lattice units by dividing by the lattice spacing $a$.
Since the Green's Function incorporates the full quantum mechanical spectrum 
of modes, the behaviour of $M_{\ro{eff}}$ is %highly %contaminated 
 strongly influenced by the excited 
states of the nucleon at small $t$. In the limit of large $t$, however, 
the ground-state nucleon 
mass can be recovered:
\eqb
M_N =\lim_{t\rightarrow\infty} \f{M_{\ro{eff}}}{a}.
\eqe

\subsection{Quenching and Computational Alternatives}
\label{subsect:qqcd}
%
%The Quenched Approximation can be then summarized by setting
% $\ro{det}[\ca{M}$] to be constant, since the 
%vacuum polarization effects of the QCD vacuum are suppressed in QQCD.
%
%The calculation of 
Computing the quantity $\ro{det}(\ca{M}[\ca{A}_\mu^a])$  
 is the most time-consuming 
operation in the calculation of the partition function in Equation 
(\ref{eqn:genfuncwithdet}). 
For this reason, calculations are %often 
performed at fermion masses 
larger than their physical value, thus decreasing the Compton wavelength 
  of a fermion and significantly reducing the computational 
resource requirement of 
the summation over all paths and the time required to execute 
all the necessary fermion matrix inversion algorithms.
%Sometimes, 
Usually, results from lattice QCD are obtained at multiple fermion masses,
so an extrapolation can be used to obtain the result at 
the physical value, or at zero mass (the %\emph
{chiral limit}).
%In the literature, 
A complementary 
computational simplification known as %\emph
{quenching} exists, 
whereby $\ro{det}(\ca{M}[\ca{A}_\mu^a])$ is set equal to a constant. 
This has the effect of removing from the theory all  vacuum polarization 
diagrams, changing the dynamics of the quantum field theory 
in a non-trivial way. For this reason, quenched QCD (QQCD) should be 
considered, in essence, a different theory from QCD. 
The results from QQCD calculations can nonetheless be interesting points of 
investigation, as they offer a unique testing ground for extrapolation schemes. 
This is because results from the unphysical QQCD calculation cannot be known 
in advance from experiment.

\begin{figure}
\centering
\includegraphics[height=135pt]{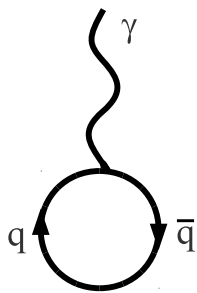}
\vspace{-3mm}
\caption{\footnotesize{An external photon coupling to a sea quark-antiquark pair. Diagrams including this kind of coupling are omitted in Valence QCD.}}
\label{fig:vqcd}
\end{figure}

Several other alternatives to quenching have been used in the literature to
 date. Sometimes, the vacuum polarizations, normally omitted in QQCD, 
are calculated for a different (usually larger) quark mass than the 
valence quarks, %ie. those that 
which couple to external sources. 
The quarks that appear in the disconnected loops are known as sea quarks. 
This distinction between %\emph
{sea} quark mass and valence 
quark mass provides some of the dynamics of QCD, albeit altered, whilst 
still ameliorating the computational intensity of the calculation 
of $\ro{det}(\ca{M}[\ca{A}_\mu^a])$. 
%Such \emph{partial} quenching...
An alternative, particularly used in electromagnetic contributions 
to QCD, is to omit diagrams that include %\emph
{indirect} couplings, that is,   
external fields coupling to 
sea quark-antiquark pairs, as shown in Figure \ref{fig:vqcd}.  
The computation of the indirect couplings to disconnected quark loops 
is by far the most time-consuming portion of the calculation of a 
diagram. 
%The calculation of amplitudes that include 
%indirect couplings is time-consuming, %and their omission provides 
%an alternative to quenching as a method for 
%This %\emph
{Valence} QCD (VQCD) therefore 
only includes diagrams where any external particles, such as incoming 
photons, couple directly to valence quarks in the relevant hadron.
Although the resulting theory differs from %\emph
{full} QCD, often 
properties of particles are calculated using an isovector combination. 
In the case of an isovector, a linear combination of isospin partners 
is formed so that the resultant combination transforms as a vector in 
isospin space. 
For example, in the case of the nucleon, the combination of the fermion 
fields: $p-n$ (proton minus neutron) 
is isovectorial with total isospin of $1$, 
and all diagrams containing indirect couplings cancel. 
This is because diagrams that contain indirect 
couplings to disconnected loops are exactly the same for the proton 
and neutron, and thus disappear in the combination: $p-n$. 
It is only the valence quark composition 
that differs between the proton 
($u\,u\,d$) 
and the neutron ($d\,d\,u$). 
Thus the distinction between full QCD and VQCD %Full and Valence QCD 
disappears for this 
observable, and the calculations of its properties are less computationally 
intensive.

\section{Lattice QCD Applicability and Issues}
%Discuss why it will always be important, even as algorthms get better,
%there are quantities that are so hard to 'get at'. Signal to noise killing
%you, etc??? --- link to baryon resonances section

It is important to identify clearly the constraints of lattice gauge theory. 
Lattice QCD is well defined over all box sizes, 
lattice spacings and quark masses, and it is also infinitely scalable. 
However, the computational cost of the 
calculation of an observable 
is generally proportional to the square of the lattice volume and 
inversely proportional to the sixth power of the lattice spacing.
To avoid major finite-volume effects, the literature suggests that
 the lattice box length should be
 about $2.5$ to $3.0$ fm \cite{Sheikholeslami:1985ij,Leinweber:1999ig,Fukugita:1994ba,Dong:1995ec,Labrenz:1996jy}. 
This is the typical  
size of most current lattice QCD calculations. Nevertheless, 
finite-volume effects 
can still be significant at these box sizes, and ought not 
to be neglected. In fact, for many observables, a box length of 
$3.0$ fm is insufficient to avoid large finite-volume corrections 
at physical quark masses. This will be demonstrated in 
Chapters \ref{chpt:intrinsic} through \ref{chpt:nucleonmagmom} for 
a variety of observables.

 %Ideally, physical quark masses could be used in lattice simulations, but 
%quarks this light exhibit non-locality and are thus sensitive to
% finite-volume effects. 
%They also critically slow down the fermion matrix inversion algorithms. 
While continued 
%Continuing 
supercomputing advances in numerical simulations of lattice QCD 
 are important, %for obtaining more accurate simulations of QCD, 
%will be vital to some extent in resolving this problem, 
one ought to recognize its limitations in providing a thorough understanding 
of the internal structure of hadrons, which can be aided, in part, 
by complementary techniques such as chiral perturbation theory ($\chi$PT). 
%questions involving  the physical value
%of the strange-quark mass, or the contributions from individual quarks within
 %a hadron, present challenges that will not diminish
%with supercomputing advances.  
For example, consider 
the effects of the mesons known as kaons, vital for understanding strangeness 
in the nucleon, which appear in the meson octet  
(see in Appendix \ref{app:fields}). 
One must either use $\chi$PT %Chiral Perturbation Theory,  
calculated to significantly high order in the relevant perturbative expansion, 
or develop new non-perturbative approaches
which utilize the non-perturbative information expressed in the
lattice simulation results.  Since the former is likely to be
compromised by the asymptotic nature of the expansion,  
attention is given to the latter approach in Chapter 
\ref{chpt:intrinsic}.

%Summary
%Now it is important to discuss the principles behind 
The computation of observables in lattice QCD %, although useful, 
%is still 
%limited in its scope with regard to an understanding of the low-energy 
%structure of QCD. 
provides great insight into the non-perturbative region of QCD. As long 
as one can account for finite-volume and momentum discretization effects, 
lattice QCD provides excellent predictions of the behaviour of quarks 
at low-energy. In simulating the interactions of hadrons, and 
demonstrating confinement, lattice QCD is a landmark achievement 
in the realm of chiral dynamics. % Nevertheless, there yet remain limitations 
% in accessing an overall understanding

The complementary methods obtained 
from  effective field theory 
offer 
guidance in the calculation of observables on the lattice. They provide  
estimates of finite-volume effects and extrapolations to physical 
quark masses,  
 and providing a deeper understanding of the applicable regions of lattice QCD.
This can serve to ameliorate the otherwise unseen difficulties encountered
 in a brute-force approach to calculation by considering symmetries, 
renormalization, power-counting, and other techniques built into 
the formalism of 
chiral effective field theory. 
Each method presents its own challenges, but also   
brings %additional 
enlightenment through the %vastly 
 significantly different approaches to a %any 
given problem.

%% file: chieft.tex
\chapter{Chiral Effective Field Theory}
\label{chpt:chieft}

\textit{``Everything can be tried, a bold abstraction of something that has succeeded elsewhere, the exploration of the faintest clue, or a leap through empty spaces$\ldots$}

\textit{Thus, the method exists, boundless, its ultimate foundation being the freedom of the mind.''}
(Omn\`{e}s, R. 2002. \textit{Quantum Philosophy: Understanding and Interpreting Contemporary Science} p.268) \cite{Omnes}

% Introduce Effective field theories
% and why they are useful
% and how they come about
% with examples!!! sigma model, representations, and QCD...

%Notation: gluon field \ca{A}, axial current A, photon field \sr{A}

%CONSIDER: AUSTRALIAN OR AMERICAN SPELLING: S OR Z? AND ARTIFACT OR ARTEFACT? 
%BEHAVIOR OR BEHAVIOUR? ETC

In an effective field theory, one identifies the relevant degrees of 
freedom at a particular energy, and encodes the behaviour of these 
degrees of freedom in a suitable Lagrangian. For a low-energy effective 
field theory corresponding to quantum chromodynamics (QCD), 
such as chiral effective field theory 
($\chi$EFT), 
effective Lagrangians may take on 
different forms to the QCD Lagrangian, but the physics of the strong 
interaction must remain the same in each case. 
That is, results for calculating elements of the S-matrix must agree among 
effective field theories, up to some order. 
In order to construct 
such a theory, terms in the effective Lagrangian are chosen so that 
they satisfy the fundamental symmetries of QCD. The coefficients 
of the terms in the effective 
Lagrangian are new coupling constants, the values 
of which are determined from experiments. %As higher energy 
%degrees of freedom, such as quarks, become irrelevant to the effective 
%theory at low-energy, it is not possible to determine the relationship

The method of effective Lagrangians provides alternative 
 machinery to lattice QCD 
for understanding the low-energy behaviour of QCD,
and physical theories in general at a specific energy level.
The dynamics of the low-energy degrees of freedom, 
such as mesons and baryons in the case of $\chi$PT, 
are incorporated directly into the Lagrangian, whereas very 
massive particles
 are treated as static sources \cite{DGH,Borasoy:2007yi}.
 Examples of important effective field theories include
the Sigma Model and its various instructive representations,
%MIT Bag and Cloudy Bag models (cite TT)
the MIT Bag Model \cite{Chodos:1974je,Johnson:1978uy}
 and Cloudy Bag Model \cite{Miller:1979kg}, 
%[add more examples here]
 as well as quantum electrodynamics (QED) and QCD themselves \cite{WQTF}.

%QCD
Recall that the QCD Lagrangian comprises a Yang-Mills term
involving  vector potentials
 $\ca{A}_\mu^a$, their field strength tensors 
$G_{\mu\nu}^a = \cd_{[\mu} \ca{A}_{\nu]}^a 
 -i\alpha_s f_{abc}\ca{A}_\mu^b \ca{A}_\nu^c$ 
 and a Dirac term
of quark spinors $\p$ corresponding to a mass matrix $\ca{M}$. 
%Note that t
The spinors and 
the mass matrix are extended to contain
the six flavour and three colour components of QCD. 
Using the slash-notation $\g^\mu D_\mu \equiv \sla{D}$, 
the %free 
 QCD 
Lagrangian may written out conveniently in matrix form\footnotemark:
\eqb
\label{eqn:qcdlag}
\cL_{\ro{QCD}} = \pb(i\Darrows - \ca{M})\p
 - \f{1}{2}\Tr[\bar{G}_{\mu\nu}\bar{G}^{\mu\nu}]\,,
\eqe
where the trace acts over colour indices for the
 matrix-valued versions of the gluon field strength tensor 
$\bar{G}_{\mu\nu}$, 
defined by summing over the generators $J^a$ of $\SU(3)$: 
\footnotetext{The double-headed arrow indicates 
the difference between the derivative acting to the right and to the left. i.e. 
$\stackrel{\leftrightarrow}{D} \quad=\quad 
\stackrel{\rightarrow}{D} - \stackrel{\leftarrow}{D}$.}
%
%\begin{align} 
\eqb
\bar{G}_{\mu\nu} %&
= J^aG_{\mu\nu}^a%\\
%&
= \cd_{[\mu}\bar{\ca{A}}_{\nu]} 
+ i\alpha_s[\bar{\ca{A}}_\mu,\bar{\ca{A}}_\nu]\,.
\label{eqn:matG}
\eqe
%\end{align}
%
% 
%This Lagrangian is simply a non-interacting version of Equation 
%(\ref{eqn:lattqcdlag}). 
The generators $J^a$ in the eight-dimensional representation of $\SU(3)$ 
are related to the Gell-Mann matrices $\la^a$, defined in Appendix 
\ref{app:struc}, by a factor of a half:
\eqb
J^a = \f{\la^a}{2}\,.
\eqe
The quark-gluon interaction vertex 
%can be constructed by imposing gauge-invariance, and 
%is satisfied by replacing $\cd_\mu$ with the covariant derivative 
is incorporated into the covariant derivative, defined as:
\eqb 
D_\mu = \cd_\mu + i\alpha_s\f{\la^a}{2}\ca{A}_\mu^a \,, 
\eqe
which acts as a parallel transport in gauge-space, so that the 
QCD Lagrangian of Equation (\ref{eqn:qcdlag}) is gauge-invariant. 
By substituting into Equation (\ref{eqn:matG}), it can be seen that the 
anti-symmetrization of the covariant derivative is the field strength tensor. 
This is a consequence of the gauge 
connection $\bar{\ca{A}}_\mu$ lacking torsion:
\eqb
\bar{G}_{\mu\nu} = \f{i}{\alpha_s}[D_\mu,D_\nu]\,.
\eqe

The fundamental symmetries of QCD are built into the QCD Lagrangian of 
Equation (\ref{eqn:qcdlag}). In particular, chiral symmetry will be 
important in the subsequent analyses of observables using $\chi$EFT. 
The consequences of chiral symmetry breaking ultimately have a profound 
effect on the behaviour of subatomic particles, their masses, magnetic moments 
and other properties. Therefore, it will be beneficial  
to describe some of the subtleties of chiral symmetry with care 
in the discussion below.  

%involving the Gell-Mann Matrices: the $\SU(3)$ generators $\la^a$.  
 
%Unlike QED, a so-called %\emph{
%$\theta$-term, 
%%}
% constructed from the product of 
%the gluon field strength $\bar{G}_{\mu\nu}$ and its dual 
%$\tilde{\bar{G}}_{\mu\nu}$, 
%can also be
%included in the general QCD Lagrangian. This is because the non-Abelian
%nature of the Yang-Mills field potential leads to 
%terms in the Lagrangian proportional to  
%$\cd_\mu\ep^{\mu\nu\rho\si}
%\Tr[\bar{\ca{A}}_\nu \bar{\ca{A}}_\rho \bar{\ca{A}}_\si]$,
%which cannot be
% discarded as a trivial divergence 
%suppressed out of the theory as a surface term by Gauss' Theorem, as in QED.
% Although such a term violates Charge-Parity (CP)-Symmetry,
% it is known that it must be
% small on experimental grounds \cite{DGH}, and so is not usually included
% when investigating phenomenological studies, except those specifically 
%relating to CP-Symmetry violation in QCD.

\section{Chiral Symmetry}
\label{sec:chisym}

% Noether's theorem BRIEF!
In general, a symmetry, or an invariance of a dynamical
quantity under a transformation of one of its parameters,
leads to important physical insights into a system.
Noether's Theorem demonstrates that a conserved current can always be 
constructed from a (non-anomalous) symmetry of a field theory.
%In the case of anomalous symmetries, such as 
%the U$(1)$ axial anomaly of QCD, Noether's 
%Theorem is not applicable because this apparent symmetry of the classical 
% Lagrangian is not preserved under quantization.

Chirality is defined as 
the handedness of the representations of the Poincar\'{e} group 
(which encodes the isometries of Minkowski spacetime) 
under which the quark spinors transform. 
It is related to the helicity of a particle: the projection of 
its spin on its direction of 
linear momentum, which is equivalent to 
 chirality if the quarks are massless. Helicity  
is not in general a Lorentz-invariant 
quantity. Its value in one frame may be flipped with respect to its value in 
a boosted frame. 

The QCD Lagrangian in Equation (\ref{eqn:qcdlag}) can be
split into separate left- and right-handed %helicity 
chiral states under the
projections $\G_{L,R} = \f{1}{2}(1 \pm \g_5)$. 
The left- and right-handed spinors are written as:
\eqb
\p_{L,R} = \G_{L,R}\p\,.
\eqe
 Note that the resultant chirality %helicity
of the quark fields is decoupled %sharp (unmixed) %and Lorentz invariant 
 only for zero mass \cite{DGH}:
\begin{align}
\label{eqn:chiralqcdlag}
\cL_{\ro{QCD}} &= \cL_L + \cL_R + \cL_{\ro{YM}} + \cL_{\ro{mass}}\nn\\
&= i \pb_{L} \Darrows\p_{L} +  i \pb_{R} \Darrows\p_{R}
- \f{1}{2}\Tr[\bar{G}_{\mu\nu}\bar{G}^{\mu\nu}] -
 (\pb_L\,\ca{M}\p_R+\pb_R\,\ca{M}\p_L)\,.
\end{align}
%
%where t
The quark fields transform under the chiral rotations $L$ and $R$, 
which are elements of the left- and right-handed Lie Algebra, respectively, 
defined for the group generators $Q_{L,R}^a$ and arbitrary, 
continuous, real parameters 
$\alpha_{L,R}^a$:
\begin{align}
L &= \ro{exp}({i\alpha^a_{L}Q_{L}^a}) \in SU(3)_{L}\,,\\
R &= \ro{exp}({i\alpha^a_{R}Q_{R}^a}) \in SU(3)_{R}\,. 
\end{align}
The transformation laws for each of the spinor fields can thus be written: 
\begin{align}
\p_L &\ra L\,\p_L = \p_L + \de\p_L\,,\\
\p_R &\ra R\,\p_R = \p_R + \de\p_R\,.
\end{align}
%
%
%Since the quarks are massless at this stage, the more 
%general notion of \emph{chirality} is completely equivalent 
%to helicity. Unlike helicity, chirality is a Loretz-invariant quantity, 
%and refers to the handedness of the representations of the Poincar\'{e}
%under which the quarks transform.
%
Noether's Theorem allows one to construct the left and right symmetry currents,
with the corresponding time-independent charges
 forming the eight unique invariants of the group.
These invariants are the generators, and are found by integrating over
a spacelike surface $\si$. Note that in the case of $\SU(3)_{L,R}$ the 
generators are %merely 
 related to the previously mentioned 
Gell-Mann matrices $\la^a$, after chiral projection
by the group elements (up to a minus sign and a factor of a half, 
by convention): 
\begin{align}
J_{L,R}^{\mu\, a} &= \f{\cd\cL_{\ro{QCD}}}{\cd\cd_\mu\pb_L}\de\pb_L = 
 \f{1}{2}\pb_{L,R}\,\g^\mu\, \f{\la^a}{2} \p_{L,R}\,,\\
Q_{L,R}^a &= \int\!\!\ud\si_\mu\, J_{L,R}^{\mu\, a} 
\,=\, \int\!\!\ud^3x\,\, J_{L,R}^{0 a} 
\,=\, -\G_{L,R}\f{\la^a}{2}\,.
\end{align}

An equivalent convention to that of left/right %helicity 
 chirality is 
the construction of vector and axial vector transformations. 
%for the group  $\SU(3)_V \otimes \SU(3)_A$.
The group action can be written out explicitly for either convention,
using the definition of a Lie group with continuous group parameters
 $\alpha^a_{V,A}$.
The charges $Q_V^a$ and $Q_A^a$ simply count the sum and the difference
of left- and right-handed fermions, respectively:
\begin{align}
V &= \ro{exp}({i \alpha_{V}^aQ_{V}^a}) \in \SU(3)_{V}\,, \\
A &= \ro{exp}({i \alpha_{A}^aQ_{A}^a}) \in \SU(3)_{A}\,,
\end{align}
\begin{align}
Q_V^a &= Q_L^a + Q_R^a = -\f{\la^a}{2}\,, \\
Q_A^a &= Q_L^a - Q_R^a = -\f{\la^a}{2}\g_5\,.
\end{align}
These sets of rotations are the most convenient for asserting 
the invocation of %\emph
an important theorem known as 
{Goldstone's Theorem}. Goldstone's Theorem, described below, 
%in regards to the axial charges $Q_A^a$ in the following way. 
 is crucial in understanding the connection between axial charges $Q_A^a$
 and the origin of mesons in QCD. 

%Goldstone's theorem

\subsection{Spontaneous Symmetry Breaking}
\label{subsect:ssb}

In QCD, particles are believed to utilize the Nambu-Goldstone
mode of spontaneous breaking of a continuous global gauge symmetry. 
This symmetry breaking occurs in flavour space, and only the lightest 
three quark flavours will be considered: up, down, strange. Since the 
up and down  
quarks have relatively low mass ($m_{u,d} \sim 2$-$6$ MeV, $m_s \sim 100$ MeV) 
compared to the other quarks ($m_c \sim 1.3$ GeV), 
they contribute the most strongly to symmetry breaking effects. 
%
%Only briefly mention, Wigner-Weyl, Higgs and anomalies
%
%This is one of the ways in which a symmetry of the Lagrangian can be
%realized in nature.
%

  Goldstone's Theorem states that the symmetry group $\SU(3)_V \otimes \SU(3)_A$
is not respected by the (no-particle) vacuum  
state $\,\mi0\rk$, even though this group is
a symmetry of the massless QCD Lagrangian.
One might %initially 
 na\"{i}vely expect that the vacuum state is invariant under the
group transformations:
\eqb
V\mi0\rk = A\mi0\rk = \,\,\,\mi0\rk\,.
\eqe
Noether's Theorem entails that the symmetry charges are 
time-independent:
\eqb
\label{eqn:heis}
\f{\ud}{\ud t}Q_{V,A}^a = 0 = [\ca{H}_\ro{QCD},Q_{V,A}^a]\,.
\eqe
This means that 
the charges should annihilate the  vacuum state $|0\rk$, since the 
QCD Hamiltonian $\ca{H}_\ro{QCD}$ annihilates the vacuum state.
In 1984, Witten and Vafa proved this result for vector charges
even without assuming chiral symmetry \cite{Vafa:1983tf}.
However, were this the case for axial charges, 
a spectrum of mass-degenerate partners with opposite parity 
would be expected to exist for all hadrons. %particles.
This is because the axial charges are odd under parity transformations, 
and any state acted on by the axial charges will also retain the same
energy eigenvalue (but with a flipped parity eigenvalue), 
 because of the commutation relation in
 Equation (\ref{eqn:heis}).
There is a stark 
lack of experimental evidence for such particles \cite{Yao:2006px}.
Thus, physical hadrons merely observe the symmetry group $\SU(3)_V$. 
%
%One arrives at this conclusion principally by the
%lack of experimental evidence for opposite parity partners of equal mass,
% which must occur for every particle obeying the 
%full, unbroken symmetry \cite{Yao:2006px}, since the axial charges
%$Q_A^a$ are odd under parity transformations.
%CONTINUE: pg 6 Borasoy
% do Goldstone's Theorem proper, with Q_A\mi0\rk != 0
%also pg 11 for Goldstones live in coset space: SU(3)_L \otimes SU(3)_R /SU(3)_V
%

%
%Explicitly, 
%Instead of preserving the vacuum state, 
%the axial charge does not annihilate the vacuum state
%and so transforms it to an element of a new Hilbert space: 
Instead of annihilating the vacuum state, the axial charges 
transform it to an element of a new Hilbert space: 
\begin{align}
Q_V^a\mi0\rk &= 0\,,\\
Q_A^a\mi0\rk &= q_A\mi\pi^a(\vec{p})\rk \ne 0\,.
\label{eqn:GT}
\end{align}
This new state (with axial eigenvalue $q_A$)
 has the same energy as the vacuum state as long as
the symmetry is not also explicitly broken by terms in the Lagrangian.
 Goldstone's Theorem states that %a cascade of 
new particles are created,
the number of which corresponds to the number of generators for the relevant
representation of $\SU(3)_A$. 
These new particles must be massless and spinless pseudoscalar
mesons, % and they are 
 called %the 
 Goldstone bosons.
%MENTION MEXICAN HAT THING WITH EXPANSION ABOUT MINIMA AND VEV
%WE CHOOSE THE FIELD CONVENTION OF SPONATEOUS SYMMETRY BREAKING

If the physical manifestation of a symmetry of a Lagrangian involves
the spontaneous breaking of one or several local continuous transformations,
the theory predicts a massive spin zero boson called a Higgs field, 
and the %\emph{
Higgs mode %} 
is said to be realized. %As of [date] it is not known
% whether such a mechanism occurs in the Standard Model, however its
Although the Higgs mode 
is not expected to occur in the strong nuclear force sector 
of the Standard Model, 
 its actualization in the electroweak sector 
would result in a mass term for the $\{W^{\pm}\!,Z\}$ 
weak gauge bosons. Such a mass is observed in experiments, 
 and also explains how the charged fermion fields gain mass, through the 
following argument. 
%This arises %comes about by 
%from 
By considering the Lagrangian for an $\SU(2)$ complex
doublet of bosons, which can be expanded about its minimum potential energy
 in the same manner as the %just as in the case of 
Goldstone bosons, %except that 
one must arbitrarily choose 
a direction in isospin space in which to expand.
Three of the Higgs degrees of freedom
combine to become the longitudinal spin modes of the three weak gauge bosons,
 and the mass of the fermions is produced by the vacuum expectation
 value of the remaining Higgs boson, which remains in the theory 
\cite{Guralnik:1964eu}.
It should be noted that the Higgs mechanism contributes only 
a small amount to the mass of hadrons in QCD, and that the dominant 
process for their mass acquisition is dynamical chiral symmetry breaking 
\cite{Gross:1974jv,Roberts:2010gh}. 
%Some of the more specific 
%details of the dynamical chiral symmetry breaking as applied to the mass
% of the 
%nucleon
% are 
A more detailed analysis of 
the consequences 
 of dynamical chiral symmetry breaking for the mass of the nucleon is 
discussed in Chapter \ref{chpt:intrinsic}, in the context of varying 
quark masses in lattice QCD results.

%\subsection{Representation Theory}
%Talk about 8\otimes8 being broken down, etc from:
%\cite{Lebed:1994ga} Deternimation of SU(6)C-G coeffs (etc)

%PCAC, ref Sidney Coleman!
\subsection{Partial Conservation of the Axial Current}
\label{subsect:pcac}
%INTRODUCE!
% 
%Since the meson octet has been identified as a set of pseudo-Goldstone boson
%fields in Section \ref{subsect:meson}, 
Before discussing the powerful techniques associated 
with effective Lagrangians, 
 a brief overview is now presented for the 
current algebra method for obtaining the 
low-energy matrix elements of pion decay. 
%It is through the current algebra approach that 
It is known that the %matrix elements
%of the 
$\SU(3)$ axial currents $J_A^{\mu\,a}$ are non-zero. But in order 
to know exactly how these matrix elements vary and how they depend on the
octet meson masses, one requires %need to use 
a current algebra technique known as
 %\emph
{Partial Conservation of the Axial Current} (PCAC).
The statement of Goldstone's Theorem in Equation (\ref{eqn:GT}) can be 
re-expressed as a matrix element:
\eqb
\label{eqn:piondecay}
\lb0\mi J_A^{\mu\,a}\mi\pi^b(\vec{p})\rk = if_\pi p^\mu \de^{ab}\,,
\eqe
from which follows the divergence:
\eqb
\label{eqn:PCAC}
\lb0\mi \cd_\mu J_A^{\mu\,a}\mi\pi^b(\vec{p})\rk = f_\pi m_\pi^2\de^{ab}\,.
\eqe
Equation (\ref{eqn:piondecay}) serves as a suitable 
definition of the pion 
decay constant $f_\pi$. Taking the value from experiment: 
$f_\pi \approx 92.4$ MeV.

Equation (\ref{eqn:PCAC}), together with the Haag Theorem, forms the principal
statement of PCAC, that either $\pi^a$ or $\cd_\mu J_A^{\mu\,a}$ can be used
equivalently, and that if the pion mass becomes zero then the axial current is
totally conserved. Thus %is allows us to write 
the following relation may be written:
\eqb
\label{eqn:PCACequiv}
\pi^a = \f{1}{f_\pi m_\pi^2}\,\cd_\mu J_A^{\mu\,a}\,.
\eqe
 This situation is a special case of the 
%\emph
{Soft-Pion Theorem} for a matrix element involving a general 
local operator $\ca{O}$ :
\eqb
\lim_{p^\mu\ra0}\lb\pi^a(\vec{p})\beta\mi\,\ca{O}\,\,\mi\alpha\rk = 
-\f{i}{f_\pi}\lb\beta\mi\,[J_A^a,\ca{O}]\,\mi\alpha\rk\,.
\eqe
While they are useful in obtaining specific information about the low-energy 
matrix  elements of pion decay, the methods of PCAC can be subtle 
in determining possible momentum dependence in an amplitude of a low-energy 
process. One must also make the assumption
 that matrix elements vary continuously 
 in taking the soft pion limit, $p^\mu\ra0$. 
The method of effective Lagrangians is less awkward in 
obtaining the appropriate momentum dependence and any quantum corrections 
to a low-energy amplitude. This is because the effective 
Lagrangians are ordered 
by a systematic expansion in momentum or mass, which encodes the relative 
importance of corrections to an amplitude in question. 
%are somewhat limited  
%in the breadth of chiral information that can be obtained from them. 
%Now, attention is turned to the method of effective Lagrangian, starting from 
%a simple model. 
%The matrix elements must be sufficiently slowly-varying in the limit 
%of soft momentum $q^\mu\rightarrow 0$.
%add a part explaining the limitations. Matrix elements must be slowly 
%varying 

\subsection{The Sigma Model}
%Sigma Models
The Linear Sigma Model \cite{GellMann:1960np} is a useful pedagogical tool,
because with it important theoretical techniques such as the construction
of symmetry currents, spontaneous symmetry breaking and changes in
parameterization can be demonstrated easily \cite{DGH}.
First, consider an $\SU(2)$   
Sigma Model Lagrangian  
consisting of a massless spinor field $\p$, a
so-called %\emph{
pion %} 
field $\vp$ spanning
 the triplet representation of $\SU(2)$
and a massive scalar field $\si$. 
The consideration of isospin symmetry in $\SU(2)$ provides a simple 
and instructive example for investigating symmetries \cite{DGH}. 
The Lagrangian takes the following form: 
\eqb
\label{eqn:lsm}
\cL_\si = \pb i\sla{\darrow}\p + 
\f{1}{2}\cd_\mu \vp \cdot \cd^\mu \vp + 
\f{1}{2} \cd_\mu \si \cd^\mu \si 
- g\pb(\si - i\vt\cdot\vp\g_5)\p
+ \f{\mu^2}{2}(\si^2 + \vp^2) 
- \f{\la}{4}{(\si^2 + \vp^2)}^2\,,
\eqe
(for constant coupling parameters $g$, $\mu$ and $\la$, and 
$\SU(2)$ Pauli Spin matrices
$\vt$ defined in Appendix \ref{app:spin}).

Spontaneous symmetry breaking occurs in the Lagrangian of 
Equation (\ref{eqn:lsm}) for $\mu^2 > 0$. In minimizing the potential: 
\eqb
V(\si,\vp) = - \f{\mu^2}{2}(\si^2 + \vp^2) 
+ \f{\la}{4}{(\si^2 + \vp^2)}^2\,,
\eqe
a ground state is found that is non-trivial (unlike the case $\mu^2 < 0$, 
for which the only ground state solution is: $\si = \vp = 0$). 
This ground state is defined by:
\eqb
\si^2 + \vp^2 = \f{\mu^2}{\la}\,.
\eqe
By redefining the $\si$ field and expanding the Lagrangian about the 
new ground state $\lb\si\rk_0 \equiv v$, the Linear Sigma Model 
exhibits spontaneous symmetry breaking, evident in the acquisition of 
mass for the $\tilde{\si}$ field:
\begin{align}
\tilde{\si} &= \si - v\,,\\
\cL_{\tilde{\si}} &= \pb (i\sla{\darrow}-gv)\p + 
\f{1}{2}\cd_\mu \vp \cdot \cd^\mu \vp + 
\f{1}{2} (\cd_\mu \tilde{\si} \cd^\mu \tilde{\si} - 2\mu^2\tilde{\si}^2 )
- g\pb(\tilde{\si} - i\vt\cdot\vp\g_5)\p\nn\\
&- \la v \tilde{\si}(\tilde{\si}^2 + \vp^2) 
- \f{\la}{4} [{(\si^2 + \vp^2)}^2-v^4]\,.
\end{align}
Nevertheless, $\SU(2)$ isospin symmetry is preserved in this Lagrangian. 

%Because this is an effective field theory, and as such
The active degrees of freedom in an effective field theory 
do not necessarily correspond to elementary 
particles of nature, and so it is expected that changes in the representation 
do not alter the outcome of physical processes. % to which amplitudes may
%correspond.
This notion is formalized in the %\emph{
Haag Theorem %} 
\cite{Haag:1958vt},
which states that for two field variables derived from (unitarily)
equivalent representations, if one is  a free field, 
then the other is also free,
regardless of how they are related and whether the associated diagrams
and Lagrangian vertices change \cite{Guenin}. As a corollary, 
an interacting quantum field theory `does not exist', in the sense that 
its fields do not transform covariantly 
under the interacting Poincar\'{e} group.
Weinberg suggested that only free fields are required to 
construct the S-matrix from the relativistic Hamiltonians in QED, 
 but in QCD one must simply resort to  
writing down the most general Lagrangian \cite{WQTF}. An alternative approach 
demonstrates that an interaction picture can be constructed consistently 
if time evolution is taken to be only locally unitarily implementable 
\cite{Guenin}.
%
%Representations eg. of Sigma models
%Introduce \Si, and U and S fields. 
%describe why they transform the way they do, in terms of cosets.
%Explain why Exponential rep, using U and S, is useful:
% integrating out heavy field S, to get a low-energy theory.
%

By redefining the scalar field in either linear or non-linear combinations
of the other involved fields, different sets of interaction vertices
can be assembled. For example, 
 using the Linear Sigma Model, two particularly instructive representations
are considered for later adaptation to low-energy QCD. 
By rewriting the heavy $\si$ field and pion triplet as a matrix quantity
$\Si \equiv \si + i \vec{\tau}\cdot \vp$,
 the resultant new field $\Si$ 
transforms as an object in the adjoint representation,
 which forms left cosets of the group  
$\SU(2)_L\otimes\SU(2)_R$, as described by Scherer \cite{Scherer:2005ri}:
% see centre of Work 4
\eqb
\Si \ra L \,\Si\, R^\da\,.
\eqe
In this representation, the Lagrangian becomes:
%
%\newpage
\begin{align}
\cL_\Si &= \pb_L i\sla{\darrow}\p_L + \pb_R i\sla{\darrow}\p_R + 
\f{1}{4}\Tr[\cd_\mu \Si \cd^\mu \Si^\da]%\nn\\
%
%&+
+ \f{1}{4}\mu^2 \Tr[\Si^\da \Si] - \f{\la}{16}\Tr[\Si^\da\Si]^2\nn\\
&-g \left(\pb_L \Si \p_R + \pb_R \Si^\da \p_L\right)\,.
\end{align}
This form is useful because it allows one to identify easily the terms
involved in spontaneous chiral symmetry breaking ($\sim \Tr[\Si^\da\Si]$). 
Terms responsible for explicit chiral symmetry breaking 
(eg. $\si \sim \Tr[\Si + \Si^\da]$) do not occur in this case, but will 
be considered in the context of $\chi$PT, in Section \ref{sec:chiptth}.

The exponential representation is  most commonly employed
for its application to low-energy QCD. By defining a matrix-valued field 
%$U \equiv e^{i\vec{\tau} \cdot \vp/v}$
 $U \equiv \ro{exp}{\left(i\vec{\tau} \cdot \vp/v\right)}$
 that transforms the same way as the previous $\Si$ field,
and a massive scalar field $S$, the Lagrangian becomes:
\begin{align}
\cL_U &=  + \pb_L i\sla{\darrow}\p_L + \pb_R i\sla{\darrow}\p_R +
\f{1}{2}\left((\cd_\mu S)^2 - 2\mu^2 S^2 \right)
+ \f{{(v+S)}^2}{4} \Tr[\cd_\mu U \cd^\mu U^\da]\nn\\
 &- \la v S^3 - \f{\la}{4}S^4
- g(v+S)\left(\pb_L U \p_R + \pb_R U^\da \p_L \right)\,,
\end{align}
for an arbitrary coupling constant $v$.
This representation combines the matrix form with a heavy scalar
degree of freedom, which can be integrated out of the theory easily 
using the prescription provided by Donoghue, Golowich and Holstein
\cite{DGH}. This is exactly the form needed to construct a
low-energy effective field theory.

\section{Chiral Perturbation Theory}
\label{sec:chiptth}
%Introduce the basic priciples, but only with a view
%to what is needed in the results chapters

% Discuss- 1: how to construct the lagrangian field theory,
% assume knowledge of level 4 QFT @ adelaide 
%          2: when referring to symmetries, only discuss Goldstone's theorem
%             or Noether's theorem. Do not branch out into chiral symmetry
%             be terse.
%          3: need to mention chiral power-counting
% Generating functional section. In previous chapter Lattice QCD? Surely,
% since lattice QCD needs it! 
The formalism of chiral perturbation theory 
($\chi$PT) will take advantage of Goldstone's Theorem 
and the study of symmetries discussed in the previous section. 
In this case, however, the global gauge group considered is flavour $\SU(3)$. 
In order for the effective field theory to emulate physical results,
one must write down the mechanics of a Lagrangian field theory
incorporating the necessary symmetries and degrees of freedom
at the observed scale. To represent particles such as pions and kaons
obeying Bose-Einstein statistics, one can write the standard massless scalar
Lagrangian:
\eqb
\cLe = \f{1}{2}\cd_\mu \pi^a \cd^\mu \pi^a + \ca{O}(\pi^4)\,,
\eqe
and interpret $\pi^a$ as the octet of Goldstone bosons (whose explicit 
form can be found in Appendix \ref{app:fields}).
 By defining a matrix-valued
function $U$, and its transformation law,
 one can collect together the interaction terms
in the exponential representation in a similar way to the Sigma Model:
\begin{align}
U &= \ro{exp} \left({\f{i}{f}\pi^a\la^a}\right)\,, \\
%U &=& e^{ {\f{i}{f}\pi^a\la^a}}\,, \\
U &\ra L\, U R^\da\,,
\end{align}
with constant $f$.
Now the effective Lagrangian can be written down as
an expansion of successive orders of momenta.
The two derivatives in the scalar Lagrangian mean that
only even chiral powers are admitted for particles such as mesons.
For the lowest-order free mesonic Lagrangian, there is only one
low-energy coupling constant, $f$:
\eqb
\label{eqn:lomeslag}
\cLe^{(2)} = \f{f^2}{4}\Tr[\cd_\mu U \cd^\mu U^\da]\,.
\eqe
(Higher-order mesonic Lagrangians can be found
in References \cite{Scherer:2005ri,Borasoy:2007yi,Bernard:2007zu}.)
The coefficient $f$ from the definition of the field $U$
appears here as a low-energy constant (LEC), 
 since it is expected that the expanded
effective Lagrangian for the pseudo-Goldstone fields
will have the standard normalization for bosons, 
$\cLe = \f{1}{2}\cd_\mu\pi^a\cd^\mu\pi^a + \ca{O}(\pi^4)$.
This LEC can further be identified with the pion decay constant
$f_\pi$ 
by first considering the Fermi weak interaction Lagrangian as a left-handed
source field and computing the decay rate from the resultant invariant
S-matrix element \cite{DGH,Scherer:2005ri}. 

The second-order  
Lagrangian of Equation (\ref{eqn:lomeslag}) 
will be the starting place for the consideration of the
low-energy meson sector of QCD.

\subsection{Meson Sector}
\label{subsect:meson}
%We shall
 In the theory of mesons, one 
considers a set of Goldstone boson fields and interprets them
as the meson sector of QCD.  %only we shall use our 
 One can use the knowledge of
%spontaneous 
 explicit symmetry breaking from Section \ref{subsect:ssb} 
to provide the fields with a (small) mass. 
Using the exponential
representation, $U(x)$ can be systematically expanded in powers
of its small momentum and mass
with respect to some energy scale $\La_\chi$.
In 1984, Manohar \emph{et al.} identified this scale of chiral symmetry 
breaking as $\La_\chi \sim 4\pi f \approx 1$ GeV \cite{Manohar:1983md}.
In renormalization, this is the  
 scale at which the
next-order loop contribution retains the same effective coupling
strength (see Section \ref{sec:renorm}). 

The total mesonic Lagrangian
can be written out in the expanded form of even chiral powers
\cite{Borasoy:2007yi}:
\eqb
\cL_\pi(U(x),\ca{M}) = \sum_{i=1}^{\infty}\cL_\pi^{(2i)}(U(x),\ca{M})\,.
\eqe
In order to quantify the extent of the chiral symmetry breaking caused
by the mass terms in the expansion,
 initially let $\ca{M}$ transform as a field 
($\ca{M} \ra L\ca{M} R^\da$), 
so that the Lagrangian will remain invariant
under global $\SU(3)_L \otimes \SU(3)_R$.
At lowest non-trivial order:
\begin{align}
\label{eqn:chiptmasslag}
\cL_\pi^{(2)} &= \cL_\pi^\ro{kin} + \cL_\pi^\ro{mass}\nn\\
&= \f{f_\pi^2}{4}\Tr[\cd_\mu U \cd^\mu U^\da] 
+ \f{f_\pi^2 B_0}{4}\Tr[\ca{M}\! U^\da +  U \ca{M}\!^\da]\,,
\end{align}
where $B_0$ is a constant (with dimensions of mass)
included for generality.
%Here, chiral symmetry breaking 
 Chiral symmetry breaking then 
results from imposing the Hermitian condition 
for the quark mass matrix $\ca{M} = \ca{M}^\da$. 
%and so 
Thus the constant $B_0$ directly corresponds to
the extent of chiral symmetry breaking 
 \cite{Scherer:2005ri,Borasoy:2007yi}.

% Discuss Weinberg Power counting here. (for mesons)

Some terms in the Lagrangians of either QCD or $\chi$PT
explicitly break chiral symmetry. 
For example, $\cL_\ro{mass}$ involving the quark mass in Equation 
(\ref{eqn:chiralqcdlag}) is invariant under an axial group action
$A = \ro{exp}({-\f{i}{2} \alpha_A^a\la^a\g_5})$.
The associated axial Noether currents $J_A^{\mu\,a}$ encountered in PCAC 
will not be conserved, but diverge according to the equation: %as: 
\eqb 
\cd_\mu J_A^{\mu\, a} = 2i\pb\ca{M}\g_5\f{\la^a}{2}\p\,.
\eqe
%
%talk about SU(3)

%GOR- keep in mind Borasoy and Ross to be terse here.
%Need to expand L_sb = B_0/2 Tr[pi^2 M],
%since Tr[M] terms are not invariant.
%The expansion, with the restrictions m_u = m_d = m, and no pi^0 eta mixing
%yields: m_\pi^2 = 2 B_0 m
%        m_K ^2  = B_0 (m + m_s)
%        m_eta ^2= 2/3 B_0 (m + 2 m_s)
%Haven't introduced B_0 yet!
%Instead, define them in terms of quark condensate <qbar q>??
%Or rearrange things !!!!****
%
To relate the meson masses to the quark masses, consider
chiral $\SU(3)$. It is expected that the vacuum expectation values %(VEV) 
of the
scalar quark densities are the same in each theory: QCD and $\chi$PT.
That is, the quark condensate
 $\lb \bar{q}q\rk$, where $q$ stands for $u$, $d$ or $s$ quarks,
 should be an observable independent of representation. 
Consider the explicit chiral symmetry breaking terms of each Lagrangian,
 namely, $\cL_\ro{mass}$ defined in Equation (\ref{eqn:chiralqcdlag}) for QCD,
 and the mass term $\cL_\pi^\ro{mass}$
 of Equation (\ref{eqn:chiptmasslag}) for $\chi$PT.
By expanding out the exponential field $U$ in $\cL_\pi^\ro{mass}$,
one obtains:
\eqb
\cL_\pi^\ro{mass} = B_0f_\pi^2\Tr[\ca{M}\!] 
-\f{1}{2}B_0\Tr[\ca{M}\pi^2]+\ca{O}(\pi^4)\,.
\eqe 
For approximate isospin symmetry 
$m_u \approx m_d \approx \f{1}{2}(m_u + m_d)\equiv \hat{m} \neq m_s$,
expanding the first of these terms yields the relations: 
\eqb
\lb\bar{q}q\rk = -\lb0\mi\f{\cd\cL_\ro{mixed}}{\cd \hat{m}}\mi0\rk
= -\lb0\mi\f{\cd\cL_\pi^\ro{mass}}{\cd \hat{m}}\mi0\rk = -B_0 f_\pi^2\,.% \\
%
%&\Rightarrow& \lb0\mi\bar{u}u + \bar{d}d + \bar{s}s\mi0\rk = -B_0 f^2\,.
\eqe
Thus there is a profound %deep 
connection between quark condensation and the process
 of dynamical chiral symmetry breaking.
The second term yields the 
%\emph
{Gell-Mann$-$Oakes$-$Renner Relations} \cite{GellMann:1968rz}
relating meson masses to quark masses:
\begin{align}
\label{eqn:GOR}
m_\pi^2 &= 2B_0 \hat{m}\,,\\
m_K^2 &= B_0 (m_d + m_s)\,,\\
m_\eta^2 &= \f{2}{3}B_0 (\hat{m} + 2m_s)\,. 
\end{align}
Just like PCAC, 
Equation (\ref{eqn:GOR}) shows that if the light quark masses are zero, then the
pion mass must also be zero, and thus chiral symmetry holds.
This leads to the %\emph
{Gell-Mann$-$Okubo Mass Relation}:
\eqb
3m_\eta^2 = 4m_K^2 - m_\pi^2\,.
\eqe

By additionally enforcing local chiral symmetry,
the set of all chiral Ward Identities become an invariant of the generating
functional encountered in Section (\ref{sec:funcmeth}),
as long as no anomalies are present \cite{Scherer:2005ri}. 
The chiral Ward Identities simply encode the statement 
of symmetry preservation and the existence of conserved quantities as a 
consequence, much like Noether's Theorem; but applied to quantum amplitudes.  
Consider QED, a $\ro{U}(1)$ gauge theory, as an example. 
The Ward Identity amounts to a statement of charge conservation, 
and the existence of a conserved electric current. 
%Consider QCD, as an example.   
%and the existence
 %In order 
In QCD, to be able to generate all the Green's Functions for the theory,
the Lagrangian must include
pseudo-scalar ($p$) and vector ($l_\mu$, $r_\mu$) source fields, which
vanish to recover the standard QCD Lagrangian in Equation (\ref{eqn:qcdlag}),
and a scalar field ($s$) that %takes on 
 assumes the role of the quark masses $\ca{M}$. This is known as 
the method of external sources. 
%Why? Scherer pdf-p38: the transformation of the external current 
%j(x) = j(x)-\cd\ep(x) cancels off the extra kinetic term in the lagrangian
%after a local transformation. ie under global U(1) sym, the gen f'n'al is
%invariant under local transf. In principle, any ward identity can be 
%obtained by taking the appropriate higher f'n'al derivs of W.
This generalization of the QCD Lagrangian is vital for calculating
the divergence of Green's Functions.
These fields obey the following transformation laws for the 
local chiral rotations
$L(x),R(x) \in \SU(3)_{L,R}\,$:
\begin{align}
l_\mu &\ra L(x) \,l_\mu\, L^\da(x) +
 i (\cd_\mu L(x)) L^\da(x)\,,   \\
r_\mu &\ra R(x)\, r_\mu\, R^\da(x) + 
i (\cd_\mu R(x)) R^\da(x)\,,   \\
(s + ip) &\ra L(x)(s + ip)R^\da(x)\,.
\end{align}
The QCD Lagrangian, invariant under local $\SU(3)_L\otimes\SU(3)_R$, 
becomes:
\begin{align}
\cL_\ro{QCD}^\ro{local} &= i\pb\sla{\darrow}\p
-\f{1}{2}\Tr[\bar{G}_{\mu\nu}\bar{G}^{\mu\nu}]
-\pb_L(s + ip)\p_R - \pb_R(s + ip)\p_L\nn\\
 &- \pb\g_\mu \G_Ll^\mu\p -
\pb\g_\mu \G_Rr^\mu\p\,.
\end{align}
In the case of the low-energy effective Lagrangian,
one must define a covariant derivative with transformation law:
%
% use D when we include e-m gauge field
%
\begin{align} 
\nabla_\mu U &= \cd_\mu U +  il_\mu U -iUr_\mu\,, \\
\nabla_\mu U &\ra L(x) \nabla_\mu U R^\da(x)\,.
\end{align}
Therefore, the lowest-order non-trivial Lagrangian for mesons obeying 
local chiral symmetry 
can now be written with mass source defined using the convention
$\chi = 2 B_0 (s + ip)$, functioning as a field, as before:
%
% cf e-m: use chi_+. also, cf. M_+ used below in tadpole lag.
%
\eqb
\label{eqn:meslag}
\cL_\pi^{(2)} = \f{f_\pi^2}{4}\Tr[\nabla_\mu U \nabla^\mu U^\da
+\chi U^\da + U\chi^\da]\,.
\eqe
%

%as reproduced in the preceding recapitulation of symmetry theory,
% Section \ref{sec:chisym}. 

%Higher order lags in \cite{Bernard:2007zu} ``Chiral PT and Baryon Properties''
%Borasoy and Scherer
%Cov deriv, and putting photon field in as per Cloet and Ha
%\cite{Ha:1998gg}... put that in baryon e-m part.

\subsection{Baryon Sector}

Since the Lagrangian of a low-energy theory can be
expanded out in a convergent series of small momenta $p/\La_\chi$,
the mass of the baryons themselves cannot be treated as an expansion
parameter, since their mass and momenta are of the same order of magnitude
as the scale $\La_\chi$; thus the perturbation theory diverges. 
That is, the mass of a baryon $M_B\sim \La_\chi$, so the expansion parameter 
$M_B/\La_\chi$ cannot be small. 
To overcome this difficulty in ordering the chiral series in the baryon sector 
of $\chi$PT, consider the heavy-baryon approximation. 

Define some alternative fields $B_{\ro{v}}(x)$ to the $\SU(3)$ 
octet baryons $B(x) = B^a(x)\la^a$, 
 with velocity 
$\ro{v}_\mu$
 largely unchanged by pion interactions
  \cite{Georgi:1990um,Jenkins:1991ne,Jenkins:1990jv}. 
%This is the `heavy-baryon' approximation.
These fields $B_{\ro{v}}(x)$ are only just off-shell by a 
small amount $k\cdot \ro{v}$:
\eqb
p^\mu_{\ro{B}} = M_B\ro{v}^\mu + k^\mu\,.
\eqe
A perturbation theory about this small momentum $k_\mu$ can now be 
constructed.
In addition, the difficult spin structure of the new fields $B_\ro{v}$
 can be handled by using the particle projection operator 
$P_\ro{v} = \f{1}{2}(1+\sla{\ro{v}})$, thus 
absorbing the effects of virtual baryon loops into higher chiral orders
of the theory:
\eqb
B_\ro{v}(x) = P_\ro{v}\, e^{iM_B\sla{v}v\cdot x}B(x)\,.
\eqe

%NOW MENTION RARITA-SCHWINGER FIELDS (DECUPLET) also from
%introduction to unitary symmetry, Coleman.
This procedure can be repeated in exact analogy for the totally symmetric
%\emph
{Rarita-Schwinger} tensor $T^{\mu abc}_\ro{v}(x)$, which represents the
spin-$3/2$ decuplet fields, as long as all spin-$1/2$ components 
 are removed ($\g \cdot T^{abc}_\ro{v} = 0$). It is defined by:
\eqb
T_\ro{v}(x) = P_\ro{v}\, e^{i(M_B+M_T)\,\sla{v}v\cdot x}T(x)\,.
\eqe
The sum of the octet and decuplet masses is used, by convention,
 in the exponential in order to avoid extra factors of mixed 
octet-decuplet fields in the final Lagrangian. This results in a positive 
term proportional to the mass splitting $\De = |M_T-M_B|$ 
\cite{Jenkins:1991ne}.
(The explicit representation of these
fields in $\SU(3)$  can be found in Appendix \ref{app:fields}).
Treating the 
mass splitting $\De \ll \La_\chi$ as a 
small perturbation, 
the new velocity-dependent fields $B_\ro{v}$ and $T_\ro{v}$ (indices
suppressed) obey a massless Dirac Equation, and a Dirac Equation
with small mass splitting, respectively:
\begin{align}
i\sla{\cd}\,B_\ro{v}(x) &= 0\,, \\
(i\sla{\cd} - \De)T_\ro{v}(x) &= 0\,.
\end{align}
%
%AND ALSO DIRAC STRUCTURE SIMPLIFICATIONS FOR B AND T FROM JENKINS
%AND CLOET
%Cov Spin operator, and Polarization sum ie. how propagators change?

To write out a completely velocity-dependent Lagrangian for baryons and
their interactions with mesons,
it now remains to rewrite all Dirac bilinears in terms of a covariant
spin operator $S^\mu_\ro{v} = -\f{1}{8}\g_5[\g^\mu,\g^\nu]\ro{v}_\nu$, which
has the useful property that its commutation and anti-commutation
rules depend only on the four-velocity $\ro{v}_\mu$.
The meson interactions are incorporated into the theory by coupling baryon
fields to the axial current encountered in PCAC (Section (\ref{subsect:ssb})), 
which is 
equivalent to the Goldstone bosons as per the Haag Theorem.
The convention is to define exponential fields $\xi^2 \equiv U$,
which follow the transformation rule \cite{Jenkins:1990jv}:
\eqb
\xi \ra L\, \xi\, H^\da(x) = H(x)\, \xi\, R^\da\,.
\eqe
%
%how does xi transform? has H, a function of the GBs. need this to
%talk about transf rules for B,T, V and A. then cov deriv for B has V in it.
%so DB -> H DB.(as needed for a cov deriv). Same for cov dervi for T.
%Talk about how xi is treated as a transformation with generators \la
%and parameters \pi. Why? Construct Vec(like a gauge field) and Axial currents.
%Worth writing out their transformation rules? for B, T and xi, V and A.
%and definition of covariant deriv?
The transformation matrix $H = H(x)$ is a spacetime dependent combination
of the chiral transformation matrices and the Goldstone bosons
themselves.
This means that the octet and decuplet fields' transformation rules
also involve $H$, and in fact, the axial current 
$A_\mu$ and the 
octet baryon field $B_\ro{v}$ are exactly analogous to the $\xi$ field
in their transformations. The additional subtlety with the decuplet field
is that each of its three indices transforms separately:
\begin{align}
B &\ra H\, B\, H^\da\,, \\
T^{abc}_\mu &\ra H^{aa'}H^{bb'}H^{cc'}T^{a'b'c'}_\mu\,.
\end{align}
Because the transformation matrix $H$ is a spacetime-dependent object,
a vectorial connection needs to be included to preserve the gauge invariance of
the Lagrangian. Similarly, an axial vector combination of exponential fields
can be defined. Under the Haag Theorem, these axial vectors are equivalent 
to the pseudo-Goldstone boson fields:
\begin{align}
\label{eqn:veccomb}
V_\mu &= \f{1}{2}(\xi\cd_\mu\xi^\da + \xi^\da\cd_\mu\xi)\,, \\
V_\mu &\ra HV_\mu H^\da - (\cd_\mu H)H^\da\,,\\
%We didn't put an 'i' in the definition here as per Jenkins and Wang.
%instead, it is absorbed into the definition of S_v as per Wang.
%Factors of 2D and 2F come from the replacing of gammas with S_v's
% and factors: pi/2f or pi'/f merely change definition: pi = 2pi'
\label{eqn:axcomb}
A_\mu &= \f{i}{2}(\xi\cd_\mu\xi^\da - \xi^\da\cd_\mu\xi)\,, \\
A_\mu &\ra HA_\mu H^\da\,.
\end{align}
Thus the covariant derivative can now be included for both octet and decuplet
fields. As before, the decuplet requires a separate connection to
act on each index:
\begin{align}
\label{eqn:covB}
\ca{D}_\mu B_\ro{v} &= \cd_\mu B_\ro{v} + [V_\mu,B_\ro{v}]\,,\\
\ca{D}_\mu T^{\alpha\,abc}_\ro{v} &= \cd_\mu T^{\alpha\,abc}_\ro{v} +
V^d_{\mu a}T^{\alpha\,dbc}_\ro{v}+
V^d_{\mu b}T^{\alpha\,adc}_\ro{v}+
V^d_{\mu c}T^{\alpha\,abd}_\ro{v}\,.
\end{align}

The most general lowest-order Lagrangian for the baryon octet and decuplet 
fields, including transition vertices, 
can be now written by identifying the relevant $\SU(3)$ invariants \cite{Jenkins:1991ne,Jenkins:1990jv,Jenkins:1991ts,Labrenz:1996jy,WalkerLoud:2004hf,Wang:2007iw}:
% good explanation of bary lag: http://arxiv.org/pdf/nucl-th/9607021
\begin{align}
\label{eqn:octdeclag}
\cL^{(1)}_{\ro{oct\&dec}} = i\,\Tr[\bar{B}_\ro{v}(\ro{v}\cdot\ca{D})B_\ro{v}] 
+2D\,\Tr[\bar{B}_\ro{v}S_\ro{v}^\mu\{A_\mu,B_\ro{v}\}]
+2F\,\Tr[\bar{B}_\ro{v}S_\ro{v}^\mu[A_\mu,B_\ro{v}]] \nn \\
-i\bar{T}_\ro{v}^\mu(\ro{v}\cdot\ca{D})T_{\ro{v}\mu}
+\ca{C}(\bar{T}_\ro{v}^\mu A_\mu B_\ro{v} +
 \bar{B}_\ro{v}A_\mu  T_\ro{v}^\mu) \nn\\
+2\ca{H}\bar{T}_\ro{v}^\mu S_{\ro{v}\alpha} A^\alpha T_{\ro{v}\mu}
+ \De \bar{T}_\ro{v}^\mu T_{\ro{v}\mu}\,.
\end{align}
%
% CHOICE: have \xi = e^{i\pi/f} and \pi = \sqrt{2}  *(...)
%     OR: have \xi = e^{2i\pi/f},   \pi = 1/\sqrt{2}*(...)
%
The so-called %\emph
{$D$-style} and %\emph
{$F$-style} couplings for the 
octet occur simply as linear combinations of the most general first-order
invariants of flavour $\SU(3)$ symmetry. The reversed sign of the 
kinetic term of the decuplet simply encodes the spacelike nature of its 
positive energy spinors ($\ca{U}^2 < 0$), and  
the Rarita-Schwinger field propagators contain 
a polarization projector that sums over these spinors \cite{Jenkins:1991ne}:
\eqb
\ca{P}_v = \sum_{i=1}^4\ca{U}_i^\mu\,\bar{\ca{U}}_i^\nu = (v^\mu\,v^\nu - g^{\mu\nu})
- \f{4}{3}S_v^\mu\,S_v^\nu.
\eqe

When considering the mass renormalization of the nucleon in 
Chapters \ref{chpt:intrinsic} and \ref{chpt:nucleonmass}, 
%contributions up to a chiral
%order of $\ca{O}(m_\pi^4\,\ro{log}\,m_\pi)$ are included. 
%In order to do this, we
%must include at least the 
contributions from the second-order octet
Lagrangian $\cL^{(2)}_\ro{oct}$ are required, 
which correspond to an $NN\pi\pi$ vertex.
This gives rise to a %\emph
{tadpole} contribution. In full
$\SU(3)$ form, the vertices required from $\cL^{(2)}_\ro{oct}$ are 
\cite{WalkerLoud:2004hf}:
%
% Put other 2nd order vertices in an appendix???
%
%\eqb
%\cL_{\ro{tad}}^{(2)} =  c_2 \Big\{\,\Tr[\bar{B}B]\Tr[\ca{M}_{+}]
%+ \Tr[\bar{B}[\ca{M}_{+},B]]
% + \Tr[\bar{B}\{\ca{M}_{+},B\}]\,\Big\}\,,
%\eqe
\eqb
\label{eqn:tadlagsu3}
\cL_{\ro{tad}}^{(2)} =  2\si_M\,\Tr[\ca{M}_{+}]\Tr[\bar{B}B]
+ 2D_M\Tr[\bar{B}\{\ca{M}_{+},B\}]
 + 2F_M\Tr[\bar{B}[\ca{M}_{+},B]]\,,
\eqe
where $\ca{M}_{+} \equiv \f{1}{2}(\xi^\da \ca{M}\xi^\da + \xi \ca{M}\xi)$
 is the Hermitian mass source constructed from the quark mass $\ca{M}$
  \cite{Gasser:1987rb,WalkerLoud:2004hf}.
%The full Lagrangian at second order is shown in Appendix \ref{app:lag}.
%In Chapters \ref{chpt:intrinsic} and \ref{chpt:nucleonmass}, 
%the combination notation $c_2$ used 
%for the coupling constant of these vertices 
%\cite{Leinweber:2005xz} 

%Note that the notation $c_2$ used 
%for the coupling constant of these vertices 
%we choose to write the coupling constant of these vertices
%
%as $c_2$ \cite{Leinweber:2005xz} 
%is chosen 
%in anticipation of 
%relating it to the %\emph
%{sigma term}
%of explicit chiral symmetry breaking, introduced in 
%Chapter \ref{chpt:intrinsic}.

Consider now the lowest-order Lagrangian
 for the nucleon-pion interaction by simplifying
Equation (\ref{eqn:octdeclag})  to
involve only the nucleon doublet field $\Psi = (p\,,\,n)^\ro{T}$
and the $\SU(2)$ pion triplet (see Appendix \ref{app:fields}).
This is a useful approach when kaon loop contributions are neglected. 
%available only in the $\SU(3)$ case, .
The axial coupling constant below is simply defined as $\gc_A = D + F$ 
\cite{Scherer:2005ri}:
\eqb
\label{eqn:Npilag}
\cL_{\pi N}^{(1)} = \bar{\Psi}\left(\slashed{\cd} - \mc_N
+ \f{\gc_A}{2\fc_\pi}\g^\mu\g_5\vec{\tau}\cdot\cd_\mu \vp \right)\Psi\,.
\eqe
The tadpole Lagrangian now takes the form:
\eqb
\label{eqn:tadlag}
\cL_{\pi N}^{(2),\ro{tad}} = c_2 \Tr[\ca{M}_{+}]\bar{\Psi}\Psi\,,
\eqe
where the coeffcient is a combination of the 
LECs $\si_M$, $D_M$ and $F_M$, labelled $c_2$ in anticipation 
of the analysis presented in Chapter \ref{chpt:intrinsic}.

A local, chirally symmetric form of Equation (\ref{eqn:Npilag})
 can be recovered simply with the 
replacement:
\begin{align}
\cd_\mu &\ra \nabla_\mu = (\cd_\mu + \G_\mu - \f{i}{2}(l_\mu + r_\mu))\,,\\
\G_\mu &= \f{1}{2}(\xi^\da(\cd_\mu - ir_\mu)\xi
+ \xi(\cd_\mu - il_\mu)\xi^\da )\,,
\end{align}
%
%vierbein obtained from orthonormal basis set for the geometry. 
%eg. for basis e(ea)/R, e(ex)/R.sina, e(ey)/(R.sina.sinx):
% e1=r.da,e2=r.sina.dx,e3=r.sina.sinx.dy
%is a dreibein for the 3-sphere.
and also by replacing the product $\vec{\tau}\cdot\cd_\mu \vp$ 
with a  more general 
object: the Hermitian 
axial combination $u_\mu \equiv i\{\xi^\da(\cd_\mu - ir_\mu)\xi
- \xi(\cd_\mu - il_\mu)\xi^\da\}$.
The values of $\mc_N$, $\gc_A$ and $\fc_\pi$ 
are taken to be the nucleon mass, the
axial coupling strength and the pion decay constant, respectively, 
in the chiral limit.
The %\emph
{Goldberger-Treiman Relation} relates the nucleon-pion interaction
strength to the axial coupling $g_A$ \cite{Goldberger:1958tr},
 and can be obtained by 
comparing the matrix elements $\lb p\mi \pi(x)\mi n\rk$ and
$\lb p\mi \cd_\mu A^{\mu}(x)\mi n\rk$
using the relation between the pion field and the axial current
 in Equation (\ref{eqn:PCAC}) as per PCAC \cite{SC}:
\eqb
\label{eqn:GTR}
g_{\pi NN} \approx g_A \f{M_N}{f_\pi}\,.
\eqe
This equation becomes exact in the chiral limit
 $g_A(m_\pi^2\ra 0) = \,\,\gc_A$.

%
%for our convention of the LEC $c_2$ corresponding with the sigma term.
%
% Discuss Weinberg Power counting here. (for baryons)

\subsection{Electromagnetic Contributions}% in $\chi$PT}

%Below is the main reference to quote on the definitions....
%MENTION: from Jenkins \cite{Jenkins:1992pi} that: 
%this is the lagrangian from which we can derive
%the Coleman-Glashow SU(3) relations \cite{Coleman:1961jn}
%AND: that ``the baryon magnetic moments were first
%predicted theoretically on the basis of SU(3) flavour symmetry.''
%
%Talk about how <B|J_\mu|B> comes about, from Zanotti
%\cite{Arrington:2006zm}
% 
%H.o.t lag. in \cite{Jenkins:1992pi}
%MUST mention how xi, meson cov deriv, vec current(bary gauge component)
%and axial current are changed to have the photon field in them. [CLOET]
%Need to mention how charge matrix Q and mass \ca{M} change (operator 
%preplacements?)
%Sachs form factor

The baryon form factors comprise a parameterization for the matrix element 
obtained from the isovector quark current $J_\mu \equiv \pb \ca{Q}\g_\mu\p$, 
where $\ca{Q}$ is the  
$\SU(3)$ quark charge matrix $\ca{Q} \equiv \ro{diag}(2/3,-1/3,-1/3)$.
To evaluate this matrix element, one must calculate the fully-amputated 
 vertex for a baryon-photon interaction, wedged between the usual 
in- and out-going fermion spinors $u^{s}(p)$ and $\bar{u}^{s'}(p')$:
\eqb
\label{eqn:matelemem}
\lb B(p')\mi J_\mu\mi B(p)\rk = \bar{u}^{s'}(p')\left\{\g_\mu\, F_1(Q^2)
+\f{i\si_{\mu\nu} q^\nu}{2M_B}\,F_2(Q^2)\right\}u^s(p)\,,
\eqe
 for the tensor quantity $\si_{\mu\nu} \equiv \f{i}{4}\{\g_\mu,\g_\nu\}$.
$Q^2$ is a positive momentum transfer $Q^2 = -(p'-p)^2$, and 
$F_{1}$ and $F_2$ are called the Dirac and Pauli form factors, respectively.
The Sachs electromagnetic form factors $G_{E,M}$ are the linear combinations:
\begin{align}
G_E(Q^2) &= F_1(Q^2) - \f{Q^2}{4M_B^2}F_2(Q^2)\,,\\
G_M(Q^2) &= F_1(Q^2) + F_2(Q^2)\,.
\end{align}
Thus, in the non-relativistic, heavy-baryon formulation: %, one obtains:
\eqb
\lb B(p')\mi J_\mu\mi B(p)\rk = \bar{u}^{s'}(p')\left\{\ro{v}_\mu\, G_E(Q^2)
+\f{i\ep_{\mu\nu\rho\si}\ro{v}^\rho\,
 S_\ro{v}^\beta\, q^\nu}{M_B}\,G_M(Q^2)\right\}u^s(p)\,.
\eqe
By considering the behaviour of the Sachs form factors at zero momentum 
transfer, one can construct moments and charge radii. Two such important 
examples that will be considered are the magnetic moment, and the electric 
charge radius of the isovector nucleon. Recall 
from Section \ref{subsect:qqcd} that the isovector nucleon 
is simply the combination $p-n$, which transforms as a vector in 
isospin space, chosen so that diagrams containing indirect couplings 
will cancel, and the computation will be less intensive. 
The magnetic moment $\mu_n^v$ is simply the value of $G_{M}^v$ 
at $Q^2 = 0$: 
\begin{align}
\label{eqn:nucmagmom}
\mu^v_n &= G_{M}^v(Q^2=0) \\
&= 1 + \kappa^v_n.
\end{align}
The first term is simply the value of the Dirac form factor of the proton 
at $Q^2 = 0$, and the second term $\kappa^v_n$ is the anomalous magnetic moment
 originating from the finite-size behaviour of the hadron interactions 
of the effective quantum field theory: the hadron cloud, %\emph{hadron cloud}, 
which surrounds 
the nucleon. 

The electric charge radius is obtained by %considering the  
%form factor as a Fourier transform of an electric charge distribution:
taking a derivative with respect to $Q^2$ in the limit that $Q^2$ equals zero: 
%
%\begin{align}
%G_E^v(Q^2) &= \int\!d^3\vec{r}\,\rho(\vec{r}\,)\,e^{i\,\vec{q}\cdot\vec{r}\,},\\
%\Rightarrow \rad_E^v &= \lim_{Q^2\rightarrow 0}-6\frac{\cd G_E(Q^2)}{\cd Q^2}.
%\label{eqn:raddefn}
%\end{align}
%
\eqb
%G_E^v(Q^2) &= \int\!d^3\vec{r}\,\rho(\vec{r}\,)\,e^{i\,\vec{q}\cdot\vec{r}\,},\\
%\Rightarrow 
\rad_E^v = \lim_{Q^2\rightarrow 0}-6\frac{\cd G_E(Q^2)}{\cd Q^2}.
\label{eqn:raddefn}
\eqe

For octet baryons, the magnetic moments obey the Coleman-Glashow $\SU(3)$ 
relations, related to the following Lagrangian of two independent terms \cite{Jenkins:1992pi,Wang:2007iw,Wang:2008vb,Wang:1900ta}:
\eqb
\cL_{\ro{oct}}^{\ro{e-m}} = \f{e}{4m_N}(\mu_D\,\Tr \bar{B}_\ro{v} \si^{\mu\nu}
\{F_{\mu\nu}^+,B_\ro{v}\}
+\mu_F\,\Tr \bar{B}_\ro{v}\si^{\mu\nu}[F_{\mu\nu}^+,B_\ro{v}])\,.
\eqe
For an electromagnetic gauge field $\sr{A}_\mu$ with field strength tensor 
$F_{\mu\nu}\equiv \cd_{[\mu}\sr{A}_{\nu]}$, 
the quantity $F_{\mu\nu}^+$ has been chosen such that it is invariant under local 
chiral symmetry transformations:
\eqb
F_{\mu\nu}^+ \equiv \f{1}{2}(\xi^\da F_{\mu\nu}\ca{Q}\xi 
+ \xi F_{\mu\nu}\ca{Q}\xi^\da)\,.
\eqe
%
%When considering 
In the case of decuplet baryons, there is only a single 
 invariant term  that can be obtained from the group product 
$\mb{\bar{10}}\otimes\mb{10}\otimes\mb{8}$ that is proportional 
to their electric charge tensor $q_{ijk}$ \cite{Jenkins:1992pi}:
\eqb
\cL_{\ro{dec}}^{\ro{e-m}} = i\f{e}{m_N}\mu_\ca{C}\, q_{ijk}
\,\bar{T}_{\ro{v},ikl}^\mu\,
 T_{\ro{v},jkl}^\nu\, F_{\mu\nu}^+\,.
\eqe
The transition Lagrangian can be written out likewise:
\eqb
\cL_{\ro{trans}}^{\ro{e-m}} = i\f{e}{2m_N}\mu_T F_{\mu\nu}
(\ep_{ijk}\,\ca{Q}_l^i\,B_{\ro{v}m}^j\,
S_\ro{v}^\mu\, T_\ro{v}^{\ro{v},klm}
+\ep^{ijk}\,\ca{Q}_i^l\, \bar{T}_{\ro{v},klm}^\mu\, 
S_\ro{v}^\nu\, B_{\ro{v}j}^m)\,.
\eqe
These electromagnetic Lagrangians are obtained simply by collecting the 
photon-baryon terms from the electromagnetic covariant derivative. 
This new covariant derivative can 
be expressed by updating Equation (\ref{eqn:covB}) 
so that the electromagnetic field is included 
in both the vector connection $V_\mu$ from Equation 
(\ref{eqn:veccomb}) and the axial combination $A_\mu$ 
from Equation (\ref{eqn:axcomb}):
\begin{align}
V_\mu &\rightarrow V_\mu + 
\f{1}{2}i\,e\sr{A}_\mu(\xi^\da\ca{Q}\xi + \xi\ca{Q}\xi^\da)\,,\\
A_\mu &\rightarrow A_\mu - 
\f{1}{2}\,e\sr{A}_\mu(\xi\ca{Q}\xi^\da - \xi^\da\ca{Q}\xi)\,.
\end{align}
The covariant derivative for the pseudo-Goldstone Lagrangian 
can be updated in a similar fashion:
\eqb
\nabla_\mu\, U \rightarrow \nabla_\mu\, U + i\,e\sr{A}_\mu\,[\ca{Q}\,,U].
\eqe

% Discuss Weinberg Power counting here. (for photons in there as well)

\section{Regularization and Renormalization}
\label{sec:renorm}

%[[[Introduce Renormalization generally, Pauli-Villars, Slavnov, 
%DR, historical context then:
% specifically DR, FRR others
%Talk about renormalization in the context of specific observables]]]
%
\subsection{Historical Overview}
The calculation and interpretation of amplitudes from a quantum field theory 
proved more subtle than other theories due to their divergent behaviour.  
Despite success in predicting hitherto unexplained phenomena, 
many quantities calculated using the relevant quantum field theory 
 become infinite, though the known experimental value is finite. 
%even though that are finite in the physical universe. 
Consideration of the Lamb Shift in the electron energy levels in 
hydrogen atoms (1947) prompted the first real insight into this problem. 
It was conceived that if a quantity were altered infinitely 
by quantum corrections so that the final result was finite, 
  the initial `bare' quantity should never have been expected to be 
finite. That is, the bare quantity becomes renormalized. %\emph{renormalized}.
For example, the bare core of an electron has certain properties, 
such as electric charge, which become altered by an 
infinite amount due to vacuum polarizations. 
This polarization cloud surrounding the unphysical, bare 
electron core contains all possible diagrams of electron-positron 
 pair-production from virtual (off-shell) photons, which 
serve to screen the electron core's infinite charge, so that the observed, 
long-range charge is $-1.6\times 10^{-19}$ Coulomb, or $-e$ (in units 
of the charge of the proton). 
%In a perturbation theory, the fact that the expansion of an observable is often 
%an asymptotic (or Poincar\'e) series is intimately connected with this 
%divergence and subsequent infinite 
%renormalization, and is, 
%in fact, a consequence of the Haag Theorem \cite{Dyson:1952tj}.
This  `running' of the electron's charge to large values under deep probing 
from hard momenta in Bhabha scattering
 was confirmed in 1997 by the TOPAZ Collaboration at TRISTAN
\cite{Achard:2005it}. 
The virtual particles of a quantum field theory are simply consequences of the 
Green's Functions of the equations of motion. The Fourier transform of a 
particle propagator integrates the whole momentum spectrum, with a pole 
on the mass shell $k^2 = m^2$ (up to factors of $c$ and $\hbar$). 
Heisenberg's Uncertainty Principle for energy and time, 
$(\De E)(\De t) \geq \hbar$, allows the extra energy of pair-production, and 
other processes, for sufficiently small time. As a corollary, %consequence?
the virtual interactions take place over a spacelike time interval.

%minimal subtraction scheme\\

%Extended on mass shell (EOMS) renormalization
\subsection{The Power-Counting Regime}

%Recall that t
The Lagrangians of $\chi$PT are constructed with the 
intention that they can be
expanded in a series of some expansion scale, 
such as small momenta or masses. 
Although, ideally,  the series is convergent for a  
sufficiently small expansion scale, 
it need not necessarily be convergent, and instead will often take the 
form of  
 an asymptotic (or Poincar\'e) series.  
Nevertheless, in a realistic calculation, which involves 
calculating the expansion series only up to some finite order, 
it is desirable to be able to ensure that the uncertainty in the truncation 
is small. 
%
%
%the perturbative expansion in $p/\La_\chi$ is convergent.
% The renormalization of observables in $\chi$PT %, such as the mass of the 
%nucleon, or its electromagnetic properties,  
%may also be written 
%as an ordered expansion, obtained from direct contributions from 
%terms in higher order Lagrangians, and from quantum amplitudes 
%in the form of loop integrals.
%The chiral expansion of an observable is generally an asymptotic 
%(or Poincar\'e) series, and 
Thus, a knowledge of the 
applicable region of the expansion is as crucial 
as knowledge of the terms of the expansion series themselves. 
The range of values of the expansion scale for which 
a chiral expansion is convergent is 
known as the power-counting regime (PCR), and the 
expansion series is generally known as the chiral expansion. 

The PCR is the region where the quark masses are small, 
and higher-order terms in
the chiral expansion are negligible beyond the order calculated. 
Within the
PCR, the truncation of the chiral expansion is reliable to some 
prescribed precision. 
A chief focus of this thesis is to establish a formal approach to
 determining the PCR of a truncated chiral expansion quantitatively. 
The chiral expansion will be examined, and 
the individual low-energy coefficients of the chiral
expansion will be analyzed.  
The approach involves the examination of these low-energy coefficients 
 as they undergo the 
process of renormalization.
This approach provides a determination of the PCR for a
truncated expansion in $\chi$EFT.
 
First, it is essential to discuss 
methods of %\emph
{regularization} in the chiral loop integrals, so that the  
renormalization can take place.
%
%An important part of this thesis will be to establish a 
% careful %rigorous 
%approach to
% determining quantitatively the PCR of a truncated chiral expansion.
%
%ETC...
%
%
In order to renormalize a quantity, one must find a way to make 
the divergent amplitudes tractable, using  a process called regularization. 
This involves solving an 
integral over propagators in such a way as to isolate the divergent piece,   
 ready for handling with a suitable renormalization scheme.
%This process is called regularization. 
There is a wide variety 
of regularization schemes available. Pauli-Villars regularization (1949) 
involves the introduction of fictitious, `auxiliary' particles, associated 
with some mass scale, into 
a Lagrangian with a quadratic interaction. The extra formal terms in the 
Lagragian vanish as the mass scale is taken to infinity, and then 
 a subtraction can %then 
take place. However, because Pauli-Villars is not a gauge-covariant scheme, 
it is not applicable directly to Yang-Mills theory.
In Slavnov's regularization scheme (1971) of higher covariant derivatives, 
 once again, additional terms are added to the Lagrangian, but these do 
not render all amplitudes finite, thereby 
requiring a Pauli-Villars or other scheme 
to be used for divergent fermion-loop Feynman diagrams. In this thesis, 
a finite-range regularization scheme is used, which has powerful benefits 
in establishing the PCR, as will become apparent in Chapter 
\ref{chpt:intrinsic}.

\subsection{Dimensional Regularization}
\label{subsec:dr}
Dimensional regularization (DR) (1972) is an important procedure whereby 
loop integrals 
are analytically continued to generalized fractional dimensions and 
shown to converge \cite{'tHooft:1972fi}. The infinitesimal four-volume box 
$\ud^4k$ is replaced with $\ud^{4-\ep}k$, and 
the limit as $\epsilon \rightarrow 0^+$ is then taken. 
For example, the integral over a single (Euclidean) 
pion propagator is easy solved 
in %$4-\ep$-dimensional 
spherical polar coordinates, 
evaluating the angular part explicitly\footnotemark:
\eqb
\int\! \! \frac{\ud^4\! k}{(2\pi)^4}  \frac{1}{k^2 + m_{\pi}^2} \rightarrow 
\lim_{\ep\rightarrow 0^+}\int_0^\infty\! \! \frac{\ud k}{(2\pi)^{4 - \epsilon}} \frac{k^{3-\epsilon}}{k^2 + m_{\pi}^2} \frac{2\pi^{2 - \epsilon/2}}{\Gamma(2 - \epsilon/2)}\,. 
\eqe
\footnotetext{The Gamma function on $\mathbb{C}$ is defined as 
 $\Gamma(z) = \int_{0}^{\infty} \! \ud s\,  e^{-s} s^{z-1}$.}
Thus the minimal subtraction scheme result is recovered correctly.

Since there is no explicit scale-dependence in the 
interaction, this minimal subtraction scheme % with no explicit scale
%dependence 
makes DR suitable for use with elementary fields, where the
 absence of new degrees of freedom at higher energies is assumed. 
This is a powerful technique by which the divergent term(s) of a loop 
integral can be obtained, and then handled using a renormalization scheme.

Nevertheless, in the case of effective field theories, 
there exists an energy scale beyond
which the effective fields are no longer the relevant degrees of
freedom, and so DR is not ideally suited. 
Selecting a hard energy scale 
in the renormalization group equation, changes the relevant degrees of freedom 
in the Lagrangian.
At high energy scales, 
the high de Broglie frequency would resolve 
the internal structure of the hadrons,
 which would be 
the quarks (and beyond, if such higher degrees of freedom exist).
However, quarks are integrated out of the low-energy $\chi$EFT 
Lagrangian by construction. 
%
%When one integrates loop contributions 
When one calculates quantum amplitudes  
over this high energy
domain, there is no guarantee that one can efficiently subtract the
model-dependent, ultraviolet physics with a finite number of
counter-terms, as is required for successful renormalization,  
unless the perturbative expansion %in $p/\La_\chi$ 
is convergent.
%As a result, the chiral expansion typically only shows reliable
%convergence properties over a narrow range of pion mass.
%
Indeed this problem of beginning with rapidly varying loop
contributions, which must then be removed with a finite number of
counter-terms, can easily be overcome. The hard momentum contributions
to the chiral loop integrals can be suppressed via the introduction of a
finite-range regulator.

\subsection{Finite-Range Regularization}
\label{subsect:frr}

One alternative to DR is finite-range regularization (FRR), in which
one introduces a functional form $u(k\,;\Lambda)$, 
known as a finite-range regulator,   
 which controls the divergent integral at 
high momentum values. 
In this case, the integral over a single pion propagator would be 
modified as follows:
\eqb
\int\! \! \frac{\ud^4\! k}{(2\pi)^4}  \frac{1}{k^2 + m_{\pi}^2} \rightarrow 
\int\! \! \frac{\ud^4\! k}{(2\pi)^4}  \frac{u^2(k\,;\La)}{k^2 + m_{\pi}^2}\,.
\eqe
%
%The aim is to find a regularization scheme which allows sensible extrapolation 
% from lattice QCD results extending \emph{outside} the chiral regime. 
FRR involves the choice of a finite-valued momentum cutoff $\Lambda$. 
Allowing hard, internal momenta to flow through a loop 
integral yields unphysical results, in the form of a divergence.  
The high de Broglie frequency would resolve 
the internal structure of the hadrons,
 which would be 
the quarks (and beyond, if such higher degrees of freedom exist).
Therefore, a finite value of $\La$ 
is suitable for an effective field theory, where 
quarks are integrated out of the Lagrangian by construction. 
The choice of parameter $\La$ determines how fast the integral 
will now converge, and the regulator function should satisfy 
$u|_{k=0} = 1$ and $u|_{k\rightarrow\infty} = 0$.
The exact functional form chosen for the regulator should be 
independent of the result of calculation, as long as the 
perturbative expansion is convergent, that is, one works within the PCR. 

FRR has already been shown to be a powerful technique in solving 
the chiral extrapolation problem and identifying the PCR. 
The infinite
 series is resummed so that leading-order terms are large and the 
series converges. 
A variety of choices of functional forms 
for the regulator have been demonstrated to agree with each other, 
and with DR, in extrapolating lattice QCD results for
 the mass of the nucleon to physical quark masses \cite{Leinweber:2005xz}. 
Thus, the results of calculations using FRR are 
consistent with %using 
DR within the PCR. 
%***put mathematical content from talk 

Consider the example of a one-pion loop contribution for a nucleon, denoted 
$\Si_N$, with constant coefficient $\chi_N$. 
(This type of calculation is considered in more detail in 
Chapter \ref{chpt:intrinsic}.) 
The chiral expansion for the mass of the nucleon in this simple case, 
with one pion loop only, takes the form:
\eqb
M_N = \{a_0 + a_2m_\pi^2 + a_4m_\pi^4 + \ca{O}(m_\pi^6)\} + \Si_N\,,
\eqe
working to chiral order $\ca{O}(m_\pi^4)$. 
The chiral expansion comprises a polynomial expansion in $m_\pi^2$ 
and the contribution from the one-pion loop. 
Each of the coefficients $a_0$, $a_2$ and $a_4$ is renormalized by the 
contributions from the loop integral, at each order. 
The result of the integral 
using DR is equivalent to a massless renormalization scheme, with no 
explicit momentum cutoff: 
\begin{align}
\Si_N  &=  
\f{2\chi_N}{\pi}\int_0^\infty\!\!\ud k \f{k^4}{k^2+m_\pi^2}\\
&=\f{2\chi_N}{\pi}\int_0^\infty\!\!\ud k 
\f{(k^2+m_\pi^2)(k^2-m_\pi^2) +  m_\pi^4}{k^2+m_\pi^2}\\
&=  \f{2\chi_N}{\pi}
\left( \int_0^\infty\!\!\ud k\, k^2 -  m_\pi^2 
\int_0^\infty\!\!\ud k \right)  +  \chi_N m_\pi^3\,. %+ \ca{O}(m_\pi^4)\,.
\end{align}
In a massless renormalization scheme, there is no explicit 
    momentum cutoff, so each of the 
 coefficients $a_i$  undergoes an infinite renormalization or none at all:
\begin{align}
c_0  &=  a_0  +  \f{2\chi_N}{\pi}\int_0^\infty\!\!\ud k\, k^2\,, \\
  c_2   &=   a_2  -  \f{2\chi_N}{\pi}\int_0^\infty\!\!\ud k\,,\\
  c_4   &=   a_4   +   0   \,, \mbox{\,\,etc.}
\end{align}
By contrast, in a FRR scheme, a 
    momentum cutoff $\La$ 
 is introduced,
    and  the chiral expansion is resummed. Using a sharp momentum cutoff $\La$:
\begin{align}
\Si_N(\La) &= \f{2\chi_N}{\pi}\int_0^{\La}
\!\!\ud k \f{k^4}{k^2+m_\pi^2}\\
&=\f{2\chi_N}{\pi} \left(\f{\La^3}{3} 
 - \La m_\pi^2 
+  m_\pi^3\,\ro{arctan}
\left[\f{\La}{m_\pi}\right]\right)\\
&=  \f{2\chi_N}{\pi}\f{\La^3}{3} 
 - \f{2\chi_N}{\pi}\La m_\pi^2 
+  \chi_N m_\pi^3 -\f{2\chi_N}{\pi}\f{1}{\La}m_\pi^4+\ca{O}(m_\pi^6)\,.
\end{align}
The result obtained from DR 
can be recovered in an FRR scheme by taking the regularization scale  
parameter $\La$ to infinity:
\begin{align}
 c_0  &= a_0  + \f{2\chi_N}{3}\La^3\,, \\
 c_2  &= a_2  - \f{2\chi_N}{\pi}\La\,, \\
 c_4  &= a_4  - \f{2\chi_N}{\pi}\f{1}{\La}\,, \mbox{\,\,etc.}
\end{align}
 Thus, DR applied outside the PCR 
could be considered equivalent to a model 
with an arguably injudicious choice 
of cutoff scheme. The polynomial expansion of hadron mass is not 
expected to converge, 
and indeed it does not, using DR $\chi$PT, 
as mentioned by Young, \textit{et al.} 
\cite{Young:2002ib}.
 Outside PCR,  %of ignoring higher order terms in the
 the  expansion 
%totally 
 breaks down % outside the chiral radius,
 since the 
 chiral expansion is truncated without an attempt to estimate 
the higher-order contributions \cite{Leinweber:2005xz,Leinweber:2005cm}.

In addition, because FRR involves the 
resummation of the higher-order terms of the chiral expansion, it affords 
an opportunity to perform a calculation beyond the PCR. 
Using FRR, one must select a value for the 
ultraviolet regularization scale $\La$. The choice in the value of $\La$ 
is irrelevant within the PCR, where the results of extrapolations are 
scheme-independent (so long as $\La$ is not chosen to be too small, as 
explained in Section \ref{sec:lowerbound}). 
Nevertheless, the principal exercise of this thesis will be to handle  
any scheme-dependence occuring in a $\chi$EFT calculation outside the PCR. 
By quantifying the scheme-dependence  
one arrives at a rigorous procedure for using FRR beyond the PCR.

\begin{figure}[tp]
\begin{center}
\includegraphics[height=0.76\hsize,angle=90]{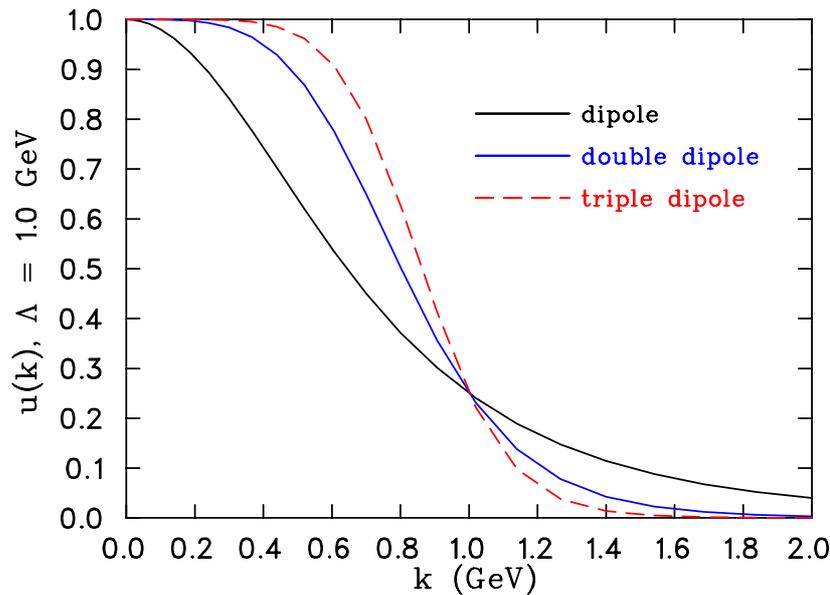}
\vspace{-12pt}
\caption{\footnotesize{Behaviour of three dipole-like regulators as a function 
of momentum $k$, for a regulator parameter $\La = 1.0$ GeV.}}
\label{fig:reg}
\end{center}
\end{figure}

In Chapter \ref{chpt:intrinsic}, %renormalization issues are considered 
%in more detail, %carefully, 
%using the nucleon mass as a test case. A 
a variety of 
finite-range regulators are used and compared.
For example, the 
Heaviside Step Function $u^2(k\,;\La) = \theta(\La - k)$ 
is an acceptable choice;  however, it is unfavorable for 
finite-volume considerations because %since
discrete lattice momenta are either fully included in the integral or
not included at all.  This results in inconvenient finite-volume 
artefacts. In the investigation of the nucleon mass, the 
%this investigation, the
family of smoothly attenuating dipole regulators will be considered.
The general multiple-dipole function of order $n$ 
takes the following form, for a
cutoff scale of $\La$:
\eqb
\label{eqn:ndipole}
u_n(k\,;\La) = {\left(1 + {\f{k^{2n}}{\La^{2n}}} \right)}^{-2}\,.
\eqe
The standard dipole is recovered for $n=1$. The cases $n=2,3$ are the
`double-' and `triple-dipole' regulators, respectively.  
 The behaviour of each of these three attenuators is shown in 
Figure \ref{fig:reg}.
%In the
%following, the notation will be to use $u(k;\La)$ to denote one of these
%regulators. 
These functional forms allow one to interpolate between the
dipole regulator and the step function (which corresponds to $n\to\infty$).
These $n-$tuple-dipole regulators generate extra non-analytic terms. 
%***show mathematical detail from mesonmass.tex
%
%As an example, consider the chiral expansion of the nucleon mass at  
% higher chiral orders. For a dipole regulator, regulator-dependent 
%non-analytic terms occur at odd 
%powers of $m_\pi$, beginning at $\ca{O}(m_\pi^5)$.
%In the case of the double-dipole, the non-analytic terms begin 
%at $\ca{O}(m_\pi^7)$, and for the triple dipole the non-analytic terms begin 
%only at $\ca{O}(m_\pi^9)$. 

It has been suggested in the literature that the only FRR scheme
 consistent with chiral symmetry uses the step function regulator 
\cite{Bernard:2003rp}. Djukanovic \textit{et al.} \cite{Djukanovic:2004px} 
have demonstrated that more general functional forms can be 
generated by proposing a scheme in which the regulator function is 
interpreted as a modification 
to the propagators of the theory, obtained from a new chiral 
symmetry-preserving Lagrangian.
Higher derivative coupling terms are 
built into the Lagrangian %in order 
to produce a regulator from the 
Feynman Rules in a symmetry-preserving manner.
%
%The regulators used in the present investigation
% are introduced in a less systematic fashion, such that chiral symmetry 
%is not automatically preserved to the order calculated. 
%The higher derivative couplings 
%of the regulator induces scheme-dependent non-analytic terms.
%To maintain chiral symmetry, one must introduce the necessary vertex 
%corrections.
%
Alternatively, one can choose the regulator judiciously such that any extra 
scheme-dependent non-analytic terms 
are removed to any chosen order. 
The regulators used in the present investigation 
follow the latter approach.  
For example, 
the $n-$tuple dipole regulators generate extra non-analytic terms 
in the chiral expansion of Equation  (\ref{eqn:mNexpansion}) in 
Chapter \ref{chpt:intrinsic} at  
 higher chiral orders. An explicit example of this is shown for the 
quenched $\rho$ meson mass in  
Section \ref{subsec:mesloop}, once the renormalization scheme has been 
introduced.

%% file: method.tex
\chapter{The Intrinsic Scale of the Nucleon}
\label{chpt:intrinsic}

\textit{``There lies the originality of our approach: \emph{to deduce common sense from the quantum premises, including its limits$-$ that is, to demonstrate also under which conditions common sense is valid, and what is its margin for error$\ldots$}}

\textit{[W]e no longer explain reality from our mental representation of it, taken for granted without question: but it is this representation,$\ldots$ that we want to \emph{explain}[.]''}
(Omn\`{e}s, R. 2002. \textit{Quantum Philosophy: Understanding and Interpreting Contemporary Science} p.165) \cite{Omnes}

\section{Renormalization Issues for the Nucleon Mass}

%%% !!! THIS PART NEEDS TO BE REWRITTEN. I NEED TO DECIDE WHICH PARTS
%       OF FRR DISCUSSION SHOULD GO HERE, AND WHICH SHOULD GO IN CH. 3.

In chiral effective field theory ($\chi$EFT), 
the nucleon mass may be written as an ordered, chiral
 expansion in 
the quark mass. The Gell-Mann$-$Oakes$-$Renner Relation 
 from Equation (\ref{eqn:GOR})  
 entails the %equivalence 
 proportionality $m_q\propto m_\pi^2$.
By considering the renormalization of the nucleon mass $\mc_N\rightarrow M_N\,$ 
from the Lagrangian in Equation (\ref{eqn:Npilag}), 
the chiral expansion will generally include a polynomial in $m_\pi^2$
 and non-analytic terms obtained from the chiral loop integrals.
 In addition, to establish a model-independent
framework in $\chi$PT, the expansion must display the properties of a
convergent series for the terms considered.  %There is a power-counting
%regime (PCR) where the quark mass is small, and higher-order terms in
%the expansion are negligible beyond the order calculated. 
 Recall that within the
power-counting regime (PCR) %, where 
the higher-order terms of the expansion may be regarded
as sufficiently small %negligible, 
 for the truncation of the chiral expansion to be reliable to a
prescribed precision. However, truncated expansions are
typically applied to a wide range of quark (or pion) masses, with
little regard to a rigorous determination of the PCR.
In the case of the nucleon mass, evidence suggests that the PCR is %may be
% quite 
small: %constrained by 
 limited to $m_\pi \lesssim 200$ MeV at $1\%$
accuracy at the chiral order $\ca{O}(m_\pi^4\,\ro{log}\,m_\pi)$
\cite{Beane:2004ks} \cite{Leinweber:2005xz}.
  This estimate of the PCR of $\chi$PT was
identified by  %specific 
%%finite-range regularization (FRR) 
%FRR 
 comparing the results of 
infrared regularization, dimensional regularization (DR) 
and a variety of finite-range regulators 
in analyzing lattice quantum chromodynamics (lattice QCD) simulation results. 
The different regularization schemes constitute 
different ways of summing higher-order terms in the chiral expansion. 
Thus, the PCR is manifest when the 
pion mass dependence of the nucleon mass is independent of the
renormalization scheme. % parameter.
%
%In a perturbation theory, 
%The fact that the expansion of an observable is often 
%an asymptotic (or Poincar\'e) series is intimately connected with this 
%divergence and subsequent infinite 
%renormalization, and is, 
%in fact, a consequence of the Haag Theorem \cite{Dyson:1952tj}.
%The asymptotic nature of the chiral expansion places the focus on the
%first few terms of the expansion.  
%
%The expansion of an observable is often 
%an asymptotic (or Poincar\'e) series, and so divergent even after 
%renormalization. 
%
%The fact that the expansion of an observable is often 
%an asymptotic series is intimately connected with 
%renormalization. %, and is, 
%in fact, a consequence of the Haag Theorem \cite{Dyson:1952tj}.
In addition, the asymptotic nature of the chiral expansion places 
the focus on the
first few terms of the expansion. 

A survey of the literature for the
baryon sector of $\chi$EFT illustrates the rarity of calculations
beyond one-loop \cite{McGovern:1998tm,McGovern:2006fm,Schindler:2006ha},
 and there are currently no two-loop calculations that 
incorporate the effects of placing a baryon in a finite volume.  With
only a few terms of the expansion known for certain, knowledge of the PCR of
$\chi$EFT  is as important as knowledge of the expansion itself. %It is within
%the PCR that higher-order terms of the expansion may be regarded
%as negligible.
Though scheme-dependent, it is %significant 
 worthwhile to note that, using
a dipole regulator with $\La = 0.8$ GeV, the 
coefficient of the induced 
$m_\pi^5$ term compares favorably with the infinite-volume two-loop calculation 
\cite{McGovern:1998tm,Leinweber:2003dg,Leinweber:2005xz,McGovern:2006fm,Schindler:2006ha}.

%Numerical simulations of QCD on a spacetime lattice are complemented
%by $\chi$EFT through the provision of a model-independent formalism
%for connecting lattice simulation results to the physical world.
%Simulations at finite volume and a variety of quark masses are related
%to infinite volume and physical quark masses through
%this formalism.  However, the formalism is accurate only if one
%works within the PCR of the truncated expansion.  Present practice in
%the field is best described as optimistic.  Truncated expansions are
%regularly applied to a wide range of quark (or pion) masses with
%little regard to a rigorous determination of the PCR.

%When considering nucleons, there is %some 
%evidence that the PCR is %may be
% quite 
%small; constrained by $m_\pi \lesssim 200$ MeV at $1\%$
%accuracy at the chiral order $\ca{O}(m_\pi^4\,\ro{log}\,m_\pi)$
%\cite{Beane:2004ks} \cite{Leinweber:2005xz}.
%  This estimate of the PCR of $\chi$PT was
%identified using specific 
%finite-range regularization (FRR) 
%FRR techniques
%to analyze lattice QCD data.   Using FRR, the regime is manifest when the 
%pion mass dependence of the nucleon mass is independent of the
%renormalization scheme parameter.

\subsection{Chiral Expansion of the Nucleon Mass}
\begin{figure}
\centering
\includegraphics[height=85pt]{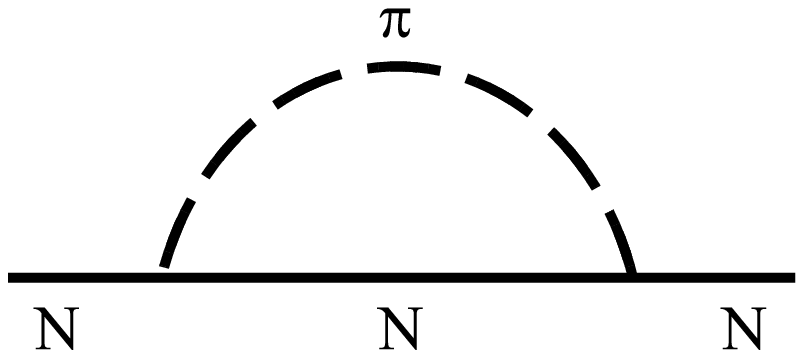}
\vspace{-3mm}
\caption{\footnotesize{The pion loop contribution to the self-energy
    of the nucleon, providing the leading non-analytic contribution to
    the nucleon mass.  All charge conserving transitions are implicit.}}
\label{fig:nucSEpiN}
%\end{figure}
%
%\begin{figure}
\centering
\vspace{6mm}
\includegraphics[height=85pt,angle=0]{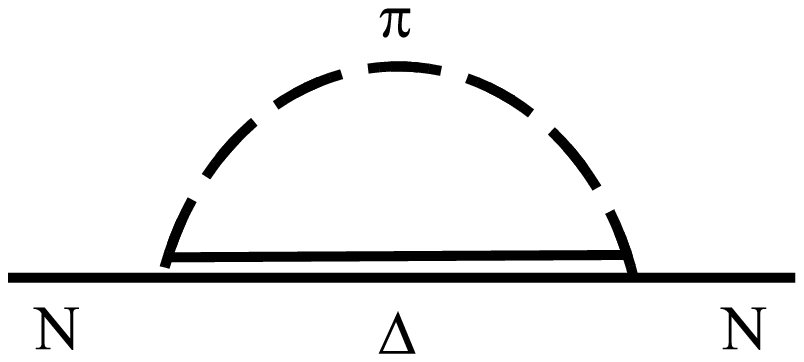}
\vspace{-3mm}
\caption{\footnotesize{The pion loop contribution to the self-energy
    of the nucleon, allowing a transition to a nearby and
    strongly-coupled decuplet baryon.}}
\label{fig:nucSEpiD}
%\end{figure}
%
%\begin{figure}
\centering
\vspace{6mm}
\includegraphics[height=85pt,angle=0]{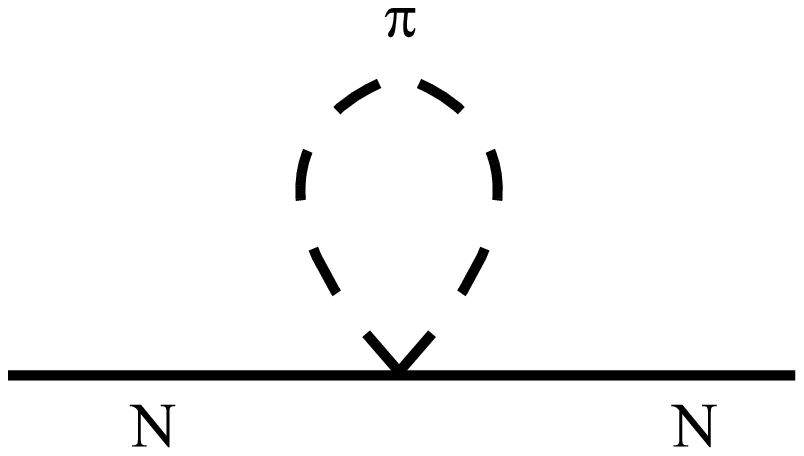}
\vspace{-3mm}
\caption{\footnotesize{The $\ca{O}(m_q)$ tadpole contribution 
to the nucleon self energy.}}
\label{fig:nucSEtad}
\end{figure}

%The discussion of renormalization is put in Ch3 so instead skip
% to the redefinition of the integrals, using FRR
The nucleon mass expansion formula can be expressed
in a form that collects the non-analytic behaviour into %up into
the loop integral contributions: % ($\Si$).
%Using the Gell-Mann$-$Oakes$-$Renner
%Relation connecting quark and pion masses (see Chapter \ref{chpt:chieft})
%$m_q \propto m_\pi^2$, we find:
%
\begin{align}
\label{eqn:mNresidB}
M_N &= \{
a_0^\La + a_2^\La\, m_\pi^2 + a_4^\La\, m_\pi^4 
+ \ca{O}(m_\pi^6)\} %\nn\\
%&& 
+ \Si_N(m_\pi^2,\La) + \Si_\De(m_\pi^2,\La) \nn\\
&+ \Si\,_{tad}(m_\pi^2,\La)\\
%
%\nn\\
%
\label{eqn:mNexpansion}
&= c_0 + c_2 m_\pi^2 + \chi_N m_\pi^3 +
c_4 m_\pi^4 %\nn\\
+ \!\Big(\!\!-\f{3}{4\pi\De}\chi_\De + \chi_\ro{t}
 \!\Big)m_\pi^4\ro{log}\,\f{m_\pi}{\mu} + \ca{O}(m_\pi^3).\quad
\end{align}
The superscript $\La$ denotes the scale-dependence of the
$a_i^\La$ coefficients. %The loop integrals are functions of the scale
%$\La$ and also $m_\pi^2$. 
The analytic terms in $m_\pi^2$ can be written as a polynomial with 
renormalized coefficients $c_i$.
The non-analytic contributions arise from the self energy integrals ($\Si$), 
which
 correspond to the diagrams in
 Figures \ref{fig:nucSEpiN} through \ref{fig:nucSEtad}. 

It is essential to note that the degrees of freedom
present in the residual series coefficients $a_i^\La$ are sufficient
to eliminate any dependence on the regularization scale parameter $\Lambda$, to
the order of the chiral expansion calculated: in this case ${\cal
O}(m_\pi^4\,\ro{log}\,m_\pi)$.  
%By definition, higher order terms in the FRR expansion
%are negligible in the PCR, and therefore FRR $\chi$EFT is
%mathematically equivalent to $\chi$PT in the PCR.
%
Any differences observed in results obtained at the same chiral order, 
but with different regularization schemes, are a direct result of
considering data that lie outside the PCR (provided that the scale  
$\La$ is not chosen so small that it introduces an unphysical 
low-energy scale).

\subsection{Chiral Loop Integrals}

Each of the loop integral contributions to the nucleon mass can be
simplified to a convenient form by
taking the non-relativistic heavy-baryon limit, and 
performing the pole integration for $k_0$. The integrals may be expanded 
out to a particular chiral order, 
in this case order $\ca{O}(m_\pi^4\,\ro{log}\,m_\pi)$, to obtain an 
analytic polynomial with coefficients $b_i^\La$, and the leading-order 
non-analytic term. 
Using a finite-range regulator $u(k\,;\La)$: 
\begin{eqnarray}
\label{eqn:NNunmod}
\Si_{N}(m_\pi^2\,;\La)  &=& \f{\chi_N}{2\pi^2}\int\!\!\ud^3\! k
\f{k^2 u^2(k\,;\La)}{k^2 + m_\pi^2} \\
 &=&  b_0^{\La,N} + b_2^{\La,N} m_\pi^2 + \chi_N m_\pi^3 + b_4^{\La,N}m_\pi^4 
+ \ca{O}(m_\pi^5)\,,
\label{eqn:NNexpnunmod}
\end{eqnarray}
\begin{eqnarray}
\label{eqn:NDeunmod}
\Si_{\De}(m_\pi^2\,;\La)  &=& \f{\chi_\De}{2\pi^2}\int\!\!\ud^3\! k
\f{k^2 u^2(k\,;\La)}{\om(k)\left(\De + \om(k)\right)} 
  \\
 &=&  b_0^{\La,\De} + b_2^{\La,\De} m_\pi^2 + b_4^{\La,\De} m_\pi^4- \f{3}{4\pi\De} 
\chi_\De m_\pi^4\,\log\f{m_\pi}{\mu} + \ca{O}(m_\pi^5)\,,\qquad
\label{eqn:NDeexpnunmod}
\end{eqnarray}
\begin{eqnarray}
\label{eqn:tadunmod}
\Si_{tad}(m_\pi^2\,;\La)  &=&  c_2 m_\pi^2\left(
\f{\chi_t}{4\pi}\int\!\!\ud^3\! k
\f{2 u^2(k\,;\La)}{\om(k)}  \right)\\
\label{eqn:tadexpnSiunmod}
 &=& c_2 m_\pi^2 \left(b_2^{\La,t} + b_4^{\La,t} m_\pi^2  + 
\chi_t m_\pi^2\,\log\f{m_\pi}{\mu}
+ \ca{O}(m_\pi^5)\right)\,,
\end{eqnarray}
where $\mu$ is an implicit mass scale from the logarithm,
 $\om(k) = \sqrt{k^2 + m_\pi^2}$ and $\De$ is the nucleon-delta
baryon mass-splitting, treated as a perturbation in
the approximate flavour symmetry. The mass of the $\De$ baryon 
is chosen to be the centre of its Breit-Wigner resonance.

The $b_i^\La$ coefficients renormalize the residual coefficients 
of the chiral expansion of Equation (\ref{eqn:mNresidB}), to obtain the 
scale-independent coefficients $c_i$. 
Though both the $a_i^\La$ coefficients and the 
$b_i^\La$ coefficients are scale-dependent,
adding them together at each order results in a scale-independent 
coefficient. These are the renormalized coefficients $c_i$. Explicitly: 
%
%The scale-dependent $a_i^\La$
%coefficients undergo a renormalization via their
%combination with the $b_i^\La$ coefficients, whose scheme-dependence
%is now explicit, reflecting the regularization of the chiral loop integrals:
%
\begin{align}
\label{eqn:c0normfrr}
c_0 &= a_0^\La + b_0^{\La,N} + b_0^{\La,\De}\,,\\
\label{eqn:c2normfrr}
c_2 &= a_2^\La + b_2^{\La,N} + b_2^{\La,\De} + b_2^{\La,t'}\,,\\
\label{eqn:c4normfrr}
c_4 &= a_4^\La + b_4^{\La,N} + b_4^{\La,\De} + b_4^{\La,t'}\,, \mbox{\,\,etc.}
\end{align}
This is simply a slight generalization from the worked example in 
Section \ref{subsect:frr}. 
Dimensional analysis reveals that the coefficients $b_i^\La$ are
proportional to $\Lambda^{(3-i)}$.  Thus it can be realized that as the cutoff
scale $\La$ tends to infinity, the result from DR, as described in Section 
\ref{subsec:dr}, is recovered.
 At any finite $\Lambda$, a partial
resummation of higher-order terms is introduced.
Previous studies indicate that extrapolation results show very little 
sensitivity to the precise functional form of the regulator 
\cite{Leinweber:2003dg}.

A modification is now made to the integrals of Equations (\ref{eqn:NNunmod}) 
through (\ref{eqn:tadexpnSiunmod}), by 
%
%Renormalization to order $\ca{O}(m_\pi^3)$ 
%is achieved by 
subtracting $b_i^\La$ terms %in the 
from their Taylor expansion, 
%of the loop integrals 
thus
absorbing them into the corresponding low-energy coefficients $c_i$. 
This achieves the renormalization to a chosen chiral order. In this case, 
only the low-energy coefficients $c_0$ and $c_2$ will be analyzed. 
The amplitudes for each process are 
thus altered: %simplified to the following:
\begin{eqnarray}
\label{eqn:NN}
\tSi_{N}(m_\pi^2\,;\La)  &=& \f{\chi_N}{2\pi^2}\int\!\!\ud^3\! k
\f{k^2 u^2(k\,;\La)}{k^2 + m_\pi^2} -  b_0^{\La,N} - b_2^{\La,N} m_\pi^2\\
 &=& \chi_N m_\pi^3 + b_4^{\La,N}m_\pi^4 + \ca{O}(m_\pi^5)\,,
\label{eqn:NNexpn}
\end{eqnarray}
\begin{eqnarray}
\label{eqn:NDe}
\tSi_{\De}(m_\pi^2\,;\La)  &=& \f{\chi_\De}{2\pi^2}\int\!\!\ud^3\! k
\f{k^2 u^2(k\,;\La)}{\om(k)\left(\De + \om(k)\right)} \nn\\
&& \quad - b_0^{\La,\De} - b_2^{\La,\De} m_\pi^2 \\
 &=&  b_4^{\La,\De} m_\pi^4- \f{3}{4\pi\De} 
\chi_\De m_\pi^4\,\log\f{m_\pi}{\mu} + \ca{O}(m_\pi^5)\,,\qquad
\label{eqn:NDeexpn}
\end{eqnarray}
\begin{eqnarray}
\label{eqn:tad}
\tSi_{tad}(m_\pi^2\,;\La)  &=&  c_2 m_\pi^2\left(
\f{\chi_t}{4\pi}\int\!\!\ud^3\! k
\f{2 u^2(k\,;\La)}{\om(k)} - b_2^{\La,t} \right)\\
\label{eqn:tadexpnSi}
 &=& c_2 m_\pi^2 \left(b_4^{\La,t} m_\pi^2  + 
\chi_t m_\pi^2\,\log\f{m_\pi}{\mu}
+ \ca{O}(m_\pi^5)\right)\qquad\\
\label{eqn:tadexpnsi}
&=& c_2 m_\pi^2 \tsi_{tad}(m_\pi^2\,;\La) \,.
\end{eqnarray}%

Note that the coefficient of the tadpole
 amplitude contains the renormalized low-energy coefficient $c_2$. 
This is because the same coefficient $c_2$ from the chiral expansion occurs 
in the tadpole Lagrangian in Equation (\ref{eqn:tadlag}). 
The tilde ($\,\tilde{\,}\,$) denotes that the
integrals are written out in renormalized form to chiral order 
 $\ca{O}(m_\pi^2)$.
  As the $b_i^\La$ 
coefficients are regulator and scale-dependent, the 
subtraction reshuffles this dependence into higher-order terms. 
The coefficients $a_0^\La$ and $a_2^\La$ 
of
the analytic terms in the chiral expansion in
Equation (\ref{eqn:mNexpansion}) automatically become the 
scale-independent renormalized
coefficients $c_0$ and $c_2$.

With the renormalized integrals specified, the finite-range regularization 
(FRR)  
modified version of the chiral expansion in Equation (\ref{eqn:mNexpansion})
 takes the form:
\begin{equation}
M_N = c_0 + c_2 m_\pi^2(1+{\tsi}_{tad}(m_\pi^2,\La))
+ a_4^\La m_\pi^4 + {\tSi}_{N}(m_\pi^2,\La) + {\tSi}_{\De}(m_\pi^2,\La)\,.
\label{eqn:mNfit}
\end{equation}
The $a_4^\La$ term is left in unrenormalized form
for simplicity. Indeed, the coefficient 
$b_4^\La$ can be evaluated by expanding out
corresponding loop integrals, such as in Reference \cite{Young:2002ib}.
However, the focus here is on the behaviour of $c_0$ and $c_2$.
%

%Introduce very briefly the sigma commutator.
\subsection{The Sigma Term}
%[Q_A,H] = 0 for chiral sym because then Q_A\mi0\rk = 0 and H\mi0\rk = 0
%We need to assume renormalization of the mass (so must come after
%Renorm section). Also, we use the important renormalized expansion
%introduced in nucleon mass. This must come in this chapter
%when discussing calculation of LECs.

%Feynman Hellmann Theorem
%Feynman$-$Hellmann Theorem \cite{Feynman:1939zz}
In addition to the mass of the nucleon in the chiral limit $c_0$, 
the low-energy constant (LEC) $c_2$ 
corresponding to the tadpole vertex  is of interest
phenomenologically because, by inspection of Equation (\ref{eqn:tadlag}),
 it is a measure of the %\emph
{explicit} chiral 
symmetry breaking of the relevant flavour symmetry group. 
That is, a sigma term can be defined for the light quarks up and down, 
and the explicit breaking of the group $\SU(2)_V \otimes \SU(2)_A$ may 
be investigated. In order to obtain a value for the sigma term 
relating to the heavier strange quark, $\chi$EFT has been used to 
 study the explicit breaking of the baryon 
octet representation of $\SU(3)$ \cite{Gasser:1980sb,Nelson:1987dg,Borasoy:1996bx,Young:2009ps}. %, and the analysis of a non

It can be seen that the term $c_2 m_q$, and higher-order terms in
the nucleon mass expansion formula of Equation (\ref{eqn:mNexpansion}),
will disappear if the chiral symmetry breaking quark mass is zero.
To investigate this, one can consider how the QCD Hamiltonian behaves
under commutation with the three-component axial charge operator of
flavour $\SU(2)$.
If the symmetry were unbroken, all quantities $Q_A^i\mi0\rk$ would vanish,
so the commutator $[Q_A^i,\ca{H}_{\ro{QCD}}]$ would also vanish. % and
%the fields commute.
In the more general case, consider two applications of the commutator,
which yield the symmetry breaking mass term $\ca{H}_{\ro{sb}}$ from 
the total Hamiltonian.
This defines the pion-nucleon sigma term
\cite{Cheng:1970mx,Leinweber:2000sa,Wright:2000gg,Holl:2005st,Young:2009ps}:
\begin{align}
\si_{\pi N} &= \f{1}{3}\lb N\mi[Q_A^i,[Q_A^i,\ca{H}_{\ro{QCD}}]]\mi N\rk \\ 
&=\lb N\mi(m_u \bar{u}u + m_d \bar{d}d)\mi N\rk = \lb N\mi\ca{H}_{\ro{sb}}
\mi N\rk\,.
\label{eqn:sigcomm}
\end{align}
Under the simplification of mass degeneracy between quark fields 
(which is approximately true under flavour $\SU(2)$),
 one can apply the Feynman$-$Hellmann Theorem \cite{Feynman:1939zz}
and recover the important result for small $m_\pi^2$:
\eqb
\si_{\pi N} = m_q\f{\cd M_N}{\cd m_q} = c_2 m_\pi^2 + \ca{O}(m_\pi^{5/2})\,.
%\cdots\,.
\eqe
That is, the value of the sigma term is dominated by the leading-order 
term with coefficient $c_2$. 
The violation of this axial symmetry is therefore important for understanding
the behaviour of hadrons, because a non-zero sigma term affects the structure 
of the interaction between hadrons and the meson cloud which surrounds them, 
and provides a small, but not statistically insignificant contribution 
to the total mass of the hadron. 

The standard result for the sigma term using $\SU(2)$ $\chi$PT, 
incorporating meson loop corrections, is: $\Si_{\pi N}=35\pm 5$ MeV 
\cite{Gasser:1980sb}. 
By analyzing data from $\pi p$ and $\pi\pi$ scattering experiments 
\cite{Hohler:1983aa}, 
an early analysis by Gasser suggests a value of $\Si_{\pi N} = 45\pm 8$ MeV 
\cite{Gasser:1990ce}.
The currently accepted value of the sigma term, due to the work of Koch,   
is larger than the theoretical value: $\Si_{\pi N} = 64\pm 8$ MeV 
\cite{Koch:1980aa,Koch:1982aa}. 
A more recent analysis of the experimental data by Pavan, 
incorporating a partial 
wave and dispersion relation analysis, suggests an even higher value of 
$\Si_{\pi N} = 79\pm 7$ MeV \cite{Pavan:2001wz}. 
In contrast, calculations from two-flavour 
dynamical quark lattice QCD comparatively underestimate the value of the 
sigma term. In a study by G\"{u}sken, it was found that  
$\Si_{\pi N} = 18\pm 5$ MeV, by direct calculation of the scalar matrix 
element in Equation (\ref{eqn:sigcomm}) 
 \cite{Gusken:1998wy}. This apparently low value for the sigma term was 
found to be a consequence of its sensitivity to chiral extrapolation, 
and large pion masses (above $500$ MeV) were used in the extrapolation   
\cite{Leinweber:2000sa,Young:2009ps}.

\subsection{Scheme-Independent Coefficients}

%chi coeffs
The chiral coefficients $\chi_{N}$, $\chi_\De$ and $\chi_t$ for each
integral are defined in terms of the pion decay constant, which is
taken to be $f_\pi = 92.4$ MeV, and the axial coupling parameters $D$,
$F$ and $\ca{C}$ which couple the baryons to the pion field, 
as shown in the Lagrangian $\cL^{(1)}_{\ro{oct\&dec}}$ 
of Equation (\ref{eqn:octdeclag}). 
The coeffcient $c_2$, which occurs in the tadpole loop integral of Equations  
(\ref{eqn:tad}) through (\ref{eqn:tadexpnsi}), is a combination of the 
LECs $\si_m$, $D_M$ and $F_M$, which occur in the tadpole Lagrangian of 
Equation (\ref{eqn:tadlagsu3}). 
Though $c_2$ is treated as a fit parameter,  
the phenomenological values for the $D$,
$F$ and $\ca{C}$ 
couplings are used, applying the $\SU(6)$
flavour-symmetry relations \cite{Jenkins:1991ts,Lebed:1994ga} to yield
$\ca{C} = -2D$, $F = \f{2}{3}D$ and the value $D = 0.76$:
\begin{align}
\chi_N   &= -\f{3}{32\pi f_\pi^2}(D+F)^2\,,\\
\chi_\De &= -\f{3}{32\pi f_\pi^2}\f{8}{9}\ca{C}^2\,,\\
\chi_t   &= -\f{3}{16\pi^2 f_\pi^2}\,.
\end{align}
%
%Because t
 These coefficients are constant and remain unaffected 
by renormalization scale or finite-volume effects. Ultimately, 
one may try to determine these directly from lattice simulation results. 
Nevertheless, because of the limited number of lattice simulation 
results currently available, this analysis will focus on the determination 
of $c_0$, $c_2$ and the nucleon mass $M_N$. 

\subsection{Finite-Volume Effects}

%mpiL and ep\regime
In lattice QCD, the introduction of finite-volume effects become significant 
for small box sizes.
The expansion parameter $1/L$ contributing to 
finite-volume effects should be of the same order of magnitude as the 
momenta for the perturbation scheme to remain valid. 
 If $L$ is small, the exponential factor $e^{-m_\pi L}$ no longer 
suppresses the finite-volume corrections \cite{Bar:2010zj}.
 As a general rule, the characteristic dimensionless quantity $m_\pi L$ 
 specifies the %\emph
{$\ep$-regime} through the condition $m_\pi L\leq 1$ 
\cite{Hansen1990685,Hansen1991201,Hasenfratz1990241}. This is a breakdown
 region in $\chi$PT, since divergences in the leading-order pion contributions 
cannot be approximated by standard perturbative techniques \cite{Bar:2010zj}.

Since the results of lattice simulations reflect the presence of
discrete momentum values associated with the finite volume of the
lattices, the formalism of $\chi$EFT 
must also take into account these finite-volume 
effects.  
$\chi$EFT is ideally suited for examining finite-volume effects, 
because of its accurate characterization of the dominant infrared physics. 
In order to accommodate the effect of the finite
volume, the continuous loop integrals occurring in the meson loop
calculations in an infinite volume are transformed into a sum over
discrete momentum values.  The difference between a loop sum and its
corresponding loop integral is defined to be 
the finite-volume correction, which
should vanish for all integrals as $m_\pi L$ becomes large
\cite{Gasser:1987zq,Beane:2004tw}.
While Equation (\ref{eqn:mNfit}) is useful in describing the pion mass
evolution of the nucleon mass, for the consideration of lattice QCD
results, one also needs to incorporate corrections to allow for the
finite-volume nature of the numerical simulations.  As the pion is the
lightest degree of freedom in the system, it is the leading-order 
 pion loop effects
 that are most sensitive to the periodic boundary
conditions.  The corrections can be determined by considering the
transformation of each loop integral in
Equations (\ref{eqn:NN}), (\ref{eqn:NDe}) and (\ref{eqn:tad})  
into a discrete sum for a given 
lattice size.  %$V = L_x L_y L_z$. %as described in Equation (\ref{eqn:discr}).
%
%Since loop integrals are now also discretized on the finite-volume lattice, 
The three-dimensional integrals %encountered in the non-relativistic 
%heavy-baryon limit
%(after the time component has been integrated out) 
can be replaced by summations over all possible momentum
 values \cite{Armour:2005mk}. 
%using the following Equation \cite{Armour:2005mk}:
%
%\eqb
%\label{eqn:discr}
%\int\! \! \ud^3k \, \,   \approx \frac{1}{N^3} \left( \frac{2 \pi}{a} \right )^%3 \sum_{k_{x},k_{y},k_{z}}\,,
%\eqe
%
%where $L = aN$ is the lattice box length, assumed equal in each spatial 
%direction. 
It is useful to define the %\emph
{finite-volume correction} to the loop 
integral, by convention, by subtracting the integral from the sum quantity.
This technique will be used to correct for finite-volume effects
 encountered in 
Chapters \ref{chpt:intrinsic} through \ref{chpt:nucleonmagmom}.

The finite-volume correction 
$\de^\ro{FVC}$ can be written as the difference between the finite sum
and the integral:
\begin{equation}
\de^\ro{FVC}_i(m_\pi^2,\La) \!\!= 
\f{\chi_i}{2\pi^2}\Big[\f{{(2\pi)}^3}{L_x L_y L_z}
\!\!\sum_{\,\,k_x,k_y,k_z}\!\!\!\!\!I_i(\vec{k},m_\pi^2,\La)\, -
  \int\!\!\ud^3\! k\,\, I_i(\vec{k},m_\pi^2,\La)\Big],\quad
\end{equation}
where $i=N,\De$, and the integrands are denoted $I_i(\vec{k},m_\pi^2,\La)$.
%as functions of vector $\vec{k}$, $m_\pi^2$ and regulator parameter $\La$.
%
The finite-volume corrections to the tadpole contribution are not 
considered in this investigation because of subtleties in their behaviour 
at large $m_\pi$. Details regarding the finite-volume behaviour of 
the tadpole amplitude are discussed in Appendix \ref{app:tadfvc}, and a 
more general discussion of its convergence properties occurs in Section  
\ref{sect:fut}. 
By adding the relevant finite-volume correction to each loop 
contribution, 
 the finite-volume nucleon mass can be parameterized:
%
%\begin{align}
\eqb
M_N^V %&
=  c_0 + c_2 m_\pi^2(1+{\tsi}_{tad}) + a_4^\La m_\pi^4 %\nn\\
+ ({\tSi}_{N}+\de_{N}^{\ro{FVC}}) %\nn\\
%&
+
 ({\tSi}_{\De}+\de_{\De}^{\ro{FVC}})\,.
\label{eqn:mNfitfin}
\eqe
%\end{align}
% 

\begin{figure}
\begin{minipage}[b]{0.45\linewidth} % A minipage that covers half the page
\centering
\includegraphics[height=1.0\hsize,angle=90]{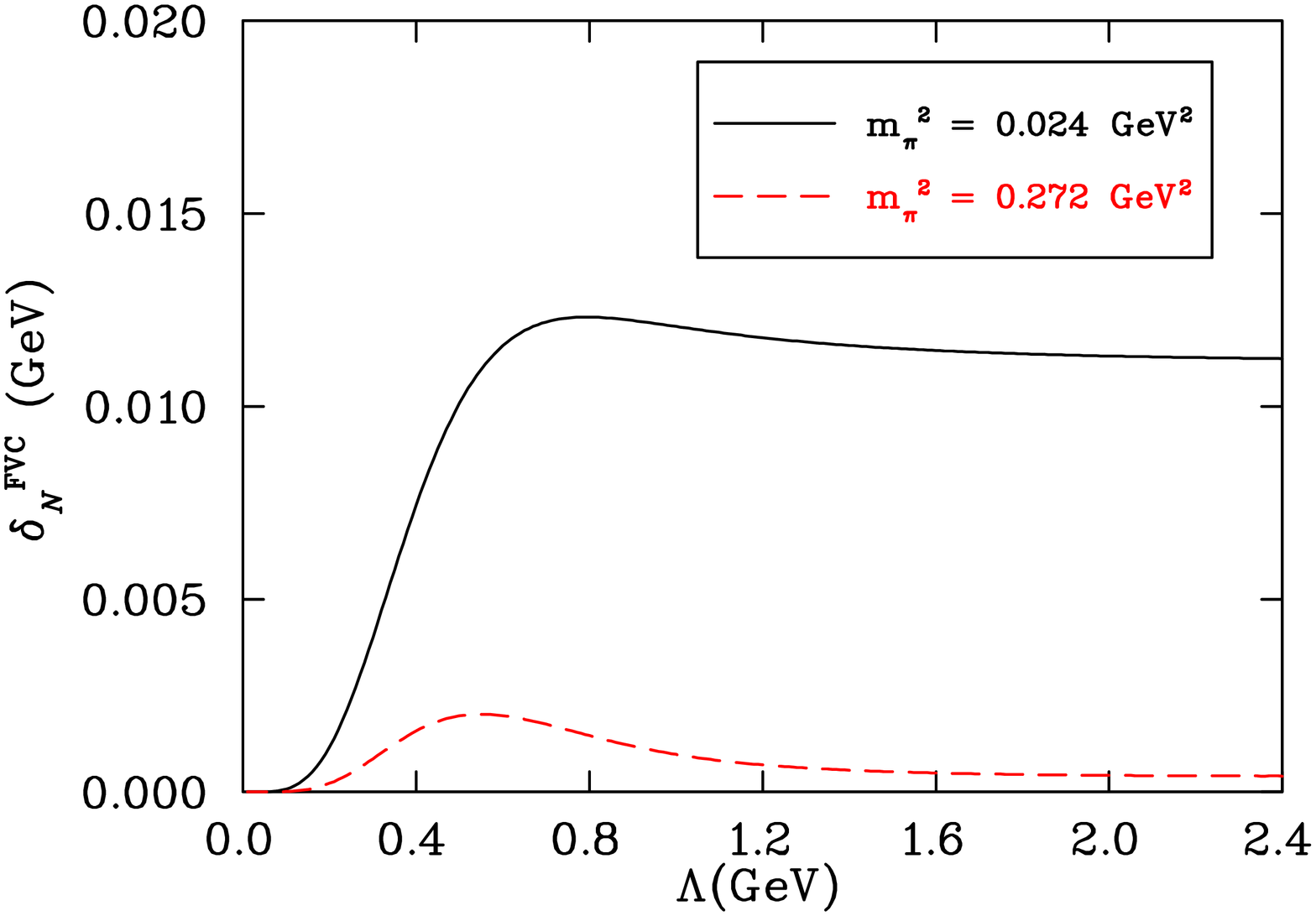}
\vspace{-12pt}
\caption{ Behaviour of the finite-volume
    corrections $\de_N^\ro{FVC}$ vs.\ $\La$ on a $2.9$ fm box using a dipole regulator. Results for two different values of $m_\pi^2$ are shown.}
\label{fig:fvcN}
\vspace{6mm}
\includegraphics[height=1.0\hsize,angle=90]{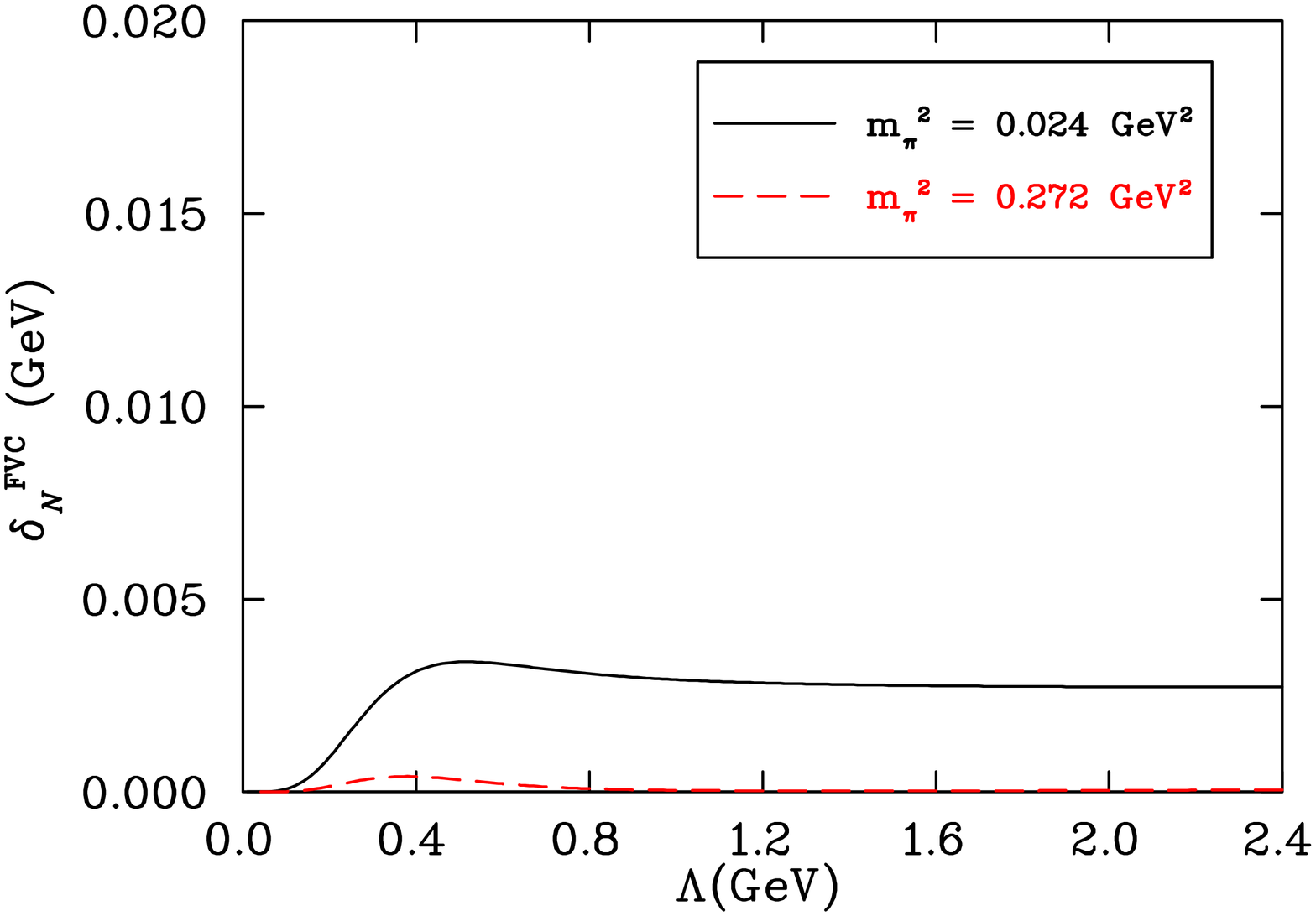}
\vspace{-12pt}
\caption{ Behaviour of finite-volume
    corrections $\de_N^\ro{FVC}$ vs.\ $\La$ on a $4.0$ fm box using a dipole regulator. Results for two different values of $m_\pi^2$ are shown.}
\label{fig:fvcN4fm}
\end{minipage}
\hspace{12mm}
\begin{minipage}[b]{0.45\linewidth}
\centering
\includegraphics[height=1.0\hsize,angle=90]{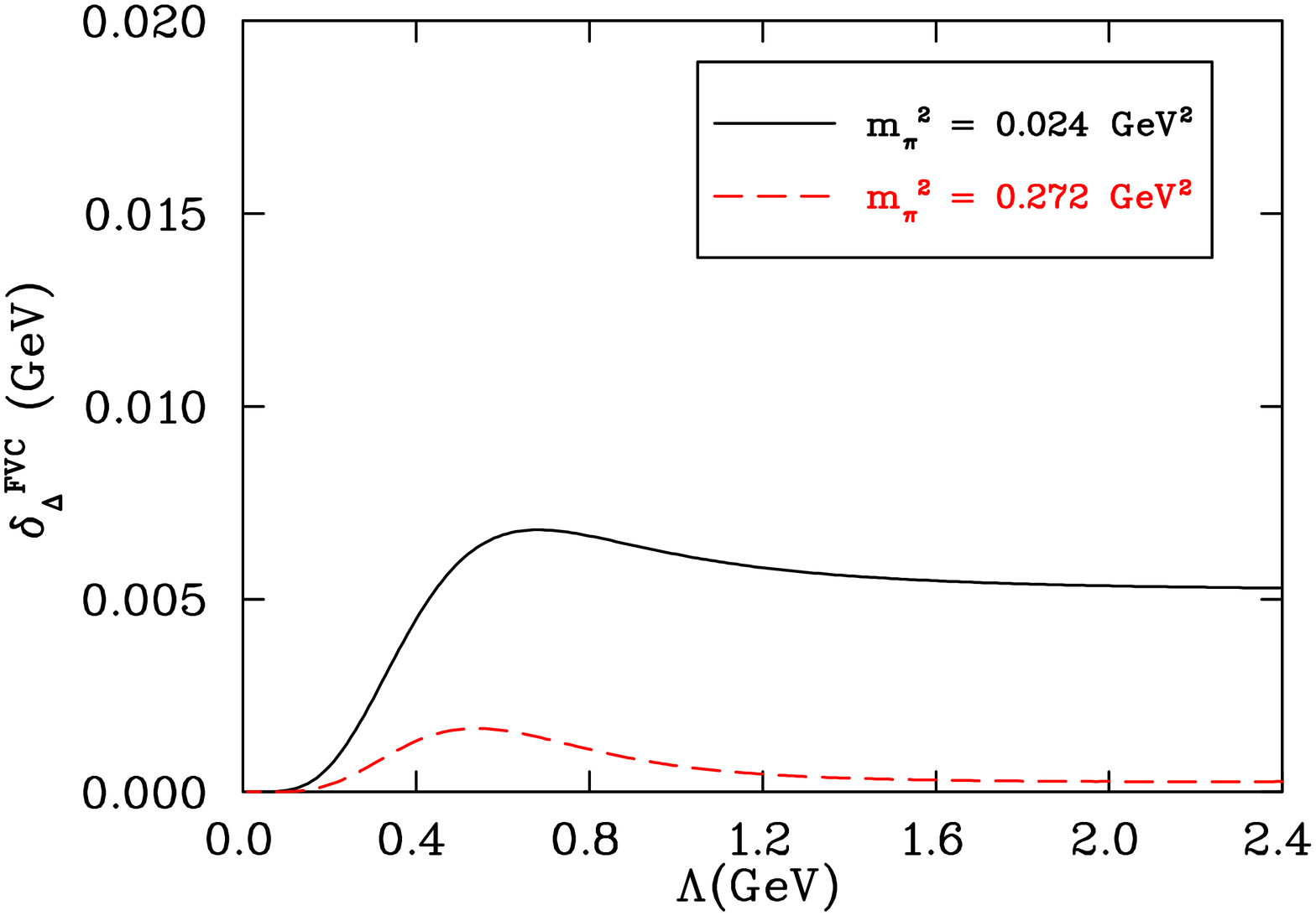}
\vspace{-12pt}
\caption{ Behaviour of finite-volume corrections
  $\de_\De^\ro{FVC}$ vs.\ $\La$ on a $2.9$ fm box using a dipole regulator. Results for two different values of $m_\pi^2$ are shown.}
\label{fig:fvcD}
\vspace{6mm}
\includegraphics[height=1.0\hsize,angle=90]{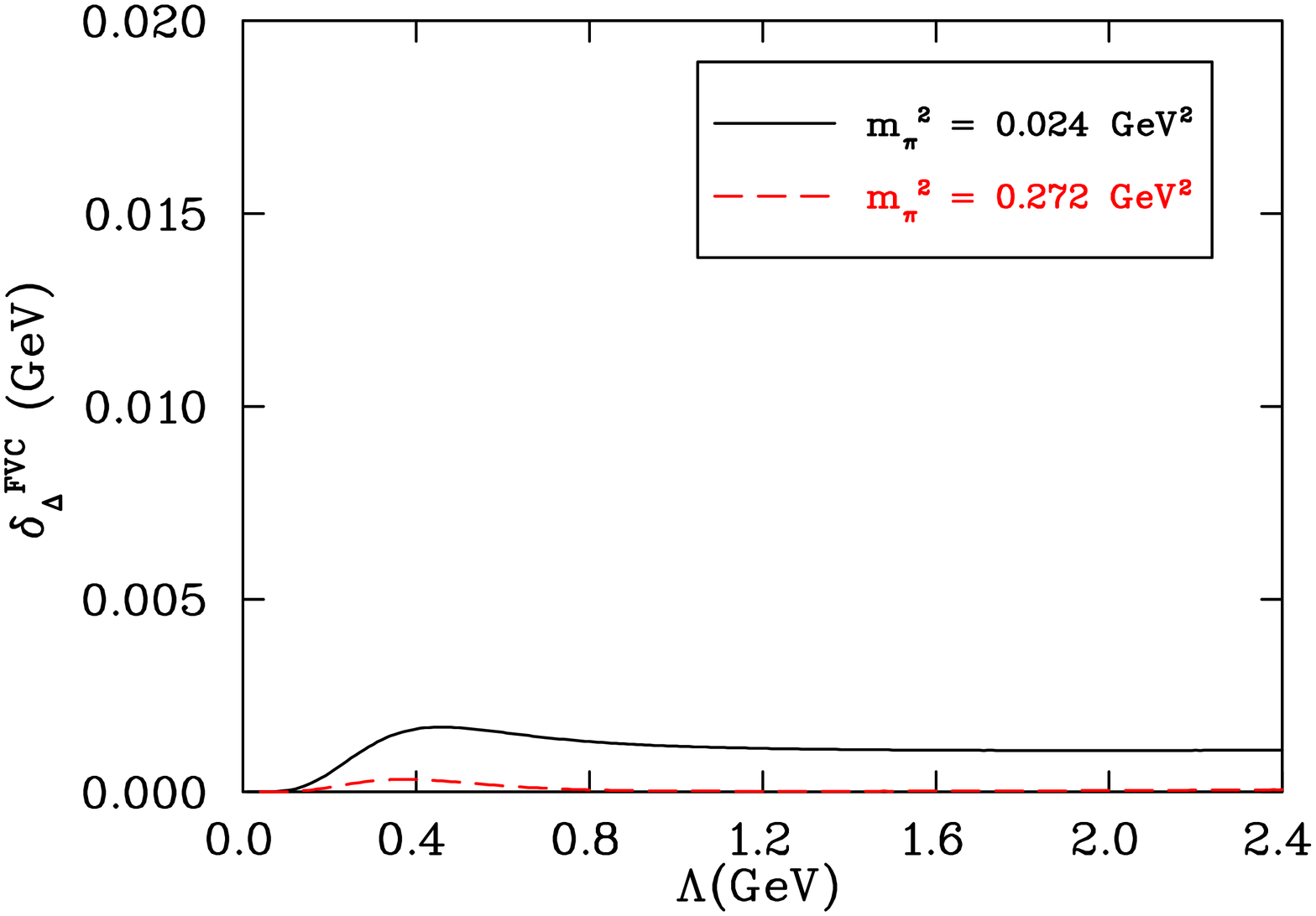}
\vspace{-12pt}
\caption{ Behaviour of finite-volume corrections
  $\de_\De^\ro{FVC}$ vs.\ $\La$ on a $4.0$ fm box using a dipole regulator. Results for two different values of $m_\pi^2$ are shown.}
\label{fig:fvcD4fm}
\end{minipage}
\end{figure}

It is also shown that the finite-volume corrections are
independent of the regularization scale $\Lambda$ in this domain. In Figures
\ref{fig:fvcN} and \ref{fig:fvcD}, the scale-dependence of
the finite-volume corrections is shown for a dipole regulator (from Equation 
(\ref{eqn:ndipole}) in Chapter \ref{chpt:chieft}) and a 
$2.9$ fm box
 (the same box size used for the PACS-CS data \cite{Aoki:2008sm}).
  It is %notable 
of note that choosing $\Lambda$
too small suppresses the very infrared physics that one is trying to
describe. Thus, caution should be exercised in choosing a 
suitable value of $\Lambda$.  %It is sensible to
% be cautious by not selecting a value of $\Lambda$ that is
%too low.
Figures \ref{fig:fvcN4fm} and \ref{fig:fvcD4fm} 
show the behaviour of the finite-volume correction 
for a $4.0$ fm box, and the corrections are much smaller, as expected. 

For large $\Lambda$, the finite-volume corrections, displayed 
in Figures \ref{fig:fvcN} through \ref{fig:fvcD4fm}, saturate to a fixed
value. %For the light pion masses, provided $\Lambda\gtrsim
%0.8$ GeV, the estimated finite-volume corrections are stable.
Provided that $\Lambda\gtrsim 0.8$ GeV, the estimated finite-volume 
corrections are stable for light pion masses.  %. 
%The 
In order to preserve the scale-independence of the finite-volume corrections, 
their asymptotic result  %of the finite-volume corrections 
will be used. This approach has been demonstrated to be
successful in previous studies \cite{AliKhan:2003cu}.  Numerically, this is
achieved by evaluating the finite-volume corrections with a parameter,
$\La' = 2.0$ GeV, $\de_i^\ro{FVC} = \de_i^\ro{FVC}(\La')$. It should be noted
that this is equivalent to the more algebraic approach outlined by Beane 
%in Ref.
\cite{Beane:2004tw}.

\section{The Intrinsic Scale: An Example by Construction}
\label{sect:intr}

\begin{figure}[tp]
\begin{center}
\includegraphics[height=0.70\hsize,angle=90]{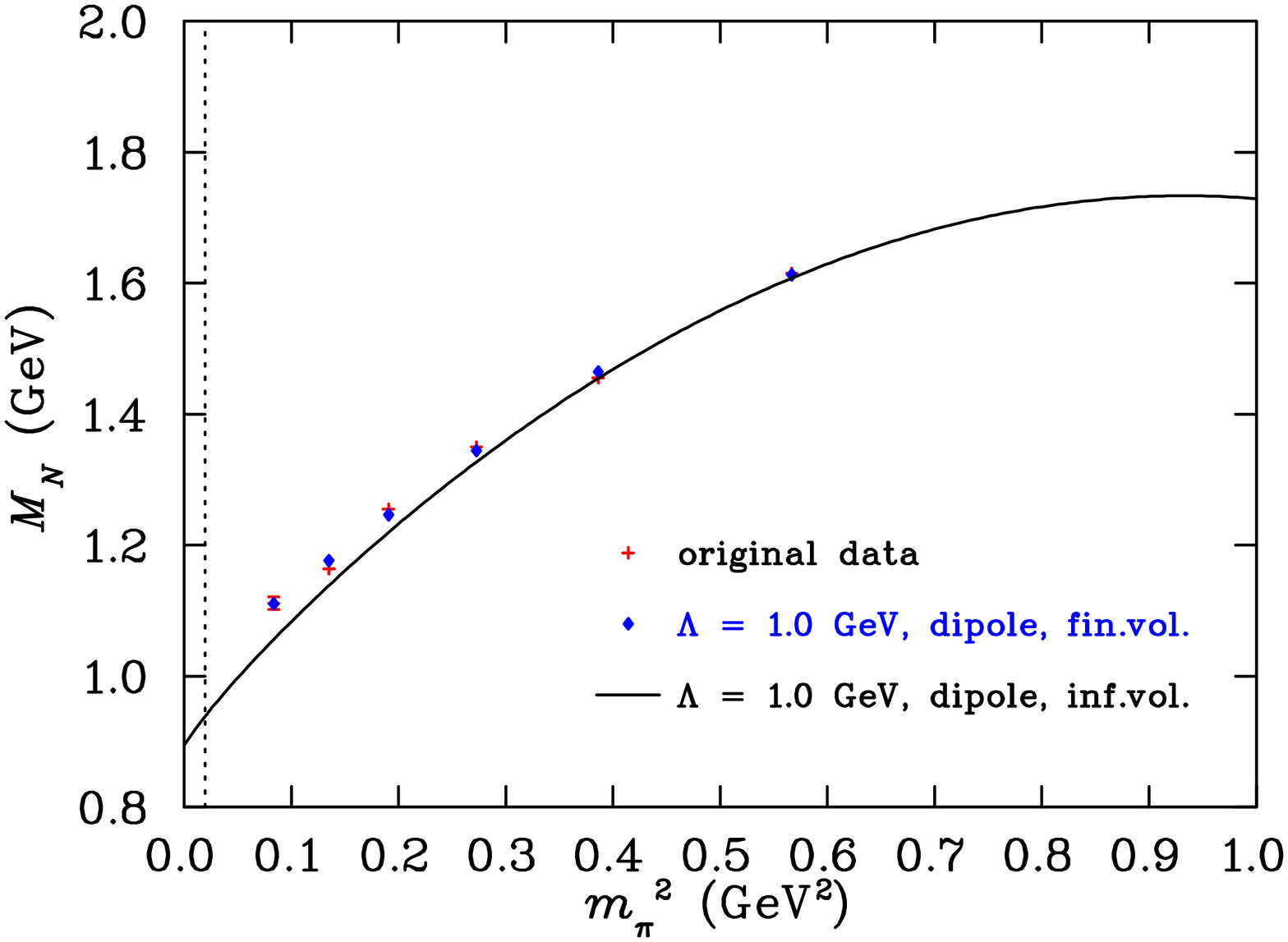}
\vspace{-12pt}
\caption{\footnotesize{ Example dipole extrapolation based on JLQCD data \cite{Ohki:2008ff}, box size: $1.9$ fm.}}
\label{fig:OhkiExt}
%\end{center}
%\end{figure}
%\vspace{5pt}
%\begin{figure}[tp]
%\begin{center}
\vspace{6mm}
\includegraphics[height=0.70\hsize,angle=90]{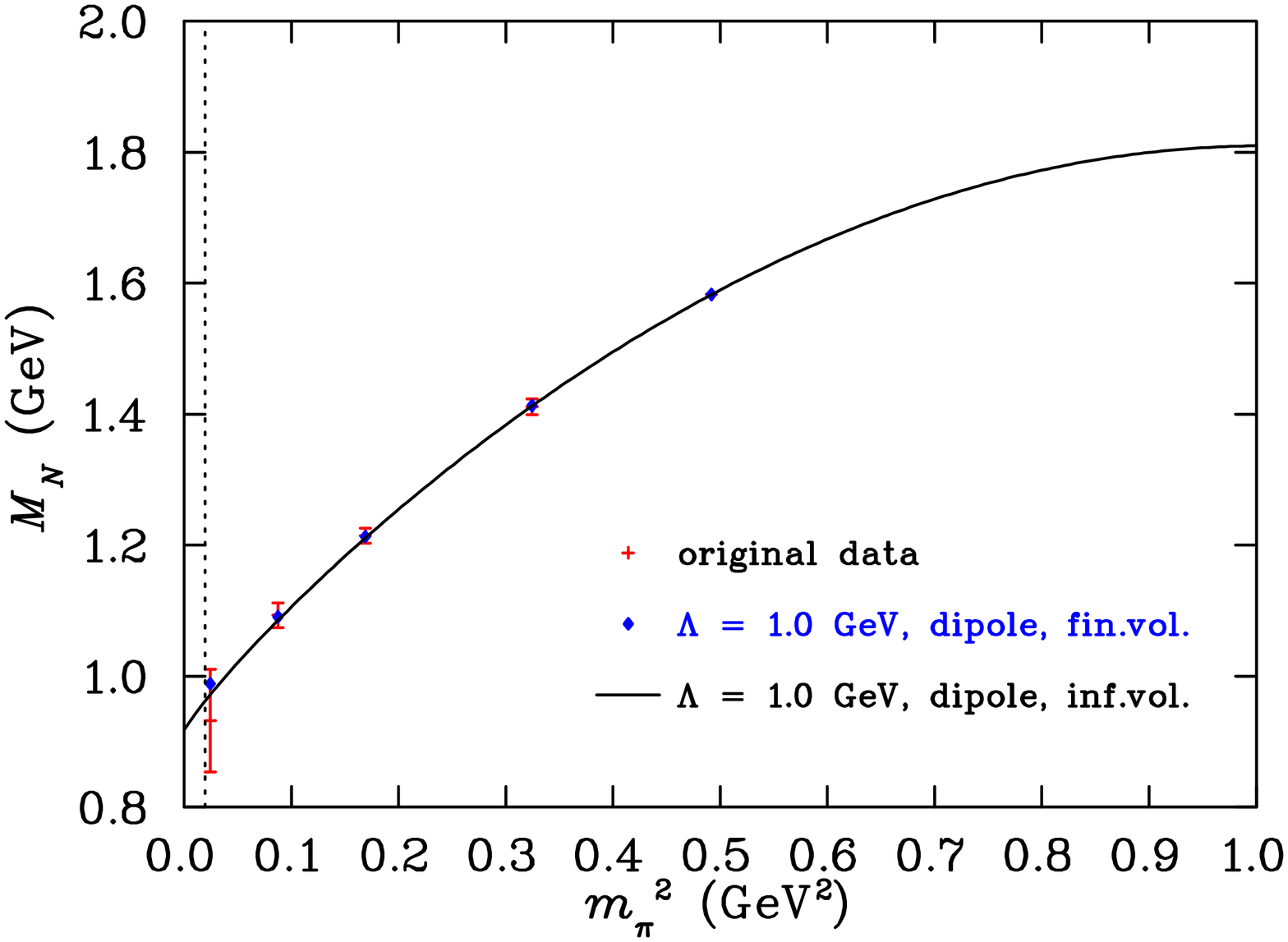}
\vspace{-12pt}
\caption{\footnotesize{ Example dipole extrapolation based on PACS-CS data \cite{Aoki:2008sm}, box size: $2.9$ fm.}}
\label{fig:AokiExt}
%\end{center}
%\end{figure}
%\vspace{5pt}
%\begin{figure}[tp]
%\begin{center}
%\end{center}
%\end{figure}
%
%\begin{figure}[tp]
%\begin{center}
\vspace{6mm}
\includegraphics[height=0.70\hsize,angle=90]{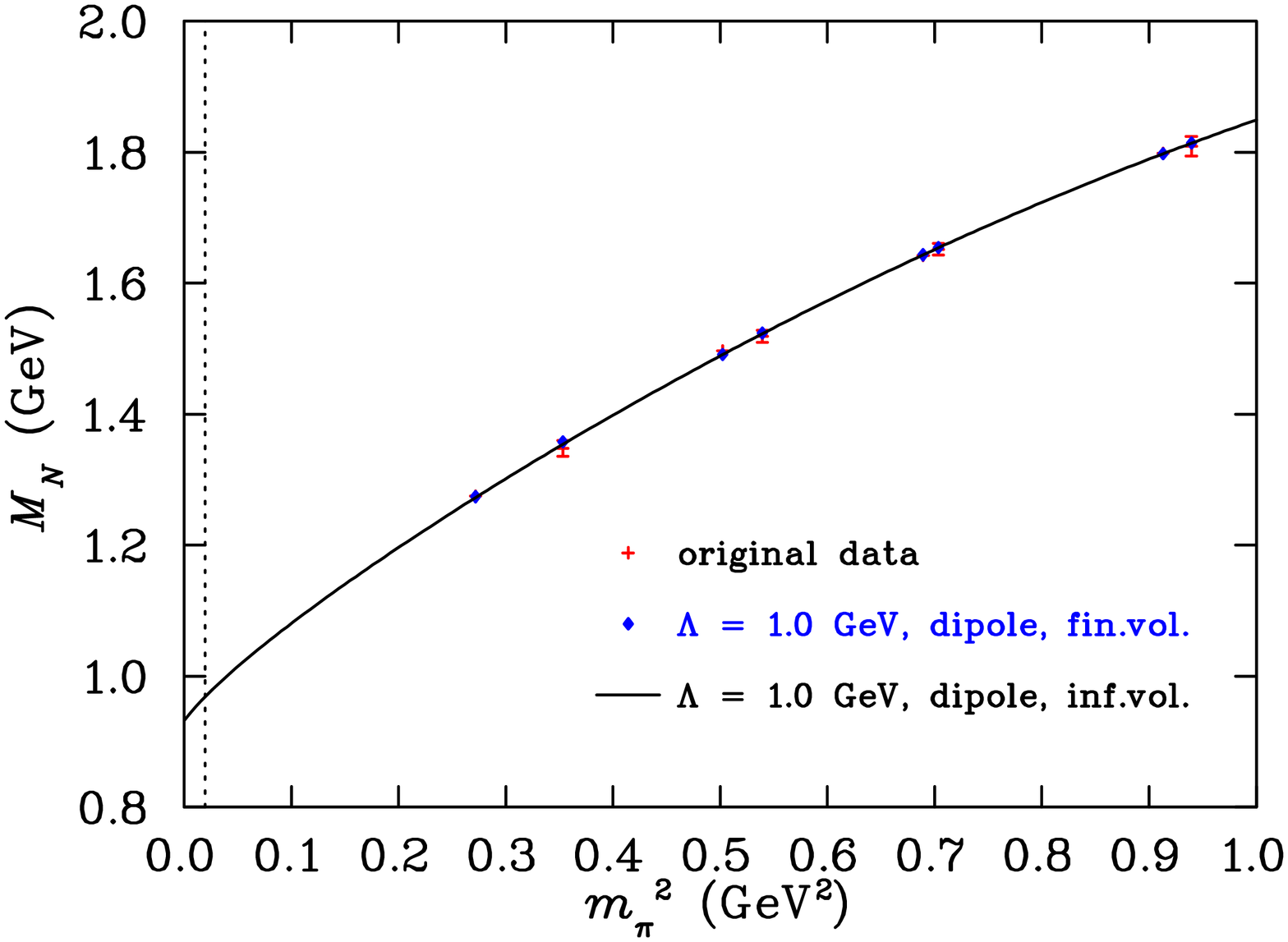}
\vspace{-12pt}
\caption{\footnotesize{ Example dipole extrapolation based on CP-PACS data \cite{AliKhan:2001tx}, lattice sizes: $2.3-2.8$ fm.}}
\label{fig:YoungExt}
\end{center}
\end{figure}

This $\chi$EFT extrapolation scheme to order 
 $\ca{O}(m_\pi^4\,\ro{log}\,m_\pi)$ 
 will be used in conjunction with 
lattice QCD data from JLQCD \cite{Ohki:2008ff}, 
PACS-CS \cite{Aoki:2008sm} and CP-PACS \cite{AliKhan:2001tx} 
to predict the nucleon mass
for any value of $m_\pi^2$. The full set of data from each of these 
collaborations is listed in Appendix \ref{chpt:appendix4}, 
Tables \ref{table:JLQCDdata} through 
\ref{table:CP-PACSdata}. 
The JLQCD data were generated using  
 overlap fermions in two-flavor QCD, but the lattice box size for each
data point is $\sim 1.9$ fm, smaller than the other two data sets.
 The PACS-CS data were generated using non-perturbatively $\ca{O}(a)$-improved
 Wilson quark action at a lattice box size of $\sim 2.9$ fm, but the data set
only contains five data points and a large statistical error in the smallest 
 $m_\pi^2$ point. The CP-PACS data were generated using 
a mean field improved clover quark 
action on lattice box sizes for each data point varying from $\sim 2.2$ fm
 to $\sim 2.8$ fm. 
The lattice data used in this analysis will be used to extrapolate 
$M_N$ to the physical point by 
taking into account the relevant curvature from the loop integrals
in Equations (\ref{eqn:NNexpn}), (\ref{eqn:NDeexpn}) and (\ref{eqn:tadexpnSi}).
As an example, a regularization scale  of $\La = 1.0$ GeV was chosen 
 for Figures \ref{fig:OhkiExt} through \ref{fig:YoungExt}, where
the finite-volume corrected effective field theory 
appears concordant with previous QCDSF-UKQCD
 results \cite{AliKhan:2003cu}.
An extrapolation or interpolation is achieved by subtracting the finite-volume 
loop integral
contributions %defined in 
%Equations 
%(\ref{eqn:NN}), (\ref{eqn:NDe}) and (\ref{eqn:tad}) 
from each data point
and then fitting the result to obtain the coefficients $c_0$, $c_2$ and 
$a_4^\La$ using Equation (\ref{eqn:mNfit}). The finite- or infinite-volume 
loop integrals are then
 added back at any desired value of $m_\pi^2$.

\begin{figure}[tp]
\begin{center}
\includegraphics[height=0.70\hsize,angle=90]{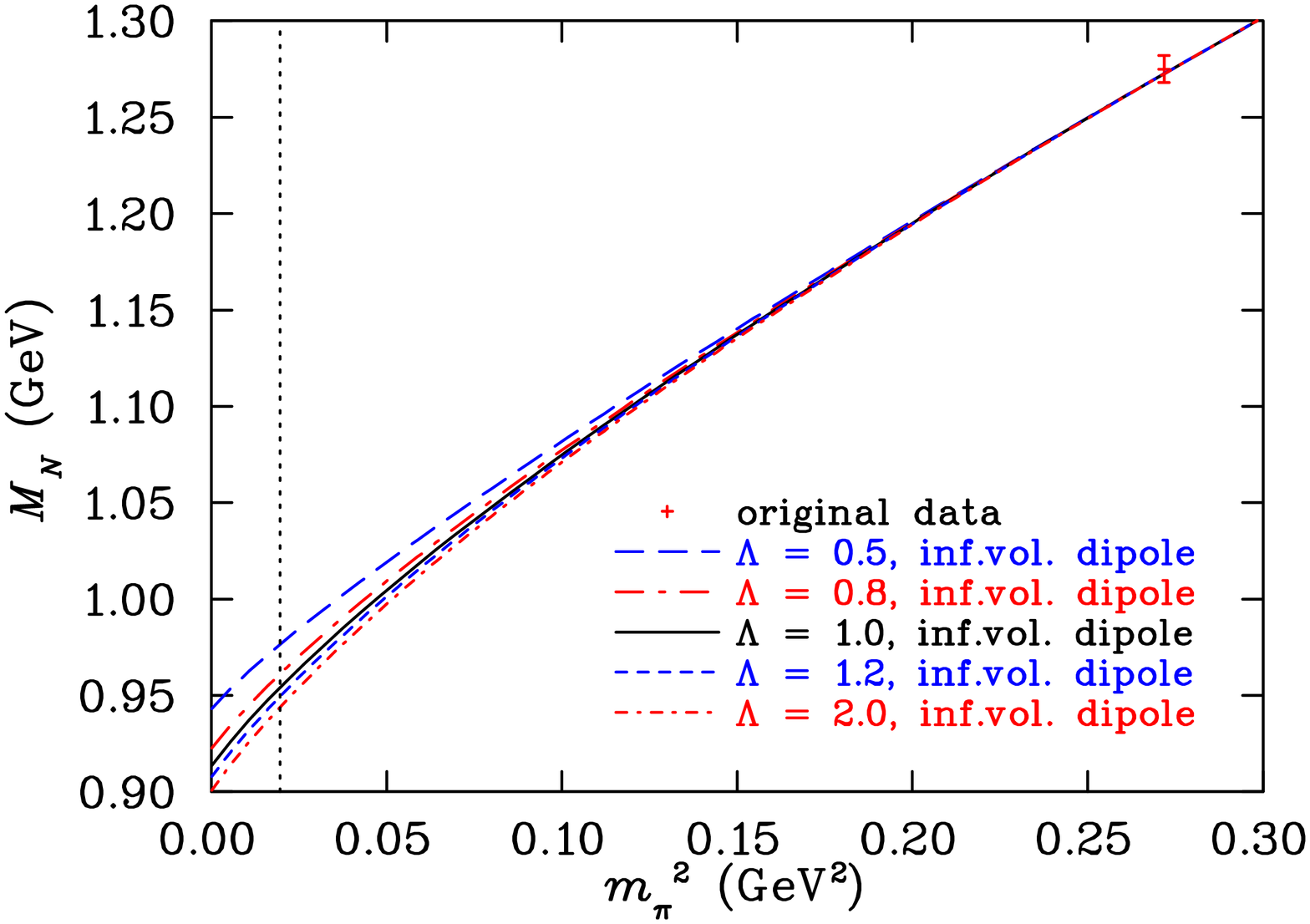}
\vspace{-12pt}
\caption{\footnotesize{ Close zoom of the regulator-dependence for dipole extrapolation based on CP-PACS data. Only the data point corresponding to the smallest $m_\pi^2$ value is shown at this scale.}}
\label{fig:YoungExtmulti}
\end{center}
\end{figure}

If the regularization scale is altered from the choice $\La = 1.0$ GeV,
 the extrapolation curve also changes. 
This signifies a scheme-dependence in the result due to using lattice QCD
data beyond the PCR.
To demonstrate this, % further, 
consider the infinite-volume extrapolation of the
CP-PACS data, as shown in Figure \ref{fig:YoungExtmulti}. 
Figure \ref{fig:YoungExtmulti} also shows that the curves overlap
exactly when $m_\pi^2$ is large, where the lattice data reside, and 
 %They 
 they diverge as the chiral regime is approached. %This section addresses
%this problem in detail.
%

\vspace{5pt}

Consider an insightful scenario, whereby a set of 
%\emph{
ideal %} 
 `pseudodata' with known low-energy coefficients is produced, 
using the formula from Equation (\ref{eqn:mNfitfin}). 
A particular regularization scale is selected and a dense and
precise pseudodata set is generated, 
which smoothly connects with %state-of-the-art 
 the lattice
simulation results. In this case, the pseudodata are converted 
to infinite-volume results in order to ensure that the following analysis 
is not simply a consequence of finite-volume effects. 
If all the data considered lie
within the PCR then the choice of regularization scale is irrelevant,
and the finite-range regularized 
chiral expansion is mathematically equivalent to scale-invariant 
renormalization schemes, including DR.  %However, the purpose here is to 
  This scenario will form the basis of the investigation of the PCR, and
ultimately, will lead to determining 
the existence of an intrinsic scale hidden within
the lattice QCD simulation results. 

The pseudodata are produced by performing an extrapolation
such as shown in Figures \ref{fig:OhkiExt} through \ref{fig:YoungExt}.
 The difference is that $100$ infinite-volume 
extrapolation points are produced close to the chiral regime.
 The exercise is to treat these pseudodata as if they were lattice QCD data.
 Clearly, a regularization scheme must be chosen 
 %to %produce 
 in generating the pseudodata. In this case,
a dipole regulator was chosen
 and pseudodata were created at $\La_\ro{c} = 1.0$ GeV.

\begin{figure}[tp]
\begin{center}
\includegraphics[height=0.70\hsize,angle=90]{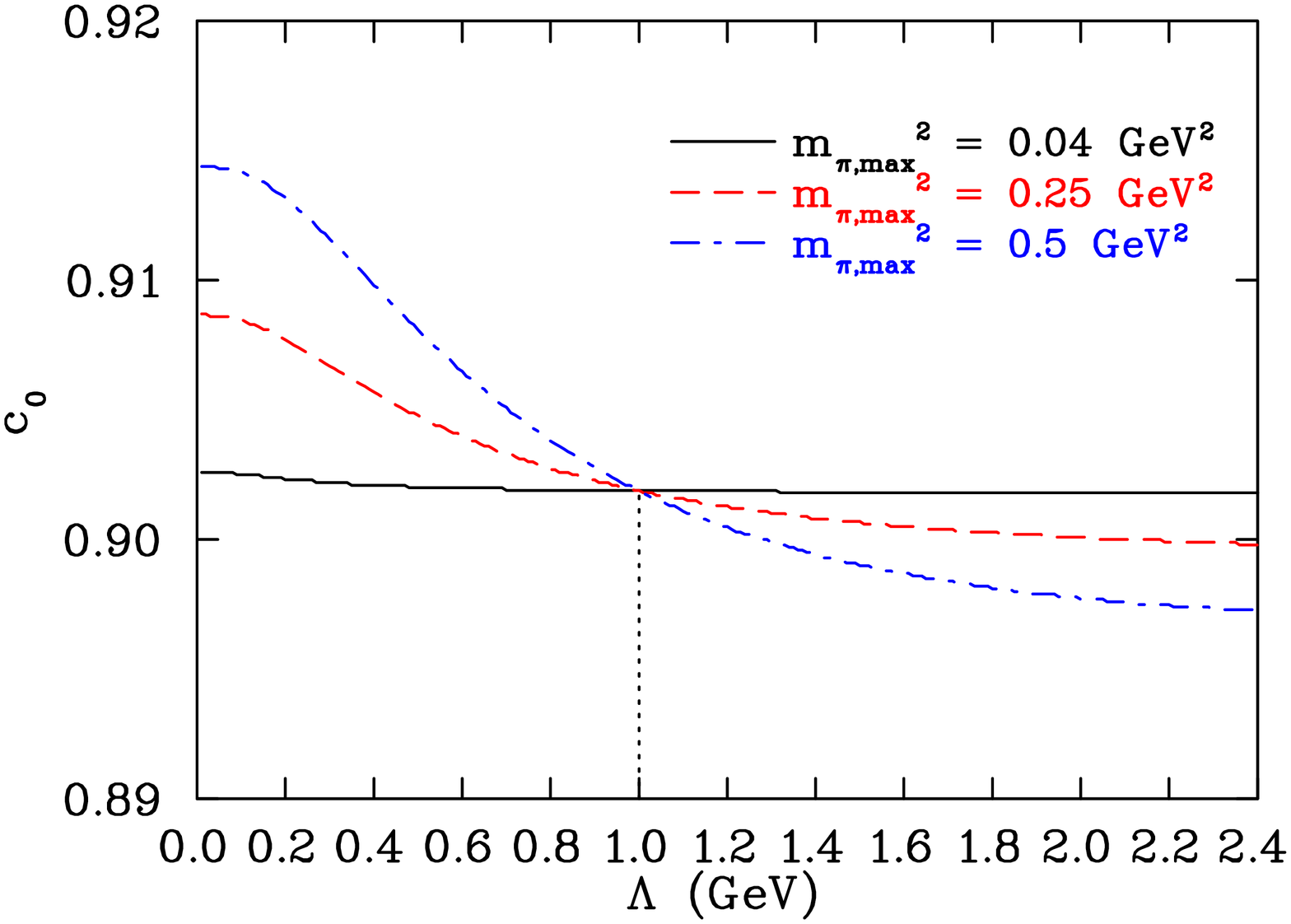}
\vspace{-12pt}
\caption{\footnotesize{ Behaviour of $c_0$ vs.\ regularization scale $\La$, based on infinite-volume pseudodata created with a dipole regulator at $\La_\ro{c} = 1.0$ GeV (based on lightest four
data points from PACS-CS). Each curve uses pseudodata with
 a different upper value of pion mass $m_{\pi,\ro{max}}^2$.}}
\label{fig:pdatac0}
%\end{center}
%\end{figure}
%\vspace{5pt}
%\begin{figure}[tp]
%\begin{center}
\vspace{12mm}
\includegraphics[height=0.70\hsize,angle=90]{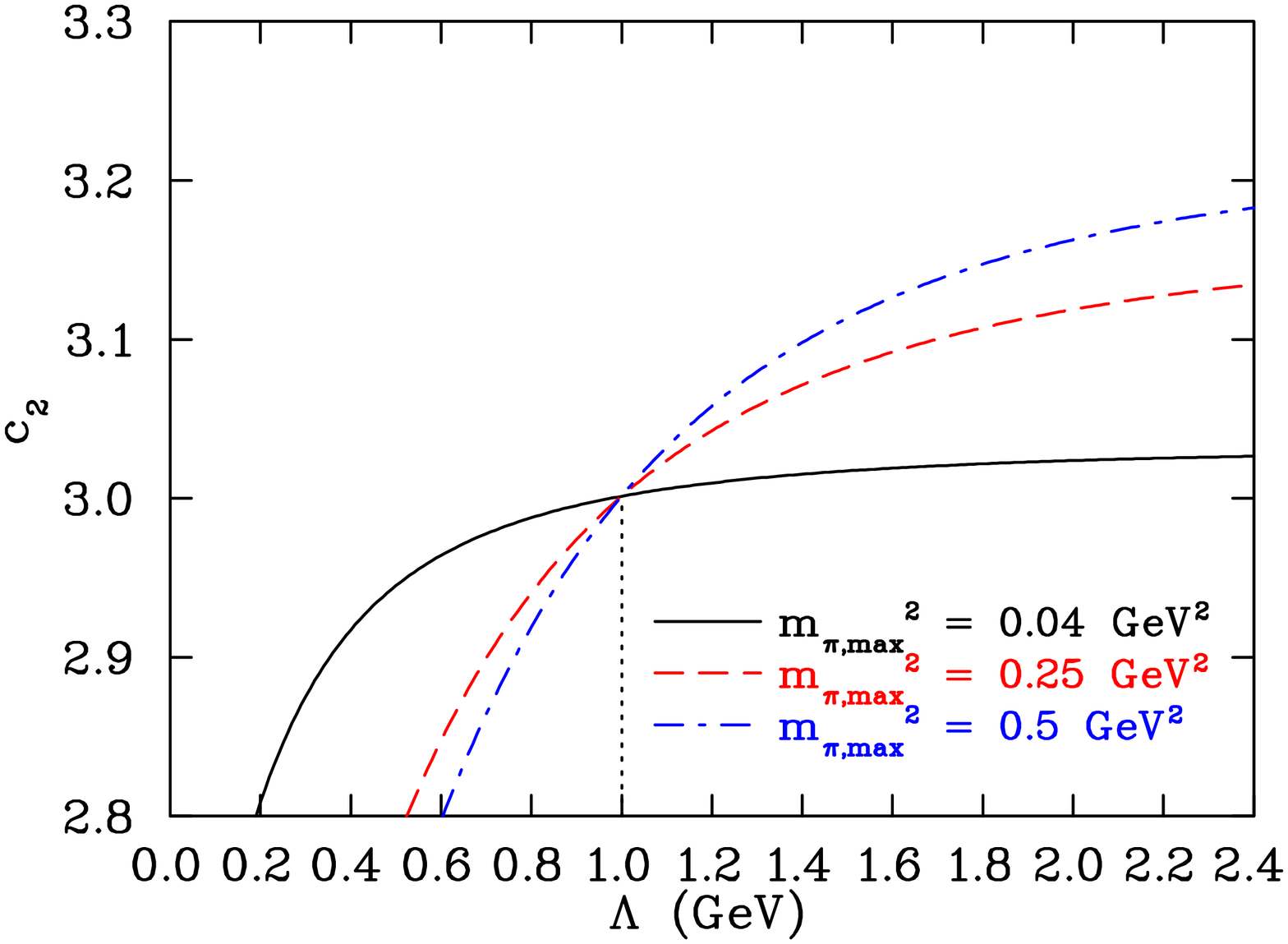}
\vspace{-12pt}
\caption{\footnotesize{ Behaviour of $c_2$ vs.\ $\La$, based on infinite-volume
pseudodata created with a dipole regulator at $\La_\ro{c} = 1.0$ GeV (based on lightest four
data points from PACS-CS). %Each curve uses pseudodata with
% a different upper value of pion mass $m_{\pi,\ro{max}}^2$.
}}
\label{fig:pdatac2}
\end{center}
\end{figure}

The regularization-dependence of the extrapolation is characterized by the 
scale-dependence of the coefficients $c_i$. These coefficients are obtained 
from fitting the pseudodata. 
Consider how $c_0$ and $c_2$ behave when analyzed with a variety of 
regularization scales in Figures \ref{fig:pdatac0} and \ref{fig:pdatac2}.
%By choosing to use pseudodata produced at infinite volume, 
By using infinite-volume pseudodata, one eliminates 
 the concern that the variation in $c_i$ with respect to $\La$ is merely 
a %behaviour of the chiral coefficients across a range of
 %regularization scales 
 %and pion masses is a 
finite-volume artefact. 
%The equivalent plots to 
%of 
%Figures \ref{fig:pdatac0} and \ref{fig:pdatac2}, but for finite-volume
% pseudodata, exhibit the same features.
%

Three pseudodata sets are compared, each with different upper bounds
on the range of $m_\pi^2$ considered in the fit. 
 An increasing regulator-dependence in $c_0$ and $c_2$ is seen as the 
data extend outside the PCR.
In Figures \ref{fig:pdatac0} and \ref{fig:pdatac2}, the behaviour  
of the fit parameters $c_0$ and $c_2$, respectively, 
are shown as functions of the 
regularization scale $\La$ for different values of $m_{\pi,\ro{max}}^2$. 
A steep line indicates 
a strong scheme-dependence in the result, 
and this occurs for data samples extending far outside the PCR.
Scheme-independence will appear as a horizontal line, as is apparent 
for $m_{\pi,\ro{max}}^2 < 0.04$ GeV$^2$,
in Figures \ref{fig:pdatac0} and \ref{fig:pdatac2}. This indicates that 
the pseudodata lie within the PCR. 

Note that in both figures  
%for both Figures \ref{fig:pdatac0} and \ref{fig:pdatac2},
 all three curves (corresponding to different values of $m_{\pi,\ro{max}}^2$)
 arrive at stable values
for $c_0$ and $c_2$ on the right-hand side of the plot, corresponding to 
 large $\La$.
To read off the values of $c_0$ and $c_2$ for large $\La$ is 
 tempting, but this does not yield the correct values of $c_0$ and $c_2$, 
which are known by construction. 
%It is known by construction that 
The correct values of $c_0$ and $c_2$
are recovered at $\La = 1.0$ GeV. %, because at that value the pseudodata were 
%created.

The analysis of the pseudodata in 
Figures \ref{fig:pdatac0} and \ref{fig:pdatac2} 
shows that even as the value of $m_{\pi,\ro{max}}^2$ is changed, 
the correct value of $c_0$ is recovered at exactly $\La = \La_\ro{c}$, where
the curves intersect.
%All that has been done in performing the fit
% is to subtract and then add back exactly the same integrals!
 The same value of $\La$ for the intersection point is obtained by analyzing 
$c_2$.
This suggests that when considering lattice QCD results extending outside the
PCR, there may be an optimal finite-range cutoff.
Physically, such a cutoff would be associated with an intrinsic 
scale reflecting the finite
size of the source of the pion dressings.
Mathematically, this optimal cutoff is reflected by an independence of 
 the fit parameters on $m_{\pi,\ro{max}}^2$.

%  New part added added JMMH

\begin{figure}[tp]
\begin{center}
\includegraphics[height=0.70\hsize,angle=90]{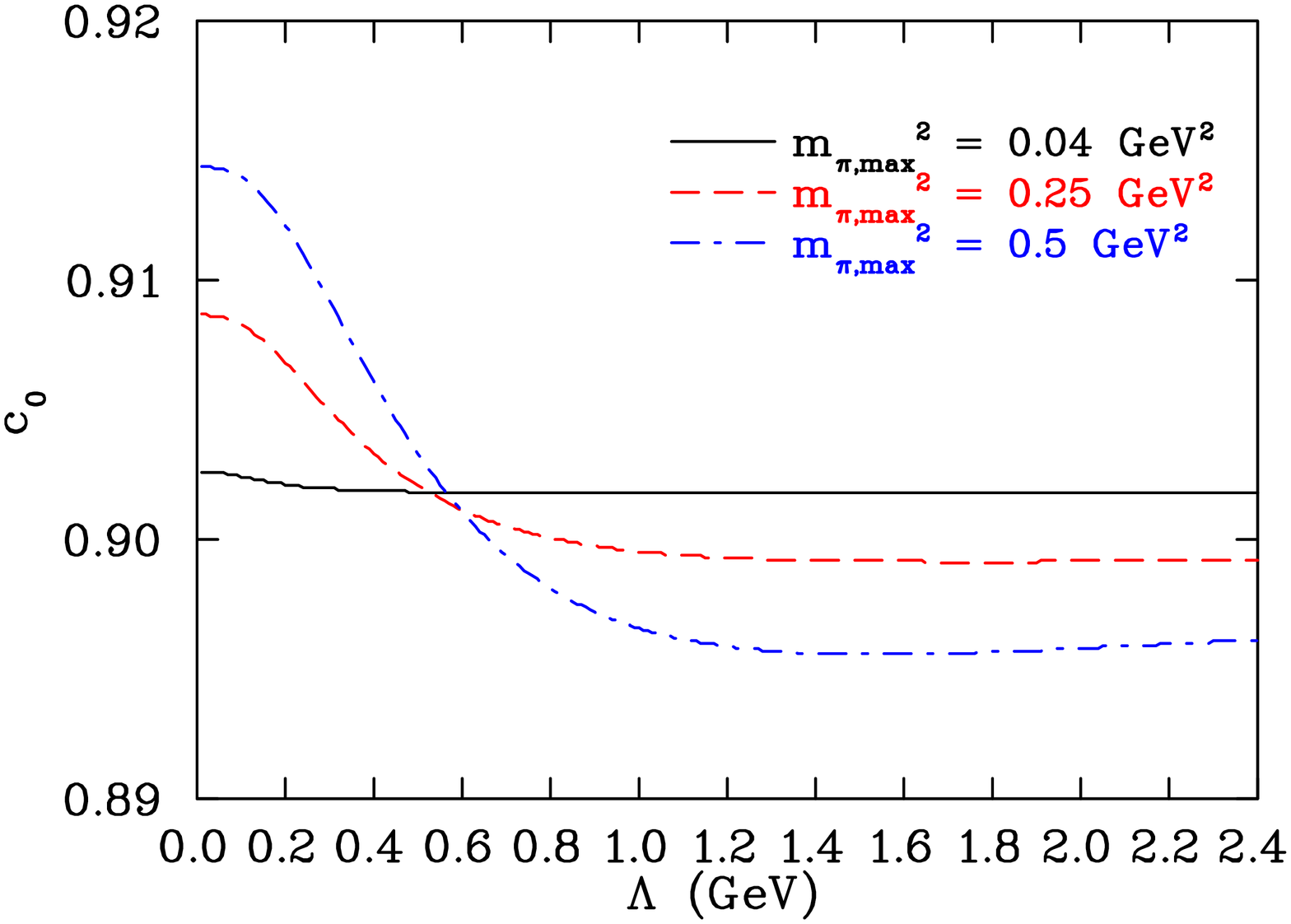}
\vspace{-12pt}
\caption{\footnotesize{ Behaviour of $c_0$ vs.\ $\La$, based on infinite-volume pseudodata created with a dipole regulator at $\La_\ro{c} = 1.0$ GeV but subsequently analyzed using a triple-dipole regulator. }}
\label{fig:pdatac0diptrip}
%\end{center}
%\end{figure}
%\vspace{5pt}
%\begin{figure}[tp]
%\begin{center}
\vspace{22mm}
\includegraphics[height=0.70\hsize,angle=90]{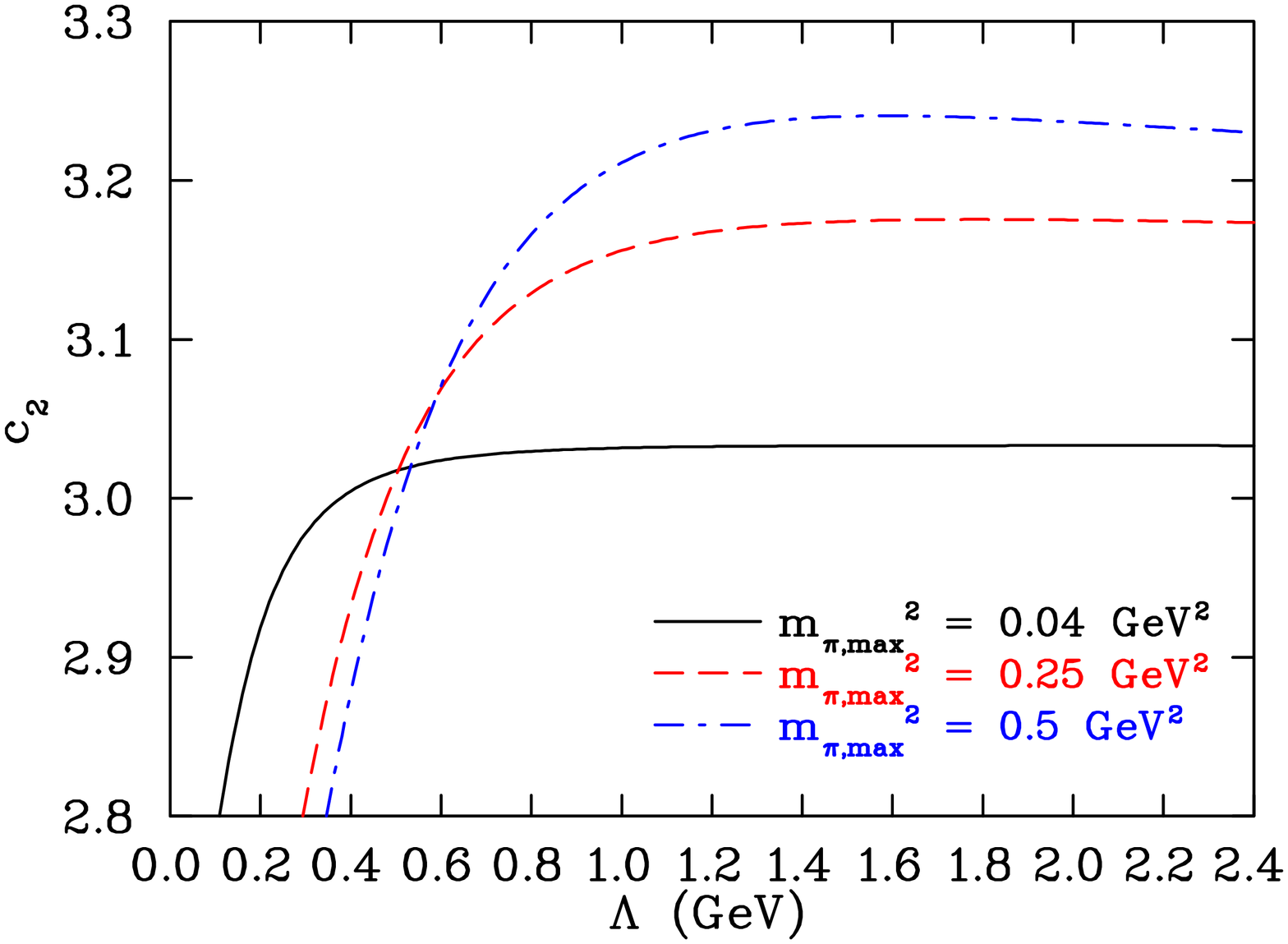}
\vspace{-12pt}
\caption{\footnotesize{ Behaviour of $c_2$ vs.\ $\La$, based on infinite-volume pseudodata created with a dipole regulator at $\La_\ro{c} = 1.0$ GeV but subsequently analyzed using a triple-dipole regulator.}}
\label{fig:pdatac2diptrip}
\end{center}
\end{figure}

%To illustrate the non-triviality of this scale of curve-intersection, 
%the pseudodata were analyzed 
By analyzing the pseudodata with a different regulator, for example,  
a triple-dipole regulator, 
Figures \ref{fig:pdatac0diptrip} and \ref{fig:pdatac2diptrip} 
show that the scale of the intersection is no longer a clear point, but a 
cluster centred about $0.5$ to $0.6$ GeV. 
%This result also supports the idea that
%there is a `best choice' of renormalization scheme.
%
The triple-dipole will of course predict
a different optimal scale, since the shape of the regulator is different from
that of the dipole used to create the pseudodata. The
 essential point of this exercise is that clustering of curve 
intersections identifies a preferred 
renormalization scale that allows one to recover the correct low-energy 
coefficients.
In this case, the crossing of the dash and dot-dash curves (from fitting) 
clearly identifies $\La^\ro{scale}_\ro{trip} = 0.6$ GeV 
as a preferred regularization scale,
 which reflects
the intrinsic scale used to create the data. Table \ref{table:pdata}
 compares the values
for $c_0$ and $c_2$ recovered in this analysis for two different regularization 
scales: 
the preferred value $\La^\ro{scale}_\ro{trip} = 0.6$ GeV, and a 
large value $\La_\ro{trip} = 2.4$ GeV reflecting
the asymptotic result recovered from DR. The input values of $c_0$ and $c_2$
used to create the pseudodata are also indicated.
\begin{table*}[tpb]
  \newcommand\T{\rule{0pt}{2.8ex}}
  \newcommand\B{\rule[-1.4ex]{0pt}{0pt}}
  \begin{center}
    \begin{tabular}{llllll}
      \hline
      \hline
      \T\B param. \!\!\!\! & input \,\,  & $\La^\ro{scale}_\ro{trip} = 0.6$  $\,\,$& $\La^\ro{scale}_\ro{trip} = 0.6$ $\,\,$& $\La_\ro{trip}=2.4$ $\,\,$& $\La_\ro{trip}=2.4$  \\
\!\!\!\! & \,\,& $m_{\pi,\ro{max}}^2 = 0.25$ $\,\,$& $m_{\pi,\ro{max}}^2 = 0.5$ $\,\,$& $m_{\pi,\ro{max}}^2 = 0.25$  $\,\,$&  $m_{\pi,\ro{max}}^2 = 0.5$  \\
      \hline
      $c_0$  \!\!\!\! &\T $0.902$ & $0.901$ & $0.902$ & $0.899$ & $0.896$ \\
      $c_2$  \!\!\!\! &\T $3.00$ & $3.07$ & $3.07$ & $3.17$ & $3.23$ \\
      \hline
    \end{tabular}
  \end{center}
\vspace{-6pt}
  \caption{\footnotesize{A comparison of the parameters $c_0$ (GeV) and $c_2$ (GeV$^{-1}$) at their input value (pseudodata created with a dipole at $\La_\ro{c} = 1.0$ GeV) with the values when analysed with a triple-dipole regulator. Different values of $\La_\ro{trip}$ (GeV) and $m_{\pi,\ro{max}}^2$ (GeV$^2$) are chosen to demonstrate the scheme-dependence of $c_0$ and $c_2$ for data extending outside the PCR. Note: the values of $c_0$ and $c_2$ are calculated from an ideal model and thus they are exact; there are no statistical uncertainties.}}
  \label{table:pdata}
\end{table*}

\subsection{Lower Bounds for the Regularization Scale}
\label{sec:lowerbound}

Figures \ref{fig:pdatac2} and \ref{fig:pdatac2diptrip} clearly indicate that 
the finite-range renormalization scheme
 breaks down if the FRR scale is too small. 
This is because $\La$ must be large enough to include
the chiral physics being studied. 
 The exact value of a sensible lower bound in the 
 FRR scale will depend on the functional form chosen
as the regulator. %This is estimated for three dipole-like regulators
% in Section (\ref{sect:scales}).
%

Figure \ref{fig:pdatac2} shows that 
 the renormalization for $c_2$ breaks down for small 
values of  $\La$. FRR breaks down for a value of $\La_\ro{dip}$ 
much 
below $0.6$ GeV, simply 
because the coefficients $b_i^\La$ 
of the loop integral expansion
in Equations (\ref{eqn:NNexpn}), (\ref{eqn:NDeexpn}) and (\ref{eqn:tadexpnSi})
 are proportional to $\La^{(3-i)}$. For higher-order terms with large $i$, 
the coefficients  will become large when $\La$ is small. 
In theory, these very large terms add up to zero, and so the limit
 $\La\rightarrow 0$ amounts to neglecting the infrared physics of the hadron. 
In practice, the finite curvature and higher-order terms of the residual series 
are not large enough to cancel the small-$\La$ behaviour of the $b_i^\La$ 
coefficients, which dominate.  
This adversely affects 
the convergence properties of the chiral expansion. On the other hand, 
one obtains a residual
series expansion with good convergence properties when $\La$ reflects
the intrinsic scale of the source of the pion dressings of the hadron 
in question. 

 The pseudodata analysis provides a good indication of a lower bound
 for $\La$ using a dipole 
regulator: $\La_{\ro{dip}} \gtrsim 0.6$ GeV. Similarly, 
Figure \ref{fig:pdatac2diptrip} suggests a lower bound for the triple-dipole
 regulator: $\La_{\ro{trip}} \gtrsim 0.3$ GeV. The same analysis can be 
repeated for the double-dipole regulator 
to obtain $\La_{\ro{doub}} \gtrsim 0.4$ GeV.

One can also estimate %constrain 
the lowest reasonable value of $\La$ 
%that $\La$ should take 
by considering arguments from phenomenology.
Based on the physical values of the sigma
commutator and the nucleon mass, a pion mass of
 $m_\pi \approx 0.5$ GeV is a suitable upper bound 
%bounds 
 for the radius of convergence 
\cite{Borasoy:2002jv,Young:2009ub,Young:2009zb}. 
This follows from the 
estimate of the two-flavour pion-nucleon sigma term due to Gasser 
\cite{Gasser:1990ce}. Using the Gell-Mann$-$Oakes$-$Renner Relation
$m_q\propto m_\pi^2$: 
\eqb
\Si_{\pi N} = m_\pi^2\f{\cd M_N}{\cd m_\pi^2} 
= c_2 m_\pi^2 + \chi_Nm_\pi^3 + c_4m_\pi^4 + \ca{O}(m_\pi^5) 
\approx 45 \,\,\mbox{MeV} \,.
\eqe
%
%Since the $c_2$ term of the chiral expansion of the nucleon dominates the 
%sigma term,
For good convergence, it is expected that the sigma term is dominated by the 
leading-order $c_2$ term.  
The second and third terms in the expansion are as large as the leading-order 
$c_2$ term  for $m_\pi \approx 0.5$ GeV.
Therefore, 
%Thus, 
in order to maintain good convergence of the chiral expansion 
whilst ensuring the inclusion of important contributions
to the chiral physics, one should choose a scale 
$\La_{\ro{sharp}} \sim 0.5$ GeV for a sharp cutoff (step function) 
regulator.  
 To compare this estimate 
for the sharp cutoff to that of dipole-like regulators, 
one can calculate the regularization scale required
such that $u^2_{n}(k\,;\La) = 1/2$ when the momentum takes the 
energy scale of $\La_\ro{sharp}$.
This results in a rough estimate for a sensible value 
for the dipole, double-dipole and triple-dipole regulators.
These values are $\La_{\ro{dip}} \sim 1.1$ GeV, 
$\La_{\ro{doub}} \sim 0.76$ GeV and 
$\La_{\ro{trip}} \sim 0.66$ GeV, respectively.
%In any event, a 

In the forthcoming chapter, 
a range of regularization scales will be considered, and
the intersections of the curves for the 
low-energy coefficients will be used 
%in order 
to construct fits that include data sets that extend outside the PCR.
This is done in order to identify the presence of an intrinsic scale
for the pion source and an associated preferred regularization scale.

%To find when u = 1/2: plug (0.5^n/(Sqrt(2)-1))^(1/n) into Calc. n=2(dip),4(doub),6(trip)
%

%To find when u^2 = 1/2, take the 2nth root of   (-2k^2n +- sqrt(4k^4n-4(1-sqrt(2))k^4n))/(2(1-sqrt(2)))
%These new lower bounds are even larger at mu_dip = 1.15, mu_doub = 0.76 and
%mu_trip = 0.66

%% file: nucleonmass.tex
\chapter{Results for the Mass of the Nucleon}
\label{chpt:nucleonmass}

\textit{``The datum is a classical property concerning only the instrument; it is the expression of a fact. The result concerns a property of the quantum world. The datum is an essential intermediary for reaching a result.''}
(Omn\`{e}s, R. 2002. \textit{Quantum Philosophy: Understanding and Interpreting Contemporary Science} p.209) \cite{Omnes}

This quotation, and those introduced in Chapters \ref{chpt:mesonmass} to 
\ref{chpt:conclusion}, contain an argument that links data, results, 
theory and experience.

%Introduce
%Say: We consider 2 flavour here (for the moment)
%3 flavour later? INTRODUCE.
%ADD MORE SECTIONS... cf. 1st article.

%Chiral perturbation theory ($\chi$PT) provides a formal approach to
%counting the powers of low-energy momenta and quark masses such that
%an ordered expansion in powers of the $m_\pi^2$
%is constructed.  $\chi$PT indicates that, in general, the most singular 
%non-analytic contributions to hadron properties lie in the one-loop 
%`meson cloud' of the hadron.  For example, the leading non-analytic
%behavior of the nucleon mass is proportional to $m_\pi^3$.
%More generally, baryon masses can be written as an ordered expansion
%in quark mass or $m_\pi^2$. 

%example extrapolations
The aim of this chapter is to apply an analysis that allows a reliable 
extrapolation of the nucleon mass to the physical point by obtaining 
an optimal regularization scale, using lattice quantum chromodynamics 
(lattice QCD) 
simulation results. 
The identification of 
an optimal regularization scale, 
along with its associated systematic uncertainty, indicates the 
degree to which the lattice QCD simulation results extend beyond the 
power-counting regime (PCR). This quantifies and effectively handles 
the scheme-dependence of chiral extrapolations. 
Ultimately, the 
agreement among optimal regularization scales obtained from different 
simulation results indicates 
the existence of an 
{intrinsic scale} that characterizes
 the interaction between the pion cloud and the 
core of the nucleon. Such an agreement will be demonstrated through the 
results in this chapter, and Chapter \ref{chpt:nucleonmagmom}. 
In Chapter \ref{chpt:mesonmass}, the procedure developed in this thesis 
for analyzing the renormalization flow of the low-energy coefficients, 
obtaining a possible intrinsic scale (or a range of acceptable regularization 
scales), and performing a robust chiral extrapolation will be tested. 

%  New graphs added JMMH
In the previous Chapter, %\ref{chpt:intrinsic}, 
extrapolation of the lattice results 
was discussed in the context of finite-range regularized chiral effective 
field theory ($\chi$EFT).  
The scheme-dependence of the various extrapolations was analyzed. 
A method was developed for extracting an optimal finite-range regularization 
scale from ideal pseudodata. Since the pseudodata were generated 
at a known scale $\La_\ro{c}$, they contain an intrinsic scale by construction, 
and so 
 %It was discovered 
it was demonstrated that an optimal finite-range regularization scale 
 could be extracted from the pseudodata by analyzing the scale-dependence 
of the low-energy coefficients. %Since the pseudodata have an %\emph{
%intrinsic scale by construction: %}
% the scale $\La_\ro{c}$ at 
%which they were generated, 
This optimal scale was the same value as the intrinsic scale built into the 
pseudodata.

The pseudodata example leads the researcher
 to consider whether actual lattice QCD simulation results  
 have an intrinsic cut-off scale embedded within them. That is, by
 analyzing lattice QCD data  
in the same way as the pseudodata, can a similar intersection
point be obtained from the renormalization-scale flow of the low-energy 
coefficients? % $c_0$ and $c_2$?
If so, %this 
it would indicate that the lattice data %do 
contain information
regarding an optimal finite-range regularization scale, and thus provide  
evidence for the existence of an underlying intrinsic scale in the 
nucleon-pion interaction. %, 
%which can be calculated.

%%%%%%%%%%%%%%%%%%%%%%%%%%%%%%%%%

%*****************************************%
%
%\section{Intrinsic Scale: Lattice Results}
%\label{sect:scales}
%\section{Evidence for an Intrinsic Scale in Lattice QCD Data}
\section{Evidence for an Intrinsic Scale in the Nucleon Mass}
\label{sect:ev}

\subsection{Renormalization Flow Analysis}
\label{subsect:renormflow}
%This optimal scale will be more tightly constrained to a window of possible
%regulator values $[\La_\ro{min},\La_\ro{max}]$ as data is taken further
%from the PCR. This window is the variance of the scale $\La^\ro{scale}$.

\begin{figure}
\begin{minipage}[b]{0.5\linewidth} % A minipage that covers half the page
\centering
\includegraphics[height=1.0\hsize,angle=90]{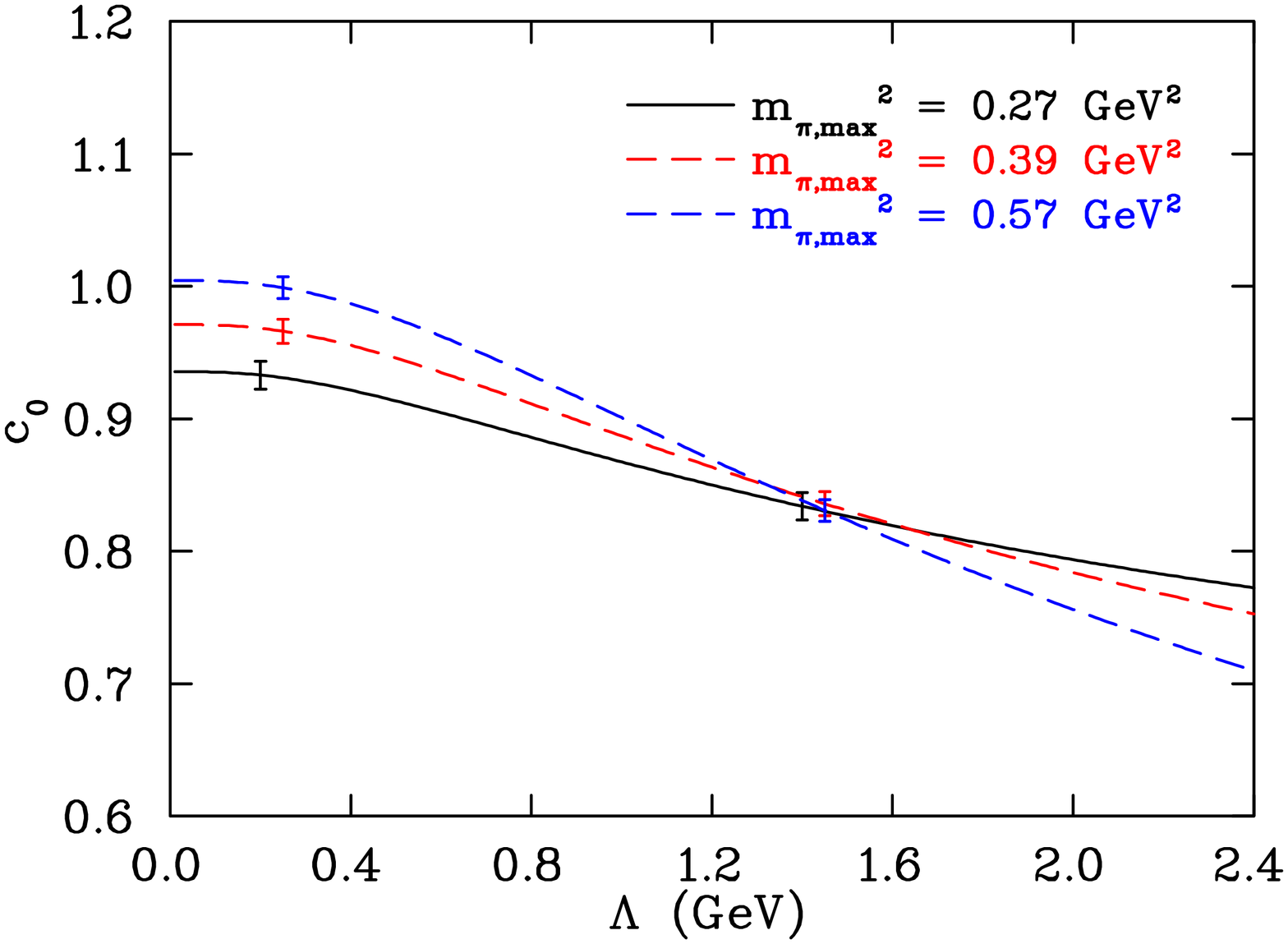}
\vspace{-3mm}
\caption{\footnotesize{ Behaviour of $c_0$ vs.\ $\La$, based on JLQCD data. The chiral expansion is taken to order $\ca{O}(m_\pi^3)$ and a  dipole regulator is used. A few points are selected to indicate the general size of the statistical error bars.}}
\label{fig:Ohkic0truncDIP}
\vspace{6mm}
\includegraphics[height=1.0\hsize,angle=90]{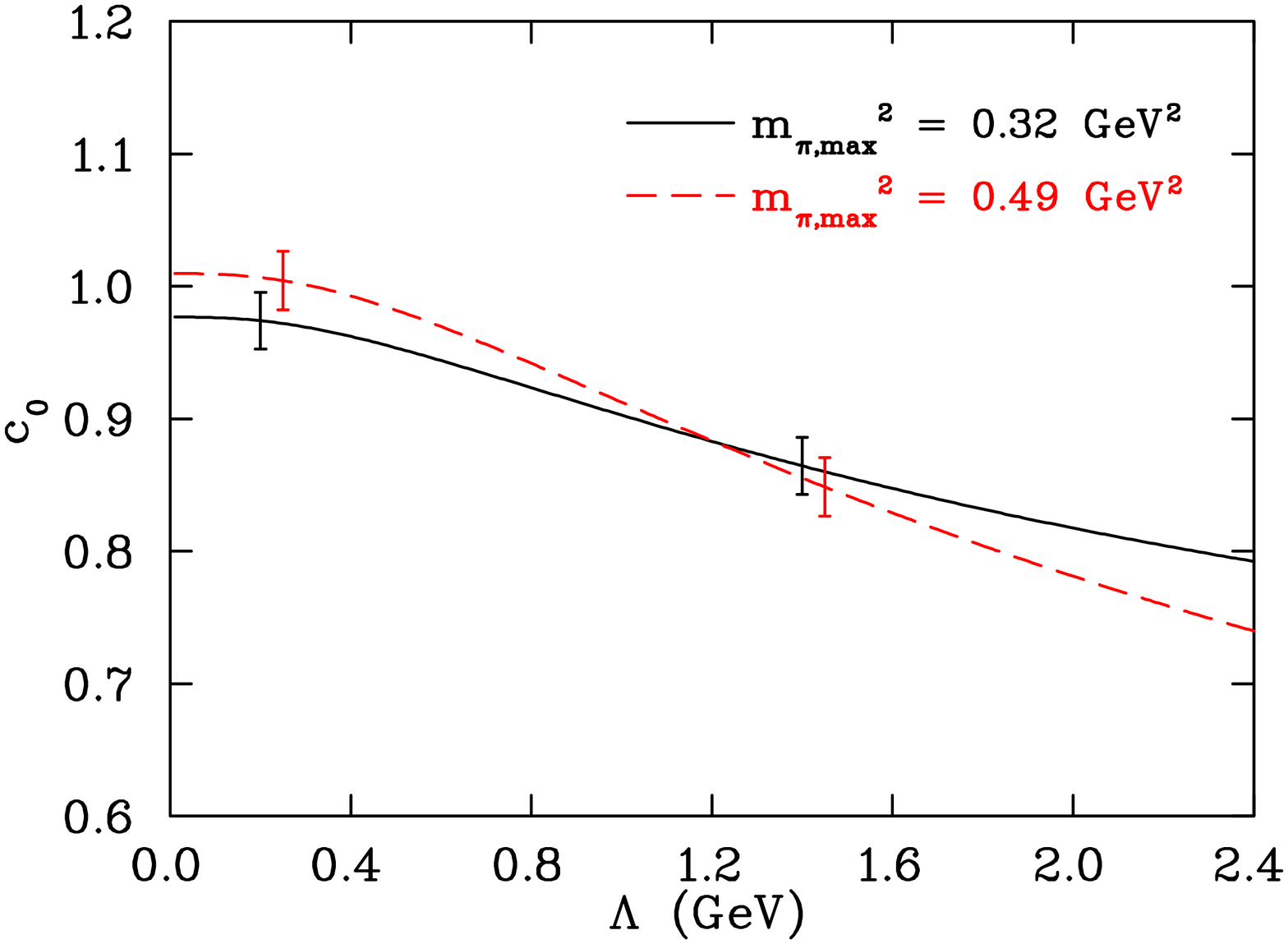}
\vspace{-3mm}
\caption{\footnotesize{ Behaviour of $c_0$ vs.\ $\La$, based on PACS-CS data. The chiral expansion is taken to order $\ca{O}(m_\pi^3)$ and a  dipole regulator is used. A few points are selected to indicate the general size of the statistical error bars.}}
\label{fig:Aokic0truncDIP}
\vspace{6mm}
\includegraphics[height=1.0\hsize,angle=90]{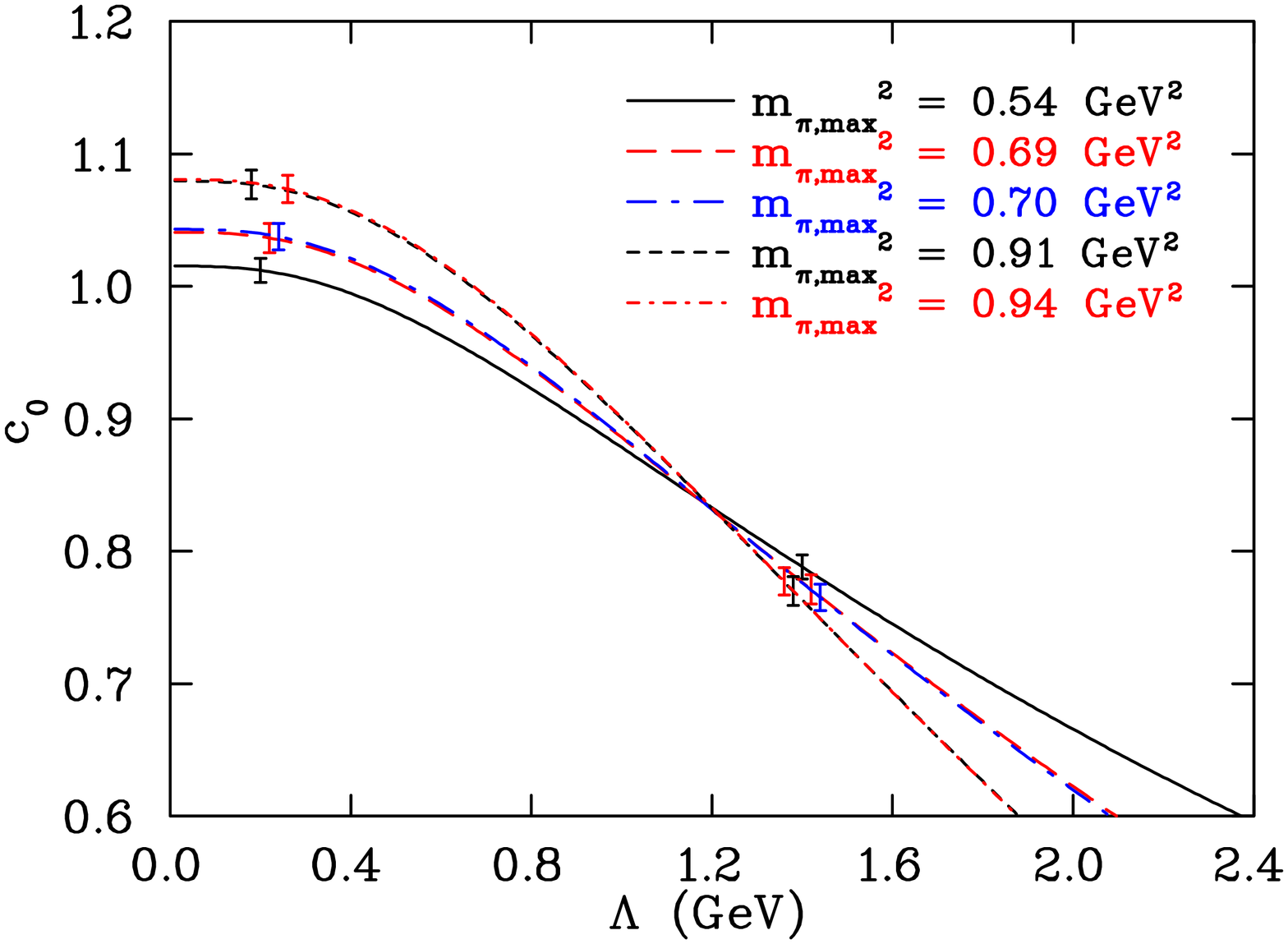}
\vspace{-3mm}
\caption{\footnotesize{ Behaviour of $c_0$ vs.\ $\La$, based on CP-PACS data. The chiral expansion is taken to order $\ca{O}(m_\pi^3)$ and a  dipole regulator is used. A few points are selected to indicate the general size of the statistical error bars.}}
\label{fig:Youngc0truncDIP}
\end{minipage}
\hspace{12mm}
\begin{minipage}[b]{0.5\linewidth}
\centering
\includegraphics[height=1.0\hsize,angle=90]{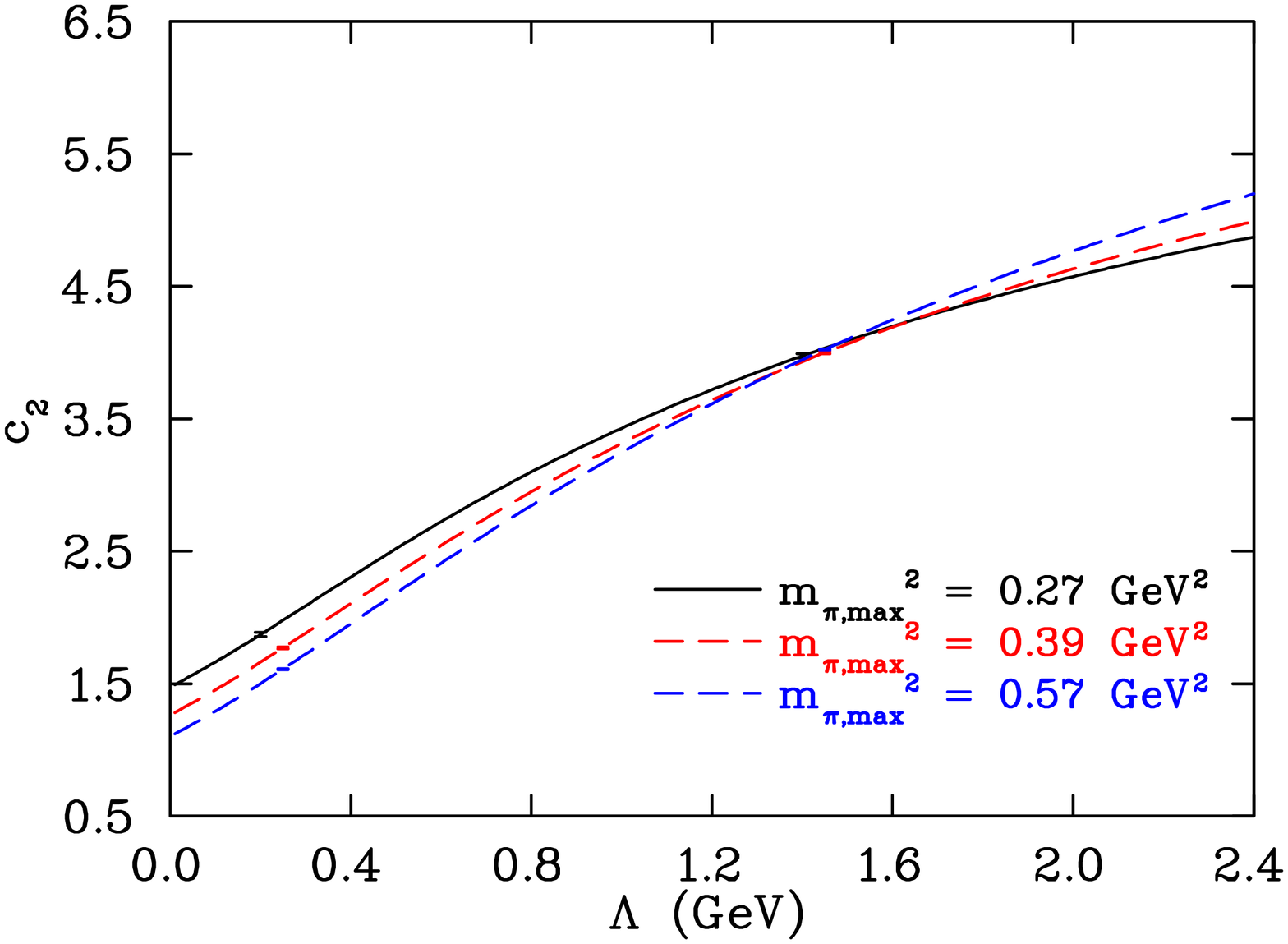}
\vspace{-3mm}
\caption{\footnotesize{ Behaviour of $c_2$ vs.\ $\La$, based on JLQCD data. The chiral expansion is taken to order $\ca{O}(m_\pi^3)$ and a  dipole regulator is used. A few points are selected to indicate the general size of the statistical error bars.}}
\label{fig:Ohkic2truncDIP}
\vspace{6mm}
\includegraphics[height=1.0\hsize,angle=90]{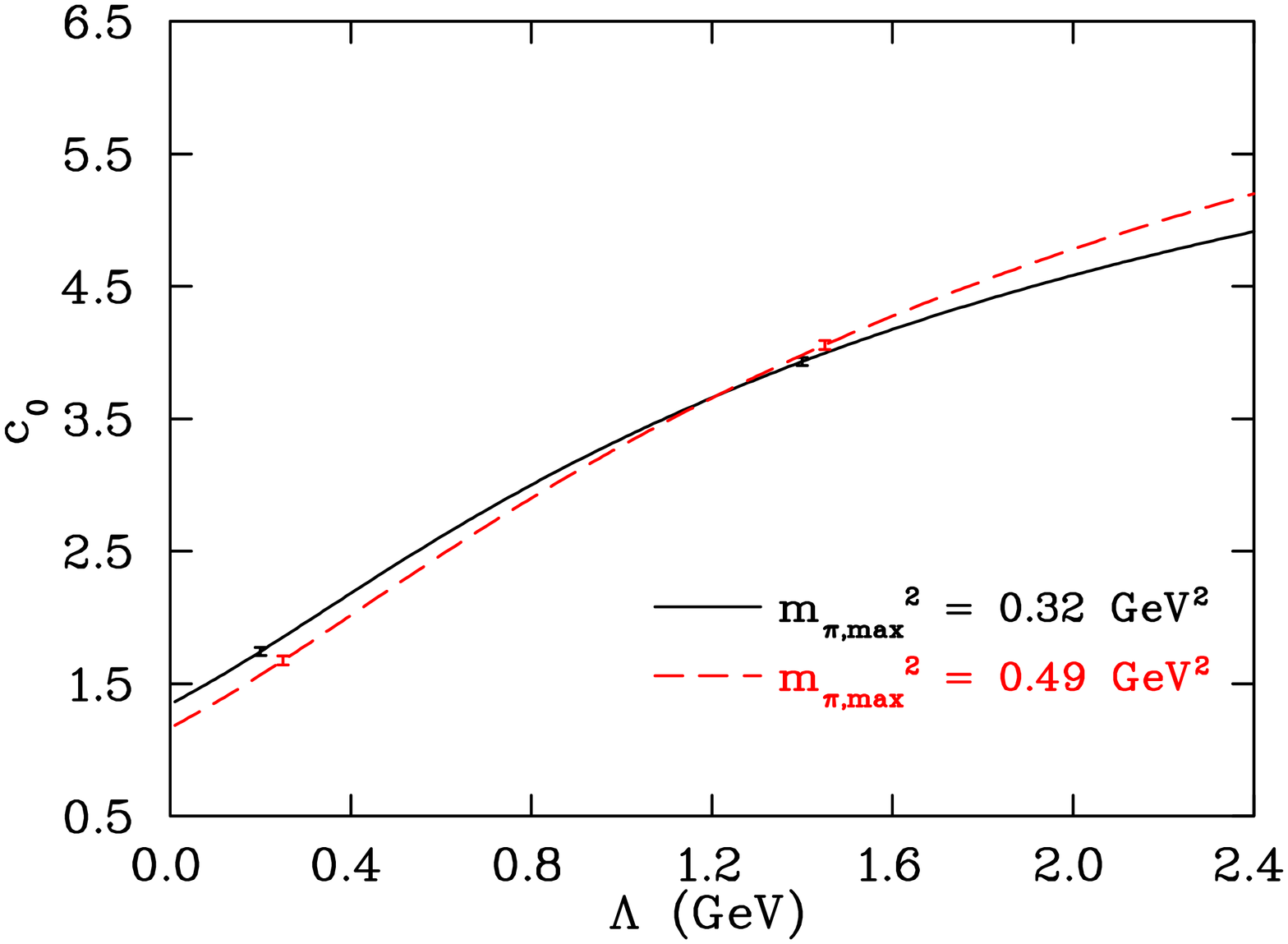}
\vspace{-3mm}
\caption{\footnotesize{ Behaviour of $c_2$ vs.\ $\La$, based on PACS-CS data. The chiral expansion is taken to order $\ca{O}(m_\pi^3)$ and a  dipole regulator is used. A few points are selected to indicate the general size of the statistical error bars.}}
\label{fig:Aokic2truncDIP}
\vspace{6mm}
\includegraphics[height=1.0\hsize,angle=90]{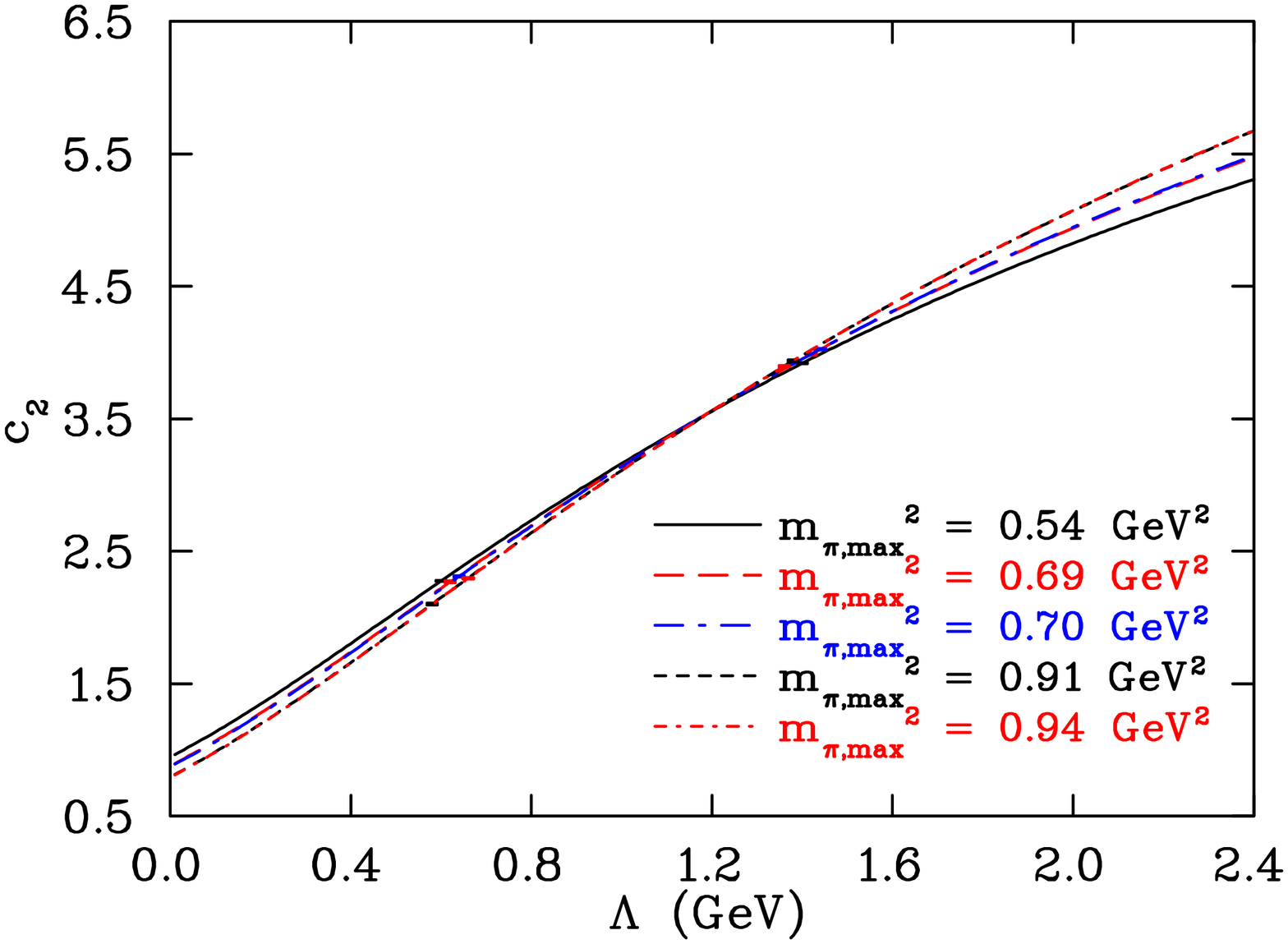}
\vspace{-3mm}
\caption{\footnotesize{ Behaviour of $c_2$ vs.\ $\La$, based on CP-PACS data. The chiral expansion is taken to order $\ca{O}(m_\pi^3)$ and a  dipole regulator is used. A few points are selected to indicate the general size of the statistical error bars.}}
\label{fig:Youngc2truncDIP}
\end{minipage}
\end{figure}

\begin{figure}
\begin{minipage}[b]{0.5\linewidth} % A minipage that covers half the page
\centering
\includegraphics[height=1.0\hsize,angle=90]{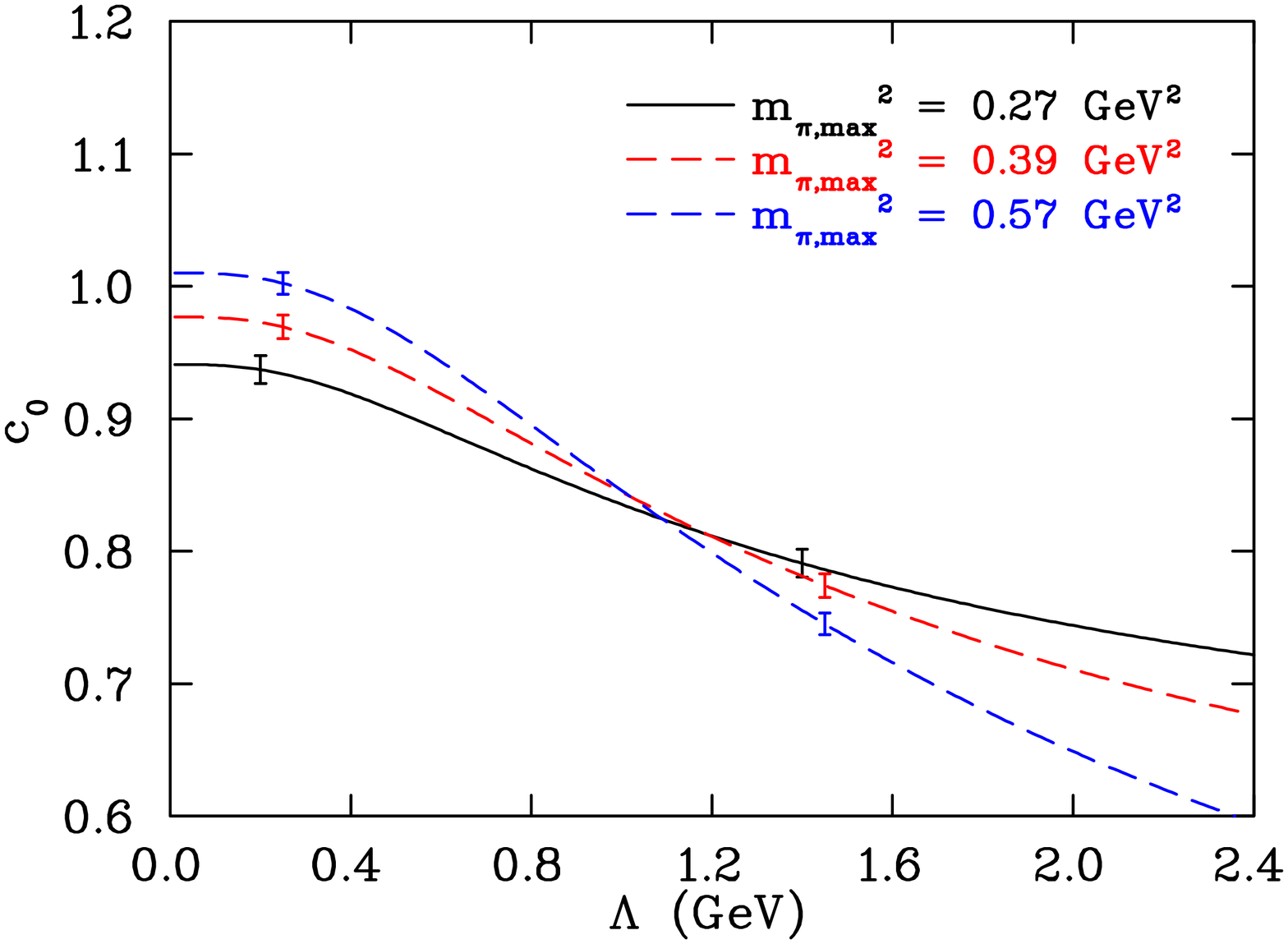}
\vspace{-3mm}
\caption{\footnotesize{ Behaviour of $c_0$ vs.\ $\La$, based on JLQCD data. The chiral expansion is taken to order $\ca{O}(m_\pi^3)$ and a double-dipole regulator is used. A few points are selected to indicate the general size of the statistical error bars.}}
\label{fig:Ohkic0truncDOUB}
\vspace{6mm}
\includegraphics[height=1.0\hsize,angle=90]{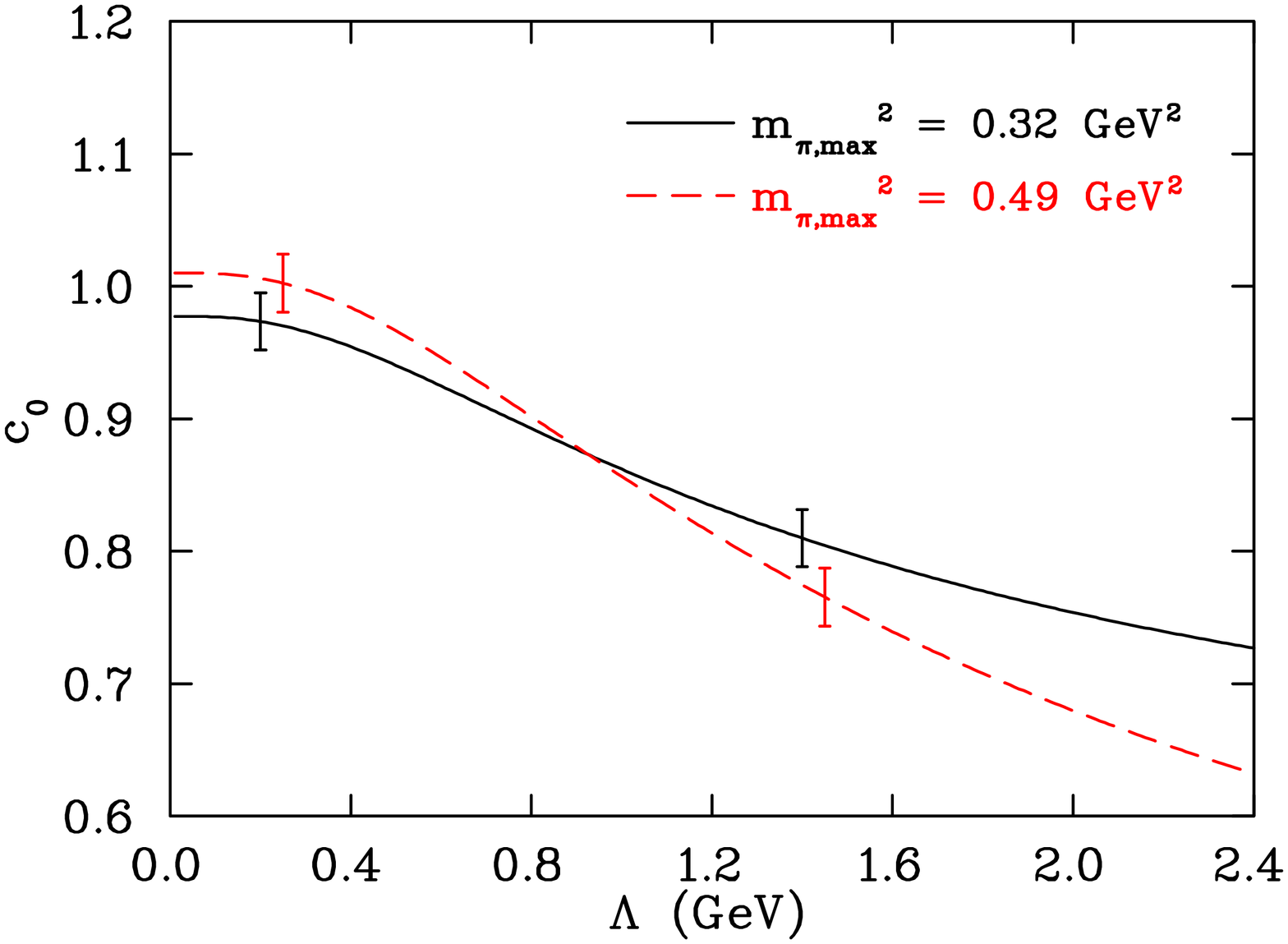}
\vspace{-3mm}
\caption{\footnotesize{ Behaviour of $c_0$ vs.\ $\La$, based on PACS-CS data. The chiral expansion is taken to order $\ca{O}(m_\pi^3)$ and a double-dipole regulator is used. A few points are selected to indicate the general size of the statistical error bars.}}
\label{fig:Aokic0truncDOUB}
\vspace{6mm}
\includegraphics[height=1.0\hsize,angle=90]{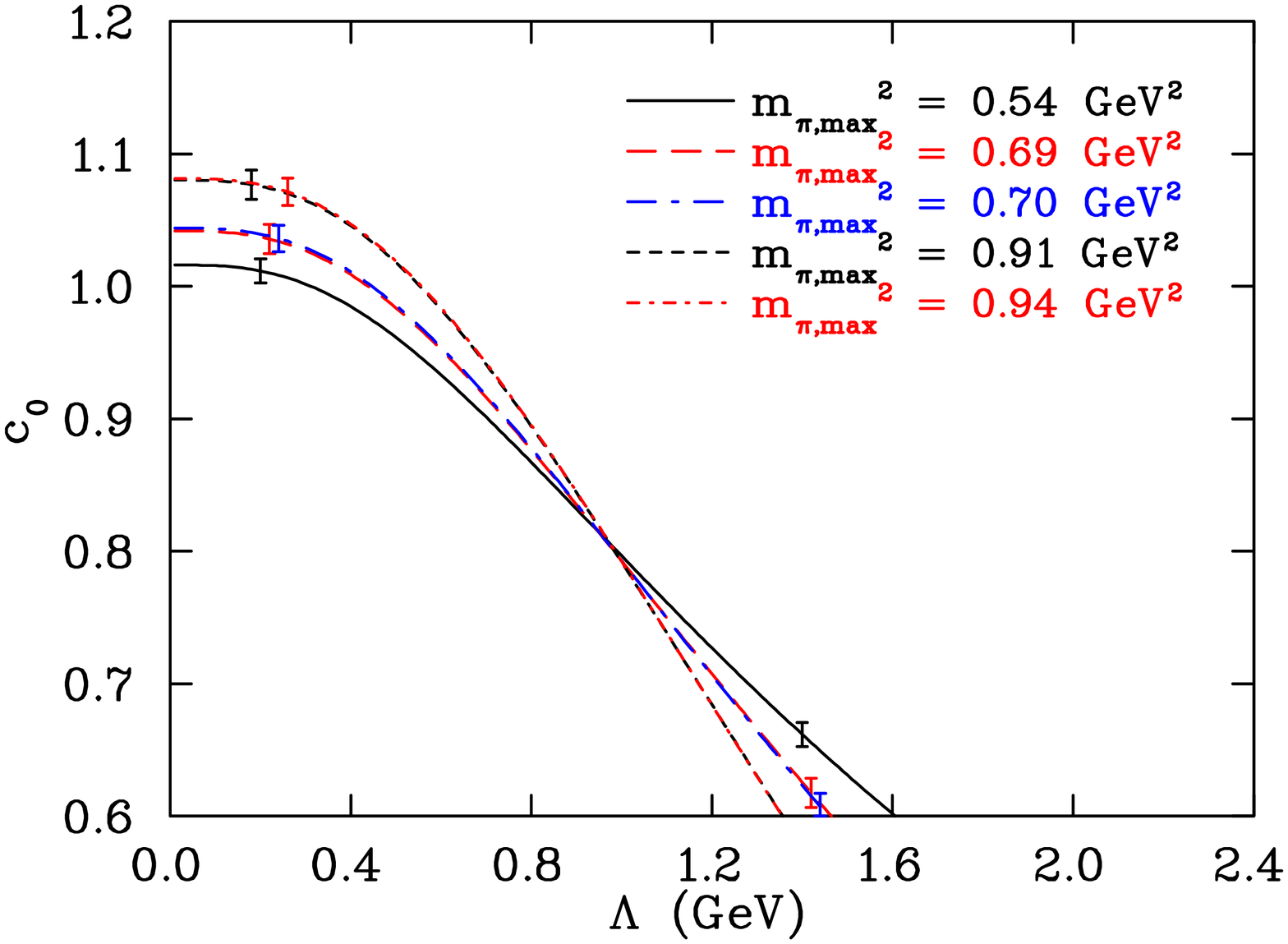}
\vspace{-3mm}
\caption{\footnotesize{ Behaviour of $c_0$ vs.\ $\La$, based on CP-PACS data. The chiral expansion is taken to order $\ca{O}(m_\pi^3)$ and a double-dipole regulator is used. A few points are selected to indicate the general size of the statistical error bars.}}
\label{fig:Youngc0truncDOUB}
\end{minipage}
\hspace{12mm}
\begin{minipage}[b]{0.5\linewidth}
\centering
\includegraphics[height=1.0\hsize,angle=90]{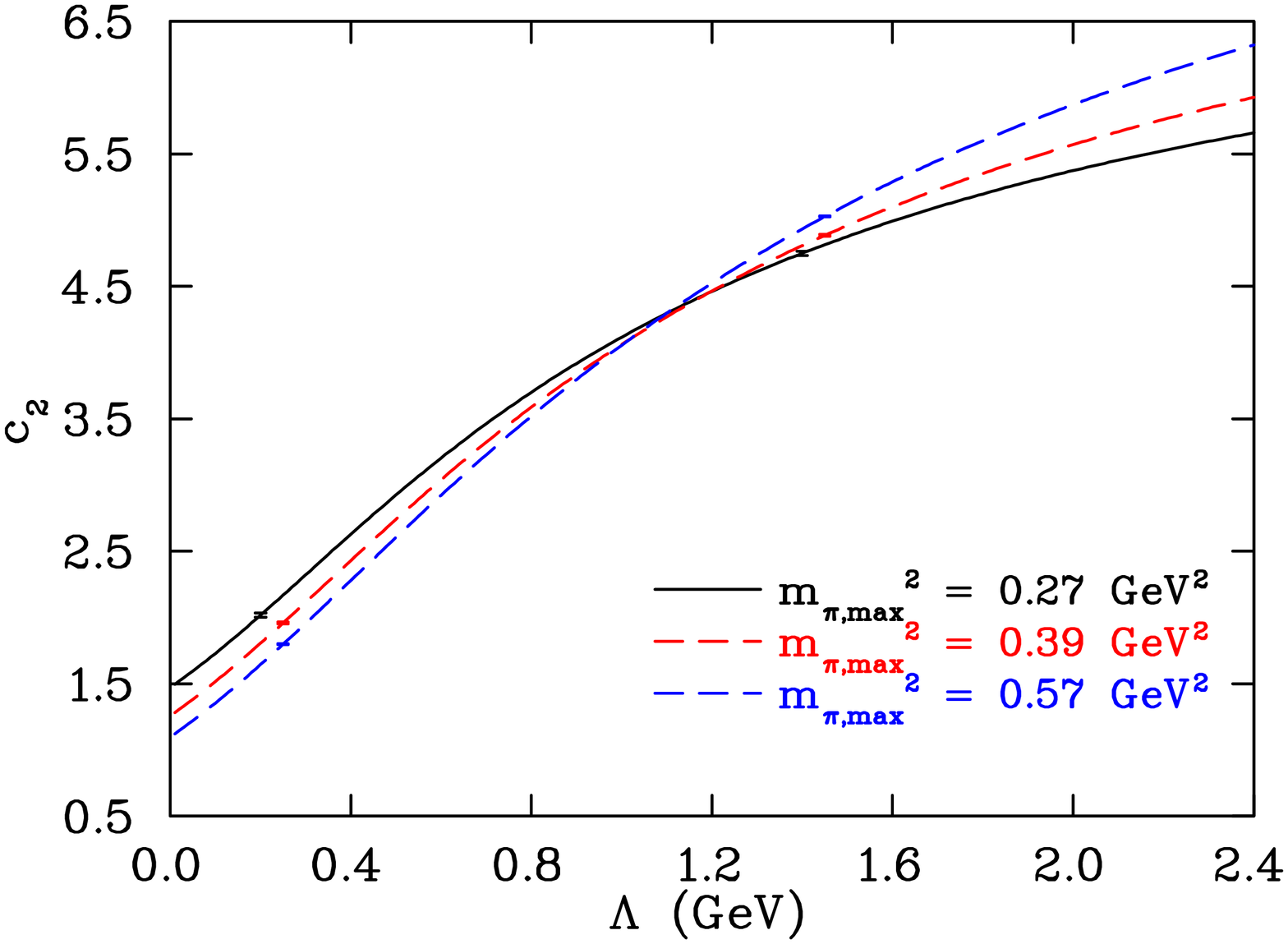}
\vspace{-3mm}
\caption{\footnotesize{ Behaviour of $c_2$ vs.\ $\La$, based on JLQCD data. The chiral expansion is taken to order $\ca{O}(m_\pi^3)$ and a double-dipole regulator is used. A few points are selected to indicate the general size of the statistical error bars.}}
\label{fig:Ohkic2truncDOUB}
\vspace{6mm}
\includegraphics[height=1.0\hsize,angle=90]{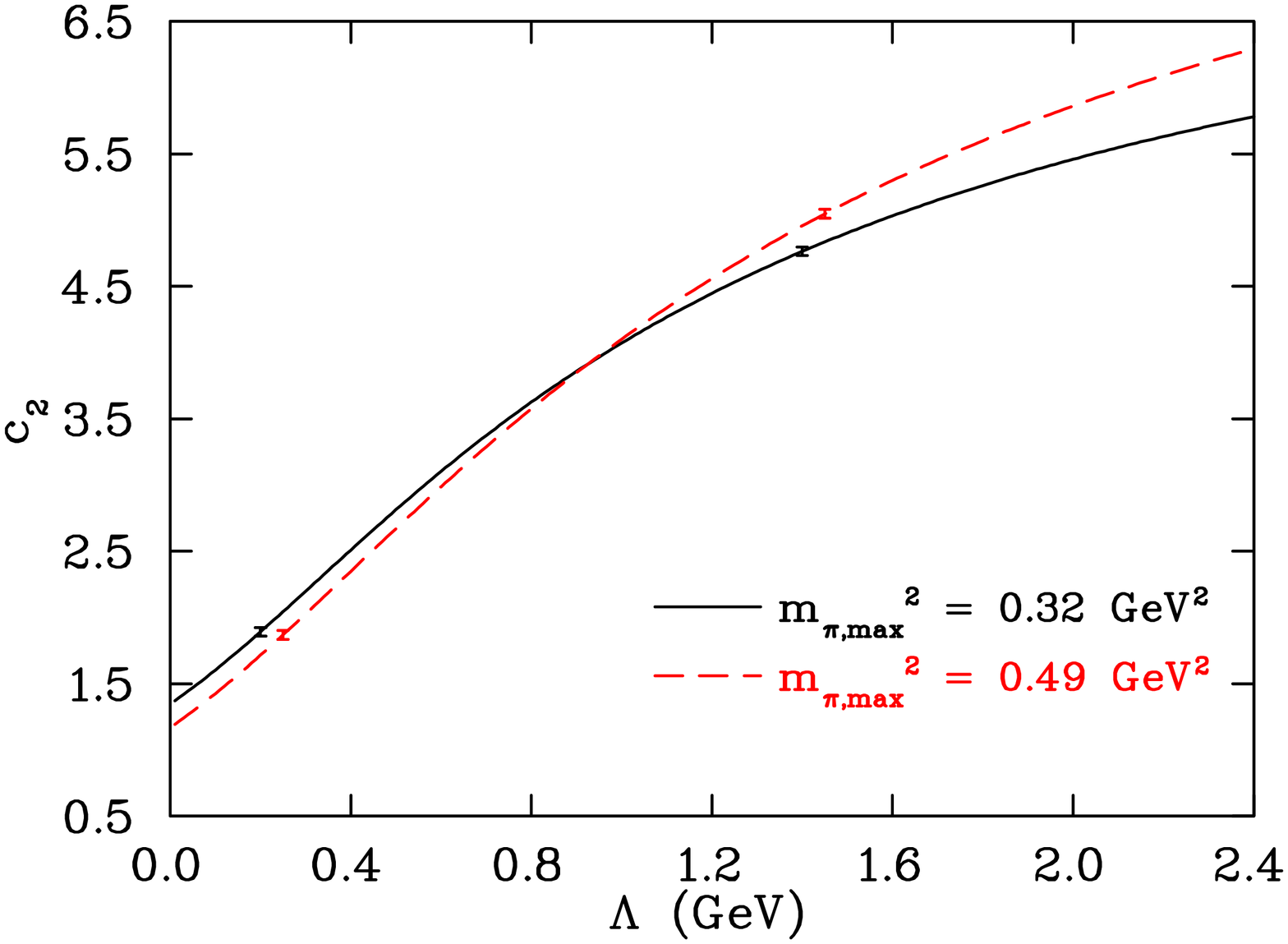}
\vspace{-3mm}
\caption{\footnotesize{ Behaviour of $c_2$ vs.\ $\La$, based on PACS-CS data. The chiral expansion is taken to order $\ca{O}(m_\pi^3)$ and a double-dipole regulator is used. A few points are selected to indicate the general size of the statistical error bars.}}
\label{fig:Aokic2truncDOUB}
\vspace{6mm}
\includegraphics[height=1.0\hsize,angle=90]{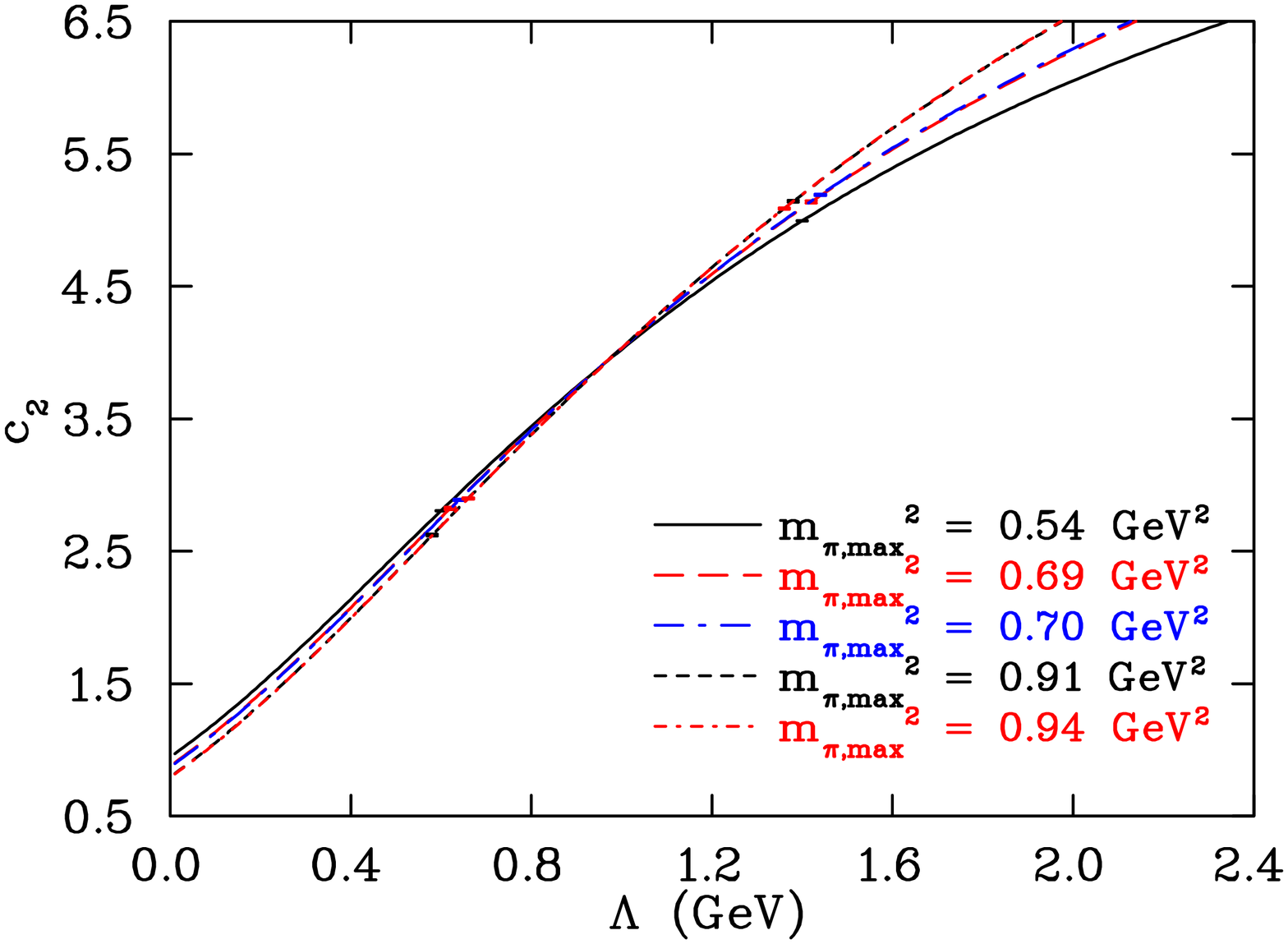}
\vspace{-3mm}
\caption{\footnotesize{ Behaviour of $c_2$ vs.\ $\La$, based on CP-PACS data. The chiral expansion is taken to order $\ca{O}(m_\pi^3)$ and a double-dipole regulator is used. A few points are selected to indicate the general size of the statistical error bars.}}
\label{fig:Youngc2truncDOUB}
\end{minipage}
\end{figure}

\begin{figure}
\begin{minipage}[b]{0.5\linewidth} % A minipage that covers half the page
\centering
\includegraphics[height=1.0\hsize,angle=90]{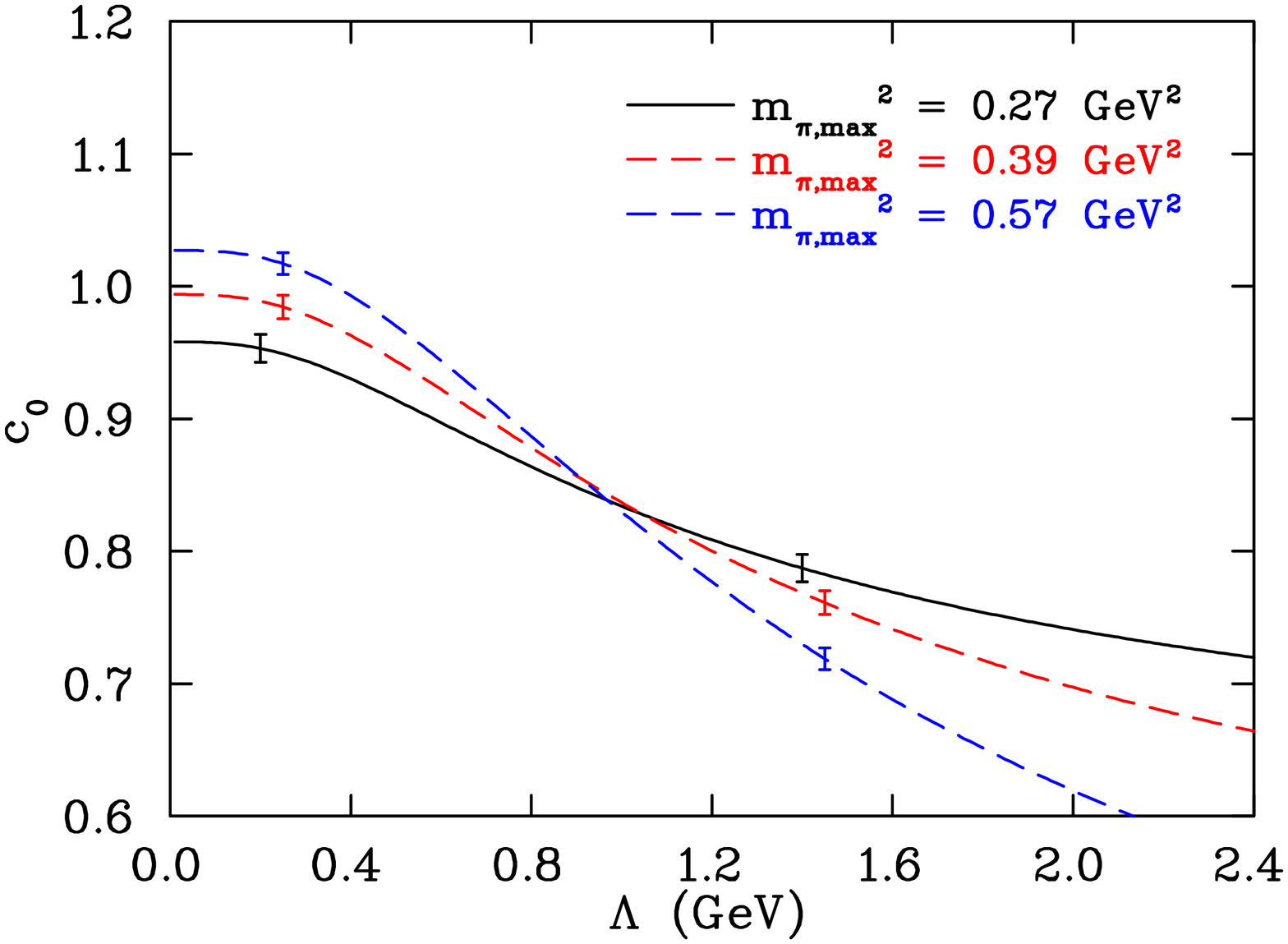}
\vspace{-3mm}
\caption{\footnotesize{ Behaviour of $c_0$ vs.\ $\La$, based on JLQCD data. The chiral expansion is taken to order $\ca{O}(m_\pi^3)$ and a triple-dipole regulator is used. A few points are selected to indicate the general size of the statistical error bars.}}
\label{fig:Ohkic0truncTRIP}
\vspace{6mm}
\includegraphics[height=1.0\hsize,angle=90]{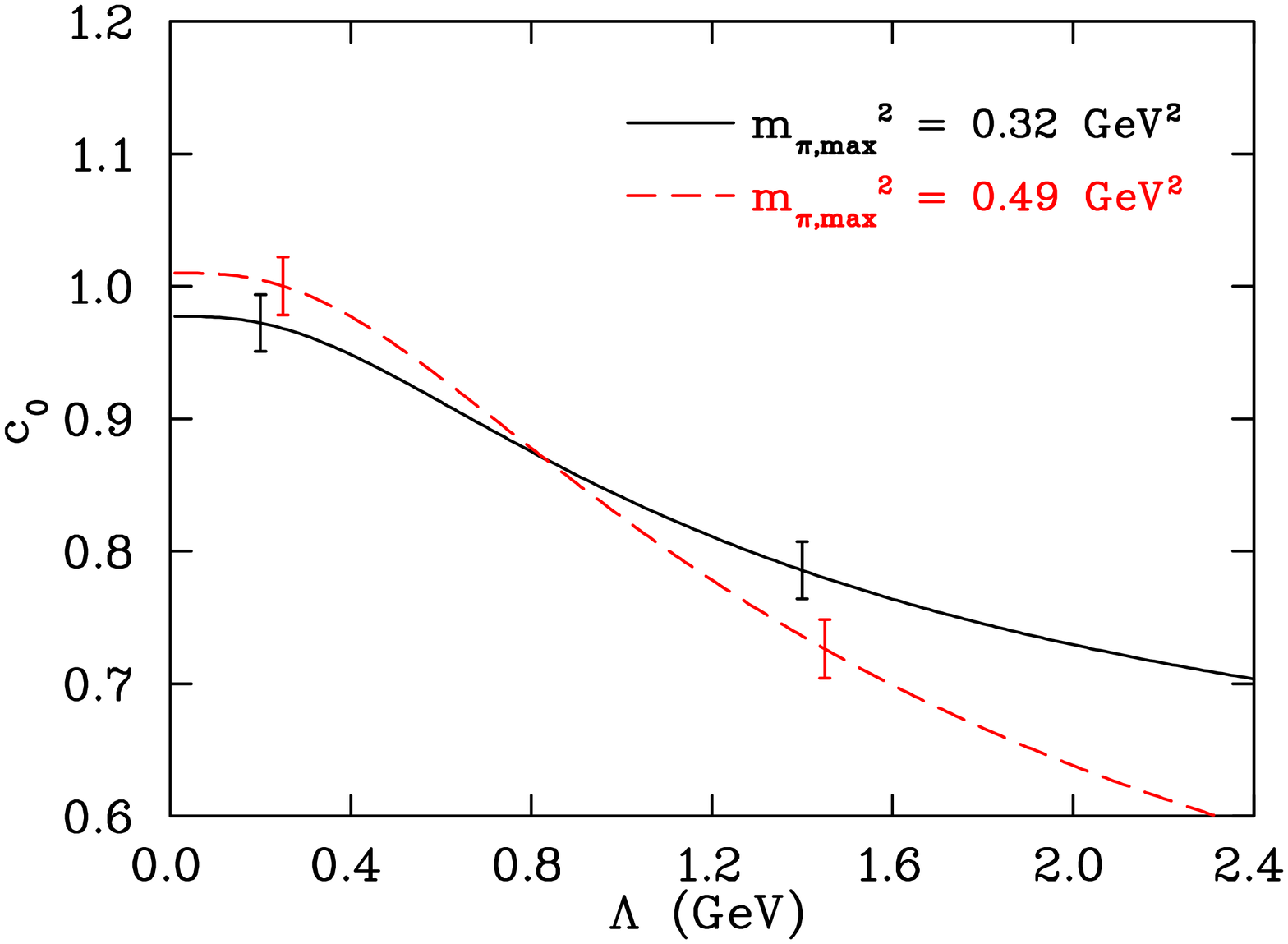}
\vspace{-3mm}
\caption{\footnotesize{ Behaviour of $c_0$ vs.\ $\La$, based on PACS-CS data. The chiral expansion is taken to order $\ca{O}(m_\pi^3)$ and a triple-dipole regulator is used. A few points are selected to indicate the general size of the statistical error bars.}}
\label{fig:Aokic0truncTRIP}
\vspace{6mm}
\includegraphics[height=1.0\hsize,angle=90]{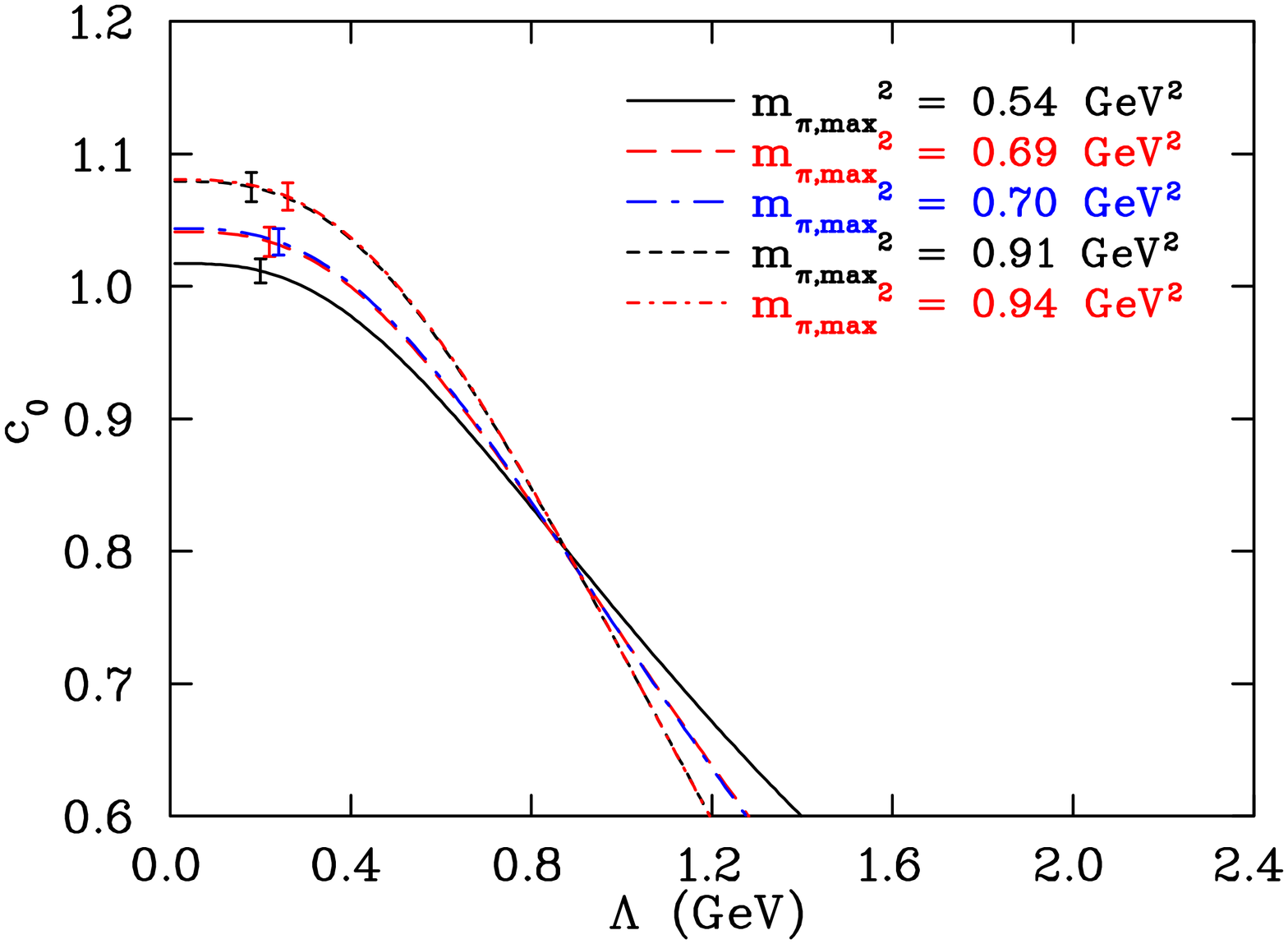}
\vspace{-3mm}
\caption{\footnotesize{ Behaviour of $c_0$ vs.\ $\La$, based on CP-PACS data. The chiral expansion is taken to order $\ca{O}(m_\pi^3)$ and a triple-dipole regulator is used. A few points are selected to indicate the general size of the statistical error bars.}}
\label{fig:Youngc0truncTRIP}
\end{minipage}
\hspace{12mm}
\begin{minipage}[b]{0.5\linewidth}
\centering
\includegraphics[height=1.0\hsize,angle=90]{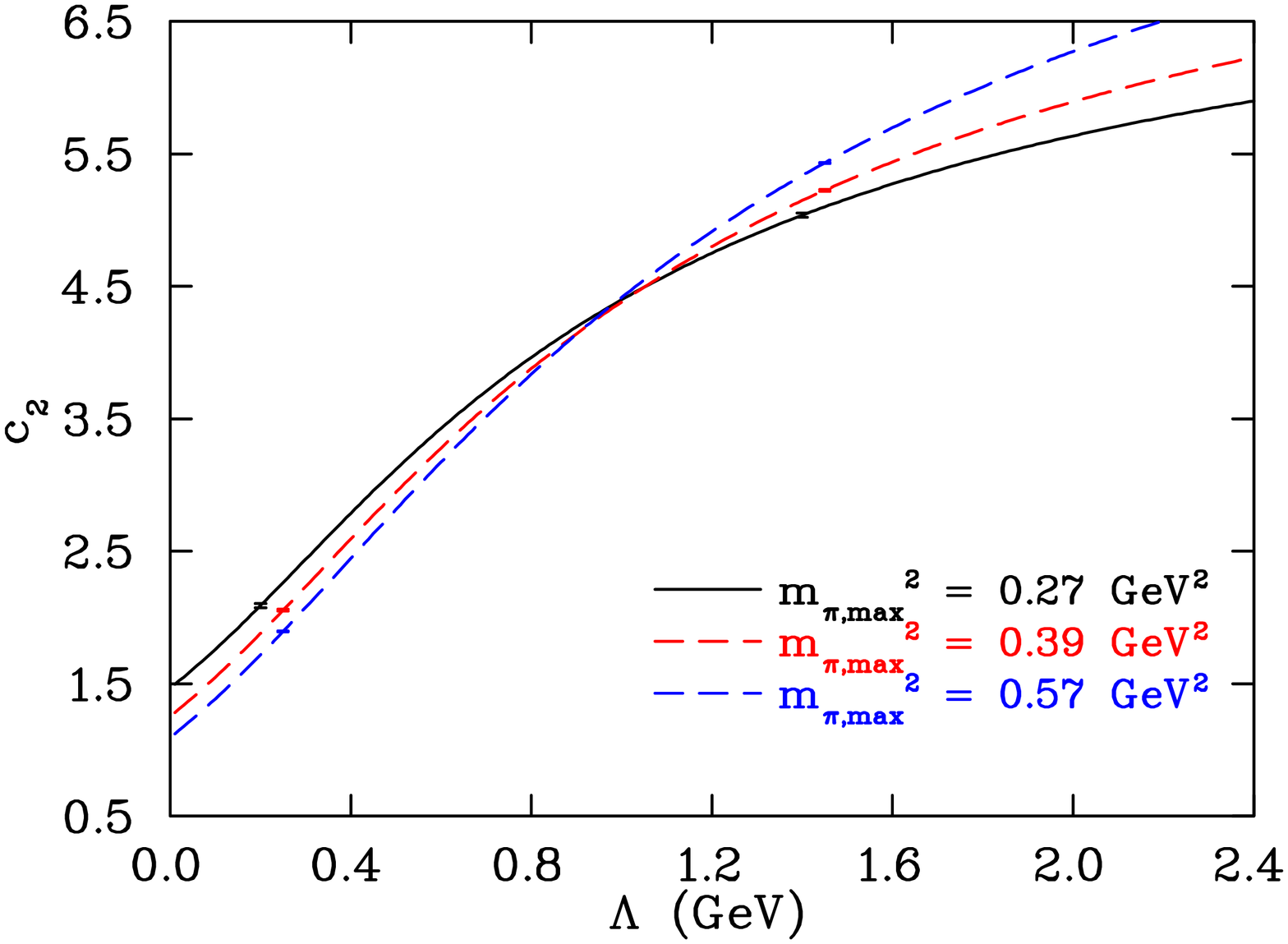}
\vspace{-3mm}
\caption{\footnotesize{ Behaviour of $c_2$ vs.\ $\La$, based on JLQCD data. The chiral expansion is taken to order $\ca{O}(m_\pi^3)$ and a triple-dipole regulator is used. A few points are selected to indicate the general size of the statistical error bars.}}
\label{fig:Ohkic2truncTRIP}
\vspace{6mm}
\includegraphics[height=1.0\hsize,angle=90]{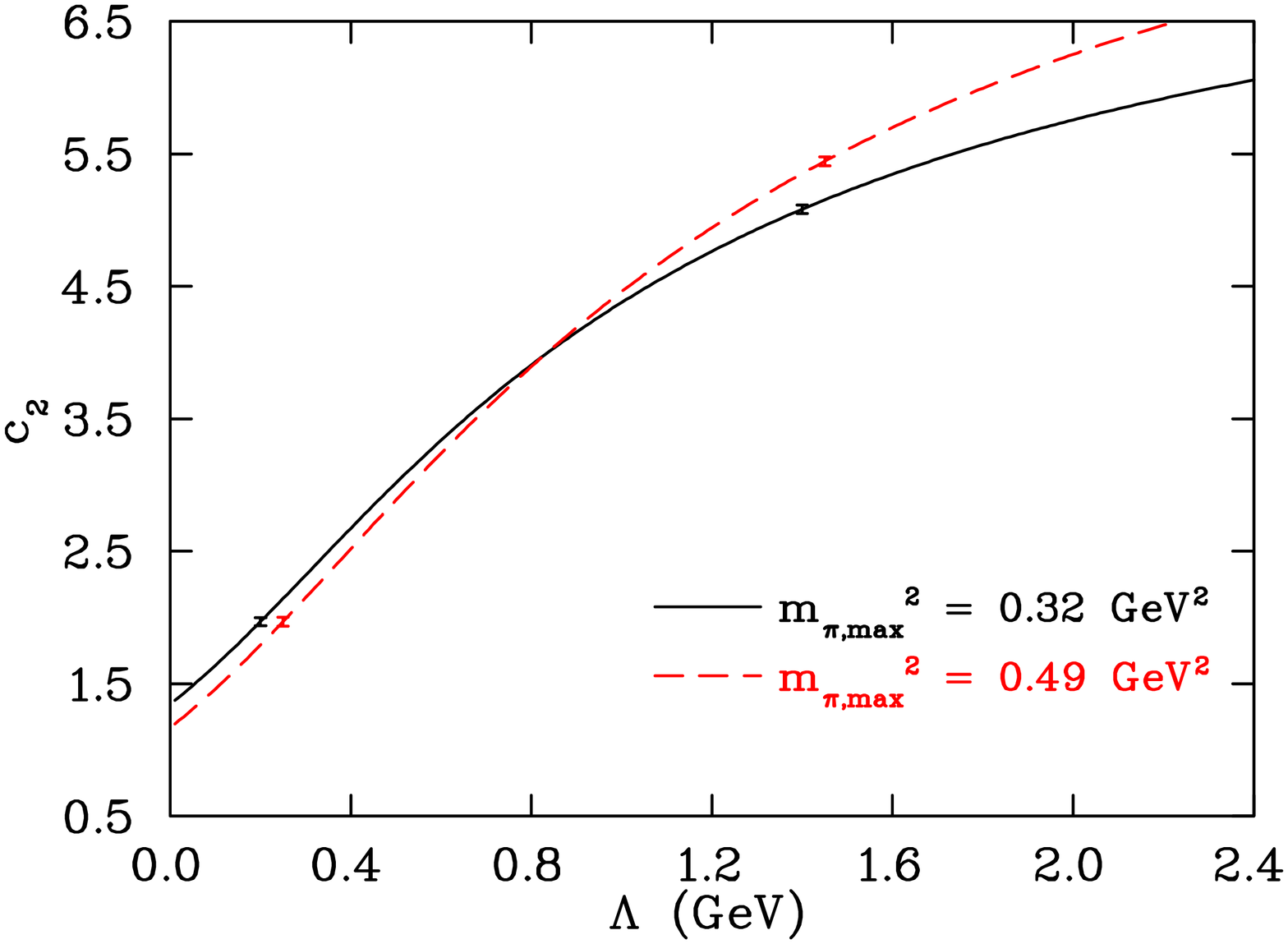}
\vspace{-3mm}
\caption{\footnotesize{ Behaviour of $c_2$ vs.\ $\La$, based on PACS-CS data. The chiral expansion is taken to order $\ca{O}(m_\pi^3)$ and a triple-dipole regulator is used. A few points are selected to indicate the general size of the statistical error bars.}}
\label{fig:Aokic2truncTRIP}
\vspace{6mm}
\includegraphics[height=1.0\hsize,angle=90]{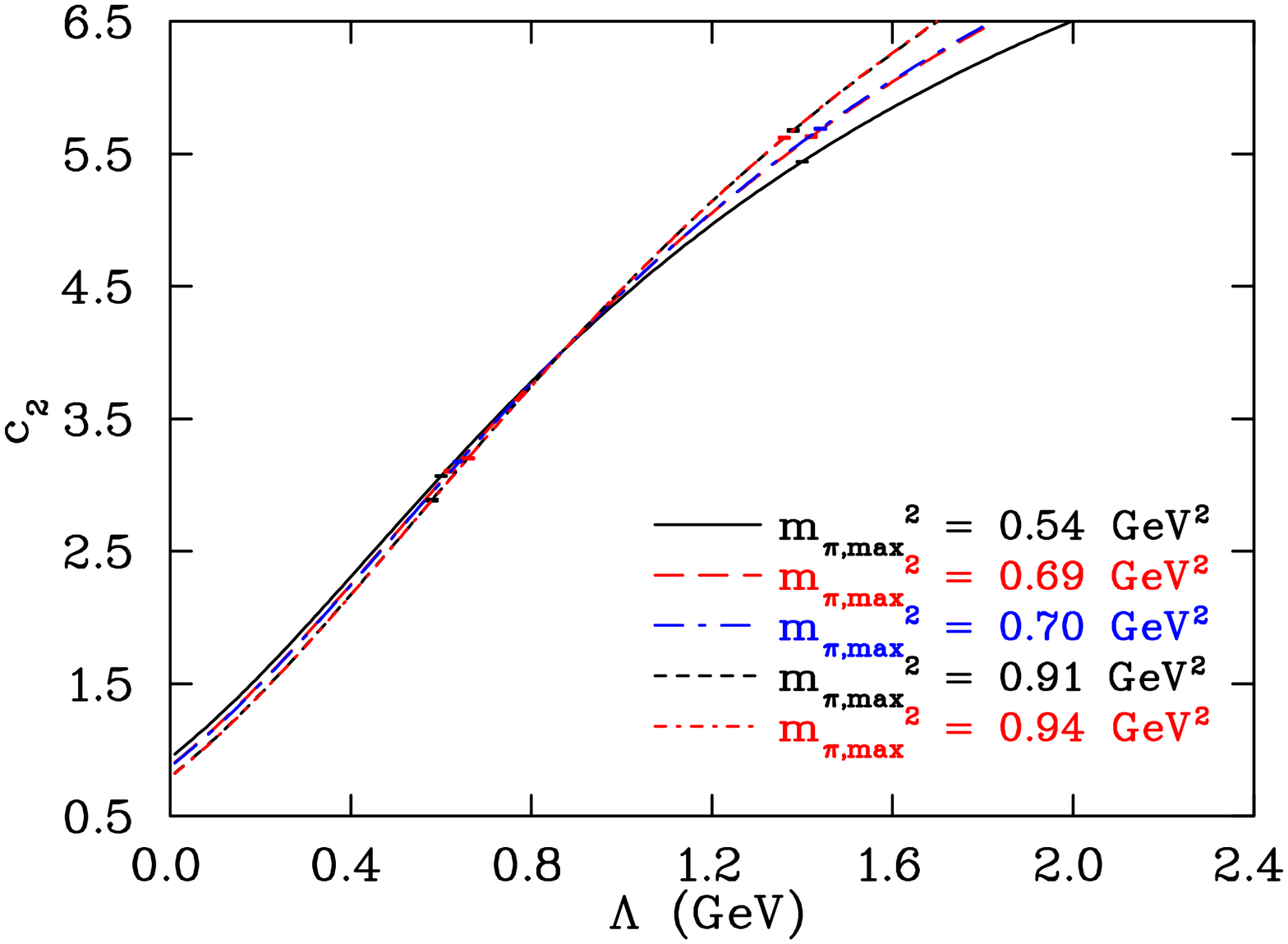}
\vspace{-3mm}
\caption{\footnotesize{ Behaviour of $c_2$ vs.\ $\La$, based on CP-PACS data. The chiral expansion is taken to order $\ca{O}(m_\pi^3)$ and a triple-dipole regulator is used. A few points are selected to indicate the general size of the statistical error bars.}}
\label{fig:Youngc2truncTRIP}
\end{minipage}
\end{figure}

Consider the mass of the nucleon as extrapolated from the results of 
lattice QCD simulations. 
The results for %the renormalization 
%of 
$c_0$ and $c_2$ as a 
function of the regularization scale $\La$ are now presented 
for  lattice QCD data from the collaborations: 
JLQCD, PACS-CS and CP-PACS. 
Initially, the chiral expansion, calculated to chiral order $\ca{O}(m_\pi^3)$, 
should be used for fitting:
\begin{equation}
M_N = c_0 + c_2 m_\pi^2(1+{\tsi}_{tad}(m_\pi^2,\La))
+ {\tSi}_{N}(m_\pi^2,\La) + {\tSi}_{\De}(m_\pi^2,\La)\,. 
\label{eqn:mNfitO3}
\end{equation}
Thus, the relevant fit parameters used in the extrapolation are $c_0$ and $c_2$ 
only. Results for the higher chiral order of $\ca{O}(m_\pi^4\,\ro{log}\,m_\pi)$ 
will be discussed in Section \ref{subsect:higher}. 
The resultant renormalization flows, using a dipole regulator, 
are shown in 
Figures \ref{fig:Ohkic0truncDIP} through \ref{fig:Youngc2truncDIP}; 
the results for the double-dipole case are shown in
Figures \ref{fig:Ohkic0truncDOUB} through \ref{fig:Youngc2truncDOUB}; 
and the results for the triple-dipole are shown in
Figures \ref{fig:Ohkic0truncTRIP} through \ref{fig:Youngc2truncTRIP}. 
On each plot of the renormalization flow  
in Figures \ref{fig:Ohkic0truncDIP} through 
\ref{fig:Youngc2truncTRIP} there are multiple curves, each corresponding
to different values of the upper bound of the fit window, 
$m_{\pi,\ro{max}}^2$. 
 %To estimate the statistical error in the renormalized
%coefficients $\de c$, a bootstrap technique of $200$ configurations of 
%nucleon mass data is used. 
%The configurations differ by the statistical error in
%the data, with values generated by a Gaussian distribution. In each plot, 
%the same configurations are used
% across the range of values of finite-range regularization scales 
%$\La$ considered. 
A few example points are selected in Figures \ref{fig:Ohkic0truncDIP} through
 \ref{fig:Youngc2truncTRIP} to indicate the general size of the statistical 
error bars.

 It should be noted that none of the curves in Figures 
\ref{fig:Ohkic0truncDIP} through
 \ref{fig:Youngc2truncTRIP} is flat to 
within $1\%$ accuracy. 
 All of the fits have lattice data included beyond the %commonly accepted 
PCR.
Clearly, there is a well-defined intersection point in each plot. 
Also, the value of $\La$ at which the
intersection point occurs is the same, even for different data sets,
and for different $c_i$. The tight groupings of the curve crossings lend 
credence to the %\emph{ 
 notion %Ansatz %} 
of an intrinsic scale that can  be interpreted as %associated with the 
 a finite
size of the source of the pion dressings of the nucleon. This is a central 
result of the analysis.

%DIP

%DOUB

%TRIP

 Using the method described in Chapter \ref{chpt:intrinsic}, the 
intersection point of the renormalization flow curves for different values of 
$m_{\pi,\ro{max}}^2$ is estimated from Figures 
\ref{fig:Ohkic0truncDIP} through
 \ref{fig:Youngc2truncTRIP}. 
As an initial estimate, by inspection, 
a mean value for the optimal regularization scale of 
$\bar{\La}^\ro{scale}_\ro{dip} \approx 1.3$ GeV
 was obtained for the dipole, a value of 
$\bar{\La}^\ro{scale}_\ro{doub} \approx 1.0$ GeV 
was obtained for the double
dipole, and a value of $\bar{\La}^\ro{scale}_\ro{trip} \approx 0.9$ GeV
 was obtained for the triple-dipole. 
These values differ because the regulators have different shapes,
as evident in Figure \ref{fig:reg}, and thus 
different
 values of $\La^\ro{scale}$ are required to create a similar suppression of
large loop momenta. In order to determine an estimate of the systematic 
uncertainty in an extrapolation due to the choice of regularization scale 
$\La^\ro{scale}$, %tt would be useful to find 
one should use a robust method for 
%determining the optimal regularization scale and for providing an 
estimating %of its 
the systematic uncertainty of $\La^\ro{scale}$ itself. 
In the following section, 
a chi-square-style analysis will be introduced 
 %in the following section. 
to fulfill this requirement.

\subsection{Analysis of Systematic Uncertainties}
\label{subsect:stat}

The optimal regularization scale  
$\La^\ro{scale}$ can be obtained from the 
renormalization flow curves using a chi-square-style analysis.
 In addition, 
 the analysis will allow the extraction of an estimate of the variance for 
$\La^\ro{scale}$.
The function $\chi^2_{dof}$ defined below allows 
 easy identification of the intersection points in the renormalization 
flow plots, and a range associated with this central regularization scale.
 This function simply measures the degree to which the renormalization 
flow curves match. 

\begin{figure}
\begin{minipage}[b]{0.5\linewidth} % A minipage that covers half the page
\centering
\includegraphics[height=1.0\hsize,angle=90]{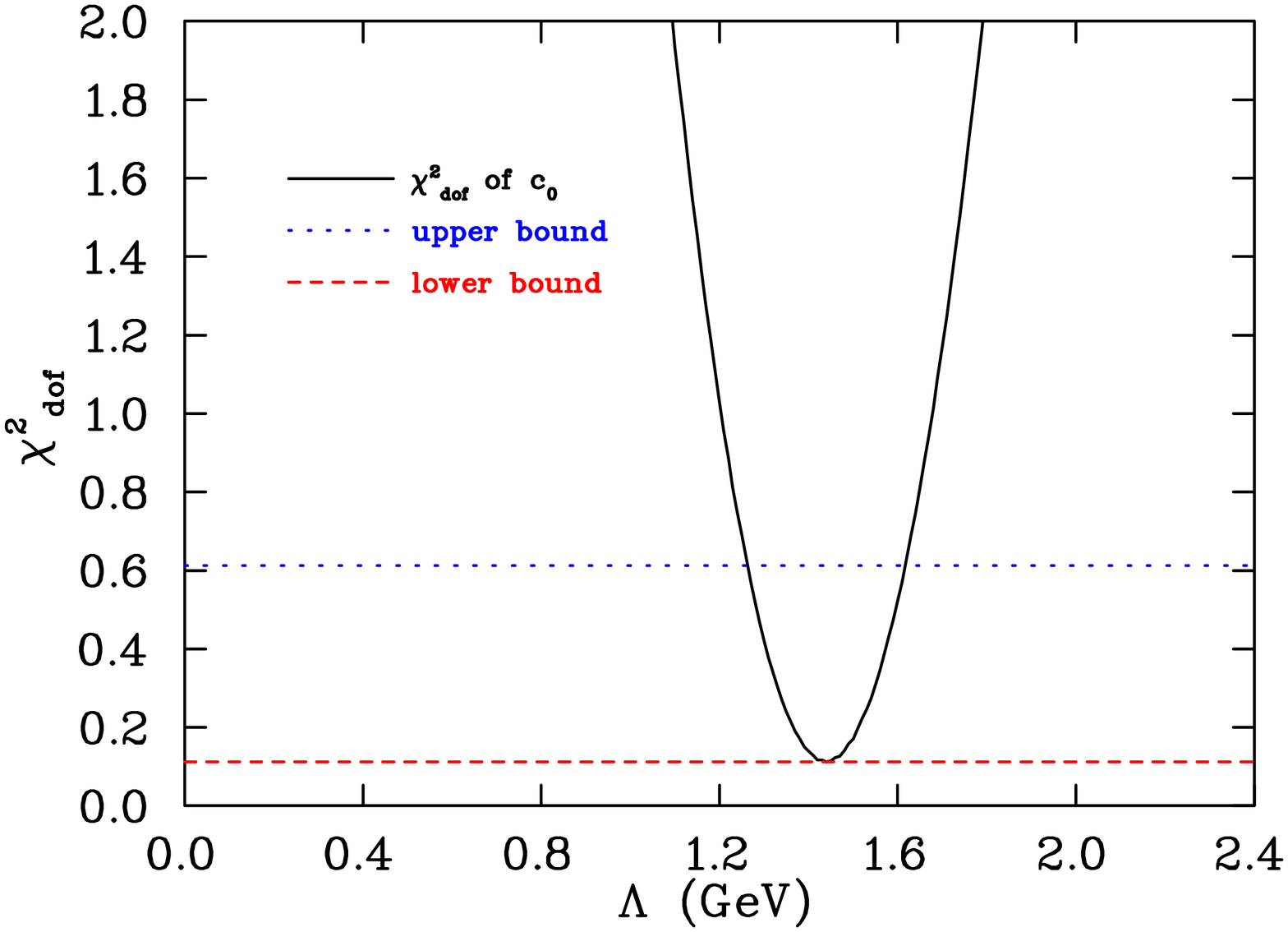}
\vspace{-3mm}
\caption{\footnotesize{ Behaviour of $\chi^2_{dof}$ for $c_0$ vs.\ $\La$, based on JLQCD data. The chiral expansion is taken to order $\ca{O}(m_\pi^3)$ and a  dipole regulator is used. }}
\label{fig:Ohkic0truncDIPchisqdof}
\vspace{6mm}
\includegraphics[height=1.0\hsize,angle=90]{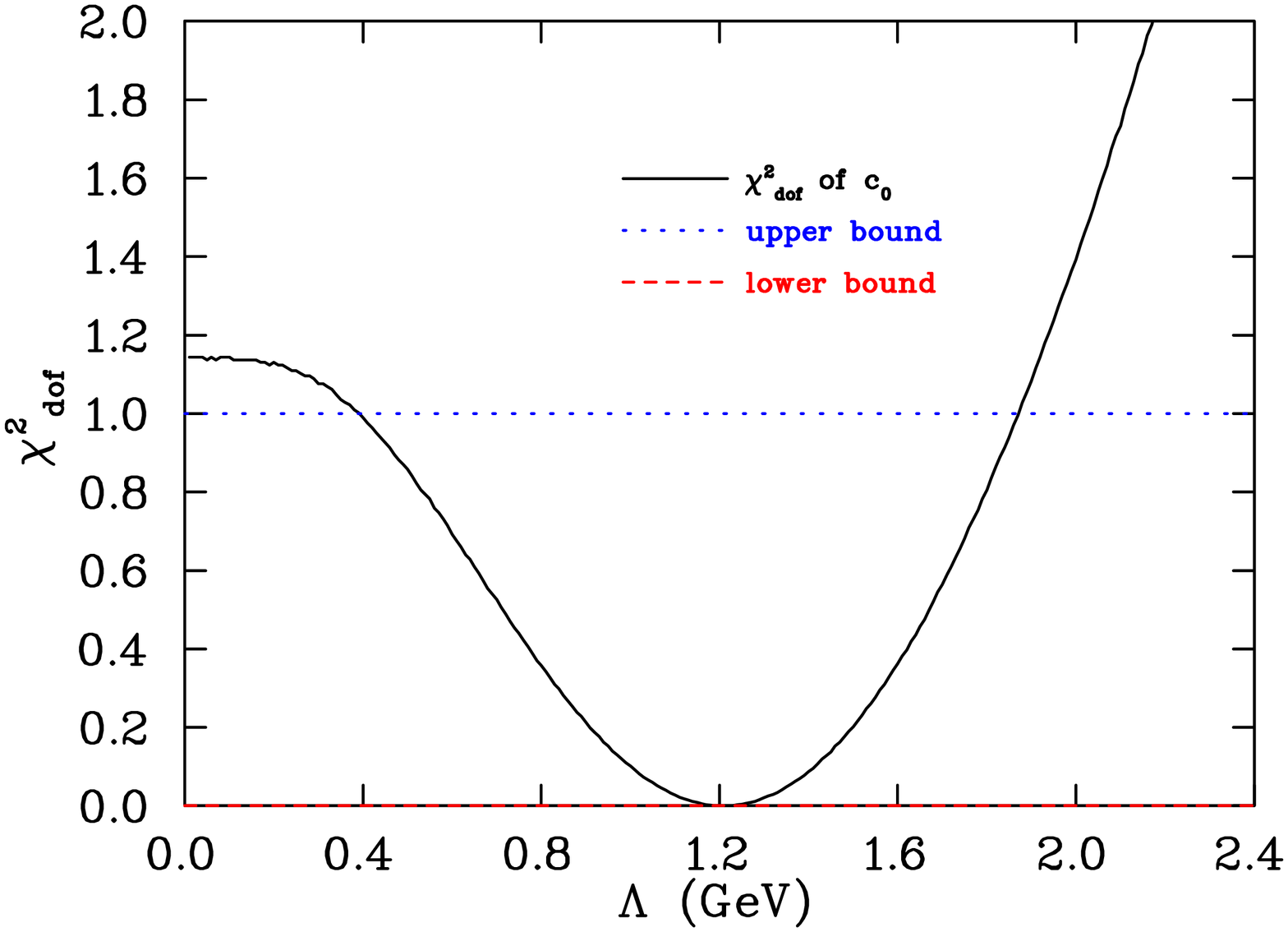}
\vspace{-3mm}
\caption{\footnotesize{ Behaviour of $\chi^2_{dof}$ for $c_0$ vs.\ $\La$, based on PACS-CS data. The chiral expansion is taken to order $\ca{O}(m_\pi^3)$ and a  dipole regulator is used. }}
\label{fig:Aokic0truncDIPchisqdof}
\vspace{6mm}
\includegraphics[height=1.0\hsize,angle=90]{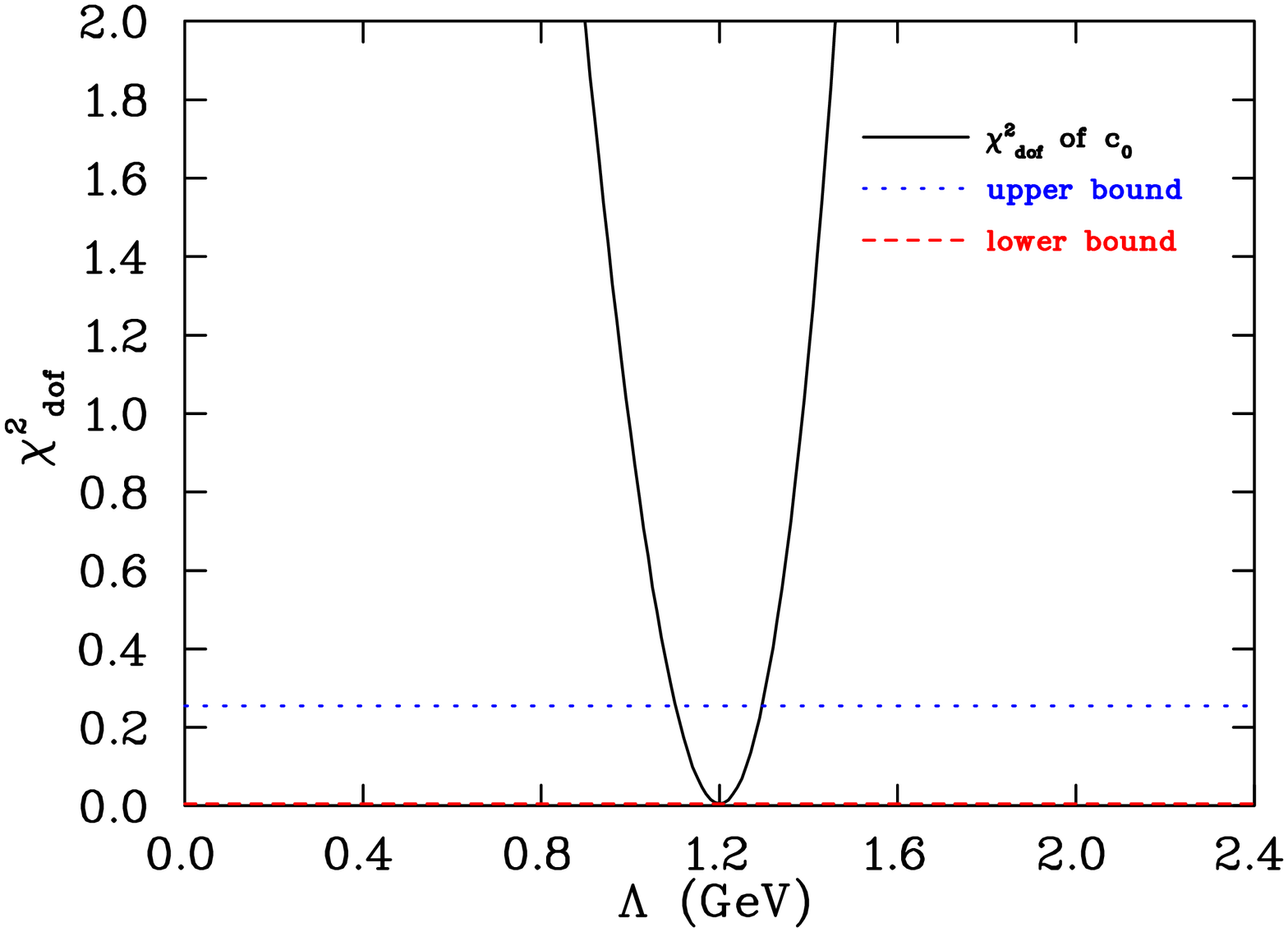}
\vspace{-3mm}
\caption{\footnotesize{ Behaviour of $\chi^2_{dof}$ for $c_0$ vs.\ $\La$, based on CP-PACS data. The chiral expansion is taken to order $\ca{O}(m_\pi^3)$ and a  dipole regulator is used. }}
\label{fig:Youngc0truncDIPchisqdof}
\end{minipage}
\hspace{12mm}
\begin{minipage}[b]{0.5\linewidth}
\centering
\includegraphics[height=1.0\hsize,angle=90]{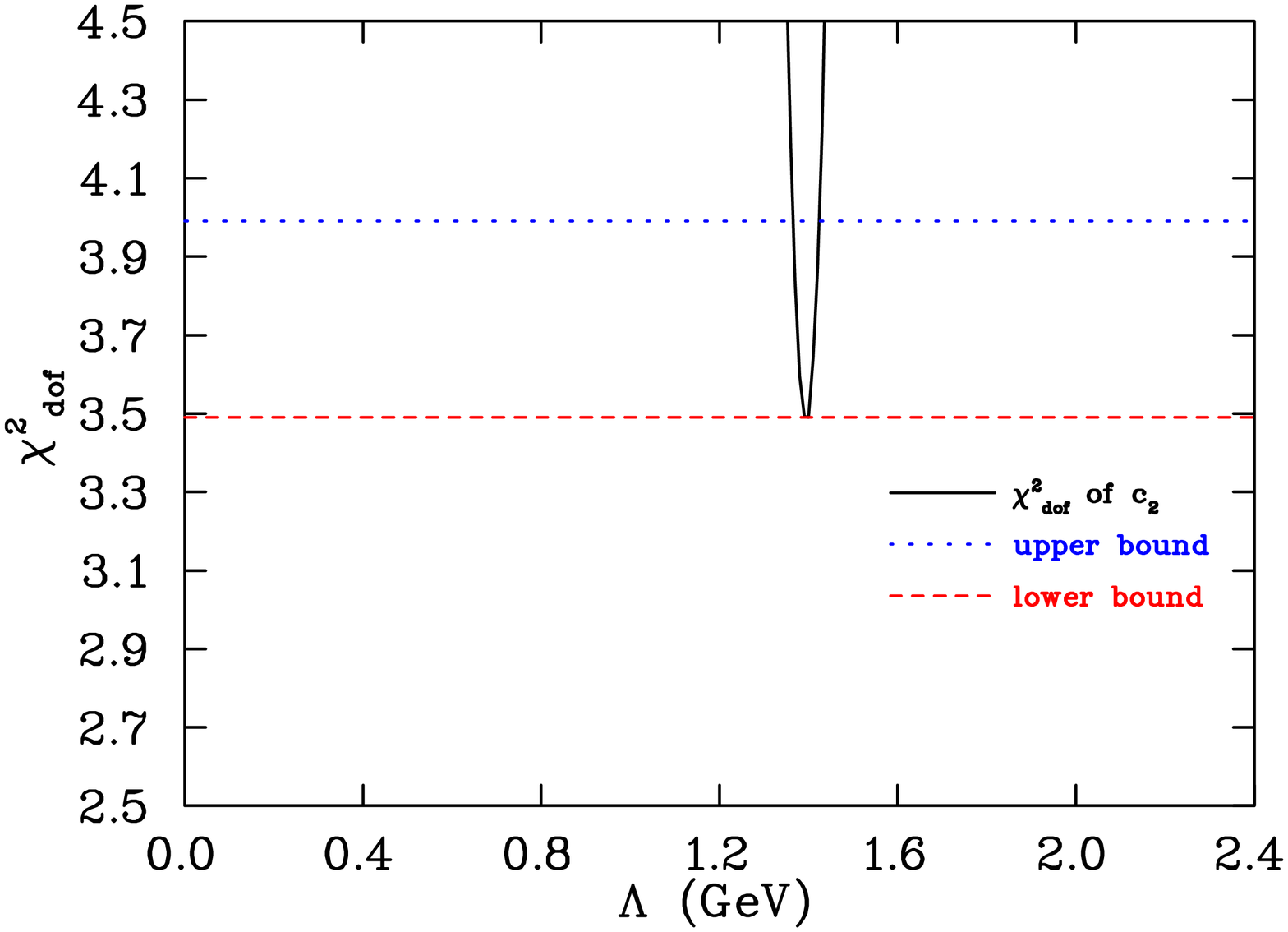}
\vspace{-3mm}
\caption{\footnotesize{ Behaviour of $\chi^2_{dof}$ for $c_2$ vs.\ $\La$, based on JLQCD data. The chiral expansion is taken to order $\ca{O}(m_\pi^3)$ and a  dipole regulator is used. }}
\label{fig:Ohkic2truncDIPchisqdof}
\vspace{6mm}
\includegraphics[height=1.0\hsize,angle=90]{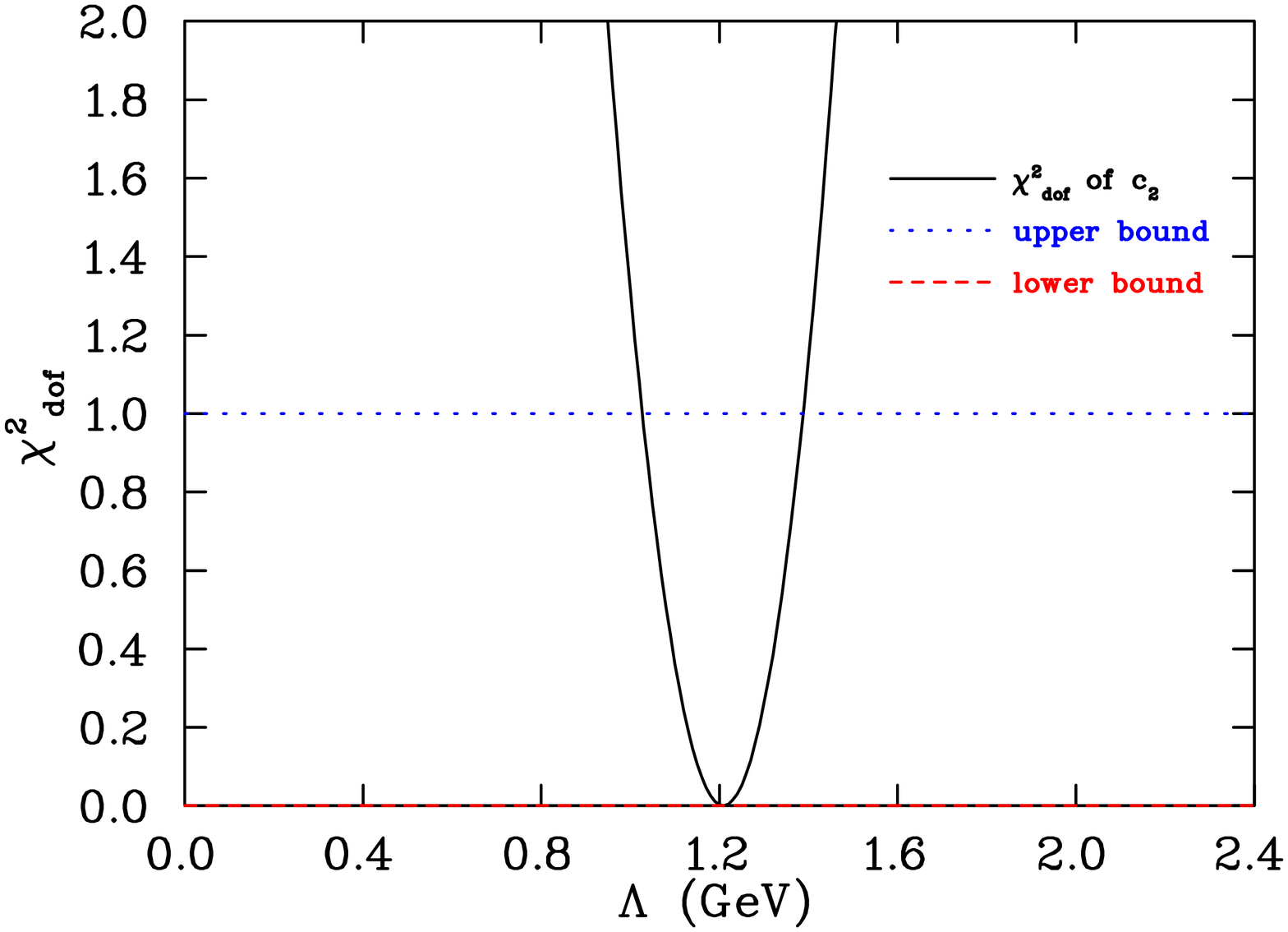}
\vspace{-3mm}
\caption{\footnotesize{ Behaviour of $\chi^2_{dof}$ for $c_2$ vs.\ $\La$, based on PACS-CS data. The chiral expansion is taken to order $\ca{O}(m_\pi^3)$ and a  dipole regulator is used. }}
\label{fig:Aokic2truncDIPchisqdof}
\vspace{6mm}
\includegraphics[height=1.0\hsize,angle=90]{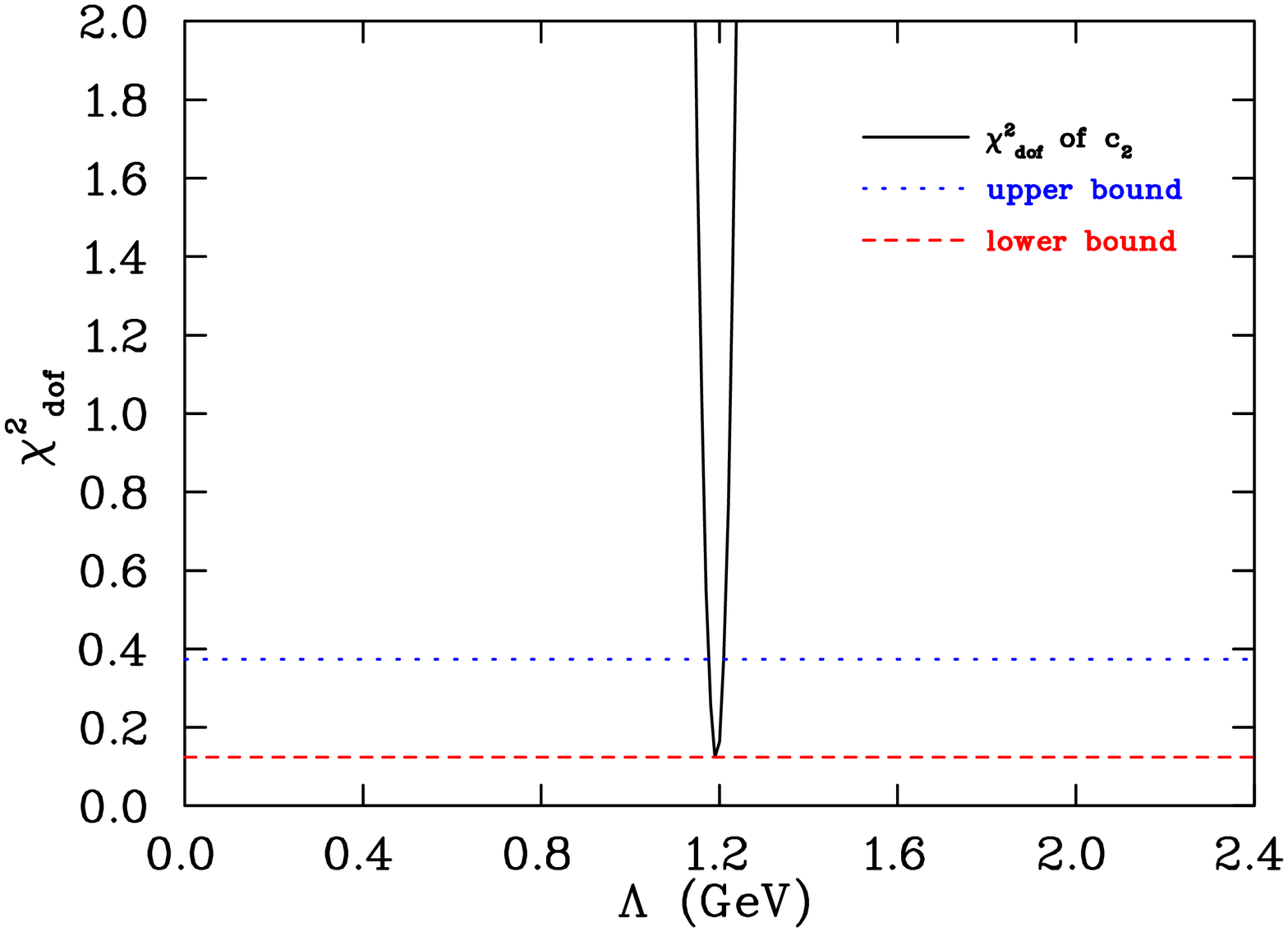}
\vspace{-3mm}
\caption{\footnotesize{ Behaviour of $\chi^2_{dof}$ for $c_2$ vs.\ $\La$, based on CP-PACS data. The chiral expansion is taken to order $\ca{O}(m_\pi^3)$ and a  dipole regulator is used. }}
\label{fig:Youngc2truncDIPchisqdof}
\end{minipage}
\end{figure}

%DOUB chi^2_{dof}
\begin{figure}
\begin{minipage}[b]{0.5\linewidth} % A minipage that covers half the page
\centering
\includegraphics[height=1.0\hsize,angle=90]{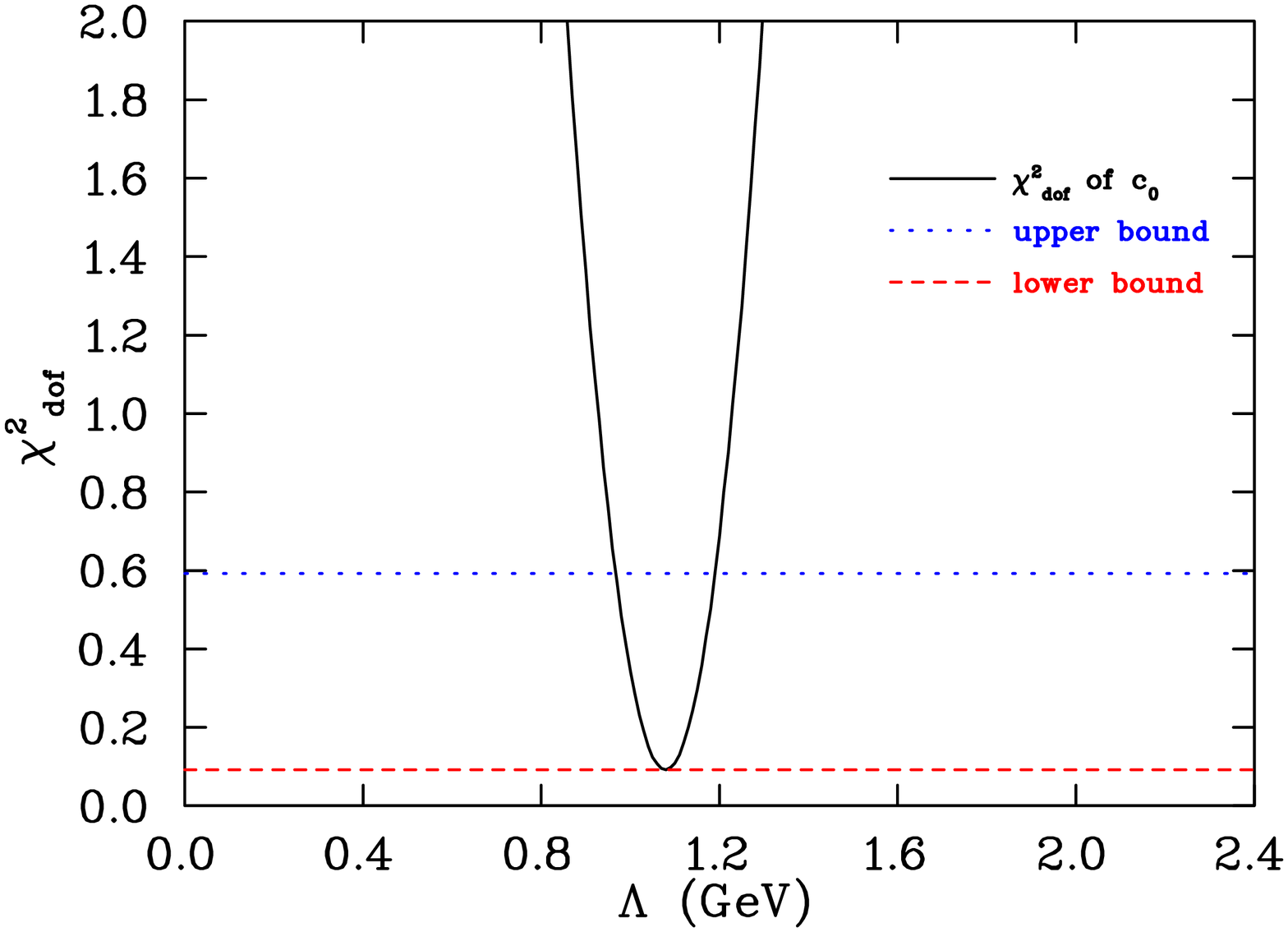}
\vspace{-3mm}
\caption{\footnotesize{ Behaviour of $\chi^2_{dof}$ for $c_0$ vs.\ $\La$, based on JLQCD data. The chiral expansion is taken to order $\ca{O}(m_\pi^3)$ and a double-dipole regulator is used. }}
\label{fig:Ohkic0truncDOUBchisqdof}
\vspace{6mm}
\includegraphics[height=1.0\hsize,angle=90]{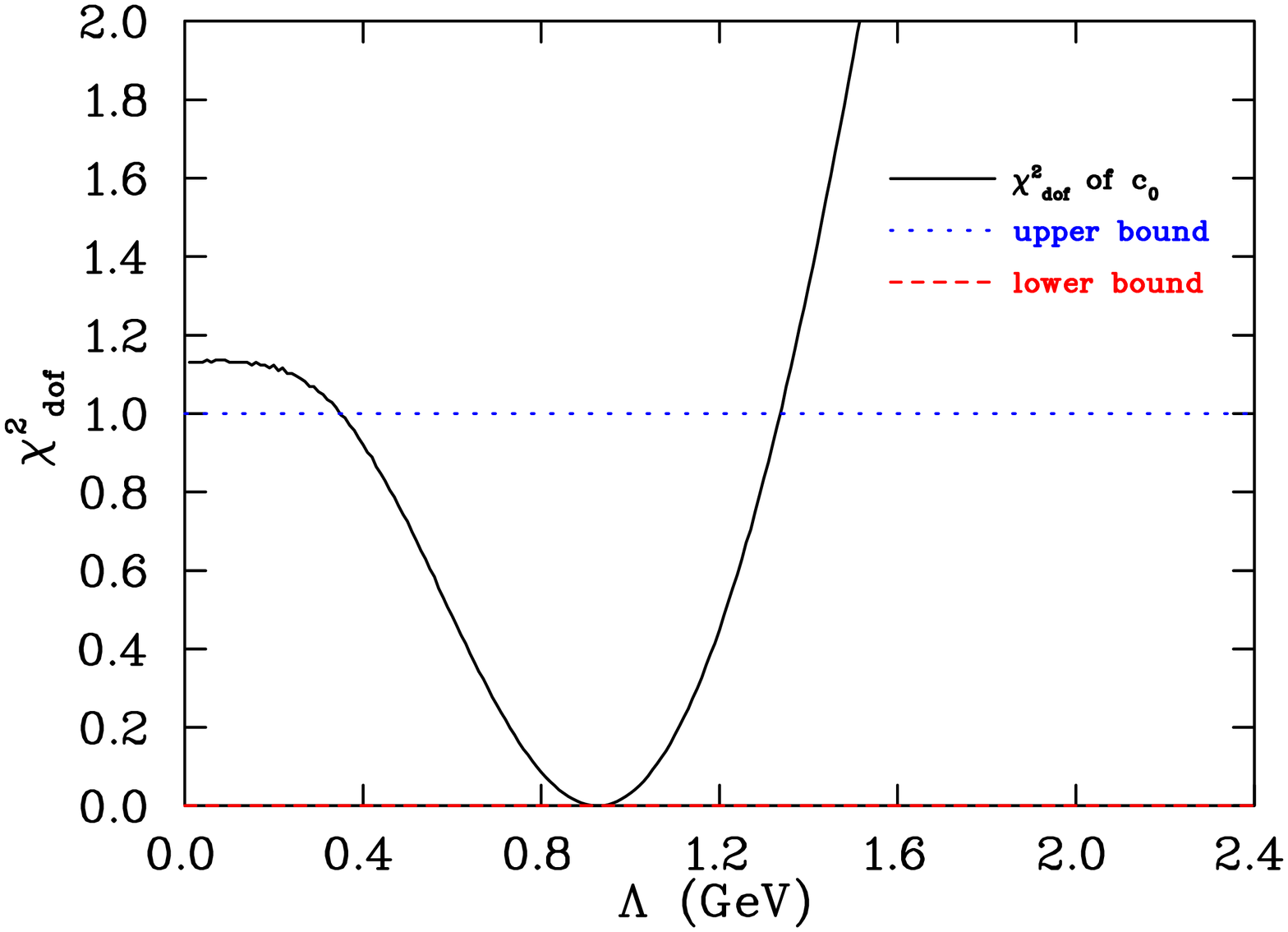}
\vspace{-3mm}
\caption{\footnotesize{ Behaviour of $\chi^2_{dof}$ for $c_0$ vs.\ $\La$, based on PACS-CS data. The chiral expansion is taken to order $\ca{O}(m_\pi^3)$ and a double-dipole regulator is used. }}
\label{fig:Aokic0truncDOUBchisqdof}
\vspace{6mm}
\includegraphics[height=1.0\hsize,angle=90]{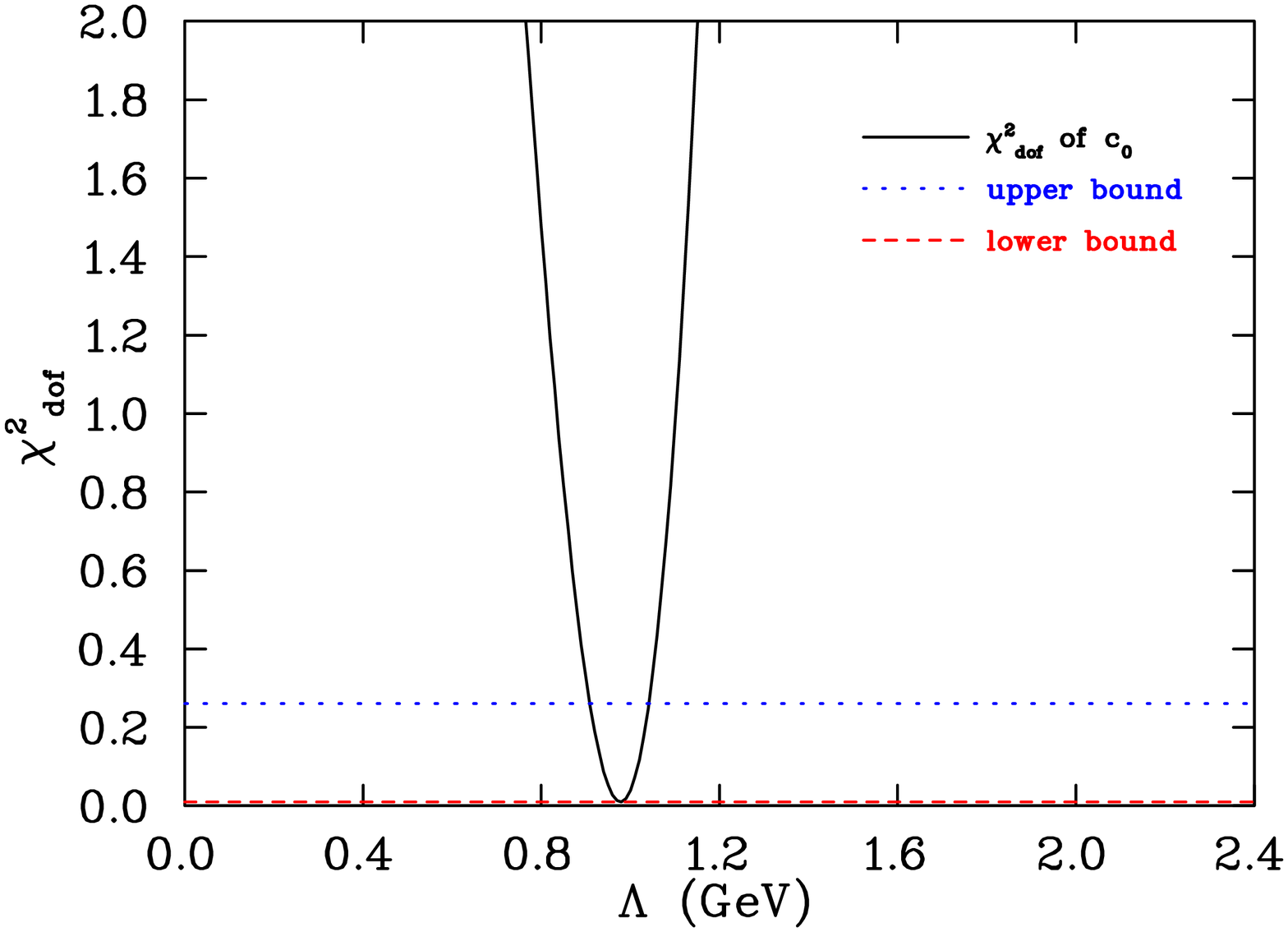}
\vspace{-3mm}
\caption{\footnotesize{ Behaviour of $\chi^2_{dof}$ for $c_0$ vs.\ $\La$, based on CP-PACS data. The chiral expansion is taken to order $\ca{O}(m_\pi^3)$ and a double-dipole regulator is used. }}
\label{fig:Youngc0truncDOUBchisqdof}
\end{minipage}
\hspace{12mm}
\begin{minipage}[b]{0.5\linewidth}
\centering
\includegraphics[height=1.0\hsize,angle=90]{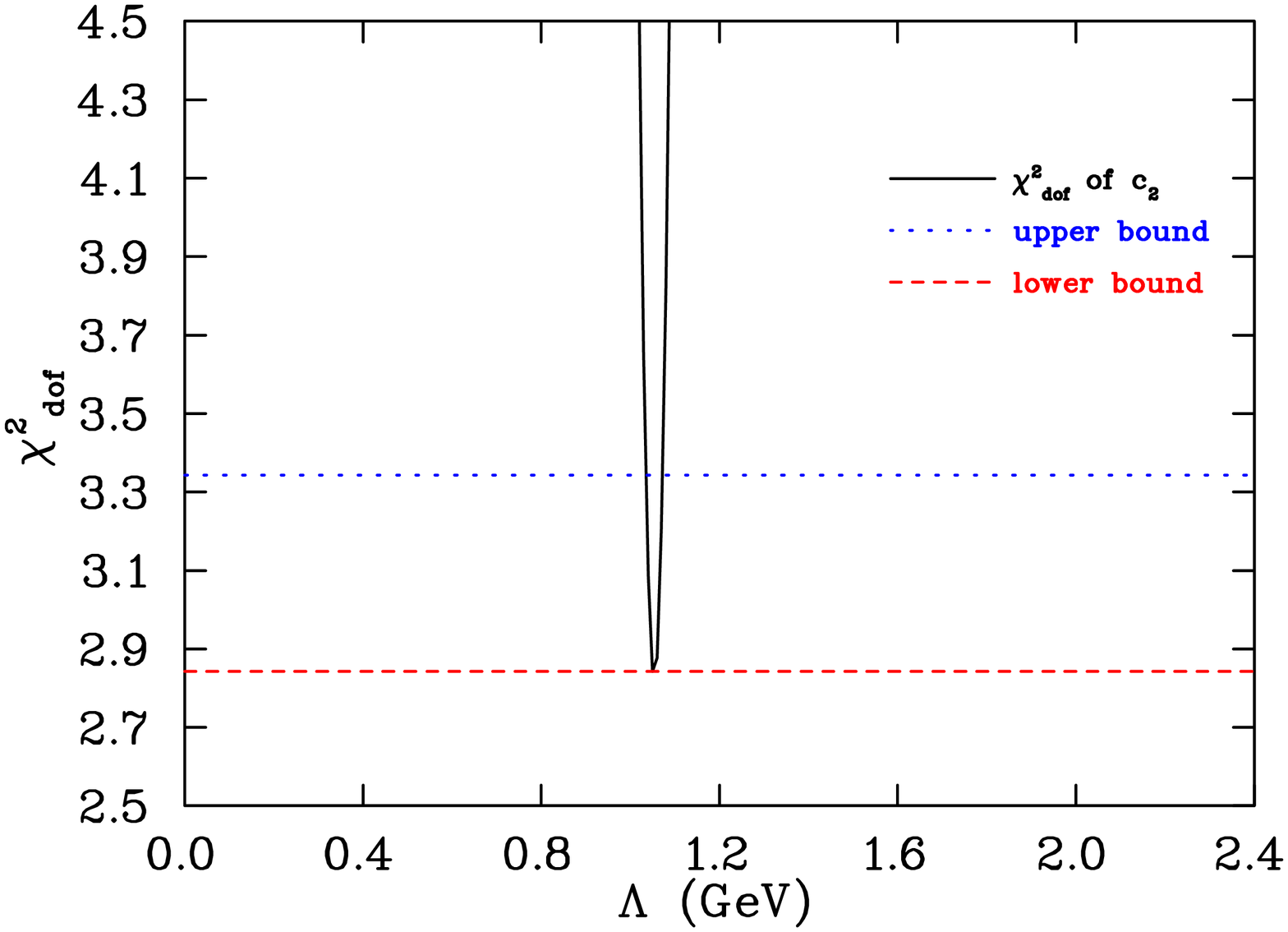}
\vspace{-3mm}
\caption{\footnotesize{ Behaviour of $\chi^2_{dof}$ for $c_2$ vs.\ $\La$, based on JLQCD data. The chiral expansion is taken to order $\ca{O}(m_\pi^3)$ and a double-dipole regulator is used. }}
\label{fig:Ohkic2truncDOUBchisqdof}
\vspace{6mm}
\includegraphics[height=1.0\hsize,angle=90]{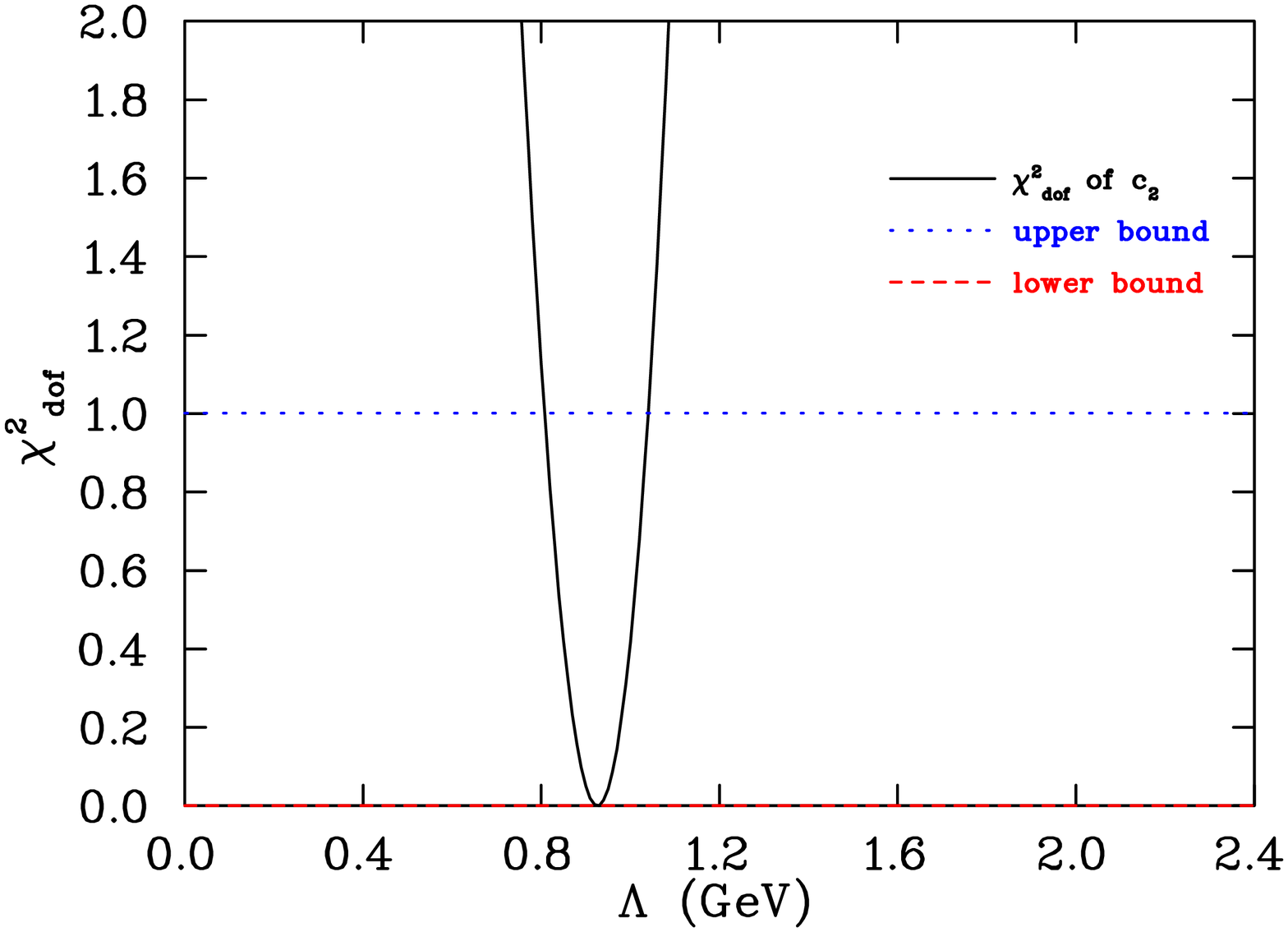}
\vspace{-3mm}
\caption{\footnotesize{ Behaviour of $\chi^2_{dof}$ for $c_2$ vs.\ $\La$, based on PACS-CS data. The chiral expansion is taken to order $\ca{O}(m_\pi^3)$ and a double-dipole regulator is used. }}
\label{fig:Aokic2truncDOUBchisqdof}
\vspace{6mm}
\includegraphics[height=1.0\hsize,angle=90]{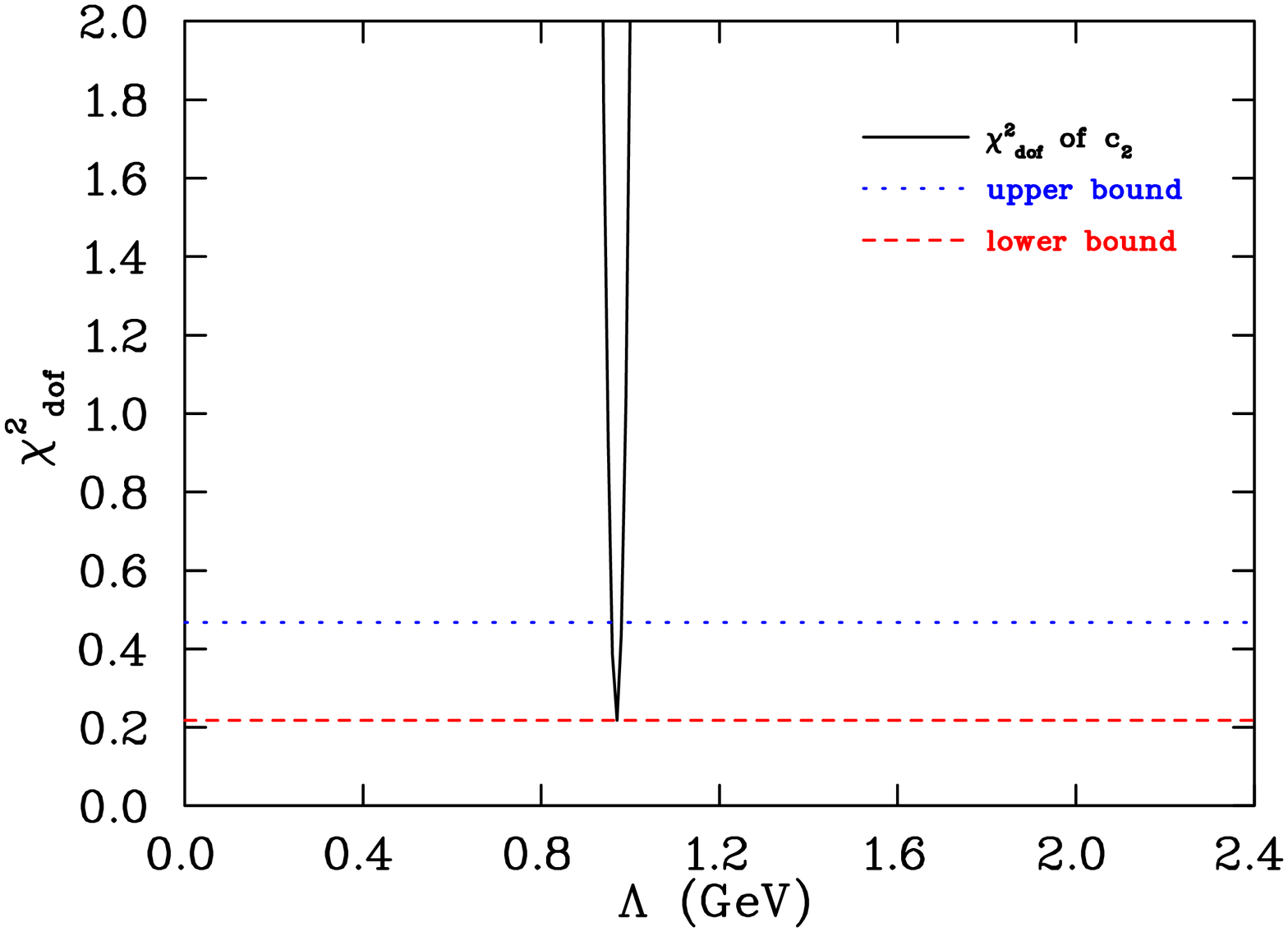}
\vspace{-3mm}
\caption{\footnotesize{ Behaviour of $\chi^2_{dof}$ for $c_2$ vs.\ $\La$, based on CP-PACS data. The chiral expansion is taken to order $\ca{O}(m_\pi^3)$ and a double-dipole regulator is used. }}
\label{fig:Youngc2truncDOUBchisqdof}
\end{minipage}
\end{figure}

%TRIP chi^2_{dof}
\begin{figure}
\begin{minipage}[b]{0.5\linewidth} % A minipage that covers half the page
\centering
\includegraphics[height=1.0\hsize,angle=90]{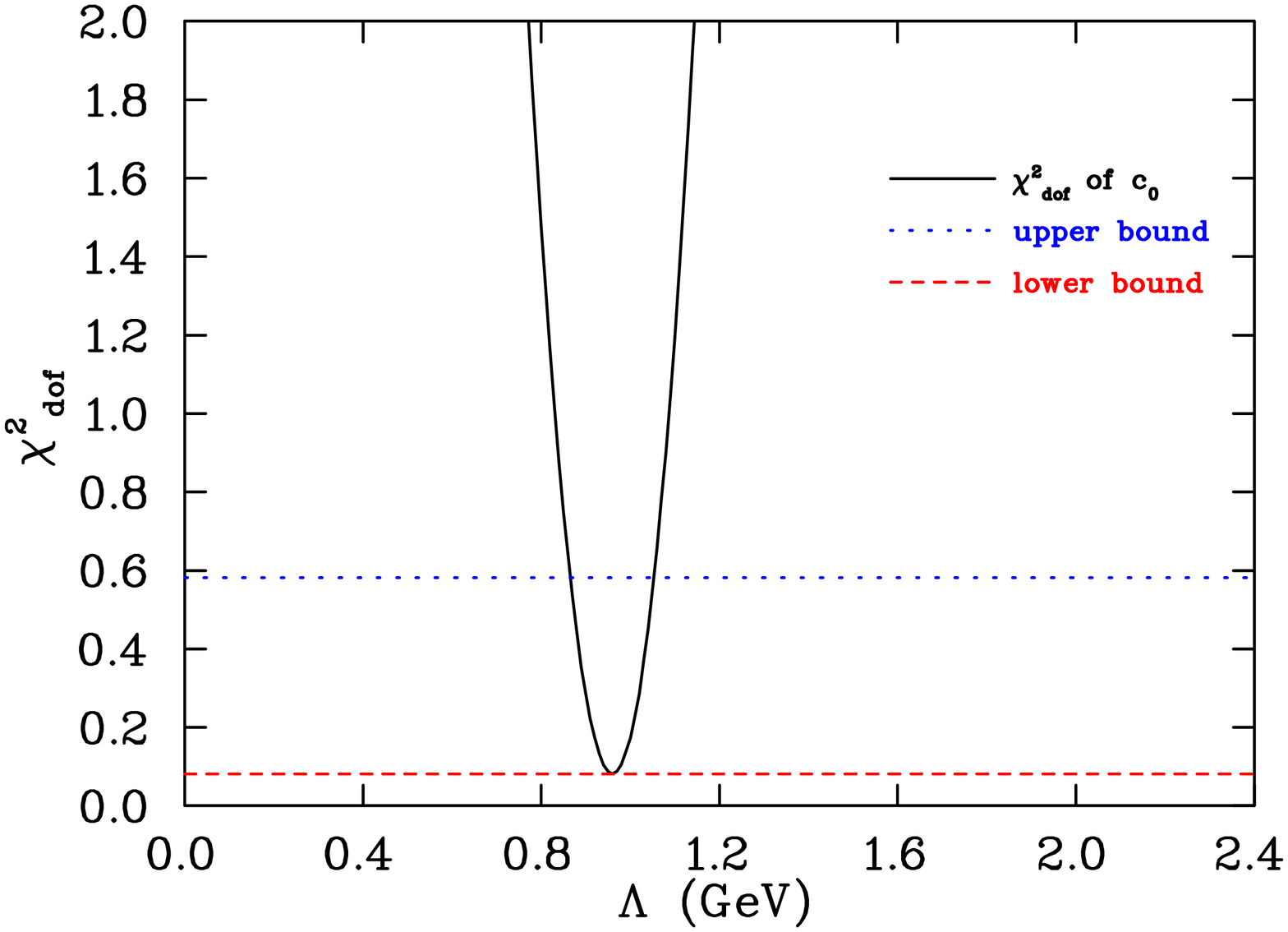}
\vspace{-3mm}
\caption{\footnotesize{ Behaviour of $\chi^2_{dof}$ for $c_0$ vs.\ $\La$, based on JLQCD data. The chiral expansion is taken to order $\ca{O}(m_\pi^3)$ and a triple-dipole regulator is used. }}
\label{fig:Ohkic0truncTRIPchisqdof}
\vspace{6mm}
\includegraphics[height=1.0\hsize,angle=90]{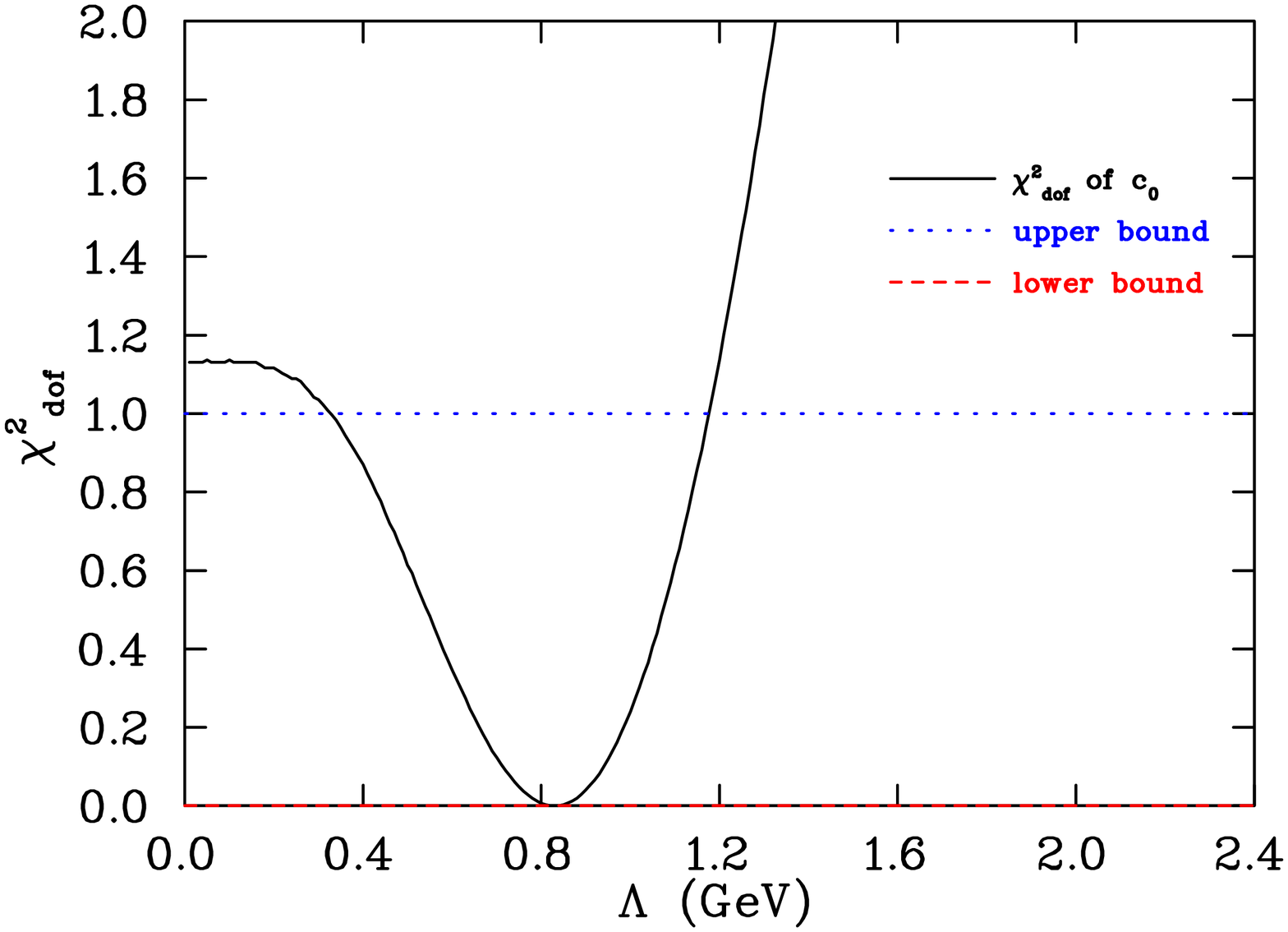}
\vspace{-3mm}
\caption{\footnotesize{ Behaviour of $\chi^2_{dof}$ for $c_0$ vs.\ $\La$, based on PACS-CS data. The chiral expansion is taken to order $\ca{O}(m_\pi^3)$ and a triple-dipole regulator is used. }}
\label{fig:Aokic0truncTRIPchisqdof}
\vspace{6mm}
\includegraphics[height=1.0\hsize,angle=90]{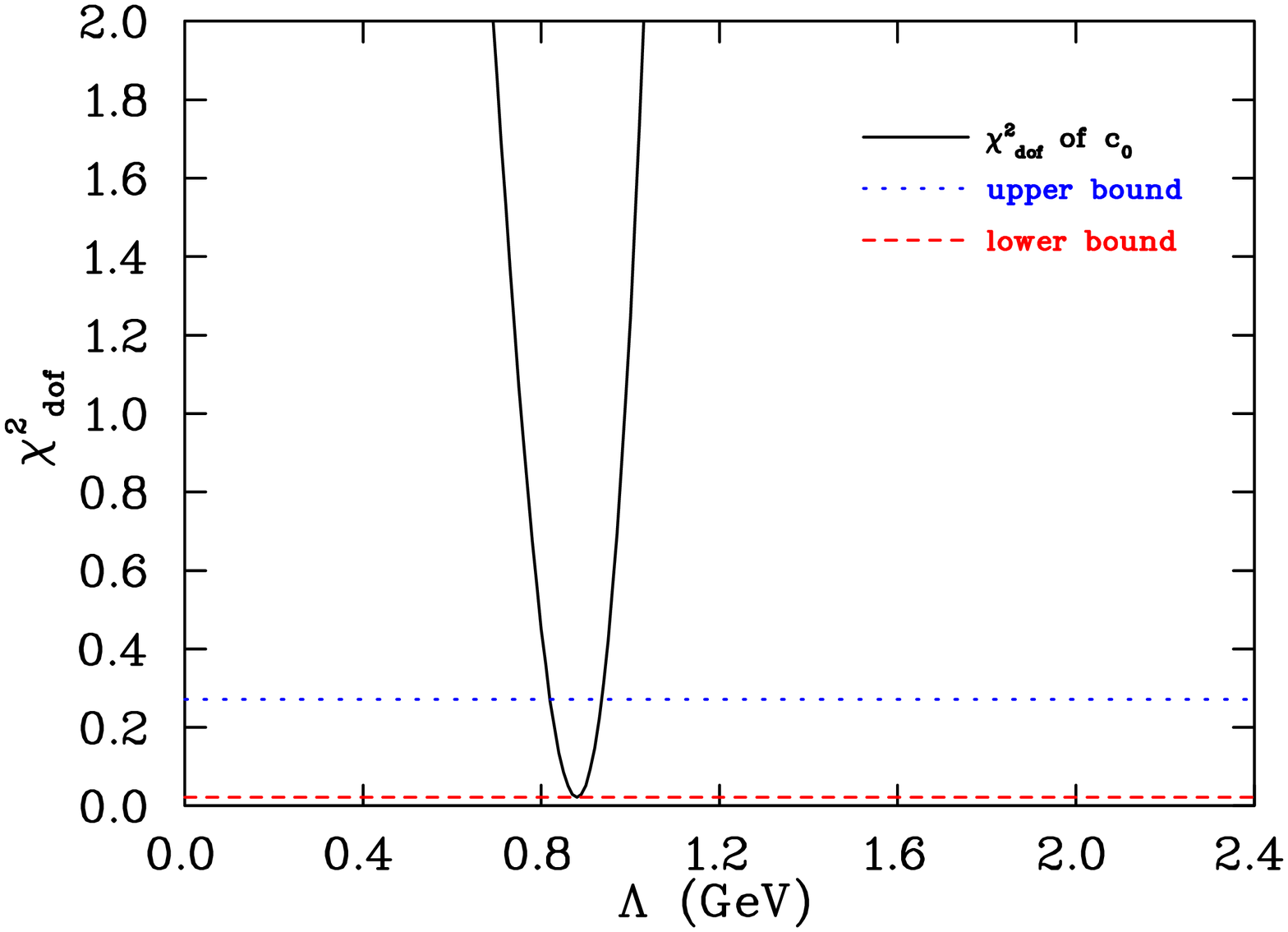}
\vspace{-3mm}
\caption{\footnotesize{ Behaviour of $\chi^2_{dof}$ for $c_0$ vs.\ $\La$, based on CP-PACS data. The chiral expansion is taken to order $\ca{O}(m_\pi^3)$ and a triple-dipole regulator is used. }}
\label{fig:Youngc0truncTRIPchisqdof}
\end{minipage}
\hspace{12mm}
\begin{minipage}[b]{0.5\linewidth}
\centering
\includegraphics[height=1.0\hsize,angle=90]{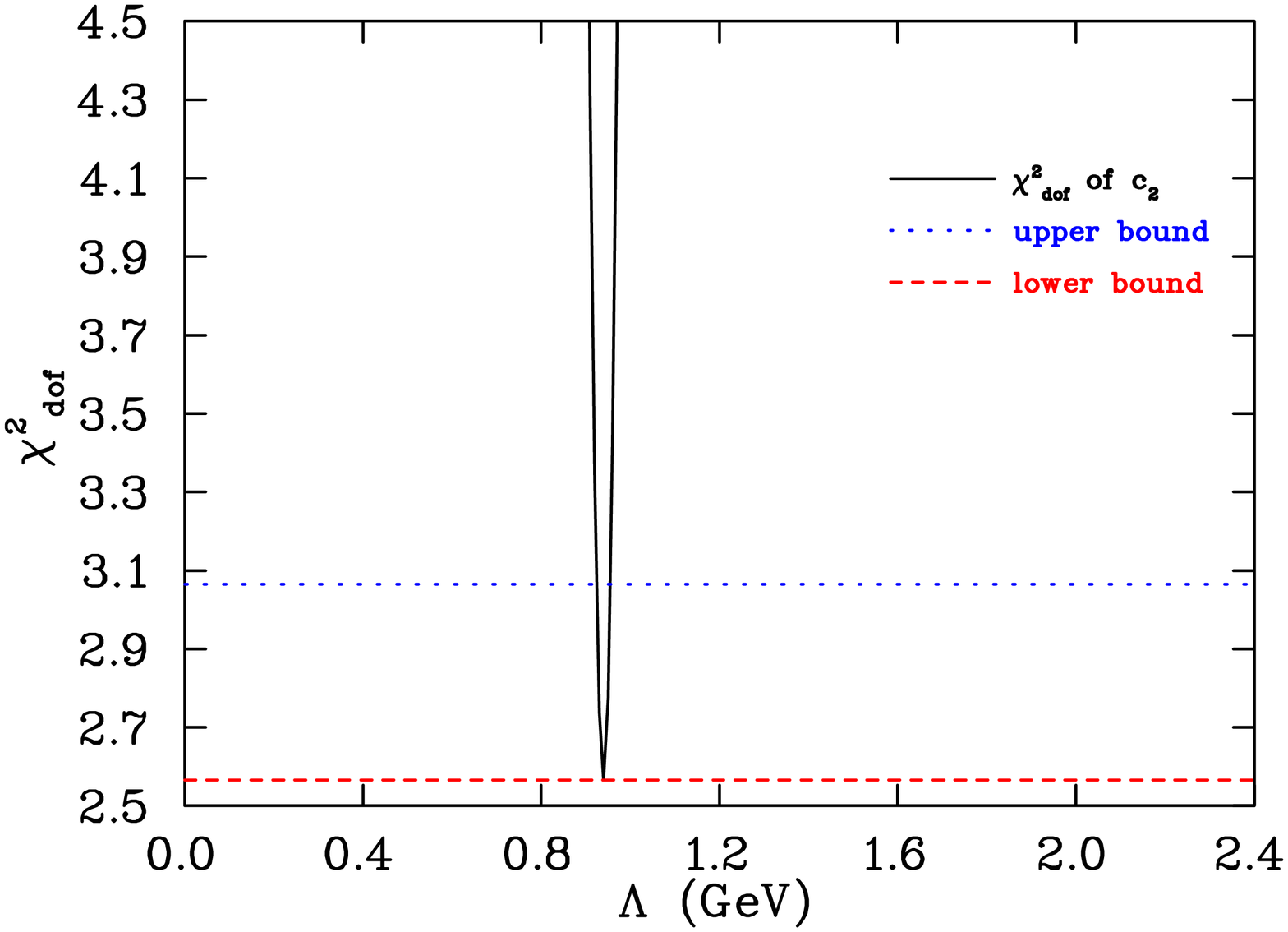}
\vspace{-3mm}
\caption{\footnotesize{ Behaviour of $\chi^2_{dof}$ for $c_2$ vs.\ $\La$, based on JLQCD data. The chiral expansion is taken to order $\ca{O}(m_\pi^3)$ and a triple-dipole regulator is used. }}
\label{fig:Ohkic2truncTRIPchisqdof}
\vspace{6mm}
\includegraphics[height=1.0\hsize,angle=90]{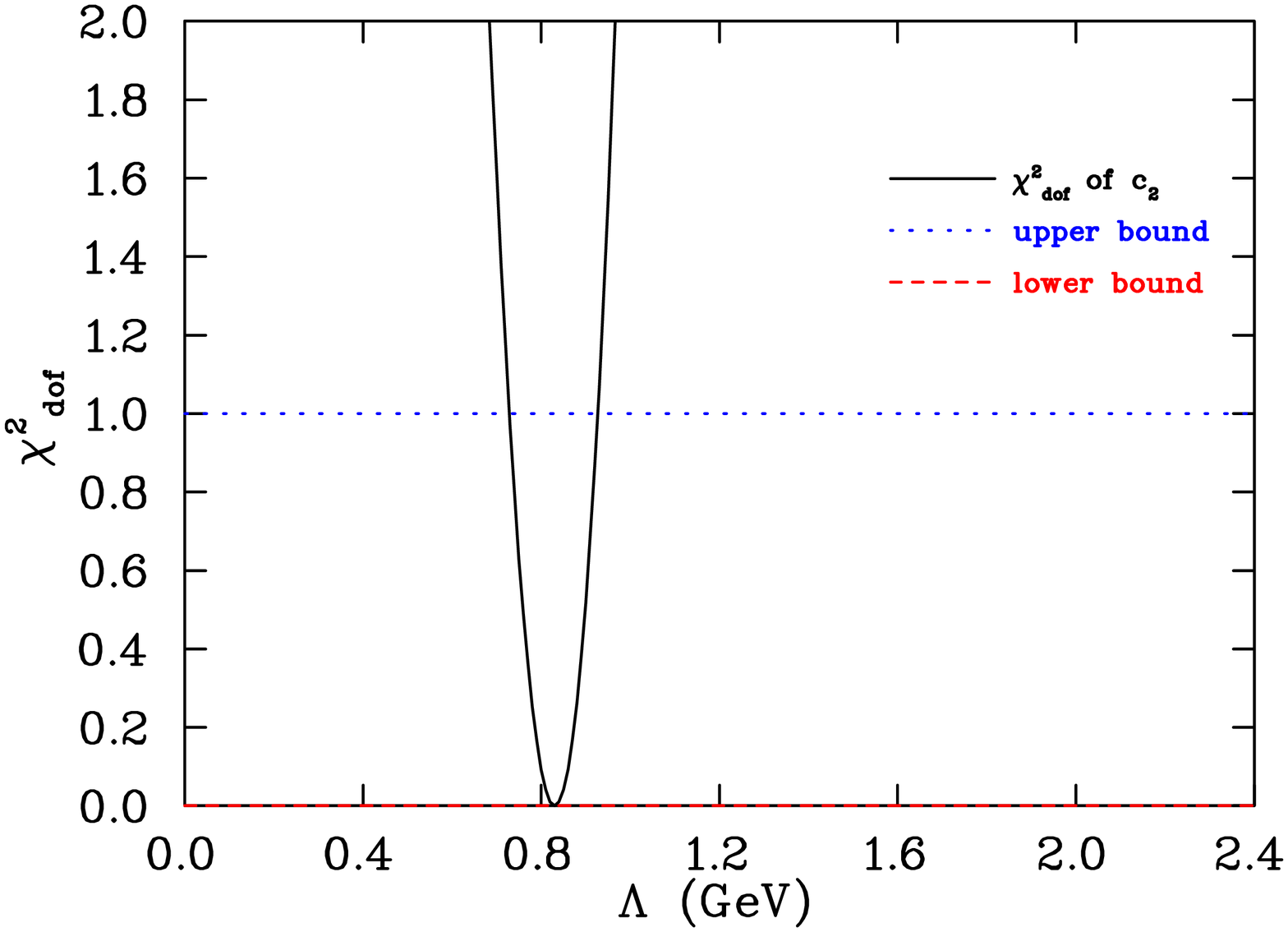}
\vspace{-3mm}
\caption{\footnotesize{ Behaviour of $\chi^2_{dof}$ for $c_2$ vs.\ $\La$, based on PACS-CS data. The chiral expansion is taken to order $\ca{O}(m_\pi^3)$ and a triple-dipole regulator is used. }}
\label{fig:Aokic2truncTRIPchisqdof}
\vspace{6mm}
\includegraphics[height=1.0\hsize,angle=90]{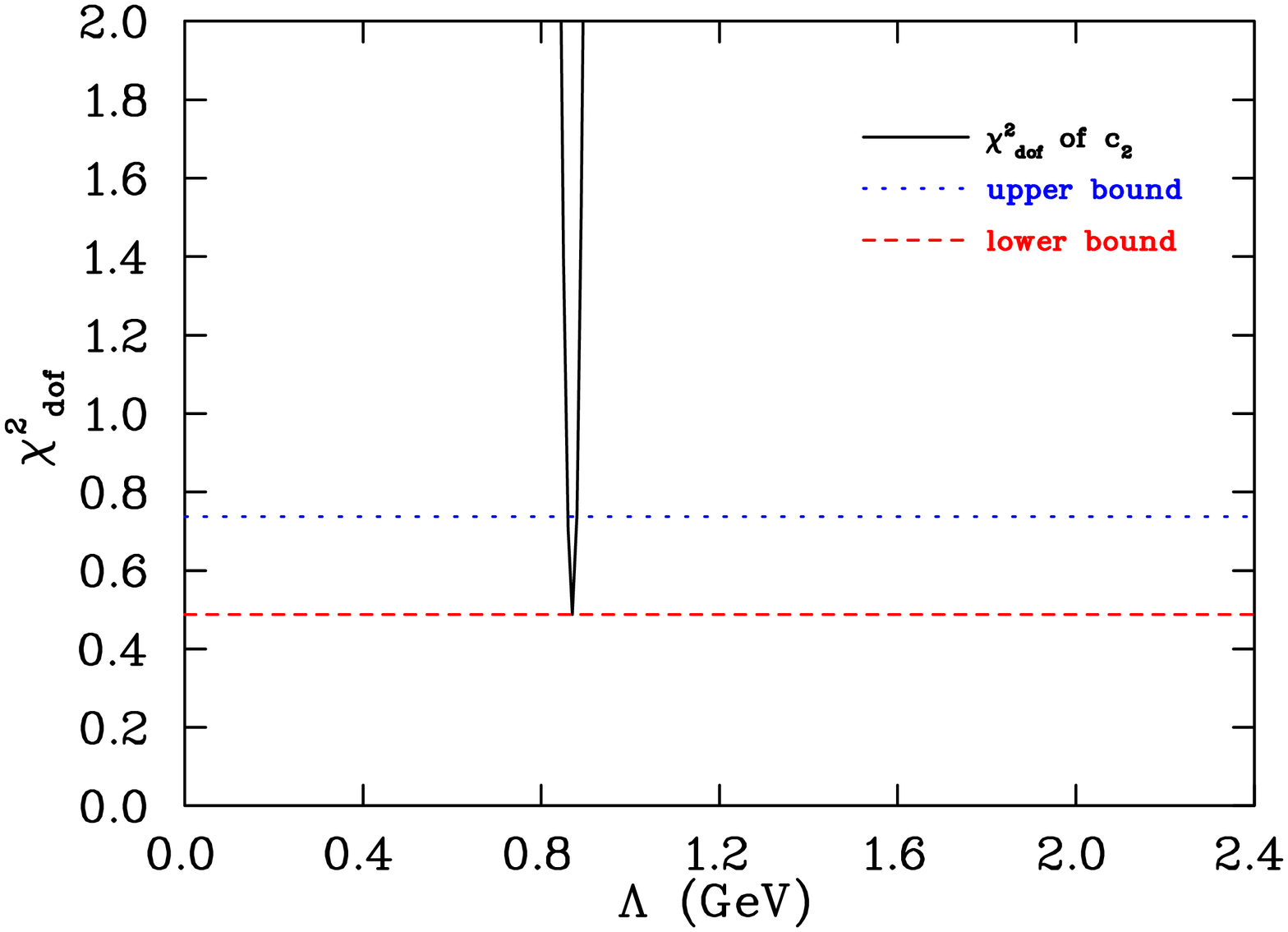}
\vspace{-3mm}
\caption{\footnotesize{ Behaviour of $\chi^2_{dof}$ for $c_2$ vs.\ $\La$, based on CP-PACS data. The chiral expansion is taken to order $\ca{O}(m_\pi^3)$ and a triple-dipole regulator is used. }}
\label{fig:Youngc2truncTRIPchisqdof}
\end{minipage}
\end{figure}

The first step is to plot $\chi^2_{dof}$ 
 against a variety of regularization scales. % $\La$. 
%It is of primary interest to
%what extent these curves match. Therefore,
% a $\chi^2_{dof}$ should be constructed, 
%where $dof$ equals the number of curves 
%on each plot minus one for the best fit value of $c_0$ or $c_2$, denoted
% by $c^T$ in the following.
% This also serves to quantify the constraint on the optimal regulator
% $\La^\ro{scale}$.
%The average value of each coefficient 
The value of 
$\bar{c}$ is given by the
weighted mean formula, evaluated separately
 for each renormalized coefficient $c$ (with error $\de c$)
 and regularization scale $\La$:
\eqb
\label{eqn:chisqwm}
\bar{c}(\La) = \f{\sum_{i=1}^{n}c(i\,;\La)/{{(\de c(i\,;\La))}^2}}
{\sum_{j=1}^{n} 1 / {(\de c(j\,;\La))}^2}\,.
\eqe
 The following $\chi^2_{dof}$ treats relevant %data 
degrees of freedom 
as the %curves 
 extracted chiral coefficients with differing values of 
 $m_{\pi,\ro{max}}^2$: 
\eqb
\label{eqn:chisq}
\chi^2_{dof} = \f{1}{n-1} \sum_{i=1}^{n} \f{{(c(i\,;\La) - \bar{c}(\La))}^2}
{{(\de c(i\,;\La))}^2}\,,
\eqe
that is,  $i$ corresponds to %data sets 
 fits with differing values of 
$m_{\pi,\ro{max}}^2$. 

The $\chi^2_{dof}$ can be calculated as a function of 
the regularization scale $\La$ for each of the renormalization plots of
 Figures \ref{fig:Ohkic0truncDIP} through 
\ref{fig:Youngc2truncTRIP}. This will indicate the spread of the extrapolated 
values at each value of $\La$. 
  In the case of the PACS-CS data,
 the minimum of the $\chi^2_{dof}$
curve will be %centered 
at the intersection point of the two curves. In the case of the 
JLQCD and CP-PACS data, 
%there appears to be a single intersection point on each plot,
%but in fact there are multiple intersections over a very small window of $\La$.
with more than two curves, there is an interaction region on each plot, 
over a narrow window of $\La$. 
The minima of $\chi^2_{dof}$ will indicate the value of $\La$ that 
obtains the best agreement among the renormalization flow curves. 
This central value of $\La$ will be taken to be the optimal regularization 
scale.
The upper and lower bounds of $\Lambda$ obey 
 the condition $\chi^2_{dof} < \chi^2_{dof, min} + 1/(dof)$. 
For each of the low-energy coefficients $c_0$ and $c_2$, 
the $\chi^2_{dof}$ curves for a dipole regulator are shown in
Figures \ref{fig:Ohkic0truncDIPchisqdof}
 through \ref{fig:Youngc2truncDIPchisqdof},
the $\chi^2_{dof}$ curves for the double-dipole case are shown in
Figures \ref{fig:Ohkic0truncDOUBchisqdof}
 through \ref{fig:Youngc2truncDOUBchisqdof}
and the $\chi^2_{dof}$ curves for the triple-dipole are shown in
Figures \ref{fig:Ohkic0truncTRIPchisqdof}
 through \ref{fig:Youngc2truncTRIPchisqdof}.
These plots indicate that there exists a statistically significant 
optimal regularization scale at this chiral order, for these data sets. 
Furthermore, for each data set and regulator functional form, 
there is an agreement between the $c_0$ and $c_2$ analyses as to the 
value of this optimal scale. This provides evidence of the 
existence of an intrinsic scale embedded in that lattice data. 
%   is remarkable, and indicative of the
% existence of an intrinsic scale in the data.

\subsection{Effects at Higher Chiral Order}
\label{subsect:higher}
\begin{figure}
\begin{minipage}[b]{0.5\linewidth} % A minipage that covers half the page
\centering
\includegraphics[height=1.0\hsize,angle=90]{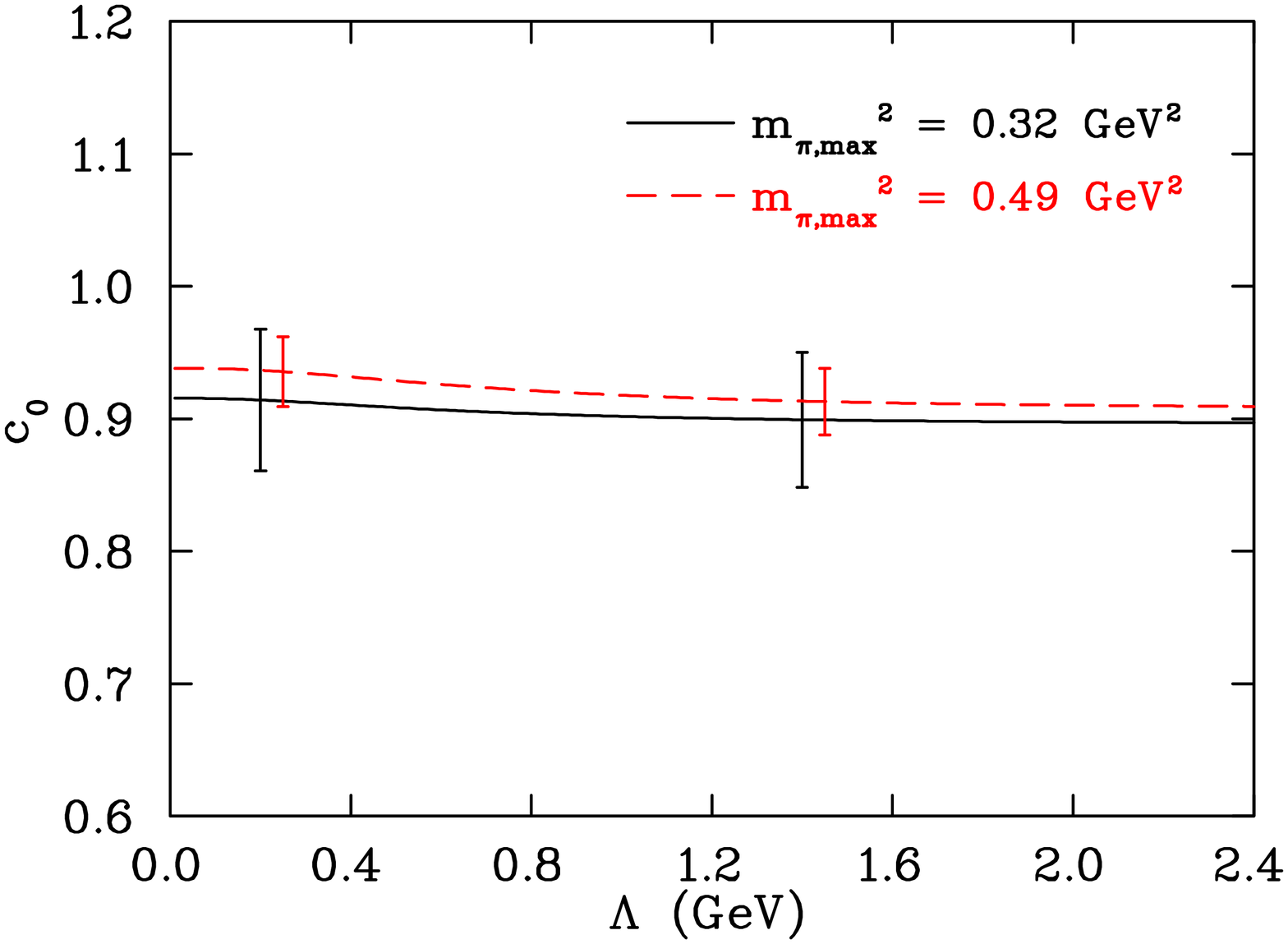}
\vspace{-3mm}
\caption{\footnotesize{ Behaviour of $c_0$ vs.\ $\La$, based on PACS-CS data. The chiral expansion is taken to order $\ca{O}(m_\pi^4)$ and a  dipole regulator is used. A few points are selected to indicate the general size of the statistical error bars.}}
\label{fig:Aokic0DIP}
\vspace{6mm}
\includegraphics[height=1.0\hsize,angle=90]{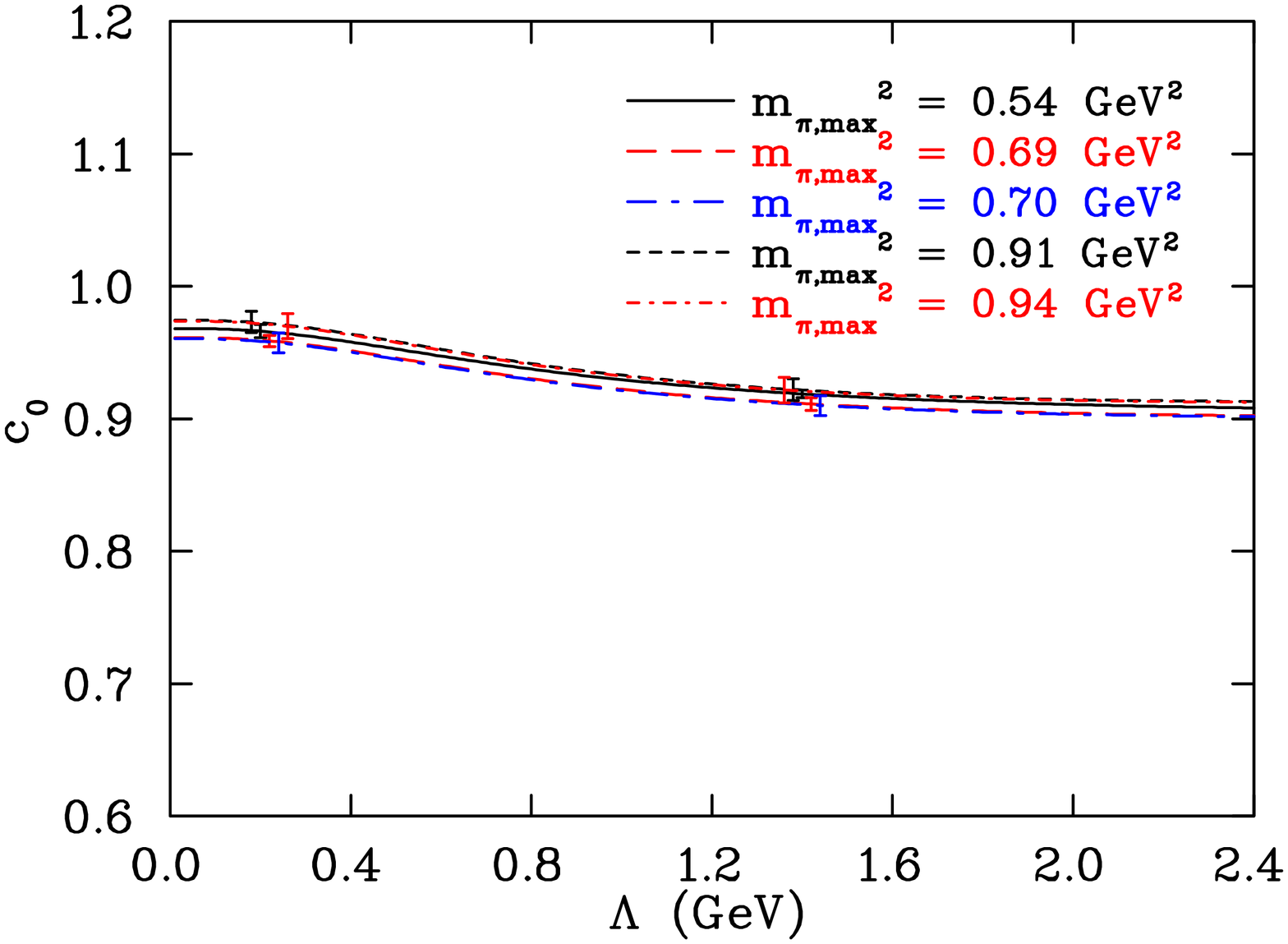}
\vspace{-3mm}
\caption{\footnotesize{ Behaviour of $c_0$ vs.\ $\La$, based on CP-PACS data. The chiral expansion is taken to order $\ca{O}(m_\pi^4)$ and a  dipole regulator is used. A few points are selected to indicate the general size of the statistical error bars.}}
\label{fig:Youngc0DIP}
\end{minipage}
\hspace{12mm}
\begin{minipage}[b]{0.5\linewidth}
\centering
\vspace{6mm}
\includegraphics[height=1.0\hsize,angle=90]{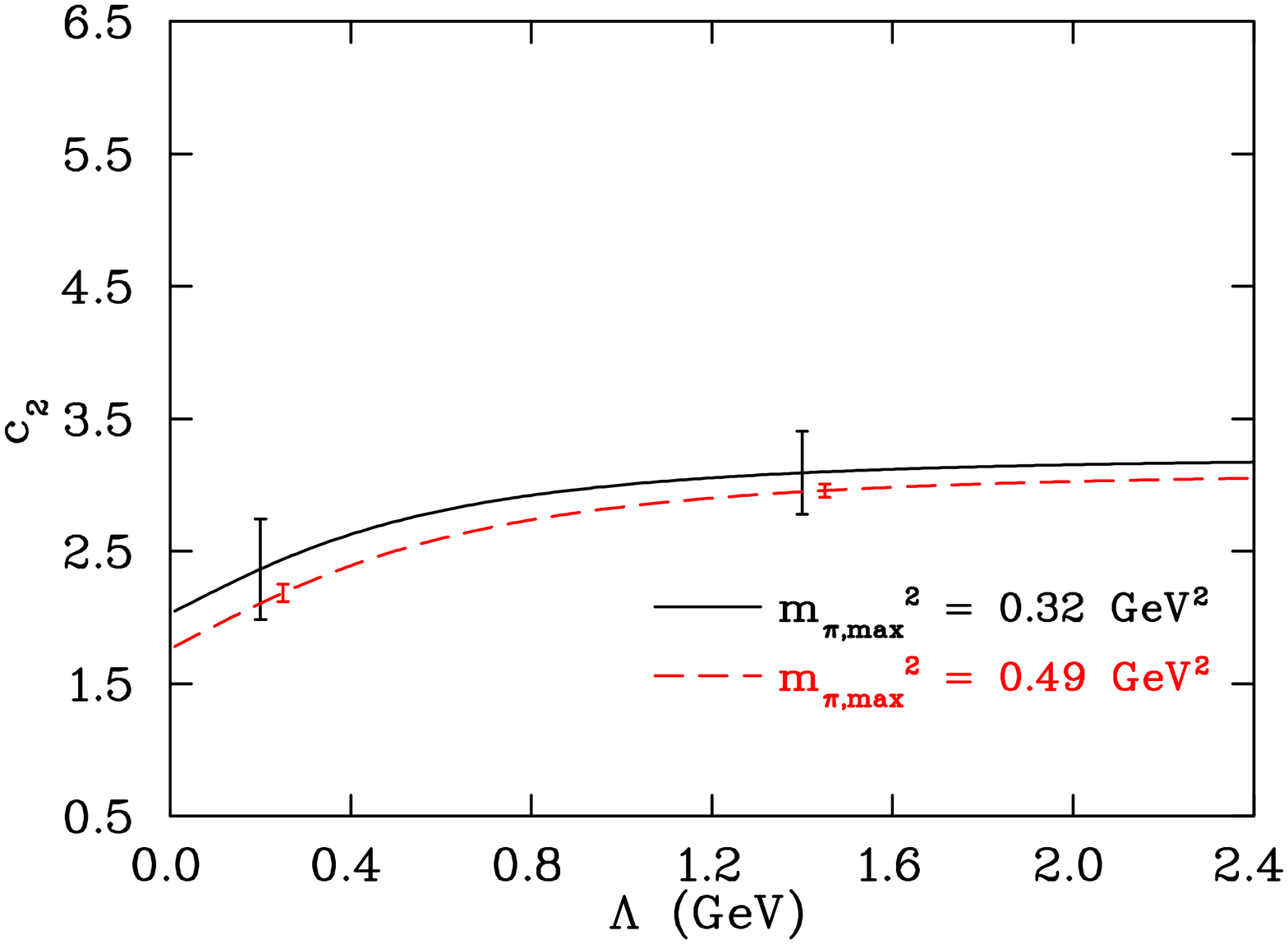}
\vspace{-3mm}
\caption{\footnotesize{ Behaviour of $c_2$ vs.\ $\La$, based on PACS-CS data. The chiral expansion is taken to order $\ca{O}(m_\pi^4)$ and a  dipole regulator is used. A few points are selected to indicate the general size of the statistical error bars.}}
\label{fig:Aokic2DIP}
\vspace{6mm}
\includegraphics[height=1.0\hsize,angle=90]{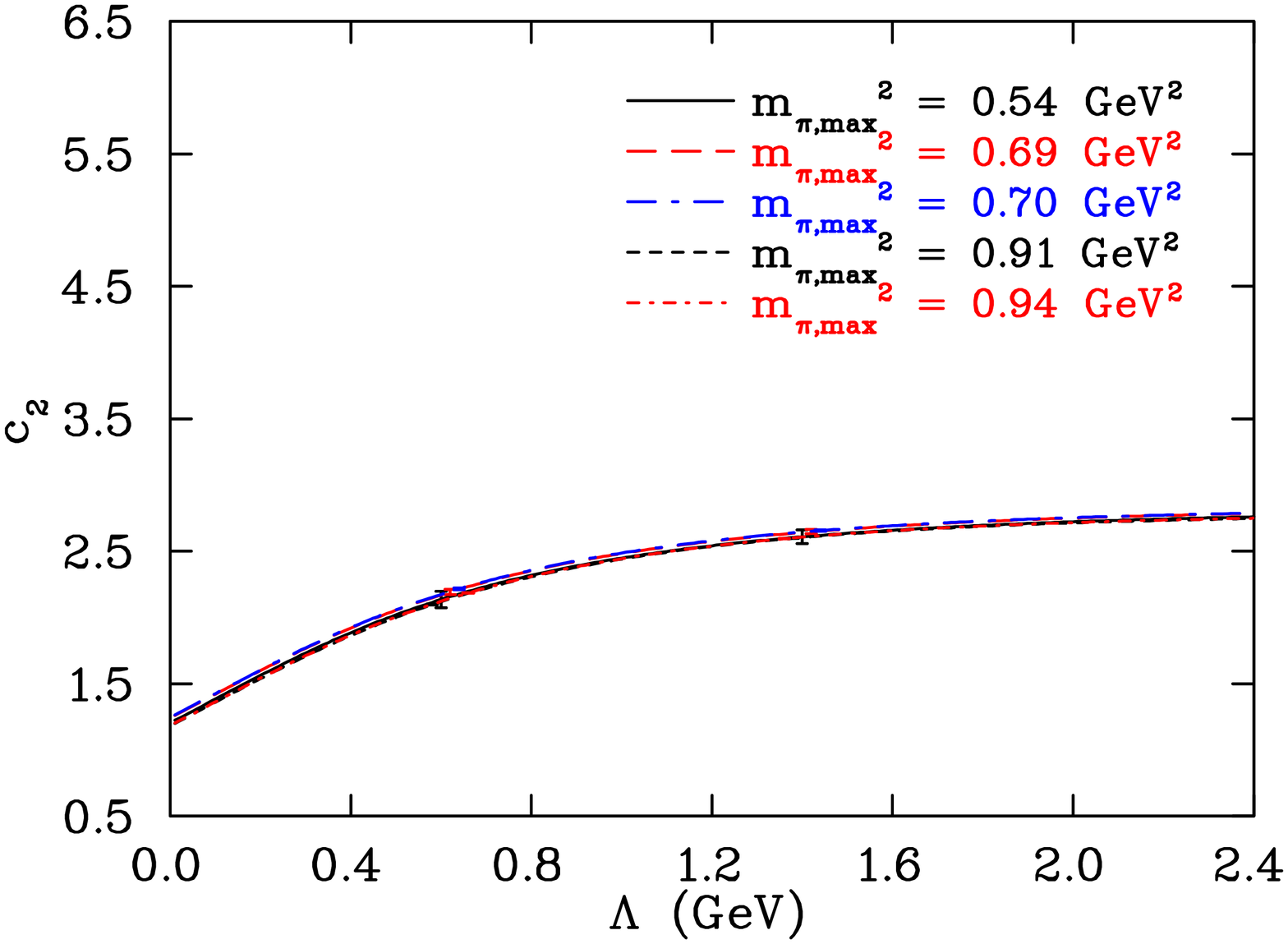}
\vspace{-3mm}
\caption{\footnotesize{ Behaviour of $c_2$ vs.\ $\La$, based on CP-PACS data. The chiral expansion is taken to order $\ca{O}(m_\pi^4)$ and a  dipole regulator is used. A few points are selected to indicate the general size of the statistical error bars.}}
\label{fig:Youngc2DIP}
\end{minipage}
\end{figure}

Consider the %renormalization of 
determination of $c_0$ and $c_2$ as a function of the 
regularization scale $\La$,
for a higher chiral order $\ca{O}(m_\pi^4\,\ro{log}\,m_\pi)$.
As an example, the results for PACS-CS and CP-PACS data are shown in 
Figures \ref{fig:Aokic0DIP} through \ref{fig:Youngc2DIP}.
In this case, no clear intersection points in the renormalization
 curves can be found, and so one is unable to specify an optimal regularization 
scale.
 This certainly should be the case 
when working with data entirely within the PCR,  
 because all renormalization procedures would be equivalent
 (to a prescribed level of accuracy) 
and so there could be no optimal scale.
 It has been demonstrated, %is known, 
however, that %this is not the case for 
the data sets used in this study extend beyond the PCR. 
 This is further verified 
 by considering the evident scale-dependence of $c_0$ and $c_2$ in 
 Figures \ref{fig:Aokic0DIP} through \ref{fig:Youngc2DIP}.
 The fact that $c_0$ and $c_2$ change over the range of $\La$ values indicates
 that the data are not inside the PCR where the renormalization must be 
scale-independent.
%%%OF NOTE
Furthermore, since no preferred scale is revealed, any choice of $\La$ appears 
 equivalent at this order. While it is encouraging that the scheme-dependence
 has been weakened by working to higher order, it must be recognized that 
there is a systematic error associated with the choice of $\La$. In 
 the case of 
the CP-PACS results shown in Figures \ref{fig:Youngc0DIP} 
and \ref{fig:Youngc2DIP}, it can be seen that the 
statistical errors are substantially smaller than the systematic error 
associated with a characteristic range, $\La_\ro{lower}<\La<\infty$, where
 $\La_\ro{lower}$ is the lowest reasonable value of $\La$, taken to be 
$0.6$, $0.4$ and $0.3$ GeV for the dipole, double-dipole and 
 triple-dipole regulator, respectively, as discussed in 
Section \ref{sec:lowerbound}. 

Since it is difficult to identify the optimal regularization scale  
at this chiral order, the results for chiral order $\ca{O}(m_\pi^3)$ will be
 chosen to demonstrate the process of handling the existence of an optimal 
regularization scale in lattice QCD data. 
%%%OF NOTE
%
%%% The results for the calculation of the optimal regularization scales 
 Values of $\La^\ro{scale}$
 for different data sets and regulators, using chiral order $\ca{O}(m_\pi^3)$,  
are given in Table \ref{table:scales}.
This table simply summarizes the central values from Figures 
\ref{fig:Ohkic0truncDIPchisqdof} through \ref{fig:Youngc2truncTRIPchisqdof}.
 %\emph{
Such excellent agreement between the $c_0$ and $c_2$ analyses
   is remarkable, and indicative of the
 existence of an intrinsic scale in the data. %}
 There is also consistency among independent data sets.
 It is important to realize that the value of 
$\La^\ro{scale}$ is always the order of $\sim 1$ GeV, not 
 $10$ GeV, nor $100$ GeV; nor is it infinity.
\begin{table}[tp]
\caption{\footnotesize{Central values of $\La$ in GeV, taken from the $\chi^2_{dof}$ analysis for $c_0$ and $c_2$, based on JLQCD, PACS-CS and CP-PACS data.}}
  \label{table:scales}
  \newcommand\T{\rule{0pt}{2.8ex}}
  \newcommand\B{\rule[-1.4ex]{0pt}{0pt}}
  \begin{center}
    \begin{tabular}{llll}
      \hline
      \hline
       \T\B                  \qquad &\multicolumn{3}{c}{regulator form} \\ 
      optimal scale \qquad & dipole $\,\,\,\,\,$& double $\,\,$& triple   \\
      \hline
      $\La^\ro{scale}_{c_0,\ro{JLQCD}}$   &\T $1.44^{+0.18}_{-0.18}$ & $1.08^{+0.11}_{-0.11}$ & $0.96^{+0.09}_{-0.09}$ \\
      $\La^\ro{scale}_{c_2,\ro{JLQCD}}$   &\T $1.40^{+0.02}_{-0.03}$ & $1.05^{+0.02}_{-0.01}$ & $0.94^{+0.01}_{-0.02}$ \\
     \hline
      $\La^\ro{scale}_{c_0,\ro{PACS-CS}}$ &\T $1.21^{+0.66}_{-0.82}$ & $0.93^{+0.41}_{-0.58}$ & $0.83^{+0.35}_{-0.50}$ \\
      $\La^\ro{scale}_{c_2,\ro{PACS-CS}}$ &\T $1.21^{+0.18}_{-0.18}$ & $0.93^{+0.11}_{-0.12}$ & $0.83^{+0.10}_{-0.10}$ \\
     \hline
      $\La^\ro{scale}_{c_0,\ro{CP-PACS}}$ &\T $1.20^{+0.10}_{-0.10}$ & $0.98^{+0.06}_{-0.07}$ & $0.88^{+0.06}_{-0.06}$ \\
      $\La^\ro{scale}_{c_2,\ro{CP-PACS}}$ &\T $1.19^{+0.02}_{-0.01}$ & $0.97^{+0.01}_{-0.01}$ & $0.87^{+0.01}_{-0.01}$ \\
      \hline
    \end{tabular}
  \end{center}
\vspace{-6pt}
\end{table}
However, in calculating the systematic 
uncertainty in the observables $c_0$, $c_2$, and the nucleon mass at the 
physical point 
  due to the optimal regularization scale  
 at order $\ca{O}(m_\pi^4\,\ro{log}\,m_\pi)$, 
two methods are provided. Firstly, the upper and lower bounds from the 
$\chi^2_{dof}$ analysis at order $\ca{O}(m_\pi^3)$ are used to constrain 
$\La$, 
and taken to be
 an accurate estimate of the systematic uncertainty
 in the contributions of higher-order terms. 
Secondly, variation of the observables  
across the aforementioned characteristic range of scale values, 
$\La_\ro{lower}<\La<\infty$ are used. %, where $\La_\ro{lower}$ takes the 
%value of $0.6$, $0.4$ and $0.3$ GeV for the dipole, double-dipole and 
% triple-dipole regulator, respectively, as discussed in 
%Chapter \ref{sec:lowerbound}. 
The results from both of these methods 
are displayed in
Table \ref{table:lerr}.
\begin{table*}[tp]
  \caption{\footnotesize{Results at $\ca{O}(m_\pi^4\,\ro{log}\,m_\pi)$ for the systematic error due to the optimal regularization scale, 
calculated using two methods,
 for the 
 values of $c_0$ (GeV), $c_2$ (GeV$^{-1}$) and the nucleon mass $M_N$ (GeV) extrapolated to the physical point ($m_{\pi,\ro{phys}} = 140$ MeV). The first number in each column is the systematic error due to the optimal regularization scale using the upper and lower bound from the $\chi^2_{dof}$ analysis at order $\ca{O}(m_\pi^3)$. The second number is the systematic error due to the intrinsic scale across the whole range of $\La$ values from the lowest reasonable value ($\La = \La_\ro{lower}$) obtained from the pseudodata analysis, to the asymptotic value ($\La = \infty$).}}
  \newcommand\T{\rule{0pt}{2.8ex}}
  \newcommand\B{\rule[-1.4ex]{0pt}{0pt}}
  \begin{center}
    \begin{tabular}{llllllll}
      \hline
      \hline
      \T\B            \quad  &\multicolumn{6}{c}{regulator form} \\
       sys. err. \quad  & \multicolumn{2}{c}{dipole} & \multicolumn{2}{c}{double} & \multicolumn{2}{c}{triple}   \\
      \hline

      $\de^\La\! c_0^\ro{JLQCD}$   &\T $0.001$ &$0.009$ \,\,\,\,\vline& $0.001$ &$0.013$ \!\quad\vline& $0.001$ &$0.016$ \\
      $\de^\La\! c_0^\ro{PACS-CS}$ &\T $0.005$ &$0.006$ \,\,\,\,\vline& $0.005$ &$0.010$ \!\quad\vline& $0.006$ &$0.012$ \\
      $\de^\La\! c_0^\ro{CP-PACS}$ &\T $0.002$ &$0.024$ \,\,\,\,\vline& $0.002$ &$0.037$ \!\quad\vline& $0.002$ &$0.045$ \\
     \hline
      $\de^\La\! c_2^\ro{JLQCD}$   &\T $0.02$ &$0.31$ \,\quad\vline& $0.03$ &$0.38$ \,\,\quad\vline& $0.01$ &$0.48$ \\
      $\de^\La\! c_2^\ro{PACS-CS}$ &\T $0.18$ &$0.25$ \,\quad\vline& $0.16$ &$0.33$ \,\,\quad\vline& $0.14$ &$0.43$ \\
      $\de^\La\! c_2^\ro{CP-PACS}$ &\T $0.02$ &$0.40$ \,\quad\vline& $0.02$ &$0.58$ \,\,\quad\vline& $0.02$ &$0.73$  \\
     \hline
      $\de^\La\! M_{N,\ro{phys}}^\ro{JLQCD}$  &\T $0.0004$ &$0.0051$ \,\vline& $0.0003$ &$0.0073$ \,\,\vline& $0.0003$ &$0.0090$  \\
      $\de^\La\! M_{N,\ro{phys}}^\ro{PACS-CS}$ &\T $0.0022$ &$0.0030$ \,\vline& $0.0025$ &$0.0046$ \,\,\vline& $0.0025$ &$0.0058$ \\
      $\de^\La\! M_{N,\ro{phys}}^\ro{CP-PACS}$ &\T $0.0012$ &$0.0175$ \,\vline& $0.0013$ &$0.0270$ \,\,\vline& $0.0014$ &$0.0326$   \\
      \hline
    \end{tabular}
  \end{center}
\vspace{-6pt}
  \label{table:lerr}
\end{table*}

The final results for the calculation of the renormalized coefficients $c_0$,
 $c_2$ and the nucleon mass $M_N$  extrapolated to the physical point 
($m_{\pi,\ro{phys}} = 140$ MeV) are summarized in Table \ref{table:cs}. 
In this table, the nucleon mass is calculated at the optimal scale 
$\La^\ro{scale}$, which is the average of $\La^\ro{scale}_{c_0}$ and $\La^\ro{scale}_{c_2}$ for each 
 data set. The extrapolations are performed at lattice sizes relevant to each data set: $L_{\ro{extrap}}^\ro{JLQCD} = 1.9$ fm, $L_{\ro{extrap}}^\ro{PACS-CS} = 2.9$ fm and $L_{\ro{extrap}}^\ro{CP-PACS} = 2.8$ fm.
 The estimate of the statistical error is quoted in the first pair of 
parentheses, and the 
systematic error, obtained from the number of $m_\pi^2$ values used, is quoted 
in the second pair of parentheses. 
Two different weighted means are calculated. One incorporates the systematic 
error in the optimal regularization scale using the upper and lower bound 
from the $\chi^2_{dof}$ analysis at order $\ca{O}(m_\pi^3)$. The other 
incorporates the systematic error due to the optimal regularization 
scale across the whole range of $\La$ values, from the lowest 
reasonable value ($\La = \La_\ro{lower}$) obtained from the 
pseudodata analysis, to the asymptotic value ($\La = \infty$).
The weighted means also include an estimate of the systematic error 
in the choice of regularization scale. All errors are added in quadrature. 
%Note that any order $\ca{O}(a)$ errors have not been incorporated into 
%the total error analysis.
 The lightest four data points from
 each of JLQCD, PACS-CS and CP-PACS lattice QCD data are used, and 
 the nucleon mass is calculated at the scale determined by 
  the data. 
\begin{table*}[tp]
  %\caption{\footnotesize{Results at $\ca{O}(m_\pi^4\,\ro{log}\,m_\pi)$ for the values of $c_0$ (GeV), $c_2$ (GeV$^{-1}$) and the nucleon mass $M_N$ (GeV) extrapolated to the physical point ($m_{\pi,\ro{phys}} = 140$ MeV). WM is the weighted mean of each row.
%}} %%%%%The error in the intrinsic scale $\de\La_\ro{scale}$ is the average of the errors from the $c_0$ analysis and the $c_2$ analysis.}}
\caption{\footnotesize{Results at $\ca{O}(m_\pi^4\,\ro{log}\,m_\pi)$ for the values of $c_0$ (GeV), $c_2$ (GeV$^{-1}$) and the nucleon mass $M_N$ (GeV) extrapolated to the physical point ($m_{\pi,\ro{phys}} = 140$ MeV). WM is the weighted mean of each row.
The nucleon mass is calculated at the optimal scale $\La^\ro{scale}$, which
is the average of $\La^\ro{scale}_{c_0}$ and $\La^\ro{scale}_{c_2}$ for each
 data set. The extrapolations are performed at lattice sizes relevant to each data set: $L_{\ro{extrap}}^\ro{JLQCD} = 1.9$ fm, $L_{\ro{extrap}}^\ro{PACS-CS} = 2.9$ fm and $L_{\ro{extrap}}^\ro{CP-PACS} = 2.8$ fm.
 The estimate of the statistical error is quoted in the first pair of parentheses, and the systematic error, obtained from the number of $m_\pi^2$ values used, is quoted in the second pair of parentheses.
Two separate weighted means are calculated for each row. WM(1) incorporates the systematic error in the intrinsic scale using the upper and lower bound from the $\chi^2_{dof}$ analysis at order $\ca{O}(m_\pi^3)$. The WM(2) incorporates the systematic error due to the intrinsic scale across the whole range of $\La$ values, from the lowest reasonable value ($\La = \La_\ro{lower}$) obtained from the pseudodata analysis, to the asymptotic value ($\La = \infty$).
The weighted means also include an estimate of the systematic error in the choice of the regulator functional form. All errors are added in quadrature. Note that any order $\ca{O}(a)$ errors have not been incorporated into the total error analysis.}}
  \newcommand\T{\rule{0pt}{2.8ex}}
  \newcommand\B{\rule[-1.4ex]{0pt}{0pt}}
  \begin{center}
    \begin{tabular}{llllll}
      \hline
      \hline
      \T\B            \quad  &\multicolumn{3}{c}{regulator form} \\
       parameter \quad  & dipole $\,\,$& double $\,\,$& triple $\,\,$& WM(1) $\,\,$& WM(2) \\
      \hline

      $c_0^\ro{JLQCD}$               &\T $0.873(18)(16)$ & $0.875(17)(16)$ & $0.891(17)(16)$ & $0.880(29)$ & $0.879(32)$ \\
      $c_0^\ro{PACS-CS}$             &\T $0.900(51)(15)$ & $0.899(51)(14)$ & $0.898(51)(14)$ & $0.899(53)$ & $0.899(55)$ \\
      $c_0^\ro{CP-PACS}$             &\T $0.924(3)(8)$ & $0.914(3)(7)$ & $0.918(3)(7)$ & $0.918(13)$ & $0.920(37)$\\
     \hline
      $c_2^\ro{JLQCD}$               &\T $3.09(9)(11)$ & $3.18(9)(12)$ & $3.20(9)(11)$ & $3.16(18)$ & $3.14(43)$\\
      $c_2^\ro{PACS-CS}$             &\T $3.06(32)(15)$ & $3.15(31)(14)$ & $3.17(31)(14)$ & $3.13(39)$ & $3.12(49)$ \\
      $c_2^\ro{CP-PACS}$             &\T $2.54(5)(4)$ & $2.70(5)(2)$ & $2.71(5)(3)$ & $2.66(18)$ & $2.61(60)$ \\
     \hline
      $M_{N,\ro{phys}}^\ro{JLQCD}$   &\T $1.02(2)(9)$ & $1.02(2)(9)$ & $1.02(2)(9)$ & $1.02(9)$ & $1.02(9)$\\
      $M_{N,\ro{phys}}^\ro{PACS-CS}$ &\T $0.967(45)(43)$ & $0.966(45)(43)$ & $0.966(45)(43)$ & $0.966(62)$ & $0.966(62)$ \\
      $M_{N,\ro{phys}}^\ro{CP-PACS}$ &\T $0.982(2)(40)$ & $0.975(2)(43)$ & $0.978(2)(42)$ & $0.979(43)$ & $0.979(50)$\\
      \hline
    \end{tabular}
  \end{center}
\vspace{-6pt}
  \label{table:cs}
\end{table*}

\section{Summary and Specific Issues for the Nucleon Mass}
%eg. not a good choice of observable
%for determining an intrinsic scale.

%Also, mention the SIGMA COMMUTATOR and its relevancce here 

%Because 
Since the chiral expansion is only convergent within the 
 PCR, a renormalization scheme such as 
finite-range regularization should be used for current lattice QCD results, 
 which typically extend beyond the PCR. It was found that 
 renormalization scheme-dependence occurs
when lattice QCD data extending outside the PCR are used in 
the extrapolation. This has provided a new quantitative test for determining 
 whether lattice QCD data lie within the PCR.

 The optimal regularization scale $\La^{\ro{scale}}$ was selected %by 
as the scale at which the renormalized 
coefficients are independent of the upper bound of the fit domain, 
$m_{\pi,\ro{max}}^2$.
 This also means that the renormalized coefficients must not be identified 
with their asymptotic values at large $\La$. 
It is also apparent that extremely low values of $\La$ cause a breakdown 
of the finite-range renormalization. The cut-off scale associated 
with an ultraviolet regulator must be 
%must have an associated scale 
large enough for the loop integral contributions to be finite,  
so that the chiral physics is not suppressed.
 
 The mean value of the optimal regularization scale 
for both the $c_0$ and $c_2$ analyses 
across each data set is $\bar{\La}^{\ro{scale}}_{\ro{dip}} \approx 1.3$ GeV for 
the dipole form, $\bar{\La}^{\ro{scale}}_{\ro{doub}} \approx 1.0$ GeV for the 
double-dipole form and $\bar{\La}^{\ro{scale}}_{\ro{trip}} \approx 0.9$ GeV for the 
triple-dipole form. Each functional form
 naturally leads to a different value of optimal regularization scale
 due to its different 
 shape of attenuation, as shown in Figure \ref{fig:reg}. The value of 
$\bar{\La}^{\ro{scale}}_{\ro{dip}}$ is of particular interest in this 
investigation. In Chapter \ref{chpt:nucleonmagmom}, the magnetic moment 
and the electric charge radius of the nucleon 
 are analyzed with the same procedure, and using a dipole regulator. 
If an optimal regularization scale 
can be obtained 
for these electromagnetic properties of the nucleon, a comparison can be made 
with the optimal regularization scale from the analysis of the nucleon mass,  
 to determine whether there exists an intrinsic scale in the nucleon. 
If the optimal regulators in each case are consistent with 
each other, this suggests the existence of a well-defined intrinsic 
energy scale in the nucleon-pion interaction. 
%In any case, 
Nevertheless, a robust method for accomplishing a chiral extrapolation 
with a reliable and systematic estimate in the uncertainty has been provided. 

In the next chapter, 
the procedure developed for obtaining an optimal regularization 
scale and performing a reliable chiral extrapolation is tested,  
by analyzing the quenched $\rho$ meson mass: an observable for which 
there does not exist an experimental value. This serves to demonstrate the 
ability of the extrapolation scheme
 to make predictions without prior bias. 
%in the next chapter summary, use the words pion cloud source.

%\section{Applications to Baryon Mass Spectrum}

%\subsection{Nucleon Mass Resonances}
%What coupling strength for the virtual processes?
%Do not know experimentally, so must use a model
%Which model? Describe the representation and model used
%very tersely, cite the person who invented it and their results
%in terms of F & D couplings
%REFER back to the rep'n section in CHAPTER: chieft.tex  !!!!

%% file: mesonmass.tex
\chapter{Results for the Mass of the Quenched $\rho$ Meson}
\label{chpt:mesonmass}

\textit{``A rigorous theory must begin by specifying the attributes that make a given experimental device into a measuring instrument.''}
(Omn\`{e}s, R. 2002. \textit{Quantum Philosophy: Understanding and Interpreting Contemporary Science} p.209) \cite{Omnes}

%Introduce

%\section{The $\rho$ Meson}
%Describe the $\rho$ (phenomenological)
%and why it is interesting, why did I study it?

The quenched $\rho$ meson mass offers 
a unique test case for the identification of an intrinsic scale, 
and subsequent extrapolation scheme. 
It serves to demonstrate the ability of the procedure 
 to make predictions with reduced phenomenological bias,
and also to highlight the difference between quenched and full 
quantum chromodynamics (QCD) in making
 extrapolations of an observable. 
By using the method developed in Chapter \ref{chpt:intrinsic}, 
 an extrapolation is performed using quenched lattice QCD data that 
extend outside the power-counting regime (PCR). 

\begin{figure}
\begin{minipage}[t]{0.5\linewidth} % A minipage that covers half the page
\centering
\includegraphics[height=85pt,angle=0]{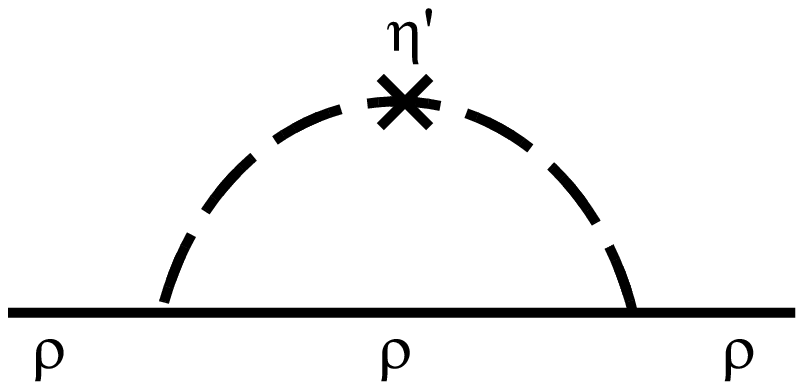}
\vspace{-3mm}
\caption{\footnotesize{Double hairpin $\eta'$ diagram.}}
\label{fig:qrhoSEdh}
\vspace{6mm}
\includegraphics[height=85pt,angle=0]{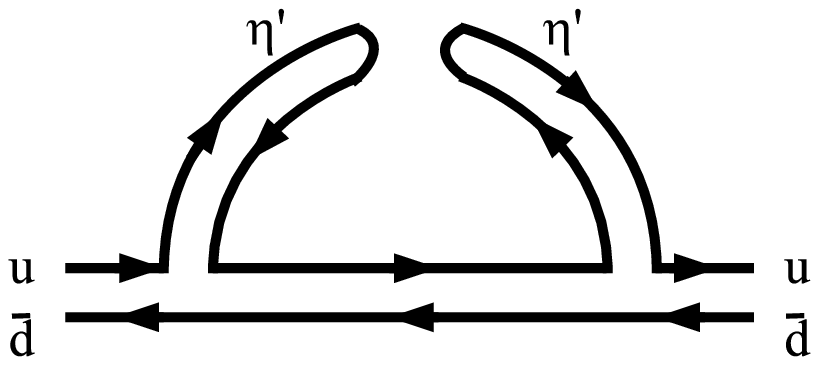}
%\vspace{-6pt}
\vspace{-3mm}
\caption{\footnotesize{Double hairpin quark flow diagram.}}
\label{fig:qrhoQFdh}
\vspace{6mm}
\includegraphics[height=85pt,angle=0]{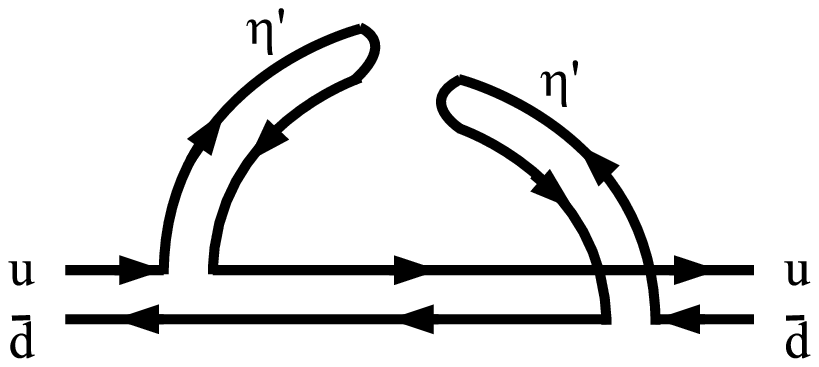}
%\vspace{-6pt}
\vspace{-3mm}
\caption{\footnotesize{Alternative double hairpin quark flow diagram.}}
\label{fig:qrhoQFdhalt}
\end{minipage}
\hspace{12mm}
\begin{minipage}[t]{0.5\linewidth}
\centering
\includegraphics[height=85pt,angle=0]{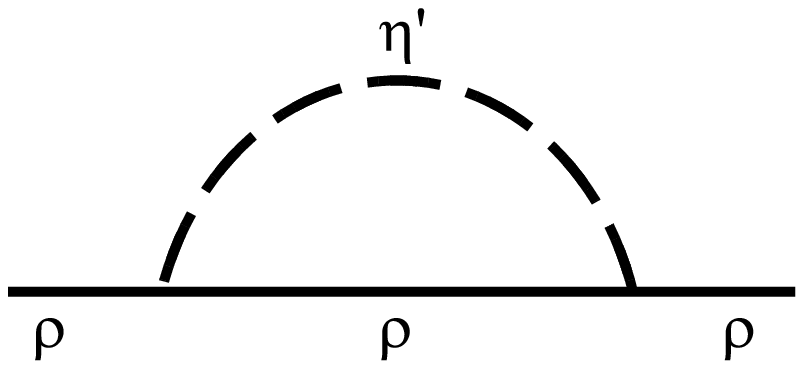}
\vspace{-3mm}
\caption{\footnotesize{Single hairpin $\eta'$ diagram.}}
\label{fig:qrhoSEsh}
\vspace{6mm}
\includegraphics[height=85pt,angle=0]{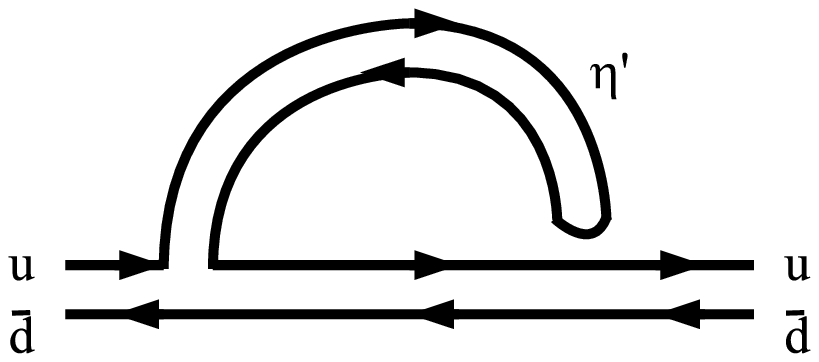}
%\vspace{-6pt}
\vspace{-3mm}
\caption{\footnotesize{Single hairpin quark flow diagram.}}
\label{fig:qrhoQFsh}
\end{minipage}
\end{figure}

In chiral effective field theory ($\chi$EFT), 
the diagrammatic formulation can be used to identify the
major contributions to the $\rho$ meson mass in quenched QCD (QQCD)
\cite{Chow:1997dw,Armour:2005mk}.  The leading-order diagrams are the
double and single $\eta'$ hairpin diagrams as shown in Figures
\ref{fig:qrhoSEdh} and \ref{fig:qrhoSEsh}, respectively.
The constant coefficients of these loop integrals are endowed with 
an uncertainty to encompass the possible effects of smaller contributions 
to order $\ca{O}(m_\pi^4)$. 
%
%\begin{figure}[tp]
%\centering
%\includegraphics[height=85pt,angle=0]{graphics/mesonmass/qrhoSEdh.eps}
%\vspace{-3mm}
%\caption{\footnotesize{Double hairpin $\eta'$ diagram.}}
%\label{fig:qrhoSEdh}
%%\vspace{10pt}
%%\centering
%\includegraphics[height=85pt,angle=0]{graphics/mesonmass/qrhoSEsh.eps}
%\vspace{-3mm}
%\caption{\footnotesize{Single hairpin $\eta'$ diagram.}}
%\label{fig:qrhoSEsh}
%\end{figure}
%
Interactions with the flavour-singlet $\eta'$ are %, surprisingly,
the most important contributions to the $\rho$ meson mass in QQCD.
This is an artefact of the quenched approximation, where the $\eta'$
also behaves as a pseudo-Goldstone boson, having a mass that is
degenerate with the pion.
%This is because, by requiring that no disconnected pion loops be present in
%any diagram, any meson self energy loop must contain a gluon coupling
%as illustrated in the quark flow diagrams in Figures 
The dressing of the $\rho$ meson by the $\eta'$ field is illustrated in
Figures \ref{fig:qrhoQFdh} through \ref{fig:qrhoQFsh}.  Since the hairpin 
 vertex must be a flavour-singlet, 
%the only possible meson that can contribute is the
%$\eta'$ meson. In QQCD, this $\eta'$ loop behaves much as a pion loop,
%yet with a slightly modified propagator.
the %only possible meson 
 mesons that can contribute are %is 
the $\eta'$ meson, and the $\omega$ meson. 
The contributions from the $\omega$ meson are insignificant 
due to OZI suppression and the small $\rho$-$\omega$ mass splitting. 
However, in QQCD, the $\eta'$ loop behaves much as a pion loop,
yet with a slightly modified propagator.
%
%\begin{figure}[tp]
%\centering
%\includegraphics[height=85pt,angle=0]{graphics/mesonmass/qrhoQFdh.eps}
%%\vspace{-6pt}
%\vspace{-3mm}
%\caption{\footnotesize{Double hairpin quark flow diagram.}}
%\label{fig:qrhoQFdh}
%%\vspace{10pt}
%\centering
%\includegraphics[height=85pt,angle=0]{graphics/mesonmass/qrhoQFdhalt.eps}
%%\vspace{-6pt}
%\vspace{-3mm}
%\caption{\footnotesize{Alternative double hairpin quark flow diagram.}}
%\label{fig:qrhoQFdhalt}
%%\vspace{10pt}
%\centering
%\includegraphics[height=85pt,angle=0]{graphics/mesonmass/qrhoQFsh.eps}
%%\vspace{-6pt}
%\vspace{-3mm}
%\caption{\footnotesize{Single hairpin quark flow diagram.}}
%\label{fig:qrhoQFsh}
%\end{figure}

In full QCD however, this would not be the case. 
The $\eta'$ masses are large compared to the pion,  
 and the propagators of the $\eta'$ meson are suppressed due to their 
large denominators.
If the $\eta'$ propagator in full QCD is expanded out, the terms 
can be summed as a geometric series and 
expressed in closed form, as a function of some massive coupling constant $M_0$
between the disconnected quark loops and pion momentum $k$,
 as argued in Allton \cite{Allton:2005fb}: 
\begin{align}
\label{eqn:etaprimeprop}
& \f{1}{k^2 + m_{\pi}^2}  - \frac{ M_0^2}{\left ( k^2 + m_{\pi}^2 \right )^2} \times \nn \\
& \left [ 1  - \f{M_0^2}{k^2 + m_{\pi}^2} + 
\left ( \f{M_0^2}{k^2 + m_{\pi}^2} \right )^2 - \cdots \right ]  \\
\nn \\
 &= \frac{1}{k^2 + m_{\pi}^2} -  
\frac{M_0^2}{ \left ( k^2 + m_{\pi}^2 \right )^2} 
 \left[1 +  \left( \frac{M_0^2}{k^2 + m_{\pi}^2}\right) \right]^{-1}   \\
\nn \\
&= \frac{1}{k^2 +  m_\pi^2 + M_0^2}
\equiv \frac{1}{k^2 +  m_{\eta'}^2 }\,.
\end{align}

\begin{figure}[bp]
\begin{center}
\includegraphics[height = 115pt,angle=0]{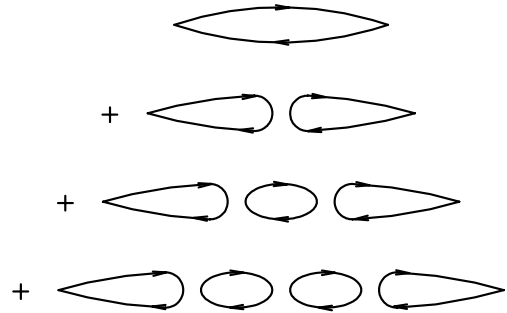}
\vspace{-6pt}
\caption{\footnotesize{Diagrammatic representation of $\eta'$ 
propagator terms.}}
\label{fig:PQetaPrime}
\end{center}
\end{figure}

However in QQCD, the first two terms of Equation (\ref{eqn:etaprimeprop}) 
form the whole $\eta'$ propagator, since they alone correspond to 
 the absence of disconnected loops, as shown in  Figure \ref{fig:PQetaPrime}.

\section{Renormalization of the Quenched $\rho$ Meson Mass}

\subsection{Chiral Expansion of the Quenched $\rho$ Meson Mass}

%Similar to the case of the nucleon mass, 
The $\rho$ meson mass extrapolation formula in QQCD can be expressed 
in a form that contains an analytic polynomial in $m_\pi^2$ plus 
the chiral loop integrals ($\Si^Q$): 
\eqb
m_{\rho,Q}^2 = a_0^\La + a_2^\La\,m_\pi^2 +a_4^\La\,m_\pi^4
 + \Si_{\eta'\!\eta'}^Q(m_\pi^2,\La)
 + \Si_{\eta'}^Q(m_\pi^2,\La) +\ca{O}(m_\pi^6)\,.
\label{eqn:mrhoexpsn}
\eqe
The coefficients $a_i^\La$ are the `residual series' coefficients, which
correspond to direct quark-mass insertions in the underlying
Lagrangian of chiral perturbation theory ($\chi$PT). 
However, the
 non-analytic behaviour of the expansion arises from the chiral loop 
integrals. 
Upon renormalization of the divergent loop
integrals, these will correspond with low-energy constants of the
quenched effective field theory. 
%Here we will demonstrate how these parameters can be
%extracted from lattice QCD results.
The extraction of these parameters %can be
%extracted 
from lattice QCD results will follow the same course as provided in Chapter  
\ref{chpt:intrinsic}. %now be demonstrated.

%LOOP INTEGRALS
By convention, the non-analytic terms from the double and single
 hairpin integrals are $\chi_1 m_\pi$ and $\chi_3 m_\pi^3$,
 respectively.
The coefficients $\chi_1$ and $\chi_3$ 
of each integral are scheme-independent constants 
that can be %obtained from experiment.
 estimated from phenomenology. %by matching the coefficient of the 
%logarithms, which are fixed by the interaction strengths.
That is, they can be expressed purely in terms of known constants 
from experiment, such as the pion decay constant $f_\pi = 92.4$ MeV, 
and a variety of parameters obtained from the underlying effective 
Lagrangian, as described in Section \ref{subsect:meschis}.
The low-order expansion of the loop contributions takes the following form:
\begin{align}
\label{eqn:doubloopexpsn}
\Si_{\eta'\!\eta'}^Q &= b^{\eta'\!\eta'}_0 + \chi_1 m_\pi + 
b^{\eta'\!\eta'}_2 m_\pi^2 + \chi_3^{\eta'\!\eta'} m_\pi^3 
+ b^{\eta'\!\eta'}_4 m_\pi^4 + \ca{O}(m_\pi^6)\,,\\
\Si_{\eta'}^Q &= b^{\eta'}_0 + b^{\eta'}_2 m_\pi^2 + \chi_3^{\eta'} m_\pi^3
 + b^{\eta'}_4 m_\pi^4 + \ca{O}(m_\pi^6)\,.
\label{eqn:singloopexpsn}
\end{align}
The coefficient $\chi_3$ is obtained by adding the contributions from 
both integrals, 
 $\chi_3 = \chi_3^{\eta'\!\eta'} + \chi_3^{\eta'}$. 
As before, each integral has a solution in the form of a polynomial expansion 
analytic in $m_\pi^2$ plus non-analytic terms, of which the leading-order 
term is of greatest interest.
%The coefficients $b_i$ are entirely scale-dependent and therefore 
%scheme-dependent. %The superscript
%$b^{(j)}$ denotes whether the coefficient corresponds to the double hairpin 
%integral ($j=1$) or the single hairpin integral ($j=3$). 
%Now, in order to achieve an extrapolation based on an optimal FRR scale,
% first the renormalized form of the low-energy coefficients must be specified.
%The renormalization 
%program of FRR combines the scheme-dependent $b_i$ coefficients from the chiral
% loops with the scheme-dependent $a_i$ coefficients from the residual series
% at each chiral order $i$. 
%The result is once again a scheme-independent coefficient $c_i$:
In order to achieve an extrapolation
based on an optimal finite-range regularization (FRR) scale, 
once again the scale-dependence of the
low-energy expansion must be removed through renormalization.  
The
renormalization program of FRR combines the scheme-dependent $b_i$
coefficients from the chiral loops with the scheme-dependent $a_i$
coefficients from the residual series at each chiral order $i$.  The
result is a scheme-independent coefficient $c_i$:
\begin{align}
c_0 &= a_0^\La + b_0^{\eta'\!\eta'} + b_0^{\eta'}\,,\\
c_2 &= a_2^\La + b_2^{\eta'\!\eta'} + b_2^{\eta'}\,,\\
c_4 &= a_4^\La + b_4^{\eta'\!\eta'} + b_4^{\eta'}\,, \mbox{\,\,etc.}
\end{align}
That is, the underlying $a_i^\La$ coefficients undergo a renormalization
from the chiral loop integrals. The renormalized coefficients $c_i$
are an important part of the extrapolation technique. In this 
chapter, a stable and
robust determination of the parameters $c_0$, $c_2$ and $c_4$ 
forms the core of the method 
for determining an optimal scale $\La^\ro{scale}$ 
of the mass of the $\rho$ meson.

\subsection{Chiral Loop Integrals}
\label{subsec:mesloop}

The loop integrals can again be
expressed conveniently %in a convenient form 
by taking the non-relativistic limit and 
performing the pole integration for $k_0$.
Renormalization is achieved by subtracting the
relevant terms in the Taylor expansion of the loop integrals, and
absorbing them into the corresponding low-energy coefficients, $c_i$:
\begin{align}
\label{eqn:doub}
\tilde{\Si}_{\eta'\!\eta'}^Q(m_\pi^2;\La) &= \f{-\chi_{\eta'\!\eta'}}
{3\pi^2}\int\!\!\ud^3\!
 k
\f{(M_0^2 k^2 +\f{5}{2}A_0 k^4)u^2(k;\La)}{{(k^2 + m_\pi^2)}^2} \nn\\
&- b_0^{\eta'\!\eta'}
 - b_2^{\eta'\!\eta'}m_\pi^2
-b_4^{\eta'\!\eta'}m_\pi^4\,,\\
\label{eqn:sing}
\tilde{\Si}_{\eta'}^Q(m_\pi^2;\La) &= \f{\chi_{\eta'}}{2\pi^2}\int\!\!\ud^3\! k
\f{k^2 u^2(k;\La)}{k^2 + m_\pi^2} - b_0^{\eta'} - b_2^{\eta'}m_\pi^2\nn\\
&-b_4^{\eta'}m_\pi^4\,.
\end{align}
The tilde ($\,\tilde{\,}\,$) denotes that the
integrals are written out in renormalized form to chiral order 
 $\ca{O}(m_\pi^4)$.
The coefficients $\chi_{\eta'\!\eta'}$ and $\chi_{\eta'}$ are related to the 
 coefficients of the leading-order non-analytic terms by the following:
\begin{align}
\chi_1 &= M_0^2\,\chi_{\eta'\!\eta'}\,,\\
\chi_3 &= \chi_3^{\eta'\!\eta'} + \chi_3^{\eta'}  
= A_0\,\chi_{\eta'\!\eta'} + \chi_{\eta'}\,.
\end{align}

\begin{figure}[tp]
\begin{center}
\includegraphics[height=0.70\hsize,angle=90]{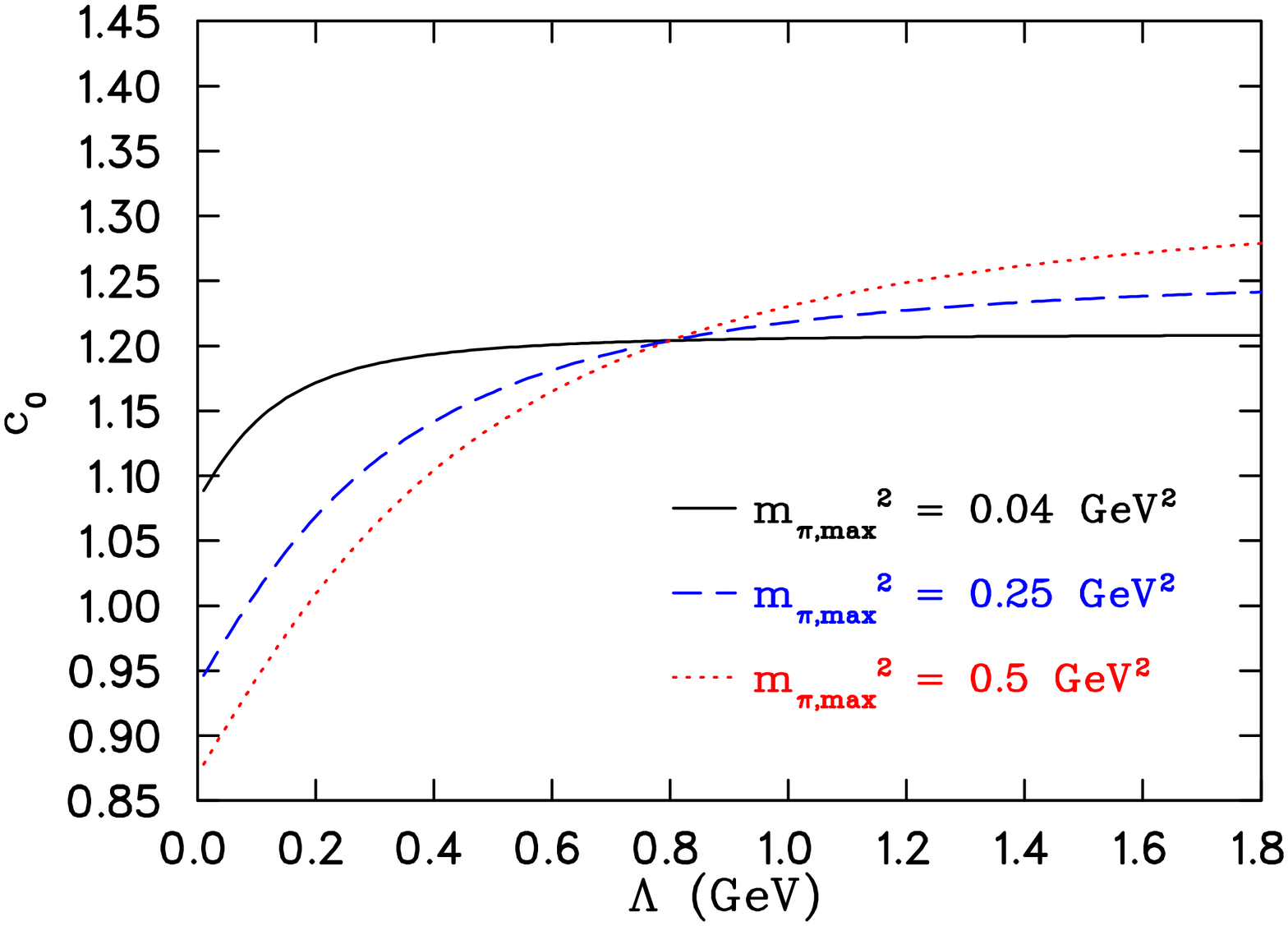}
\vspace{-13pt}
\caption{\footnotesize{ Behaviour of $c_0$ vs.\ $\La$ based on infinite-volume pseudodata created with a dipole regulator at $\La_\ro{c} = 0.8$ GeV (based on Kentucky Group data). Each curve uses pseudodata with
 a different upper value of pion mass $m_{\pi,\ro{max}}^2$.}}
\label{fig:c0dipinfpdata}
%\vspace{6mm}
%
\includegraphics[height=0.70\hsize,angle=90]{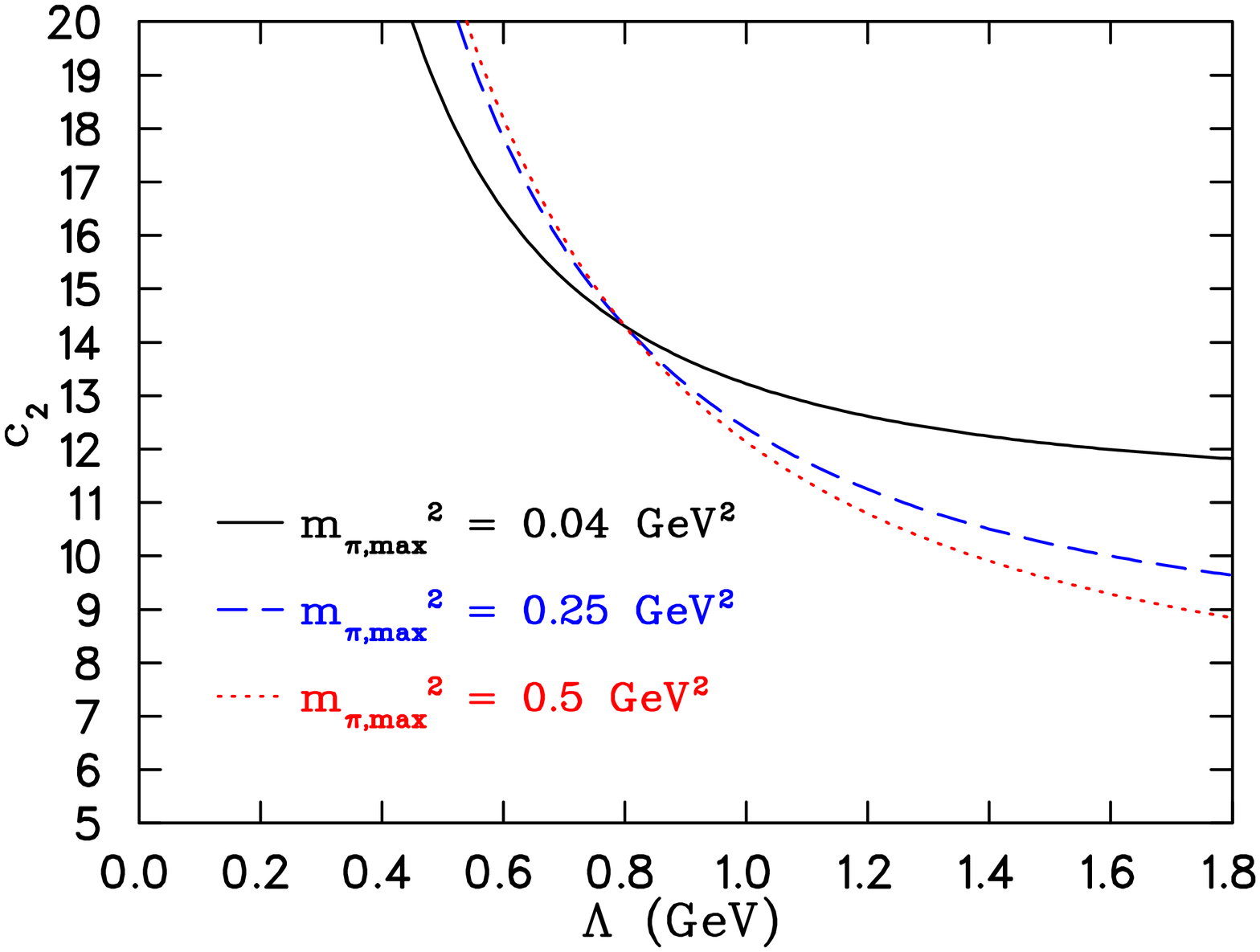}
\vspace{-13pt}
\caption{\footnotesize{ Behaviour of $c_2$ vs.\ $\La$ based on infinite-volume pseudodata created with a dipole regulator at $\La_\ro{c} = 0.8$ GeV (based on Kentucky Group data). %Each curve uses pseudodata with
% a different upper value of pion mass $m_{\pi,\ro{max}}^2$.
}}
\label{fig:c2dipinfpdata}
\vspace{1mm}
\includegraphics[height=0.70\hsize,angle=90]{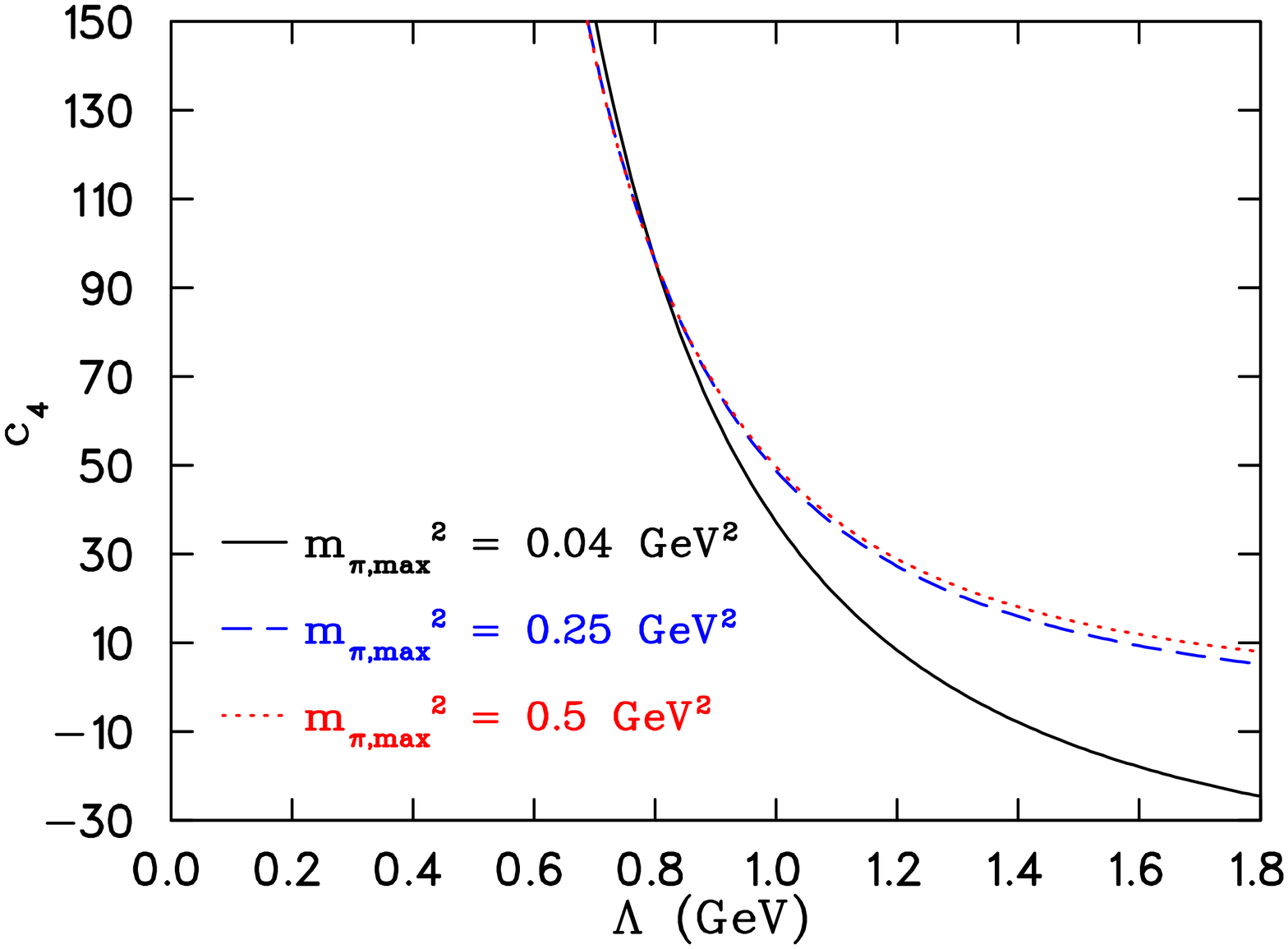}
\vspace{-13pt}
\caption{\footnotesize{ Behaviour of $c_4$ vs.\ $\La$ based on infinite-volume pseudodata created with a dipole regulator at $\La_\ro{c} = 0.8$ GeV (based on Kentucky Group data). %Each curve uses pseudodata with
% a different upper value of pion mass $m_{\pi,\ro{max}}^2$.
}}
\label{fig:c4dipinfpdata}
\end{center}
\end{figure}

\begin{figure}[tp]
\begin{center}
\includegraphics[height=0.70\hsize,angle=90]{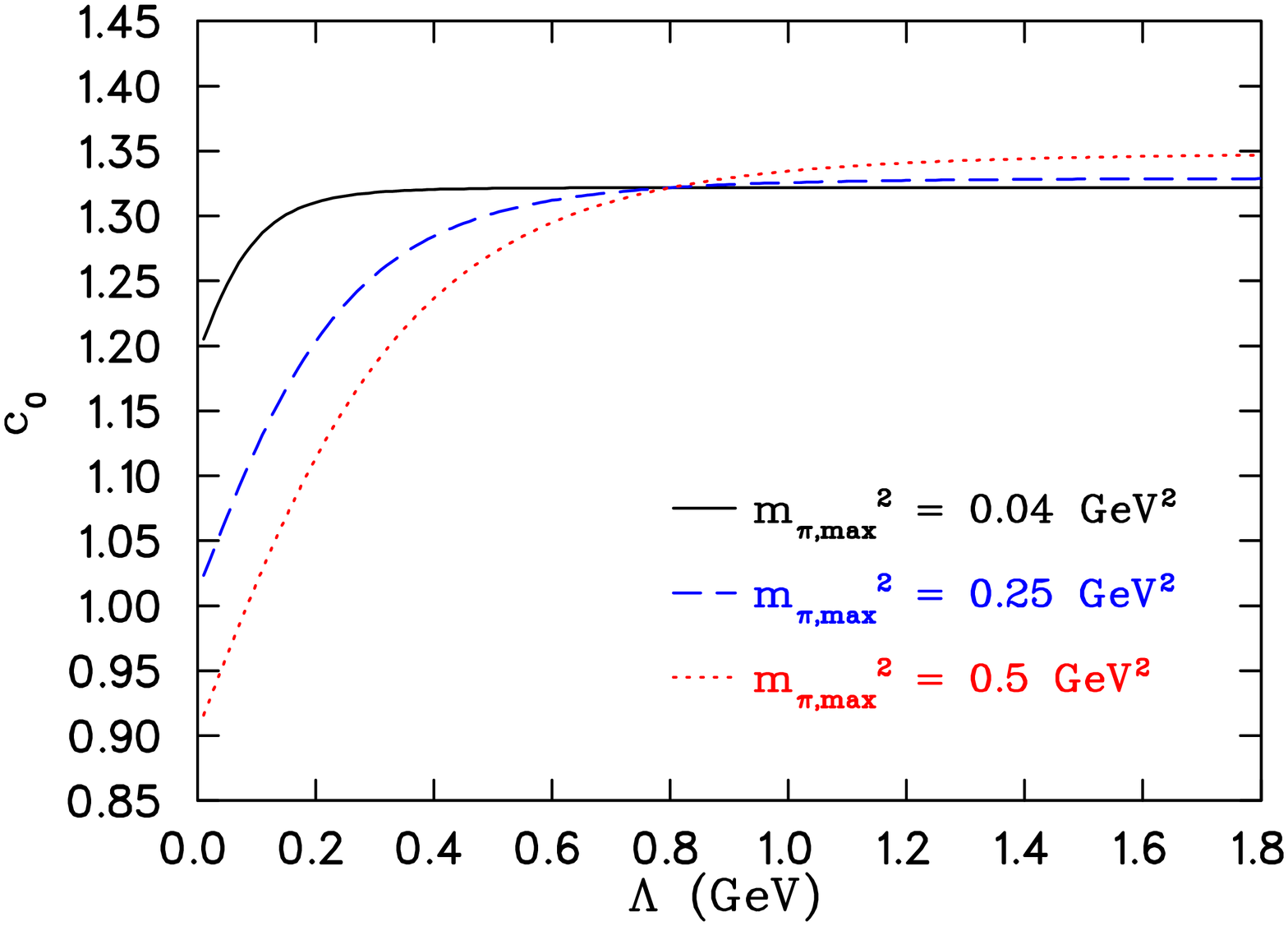}
\vspace{-13pt}
\caption{\footnotesize{ Behaviour of $c_0$ vs.\ $\La$ based on infinite-volume pseudodata created with a triple-dipole regulator at $\La_\ro{c} = 0.8$ GeV (based on Kentucky Group data). Each curve uses pseudodata with
 a different upper value of pion mass $m_{\pi,\ro{max}}^2$.
}}
\label{fig:c0diptripinfpdata}
%\vspace{6mm}
%
\includegraphics[height=0.70\hsize,angle=90]{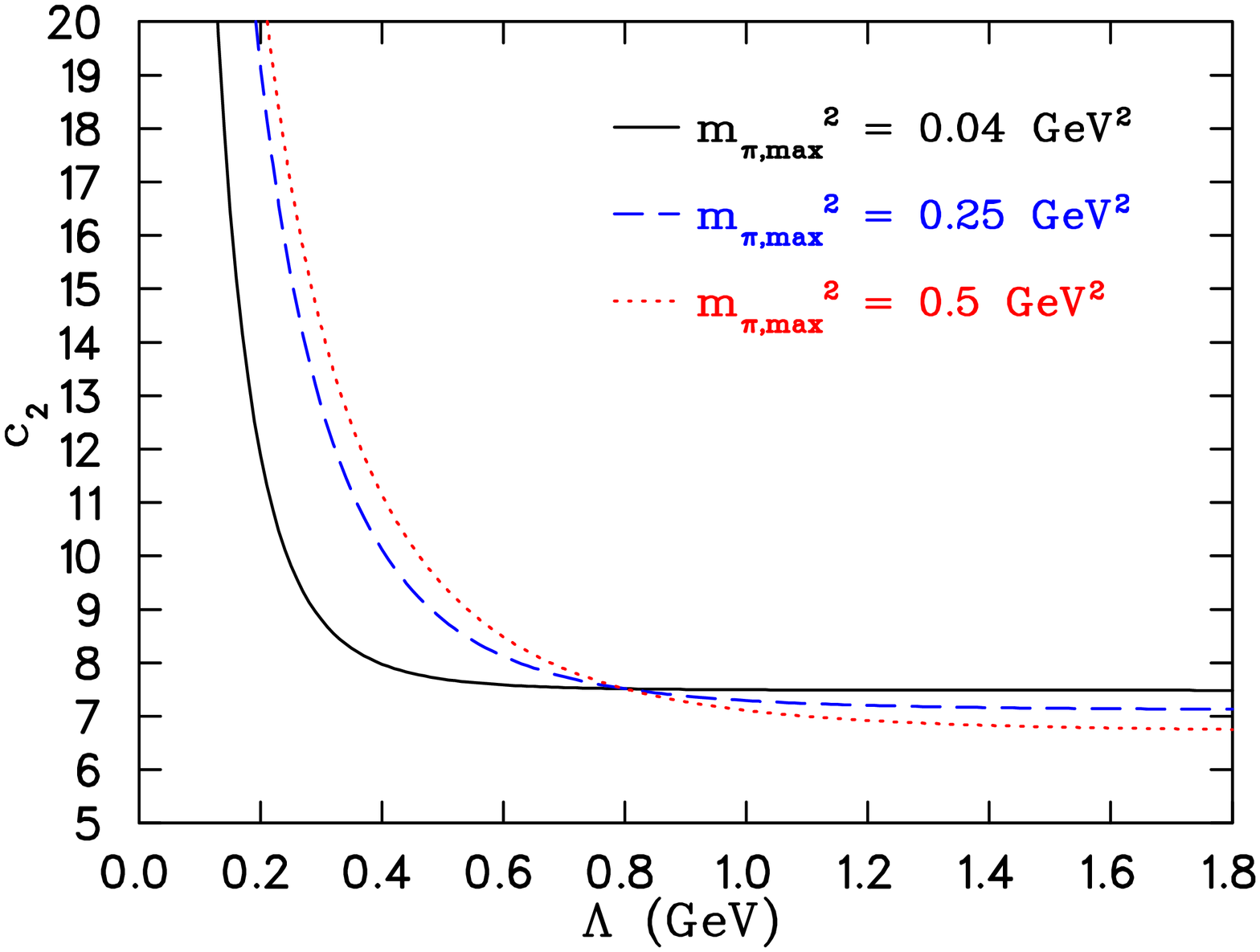}
\vspace{-13pt}
\caption{\footnotesize{ Behaviour of $c_2$ vs.\ $\La$ based on infinite-volume pseudodata created with a triple-dipole regulator at $\La_\ro{c} = 0.8$ GeV (based on Kentucky Group data). %Each curve uses pseudodata with
% a different upper value of pion mass $m_{\pi,\ro{max}}^2$.
}}
\label{fig:c2diptripinfpdata}
\vspace{1mm}
\includegraphics[height=0.70\hsize,angle=90]{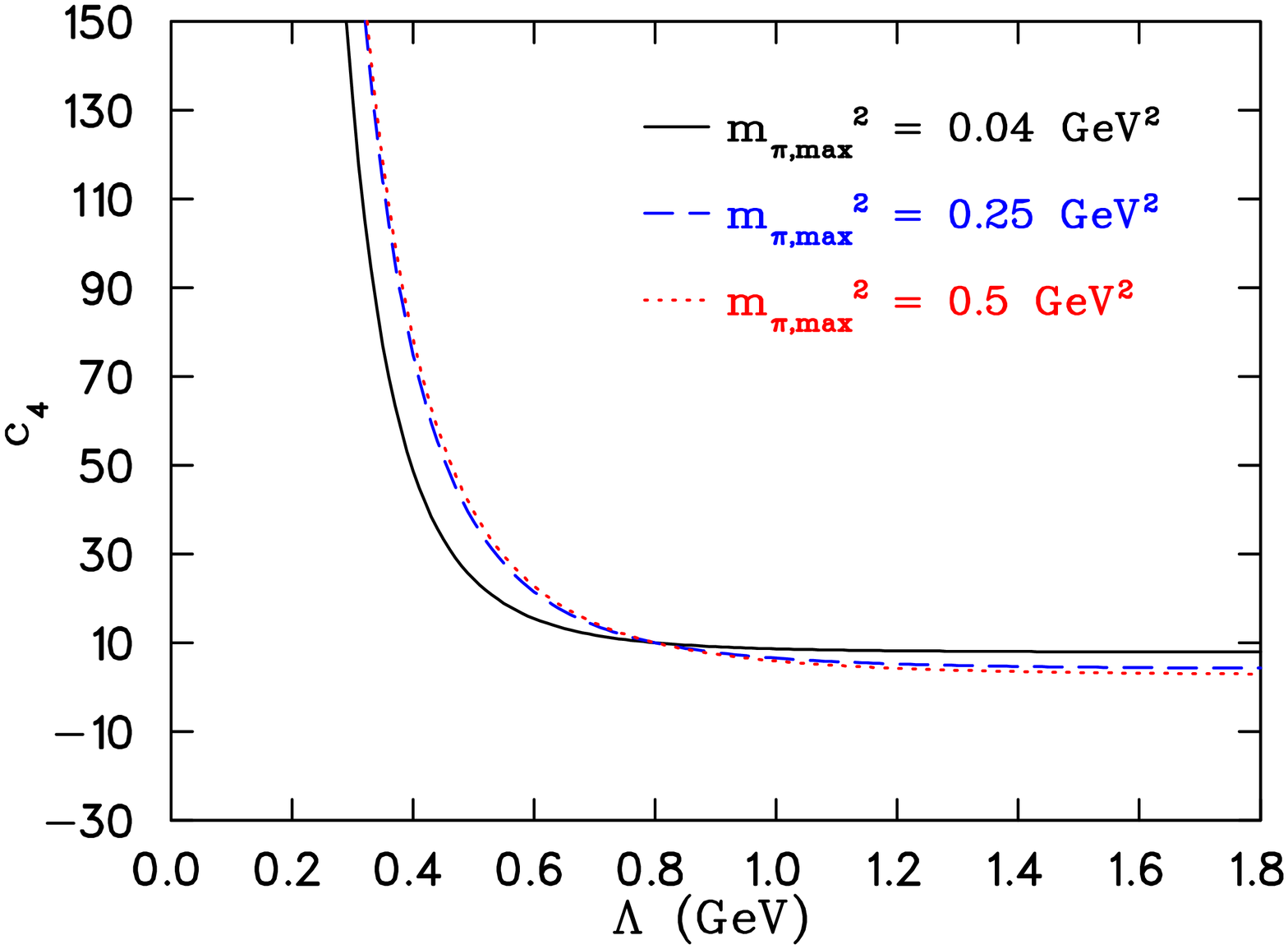}
\vspace{-13pt}
\caption{\footnotesize{ Behaviour of $c_4$ vs.\ $\La$ based on infinite-volume pseudodata created with a triple-dipole regulator at $\La_\ro{c} = 0.8$ GeV (based on Kentucky Group data). %Each curve uses pseudodata with
 %a different upper value of pion mass $m_{\pi,\ro{max}}^2$.
}}
\label{fig:c4diptripinfpdata}
\end{center}
\end{figure}

In choosing the form of regulator, one must be cautious to avoid 
 any extra scheme-dependent, non-analytic terms that might occur 
at working chiral order. 
 For example, consider pseudodata created from the lattice QCD simulations 
from the Kentucky Group, using a dipole regulator created  
using the scale: $\La_\ro{c} = 0.8$ GeV. 
 The renormalization flow curves are shown in Figures 
\ref{fig:c0dipinfpdata} through \ref{fig:c4dipinfpdata}. 
The dipole regulator induces non-analytic terms 
proportional to $m_\pi$ and $m_\pi^3$ in the loop integral expansion formulae. 
By writing out the regulator-dependence explicitly in the coefficients 
$\tilde{b}_i$, the following equations are obtained:
\begin{align}
\Si^{Q,\ro{dip}}_{\eta'\!\eta'}(k\,;\La) &= \La\, \tilde{b}_0^{\eta'\!\eta'} +
 \chi_1m_\pi +
 \f{\tilde{b}_2^{\eta'\!\eta'}}{\La}\mpisq + 
\f{\tilde{b}_3^{\eta'\!\eta'}}{\La^2}m_\pi^3  %\\
%&
+ \f{ \tilde{b}_4^{\eta'\!\eta'}}{\La^3} m_\pi^4 + 
\f{\tilde{b}_5^{\eta'\!\eta'}}{\La^4}m_\pi^5  + \ca{O}(m_\pi^6)\,, \\
%
%\\
%
\Si^{Q,\ro{dip}}_{\eta'}(k\,;\La) &= \La^3 \tilde{b}_0^{\eta'} +
 \La\, \tilde{b}_2^{\eta'}\mpisq +
 \chi_3m_\pi^3 %\\
%&
+ \f{\tilde{b}_4^{\eta'}}{\La} m_\pi^4 + 
\f{\tilde{b}_5^{\eta'}}{\La^2}m_\pi^5  + \ca{O}(m_\pi^6)\,.
\end{align}
%
%This regulator form is used from now on. 
%
\begin{figure}[tp]
\begin{center}
\includegraphics[height=0.70\hsize,angle=90]{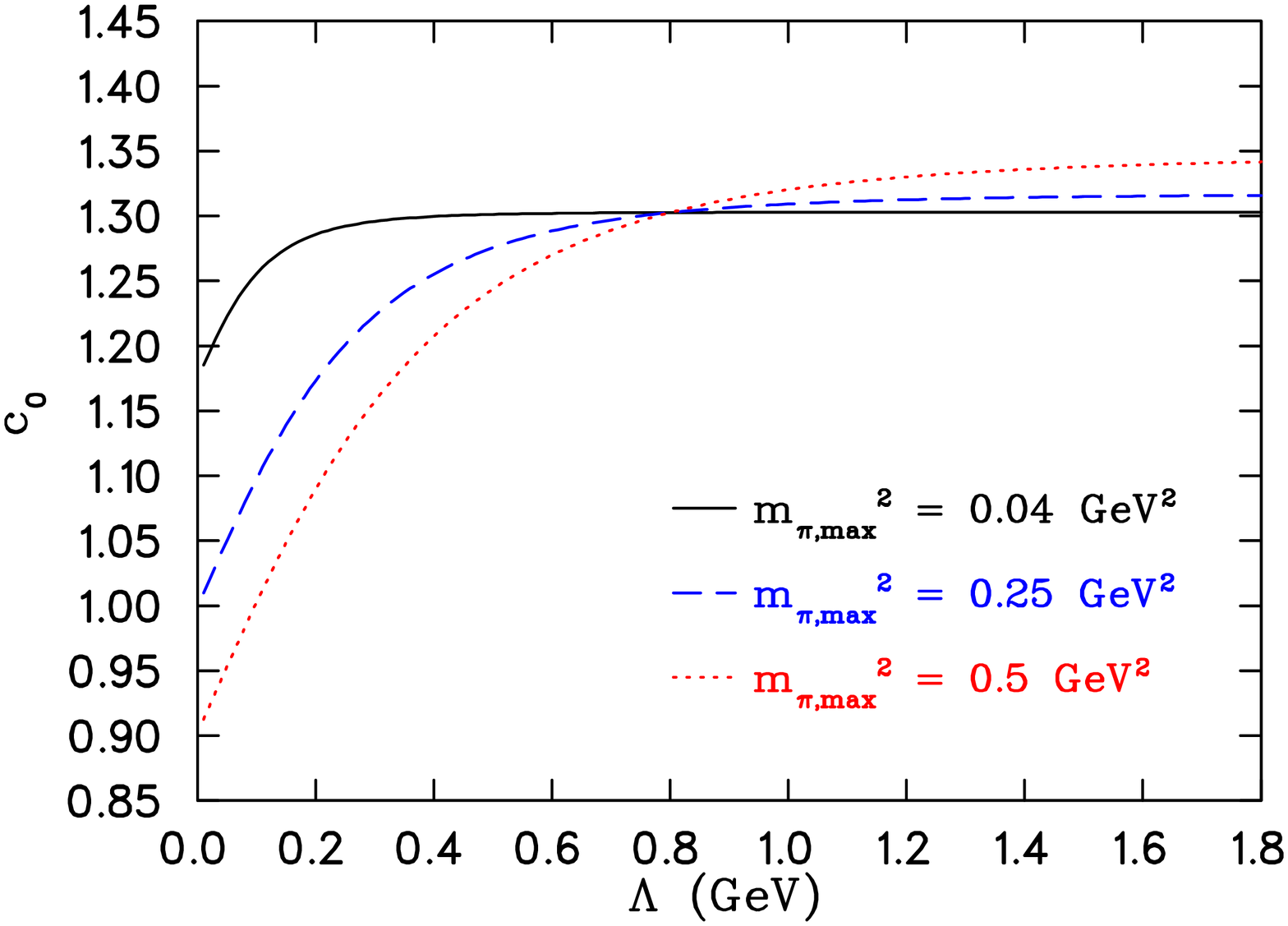}
\vspace{-13pt}
\caption{\footnotesize{ Behaviour of $c_0$ vs.\ $\La$ based on infinite-volume pseudodata created with a double-dipole regulator at $\La_\ro{c} = 0.8$ GeV (based on Kentucky Group data). Each curve uses pseudodata with
 a different upper value of pion mass $m_{\pi,\ro{max}}^2$.
}}
\label{fig:c0dipdoubinfpdata}
%\vspace{6mm}
%
\includegraphics[height=0.70\hsize,angle=90]{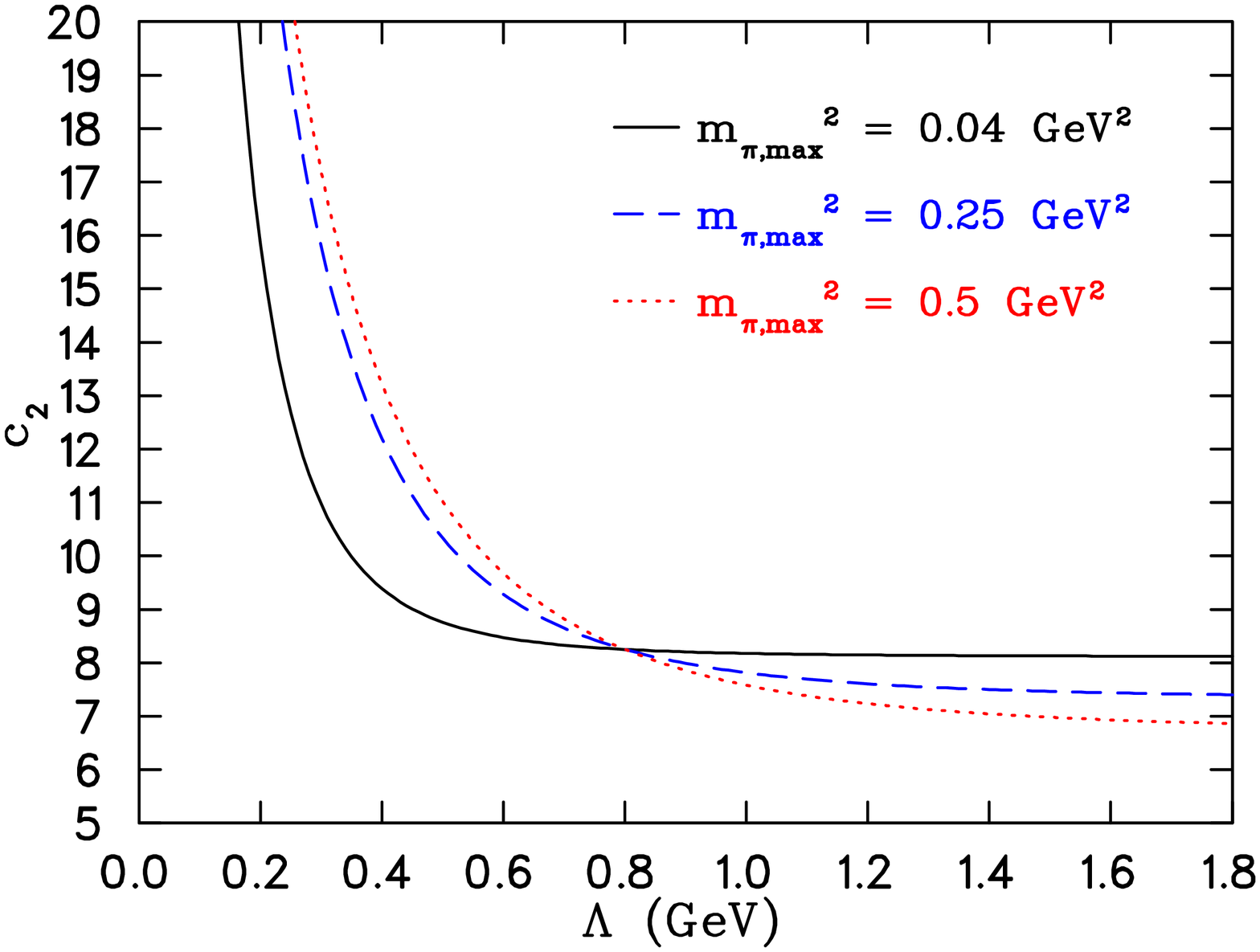}
\vspace{-13pt}
\caption{\footnotesize{ Behaviour of $c_2$ vs.\ $\La$ based on infinite-volume pseudodata created with a double-dipole regulator at $\La_\ro{c} = 0.8$ GeV (based on Kentucky Group data). %Each curve uses pseudodata with
% a different upper value of pion mass $m_{\pi,\ro{max}}^2$.
}}
\label{fig:c2dipdoubinfpdata}
\vspace{1mm}
\includegraphics[height=0.70\hsize,angle=90]{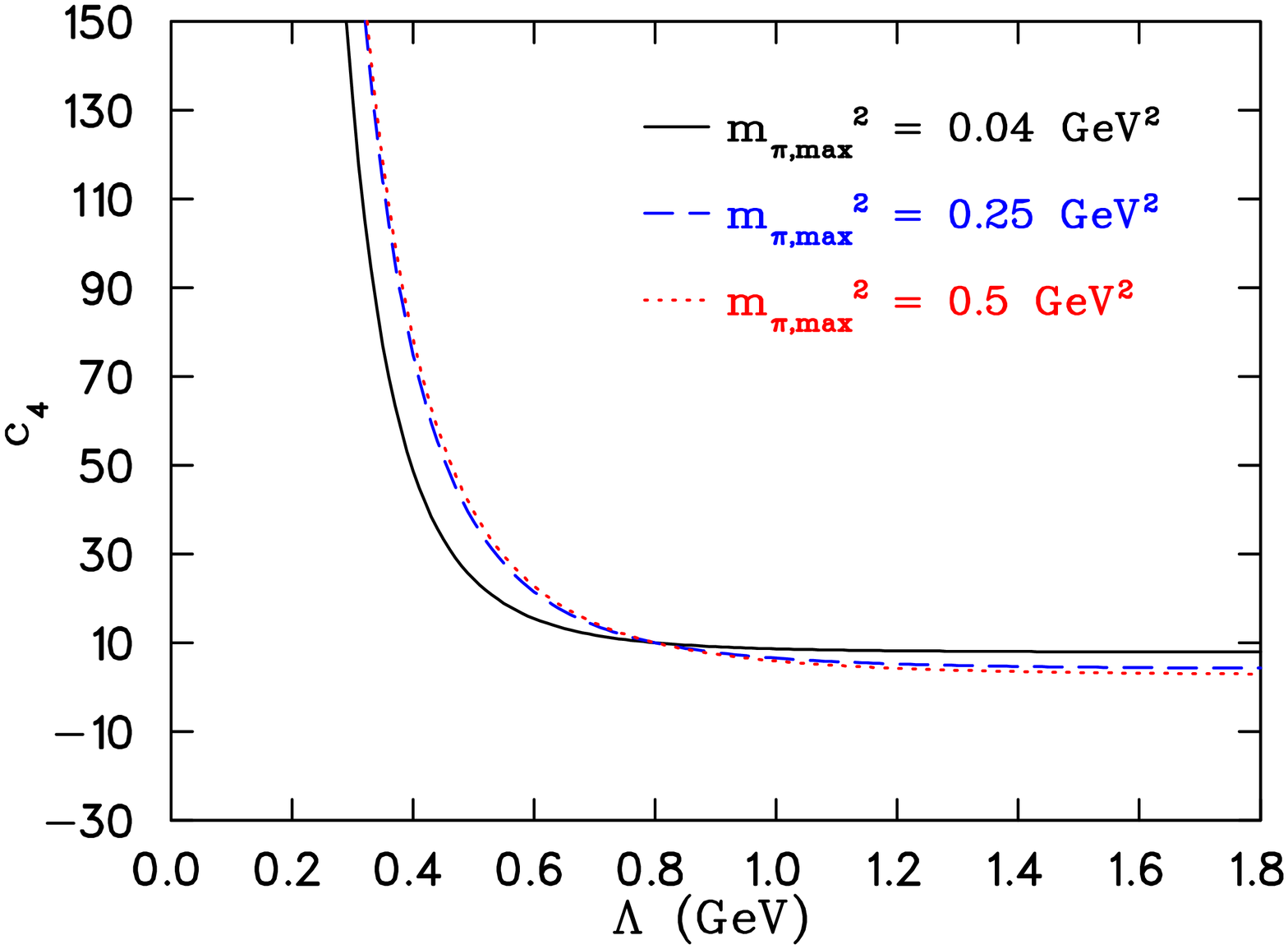}
\vspace{-13pt}
\caption{\footnotesize{ Behaviour of $c_4$ vs.\ $\La$ based on infinite-volume pseudodata created with a double-dipole regulator at $\La_\ro{c} = 0.8$ GeV (based on Kentucky Group data). %Each curve uses pseudodata with
 %a different upper value of pion mass $m_{\pi,\ro{max}}^2$.
}}
\label{fig:c4dipdoubinfpdata}
\end{center}
\end{figure}
%
%Since the chiral expansion in Equation \ref{eqn:renormexpsn2} 
%contains scheme-independent, non-analytic terms proportional to $m_\pi$ 
%and $m_\pi^3$, a successful renormalization with three fit parameters 
%requires that extra non-analytic terms below $\ca{O}(m_\pi^6)$ are suppressed. 
% %INclude plot??
%
Clearly, the renormalization flow is compromised by the extra non-analytic 
terms appearing at such a close chiral order to the fit parameters. 
Though it is possible to provide additional fit parameters $a_3^\La$ 
and $a_5^\La$ to contain the contribution from these terms, 
 there are often not enough available lattice simulation 
results to constrain all coefficients. 
Instead, a more effective approach is to 
choose a regulator functional form such that the extra non-analytic 
terms do not appear in the chiral expansion. 
By selecting an multiple-dipole regulator corresponding to 
a choice of $n>2$ in Equation (\ref{eqn:ndipole}) in Chapter 
\ref{chpt:chieft}, the suppression 
of additional non-analytic terms below the working chiral order 
$\ca{O}(m_\pi^4)$ is assured.  If one also decides to remove extra 
$m_\pi^5$ terms, a triple-dipole is sufficient to remove additional 
non-analytic terms below chiral order $\ca{O}(m_\pi^6)$. 
The renormalization flow curves for pseudodata created 
with a double-dipole are shown in Figures 
\ref{fig:c0dipdoubinfpdata} through \ref{fig:c4dipdoubinfpdata}.  
The renormalization flow curves for pseudodata created 
with a triple-dipole regulator are shown in 
Figures \ref{fig:c0diptripinfpdata} through \ref{fig:c4diptripinfpdata}. 
In both cases, the pseudodata are created using 
the scale: $\La_\ro{c} = 0.8$ GeV.

With the loop integrals specified, Equation (\ref{eqn:mrhoexpsn}) 
can be rewritten in terms 
of the renormalized coefficients $c_i$:
\begin{align}
\label{eqn:renormexpsn1}
m_{\rho,Q}^2 &= c_0 + c_2m_\pi^2 + c_4m_\pi^4 %\nn\\
%&
+ \tilde{\Si}_{\eta'\!\eta'}^Q(m_\pi^2;\La)
+ \tilde{\Si}_{\!\eta'}^Q(m_\pi^2;\La) + \ca{O}(m_\pi^6)\\
&= c_0 + \chi_1 m_\pi + c_2 m_\pi^2 %\nn\\
%
%&
+ \chi_3 m_\pi^3 + c_4 m_\pi^4 + \ca{O}(m_\pi^6)\,.
\label{eqn:renormexpsn2}
\end{align}
Equation (\ref{eqn:renormexpsn1}) is the extrapolation formula for 
$m_{\rho,Q}^2$ 
at infinite lattice volume. The fit coefficients are $c_0$, $c_2$ and $c_4$; 
and $m_{\rho,Q}$ is obtained by taking the square root of either Equation   
(\ref{eqn:renormexpsn1}) or (\ref{eqn:renormexpsn2}).

\subsection{Scheme-Independent Coefficients}
\label{subsect:meschis}
\begin{figure}[tp]
\begin{center}
\includegraphics[height=150pt,angle=0]{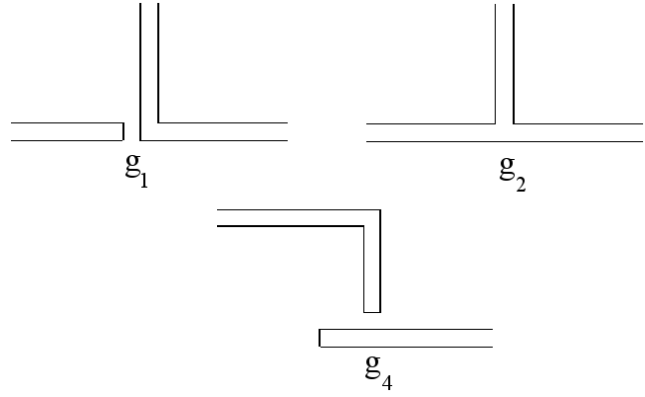}
\vspace{-6pt}
\caption{\footnotesize{Coupling types following convention 
introduced by Booth \cite{Booth:1996hk}.}}
\label{fig:coupl}
\end{center}
\end{figure}

%UP TO HERE 8/2/11
The convention used for defining the values of $\chi_1$, $\chi_3$ and
the various coupling constants that occur in each, 
follows Booth \cite{Booth:1996hk}. For the possible different
values that coupling constants can take, 
 the definitions from %Refs. 
Chow \& Rey 
\cite{Chow:1997dw}, 
Armour \textit{et.al.} 
 \cite{Armour:2005mk} and
Sharpe \cite{Sharpe:1996ih} are used. 
The types of vertices available are displayed in Figure \ref{fig:coupl},
where the couplings 
$g_2$ and $g_4$ occur explicitly in the two diagrams considered here.
Booth suggests naturalness for  $g_2 \sim 1$, and that $g_4 \sim 1/N_c$ 
\cite{Booth:1996hk}. 
These quenched coupling constants can be connected to the experimental
value of $g_{\om\rho\pi}$ as per Lublinsky \cite{Lublinsky:1996yf} by
the relation:
\eqb
g_2 = \f{1}{2}g_{\om\rho\pi}f_\pi\,,
\eqe
where $g_{\om\rho\pi} = 14\pm2$ GeV$^{-1}$ and
the pion decay constant is again taken to be $f_\pi = 92.4$ MeV. 
Thus $g_2$ is chosen to be $0.65\pm0.09$ GeV and $g_4$ is chosen to be 
 $g_2/3$. 
The coupling between the separate legs of the double hairpin
diagram are approximated by the massive constant $M_0^2 \propto m_{\eta'}^2$.
The next-order correction to $M_0$ in momentum $k$ defines the 
coupling to be $-M_0^2 + A_0 k^2$. These constants can be connected to
the full QCD $\eta'$ meson mass $m_{\eta'}$ by considering the geometric
series of terms as %previously 
illustrated earlier, in Figure \ref{fig:PQetaPrime}. 
%Thus a relation is found:
%
%\eqb
%M_0^2 = (1+A_0)m_{\eta'}^2 - m_\pi^2\,.
%\eqe
%
As a result, $M_0^2$ is taken to be $0.6\pm0.2$ GeV$^2$ and 
$A_0$ is taken to be $0\pm0.2$. The central values of each quantity 
are used in the final analysis. 

%CHIS
The coefficients $\chi_{\eta'\!\eta'}$ and $\chi_{\eta'}$ 
 can be specified in terms of the
relevant coupling constants:
\begin{align}
\chi_{\eta'\!\eta'} &= -2\mr\!\f{g_2^2}{4\pi f_\pi^2}\,,\nn\\
\chi_{\eta'} &= -2\mr\!\f{g_2 g_4}{6\pi f_\pi^2}\,,
\end{align}
where the couplings are defined relative to $\mr$, representing the $\rho$ 
meson mass in the chiral limit, which 
is taken to be $770$ MeV.

%\subsection{Finite-Volume Effects for the $\rho$ Meson}

%Since any data analyzed is
% obtained from a lattice of discretized momenta,
%any extrapolations performed 
%should take into account finite-volume effects.
%In order to achieve this,
%each of the 3-dimensional integrals can be
% transformed to its form on the lattice
%using the discretization procedure as per Allton \textit{et.al.}
%\cite{Armour:2005mk}:
%%
%\eqb
%\int\!\!\ud^3\! k \ra \f{{(2\pi)}^3}{L_x L_y L_z} \sum_{k_x,k_y,k_z}\,. 
%\eqe
%%
%Each momentum component is quantized in units of $2\pi/L$, that
%is $k_i=n_i2\pi/L$ for integers $n_i$.  finite-volume corrections 
%$\de^{\ro{FVC}}$
% can be written simply as the difference between the finite sum
%and the corresponding integral. It is known that the finite-volume 
%corrections saturate to a fixed result for large regulator values 
%\cite{Hall:2010ai}.
%Following the example set by this article, the value $\La'= 2.0$ GeV is chosen
%to evaluate all finite-volume corrections independent of the integral cutoff
% scale $\La$ in Equations (\ref{eqn:doub}) and (\ref{eqn:sing}).
The finite-volume version of Equation (\ref{eqn:renormexpsn1})
 can thus be expressed:
\begin{align}
m_{\rho,Q}^2 &= c_0 + c_2m_\pi^2 + c_4m_\pi^4 
+ (\tilde{\Si}_{\eta'\!\eta'}^Q (m_\pi^2;\La)
+ \de^\ro{FVC}_{\eta'\!\eta'}(m_\pi^2;\La')) \nn\\
&+ (\tilde{\Si}_{\!\eta'}^Q (m_\pi^2;\La) + \de^\ro{FVC}_{\eta'}(m_\pi^2;\La'))
 + \ca{O}(m_\pi^6)\,.
\label{eqn:renormexpsnfin}
\end{align}

\section{Extrapolating the Quenched $\rho$ Meson Mass}

\subsection{Renormalization Flow Analysis}
\label{subsect:curves}

%*****************************************%
%
%DATA ANALYSIS

\begin{figure}[tp]
\centering
\includegraphics[height=0.70\hsize,angle=90]{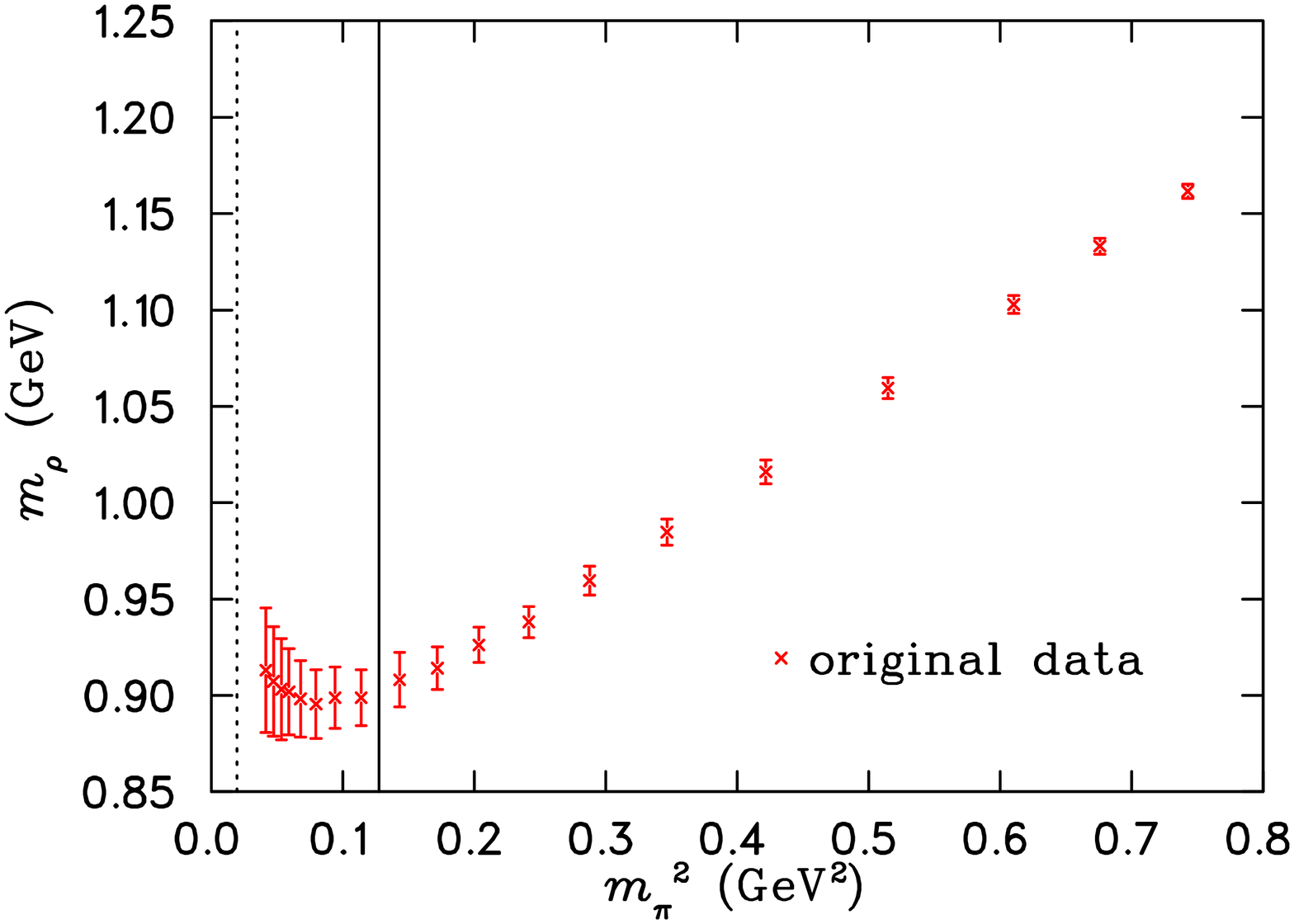}
\vspace{-13pt}
\caption{\footnotesize{ Quenched lattice QCD data for
    the $\rho$ meson mass provided by the Kentucky Group. The dashed
    vertical line indicates the physical pion mass and the solid
    vertical line shows how the data set is split into two parts.}}
\label{fig:rhodata}
%\end{figure}
\vspace{12mm}
%\begin{figure}[tp]
%\centering
\includegraphics[height=0.70\hsize,angle=90]{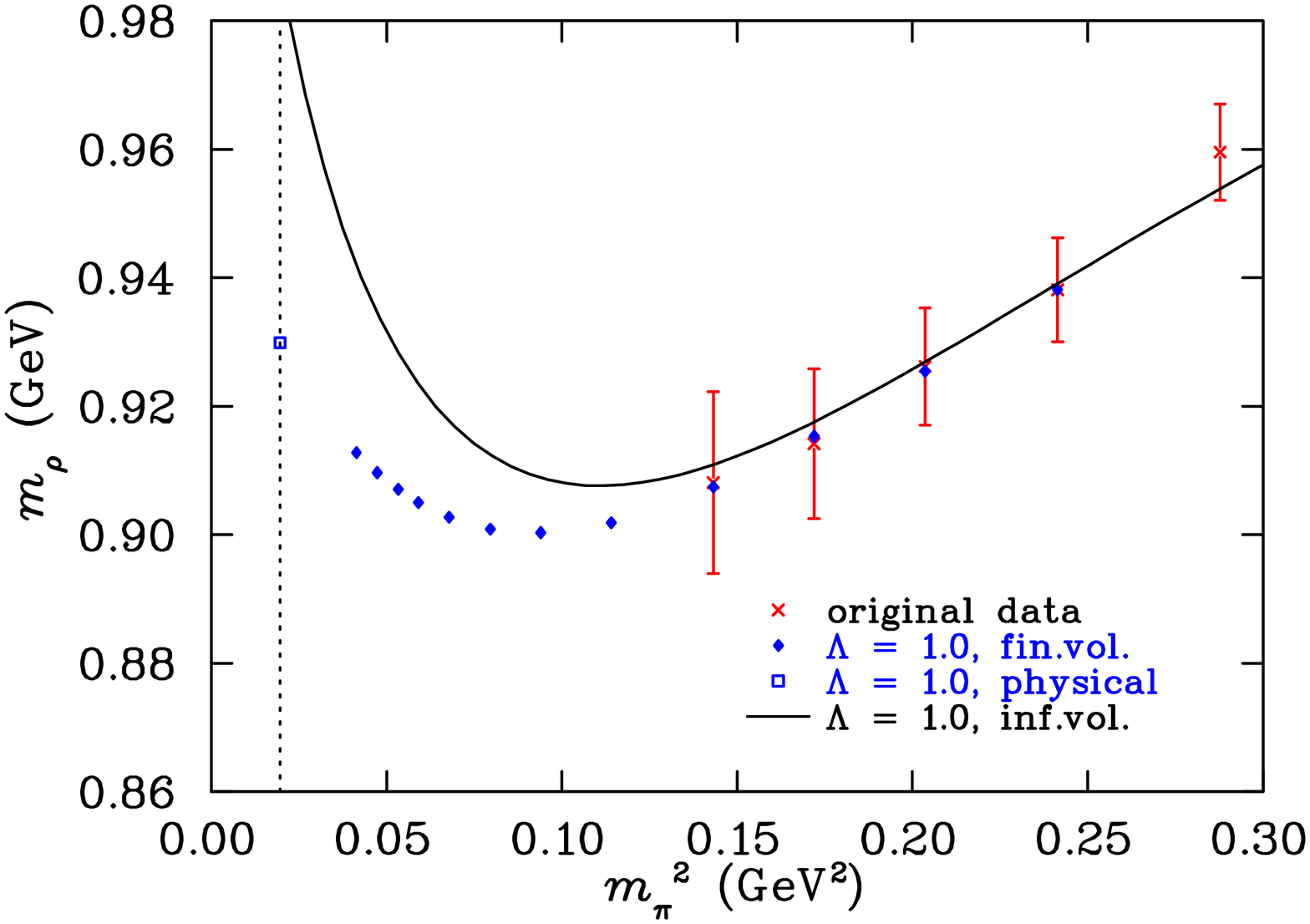}
\vspace{-13pt}
\caption{\footnotesize{ A test extrapolation based on the four original data points shown.  Both the finite- and infinite-volume results are shown for a triple-dipole regulator at $\La = 1.0$ GeV. The dashed vertical line indicates the physical pion mass.}}
\label{fig:testextrap}
\end{figure}

The data displayed in Figure \ref{fig:rhodata} 
are split into two parts. All the data points to the 
left of the solid vertical line are unused in the extrapolation and 
kept in reserve. This is so that the extrapolation can be checked against
 these known data points. The data points to the right of the solid vertical 
line are used for extrapolation. The full set of data is also listed in 
Appendix 
\ref{chpt:appendix4}, Table \ref{table:rhodata}. Note that 
in QQCD, the simulation 
results are correlated. The correlations have been taken into account 
in all fits and extrapolations. 
%
%THE DATA

In order to produce an extrapolation to each test value of $m_\pi^2$,
 an FRR scale $\La$ must be selected.
As an example, one can choose a triple-dipole regulator at $\La = 1.0$ GeV.
% For each data point used in the 
%extrapolation at a particular $m_\pi^2$ value,
% the finite volume loop sums are subtracted. Then the values of the 
%parameters $a_0$, $a_2$ and $a_4$ are obtained by fitting the data. Finally, 
%the extrapolations are made by calculating Eq.~(\ref{eqn:renormexpsn1}),
% adding 
%back in the loop contributions (at finite or infinite volume)
% at the desired value of $m_\pi^2$.
 By using Equation (\ref{eqn:renormexpsnfin}), finite- and infinite-volume 
extrapolations are shown in Figure 
\ref{fig:testextrap}. The values of $m_\pi^2$ selected for the 
 finite-volume extrapolations 
exactly correspond with the missing low-energy data points set aside
 earlier.  
The physical point $m_\pi^2 = 0.0196$ GeV$^2$ is included as well.
%

%c0,c2,c4 vs Lambda:
\begin{figure}[tp]
\centering
\includegraphics[height=0.70\hsize,angle=90]{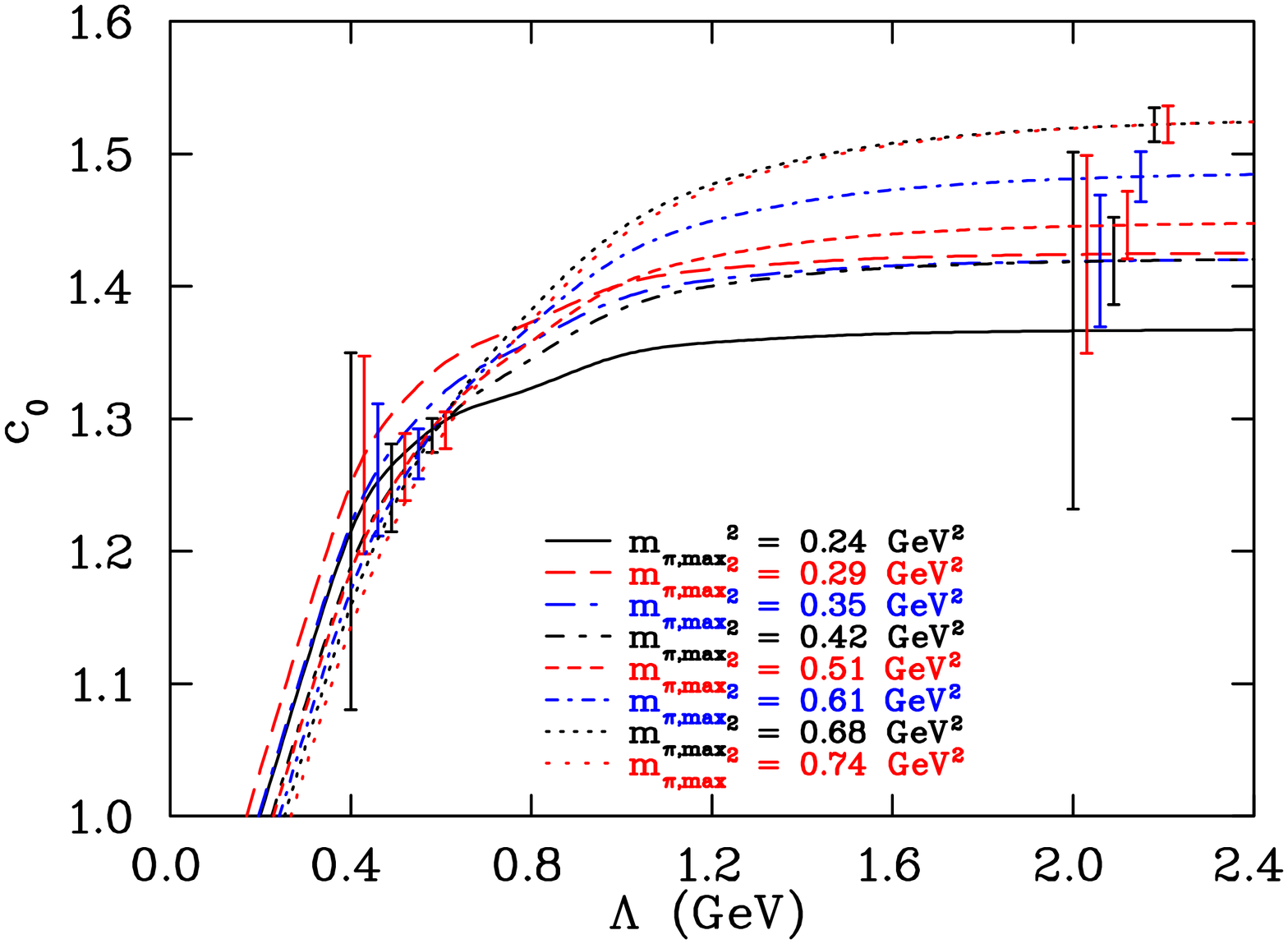}
\vspace{-13pt}
\caption{\footnotesize{ Behaviour of $c_0$ vs.\ $\La$ based on Kentucky Group data. A triple-dipole regulator is used. A few points are selected to indicate the general size of the statistical error bars.}}
\label{fig:Kehfeic0}
\vspace{5pt}
\includegraphics[height=0.70\hsize,angle=90]{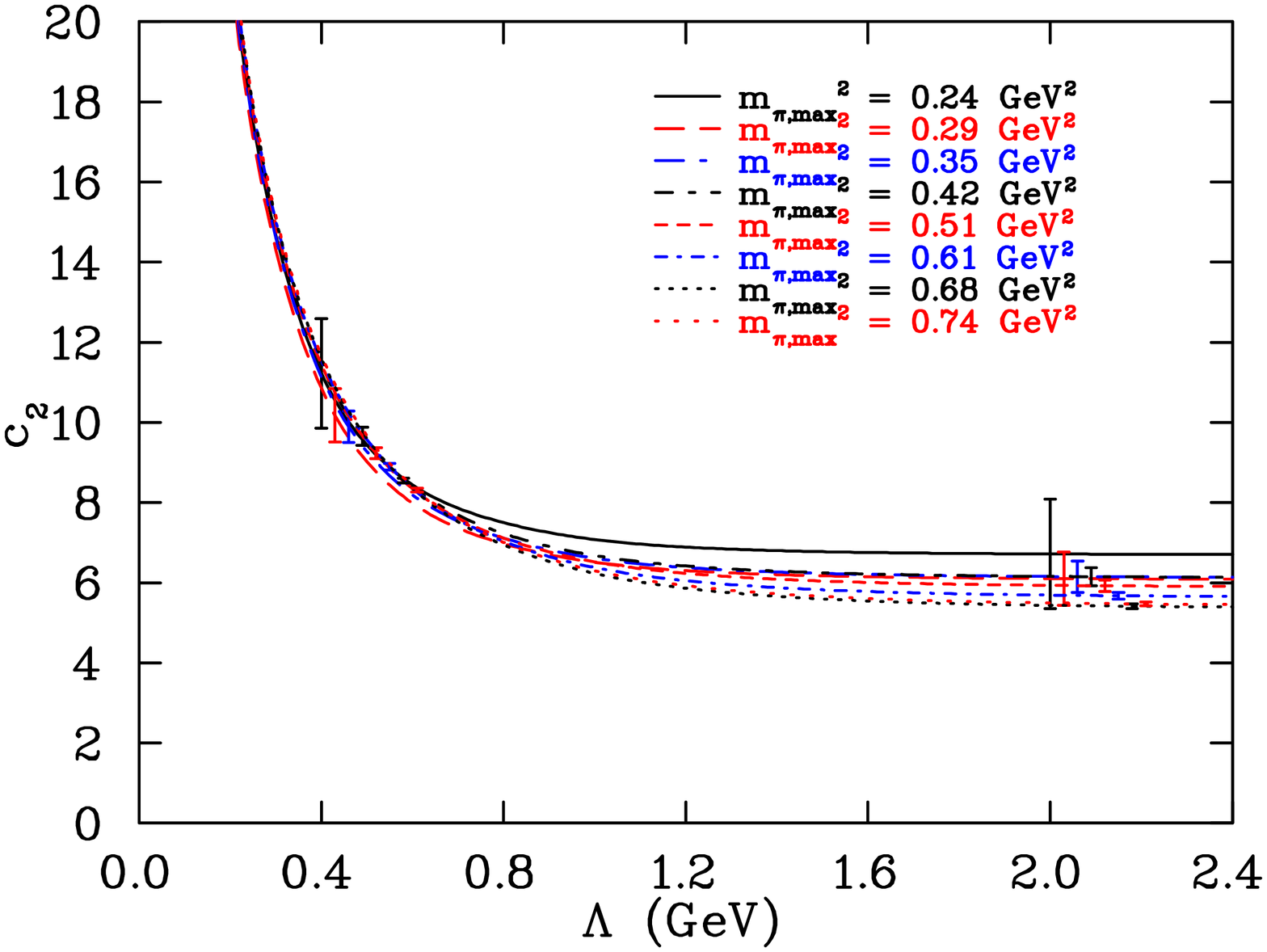}
\vspace{-13pt}
\caption{\footnotesize{Behaviour of $c_2$ vs.\ $\La$ based on Kentucky Group data. A triple-dipole regulator is used. A few points are selected to indicate the general size of the statistical error bars.}}
\label{fig:Kehfeic2}
\vspace{5pt}
\includegraphics[height=0.70\hsize,angle=90]{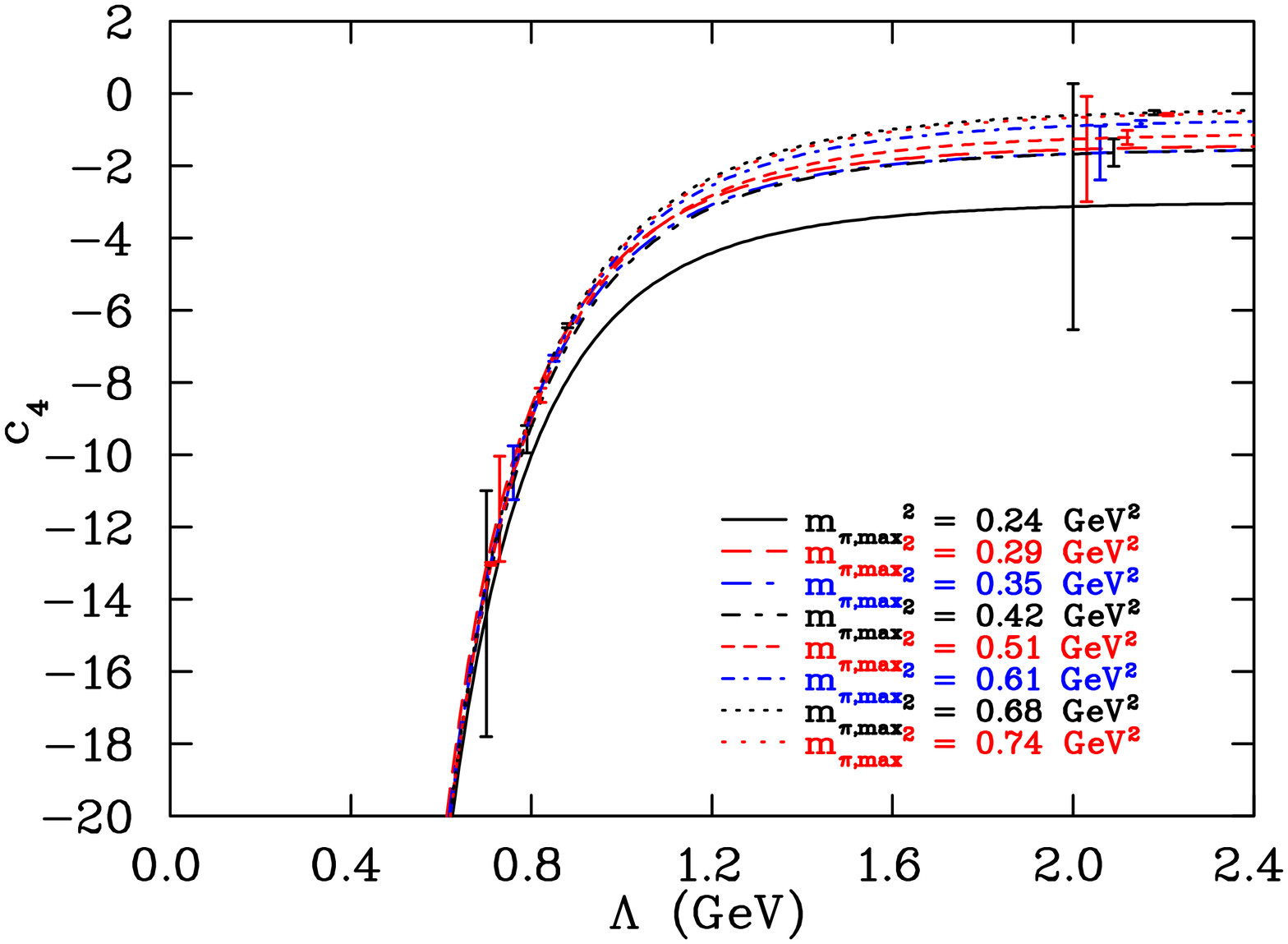}
\vspace{-13pt}
\caption{\footnotesize{Behaviour of $c_4$ vs.\ $\La$ based on Kentucky Group data. A triple-dipole regulator is used. A few points are selected to indicate the general size of the statistical error bars.}}
\label{fig:Kehfeic4}
\end{figure}

Now the regularization 
scale-dependence of low-energy coefficients $c_0$, $c_2$ and 
$c_4$ is investigated for various upper limits of range of pion masses.
The renormalization of these low-energy coefficients 
 is considered across a range of $\La$ values.
Each renormalization flow curve corresponds to a
different value of maximum pion mass, $m_{\pi,\ro{max}}^2$. 
Thus the behaviour of the renormalization of the low-energy coefficients 
can be examined as the lattice data set is extended further outside the PCR.
%For example, the first renormalization 
%flow curve will have $m_\pi^2 = 0.24$ GeV$^2$, and so on.
Figures \ref{fig:Kehfeic0} through \ref{fig:Kehfeic4}
 show the renormalization flow curves for each of $c_0$, $c_2$ and $c_4$.
Each data
point plotted has an associated error bar, but for the sake of clarity 
only a few points are selected to indicate the general size of the 
statistical error bars.
Using the procedure described in Chapter \ref{chpt:intrinsic}, 
 the optimal regularization scale %will be 
 is identified by the value of the regularization scale 
%$\La$ 
that minimizes the discrepancies among 
the renormalization flow curves. This 
indicates the scale at which the renormalization of %$c_0$, $c_2$ 
%and $c_4$ 
each $c_i$ is least sensitive to truncation of the data. Physically, 
this value of $\La$ can be associated with an intrinsic scale related 
to the size of the source of the pion cloud. 
%
%value $\La^\ro{scale}$ is obtained by plotting the renormalization of the 
%constants $c_0$, $c_2$ and $c_4$ for a variety of values of $\La$. This 
%produces a renormalization flow curve.
%The idea is to obtain many renormalization flow curves corresponding to 
%different values of maximum pion mass, $m_{\pi,\ro{max}}^2$.
%Note that at least four data points must be used, since there are three 
%fit parameters involved: $a_0$, $a_2$ and $a_4$. The first renormalization 
%flow curve will have $m_\pi^2 = 0.24$ GeV$^2$, and so on. 
%Figures \ref{fig:Kehfeic0} through \ref{fig:Kehfeic4}
% show the renormalization flow curves plotted on the 
%same graph, for each of $c_0$, $c_2$ and $c_4$.
%The optimal regulator value 
%will be the value of $\La$ where the most 
%intersections occur among the renormalization flow curves. This will indicate 
%the value of $\La$ where the correct value of $c_0$, $c_2$ and $c_4$ are 
%obtained. In order to find the best overall agreement among the curves
% for each graph, a simple chi-square analysis will be used.

%

By examining Figures \ref{fig:Kehfeic0} through \ref{fig:Kehfeic4}, 
 increasing $m_{\pi,\ro{max}}^2$ %enhances 
 leads to greater scale-dependence in the 
renormalization, since the data sample lies %naturally 
further from the PCR. 
%Complete scale-independence would be indicated by a horizontal line at the 
% physical point.
%and so this 
% necessarily increases any possible scheme-dependence in the extrapolation. 
%This is evident in the behaviour of the renormalized constants
% $c_0$, $c_2$ and $c_4$ with respect to the regulator parameter $\La$.
%A steep line indicates a strong scheme-dependence in the result, 
%and naturally occurs for data samples far outside the PCR, and also
%
 Since the effective field theory is calculated to a 
finite chiral order, complete scale-independence across all possible 
 $\La$ values will not occur in practice.
 An asymptotic value is usually  
observed in the renormalization flow as $\La$ becomes large, indicating that 
%no more chiral physics is included. 
 the higher-order terms of the chiral expansion are effectively zero. 
 However, these asymptotic values of the coefficients 
are  poor estimates 
of their correct values, as previously demonstrated in the pseudodata 
analysis in Chapter \ref{chpt:intrinsic}. %\cite{Hall:2010ai}. 
Instead, the best estimates of the low-energy
 coefficients lie in the identification of the intersection point of the 
renormalization flow of these coefficients.
It is also of note that, 
for small values of $\La$, the FRR scheme breaks down, as observed 
for the nucleon mass in Section \ref{sec:lowerbound}. 
The regularization scale must be at least large enough to include
the chiral physics being studied. 

\subsection{Intrinsic Scale and Systematic Uncertainties}
\label{subsect:err}

The optimal regularization scale $\La^\ro{scale}$ can be obtained from the 
renormalization flow curves using a chi-square-style analysis.
 In addition, 
 the analysis will allow the extraction of a variance for 
$\La^\ro{scale}$. Knowing how the data are correlated, the systematic 
errors from the coupling constants and $\La^\ro{scale}$ will be combined
 to obtain an error estimate for each extrapolation point. 
%, in particular, the 
%value of $m_{\rho,Q}^2$ at the physical point.
Of particular interest are the values of $m_{\rho,Q}$ at those values of 
$m_\pi^2$ that are explored in the lattice simulations,  
but are excluded in the chiral 
extrapolation. 
The function $\chi^2_{dof}$ is constructed in the same way as 
Equations (\ref{eqn:chisqwm}) and (\ref{eqn:chisq}).

%CHISQDOF
%graph of chisq/dof for c0,c2,c4
\begin{figure}[tp]
\centering
\includegraphics[height=0.70\hsize,angle=90]{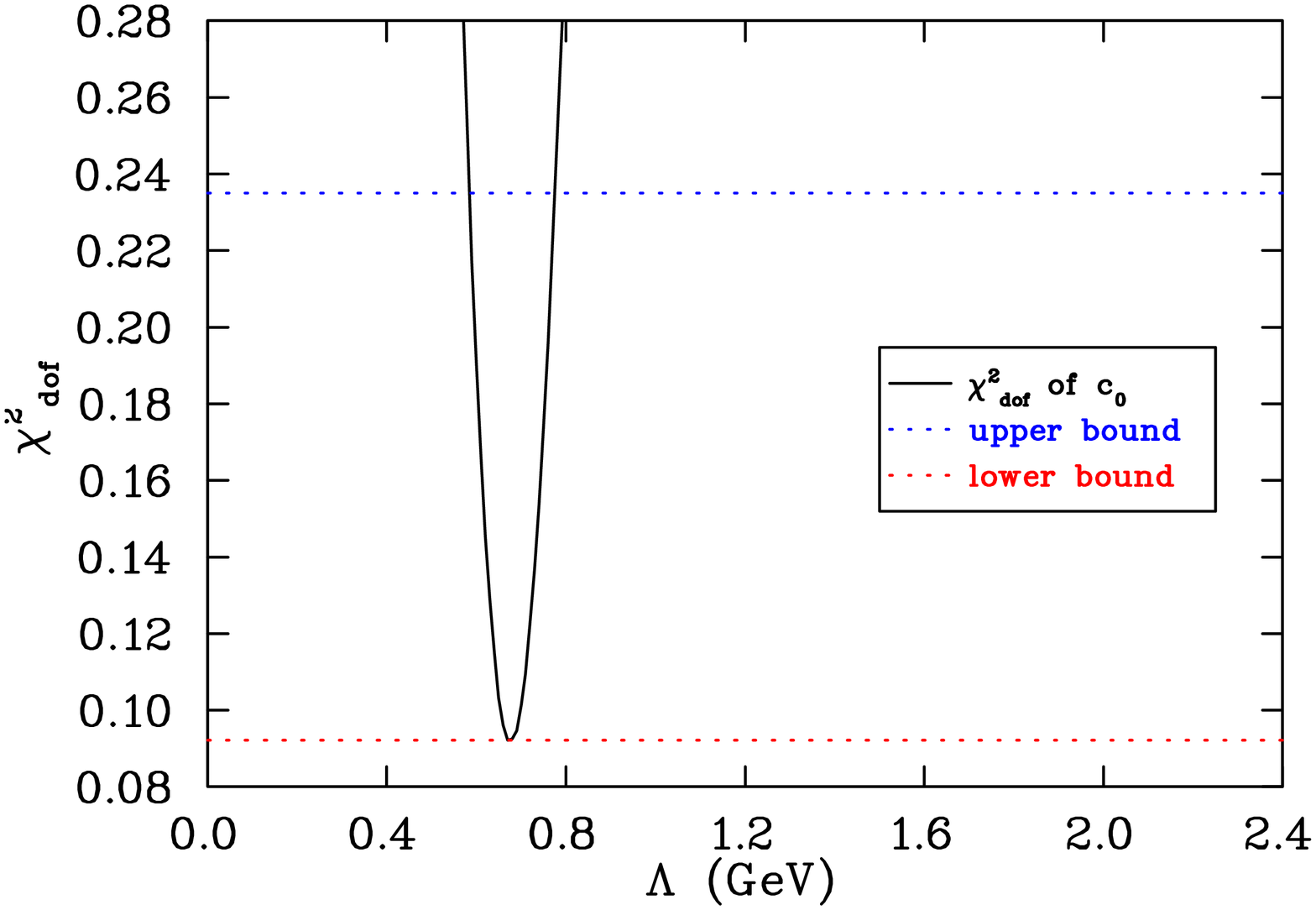}
\vspace{-11pt}
\caption{\footnotesize{$\chi^2_{dof}$ for $c_0$ versus $\La$, corresponding to the renormalization flow curves displayed in Figure \ref{fig:Kehfeic0} based on Kentucky Group data. A triple-dipole regulator is used.}}
\label{fig:Kehfeichisqdofc0}
\vspace{5pt}
\includegraphics[height=0.70\hsize,angle=90]{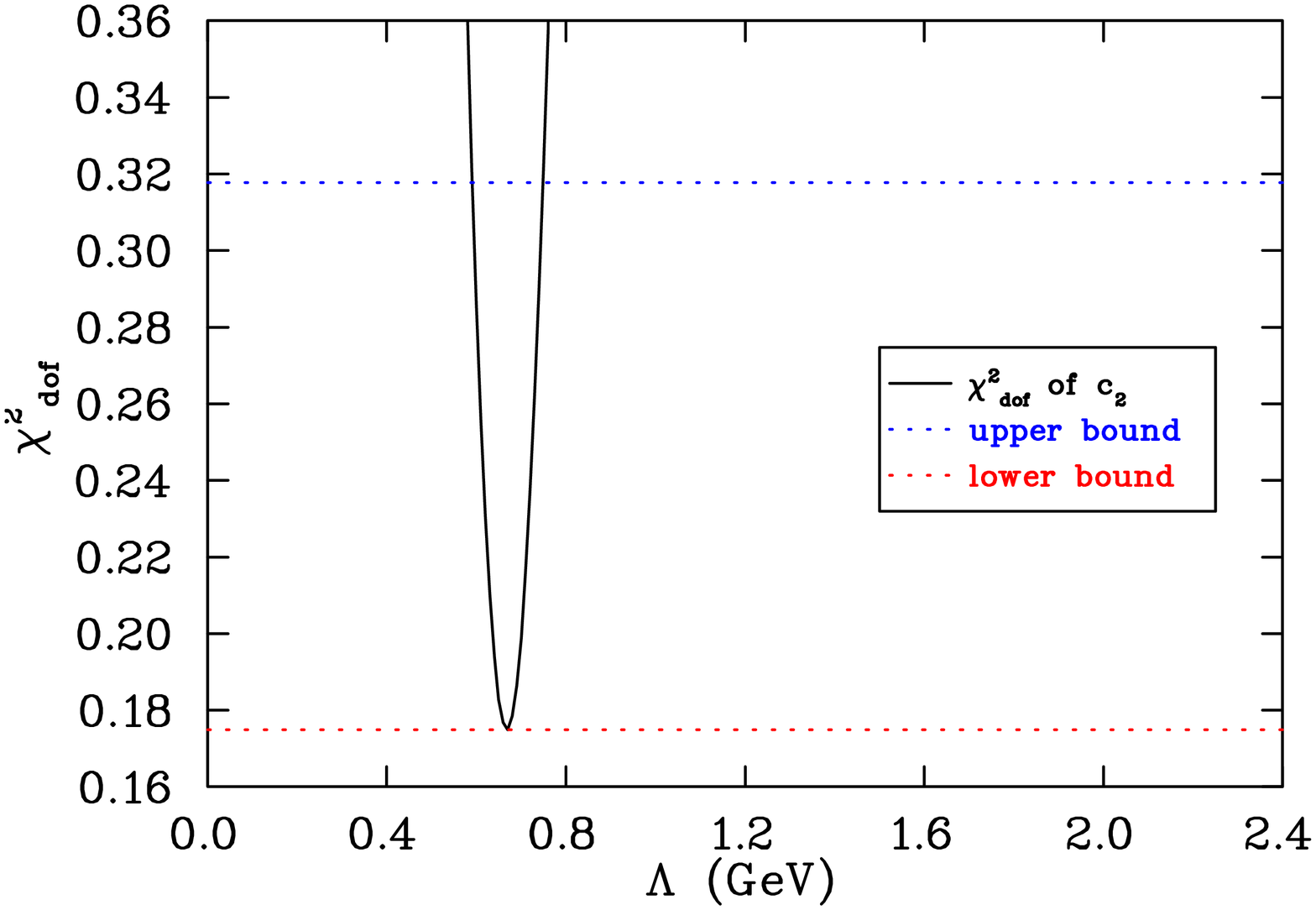}
\vspace{-11pt}
\caption{\footnotesize{$\chi^2_{dof}$ for $c_2$ versus $\La$, corresponding to the renormalization flow curves displayed in Figure \ref{fig:Kehfeic2} based on Kentucky Group data. A triple-dipole regulator is used.}}
\label{fig:Kehfeichisqdofc2}
\vspace{5pt}
\includegraphics[height=0.70\hsize,angle=90]{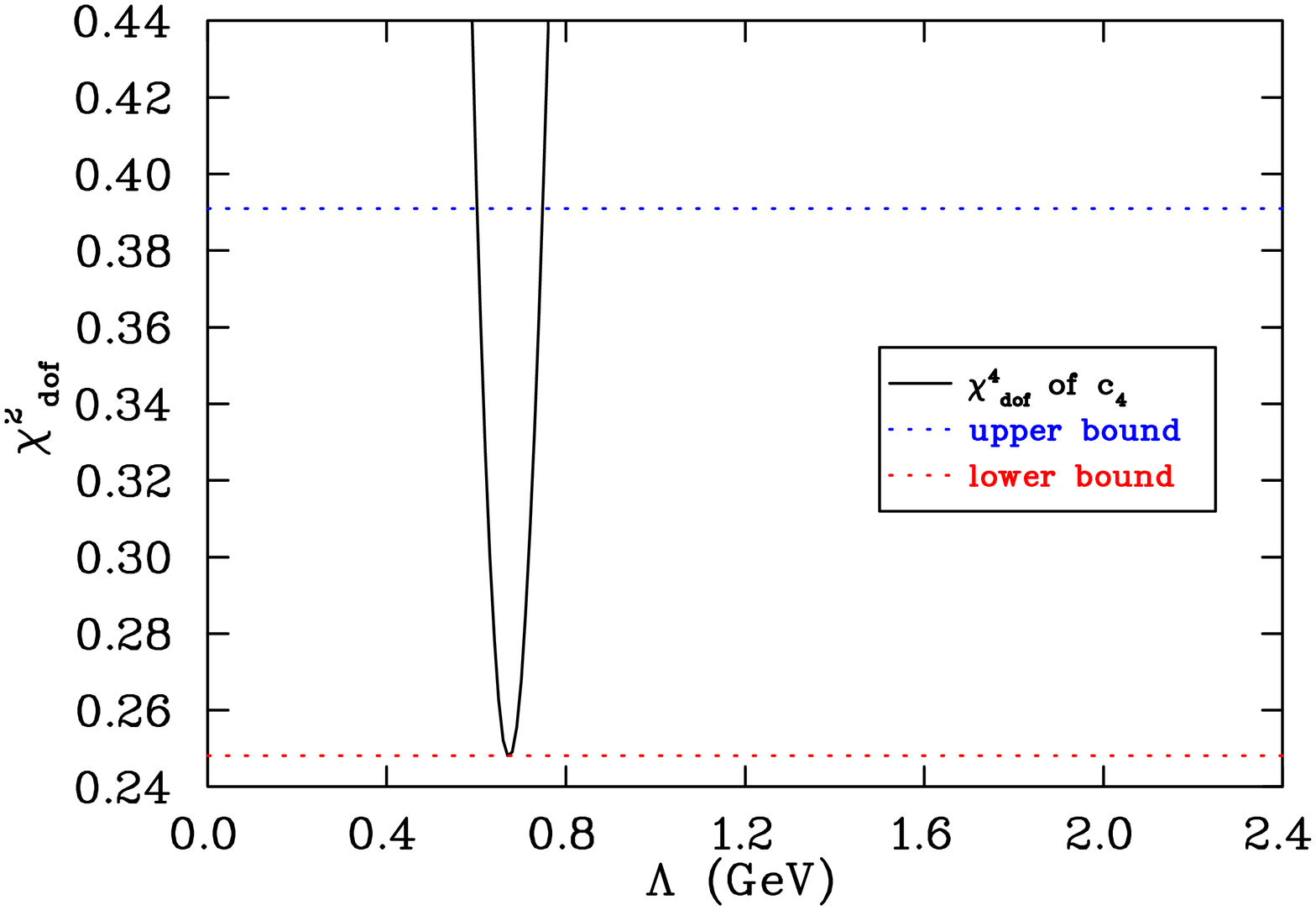}
\vspace{-11pt}
\caption{\footnotesize{(color online). $\chi^2_{dof}$ for $c_4$ versus $\La$, corresponding to the renormalization flow curves displayed in Figure \ref{fig:Kehfeic4} based on Kentucky Group data. A triple-dipole regulator is used.}}
\label{fig:Kehfeichisqdofc4}
\end{figure}

%To obtain a measure of the uncertainty associated with %the identification of 
%an optimal regulator, %(as identified from the intersection
%a $\chi^2_{dof}$ function is constructed. This function should allow 
% easy identification of the intersection points in the renormalization 
%flow curves, and a range associated with this central regulator value.
%The first step is to plot $\chi^2_{dof}$
% against a variety of regulator values $\La$. 
%The relevant degrees of freedom are the curves with differing values of 
%$m_{\pi,\ro{max}}^2$.
%
% A plot of $\chi^2_{dof}$ is constructed separately
% for each renormalized constant $c$ (with uncertainty $\de c$):
%%
%\eqb
%\chi^2_{dof} = \f{1}{n-1} \sum_{i=1}^{n} \f{{(c_i(\La) - c^T(\La))}^2}
%{{(\de c_i(\La))}^2},
%\eqe
%
%for $i$ corresponding to data sets with differing 
%$m_{\pi,\ro{max}}^2$ ($n = 8$). The theoretical value $c^T$ is given by the
%weighted mean:
% 
%\eqb
%c^T(\La) = \f{\sum_{i=1}^{n}c_i(\La)/{{(\de c_i(\La))}^2}}
%{\sum_{j=1}^{n} 1 / {(\de c_j(\La))}^2}.
%\eqe
% 
The $\chi^2_{dof}$ plots using a triple-dipole regulator are shown in 
Figures \ref{fig:Kehfeichisqdofc0} through \ref{fig:Kehfeichisqdofc4}. 
The optimal regularization scale $\La^\ro{scale}$ 
is taken to be the central value $\La^\ro{central}$ of each plot. 
The upper and lower bounds of $\Lambda$ obey 
 the condition $\chi^2_{dof} < \chi^2_{dof, min} + 1/(dof)$. 
The results for the optimal regularization scales obtained from 
analyzing each low-energy coefficient, and their associated  
 upper and lower bounds, are shown in 
Table \ref{table:messcales}. 
%
%It is truly remarkable that each low-energy coefficient leads to 
%the same optimal value of $\La$, that is, $\La^{\ro{central}} = 0.64$ GeV. 
% By averaging the results among $c_0$, $c_2$,
% and $c_4$, the optimal regularization scale $\La^{\ro{scale}}$
%for the quenched $\rho$ meson mass can be calculated for this data set: %; 
%% Using a triple dipole regulator, 
%  %$ \La_{\ro{scale}} = 0.64$ %($\stackrel{+0.08}{-0.07}$) GeV.
% %$\Big\{\begin{matrix}+0.08\\-0.07 \end{matrix}$ GeV.
% $\La^{\ro{scale}} = 0.64^{+0.08}_{-0.07}$ GeV.
%
   It is remarkable that each low-energy coefficient leads to 
the same optimal value of $\La$, i.e. $\La_{\ro{central}} = 0.67$ GeV. 
 By averaging the results among $c_0$, $c_2$,
 and $c_4$, the optimal regularization scale $\La_{\ro{scale}}$
for the quenched $\rho$ meson mass can be calculated for this data set: 
% Using a triple dipole regulator, 
  %$ \La_{\ro{scale}} = 0.64$ %($\stackrel{+0.08}{-0.07}$) GeV.
 %$\Big\{\begin{matrix}+0.08\\-0.07 \end{matrix}$ GeV.
 $\La_{\ro{scale}} = 0.67^{+0.09}_{-0.08}$ GeV. %  
%

%  %Paragraph explaining the renormalization flow for the excluded low-energy data. Comparing the result with the new data set provides a check to see if we are closer to the PCR.]]
%[[explanation for the wiggle: behaviour of $a_0$ and $b_0^{\eta'\eta'}$.]]
%[[the chisqdof curves for them?? why? I don't need these, surely.]]
%[[table showing comparison of central values of $c_{0,2,4}$, using best mpisq.]]
%[[plot of $m_\rho$ vs.\ $m_{\pi,\ro{max}}^2$. Explain the new optimal no. of data points.

\begin{figure}[tp]
\begin{center}
\vspace{-5mm}
\includegraphics[height=0.70\hsize,angle=90]{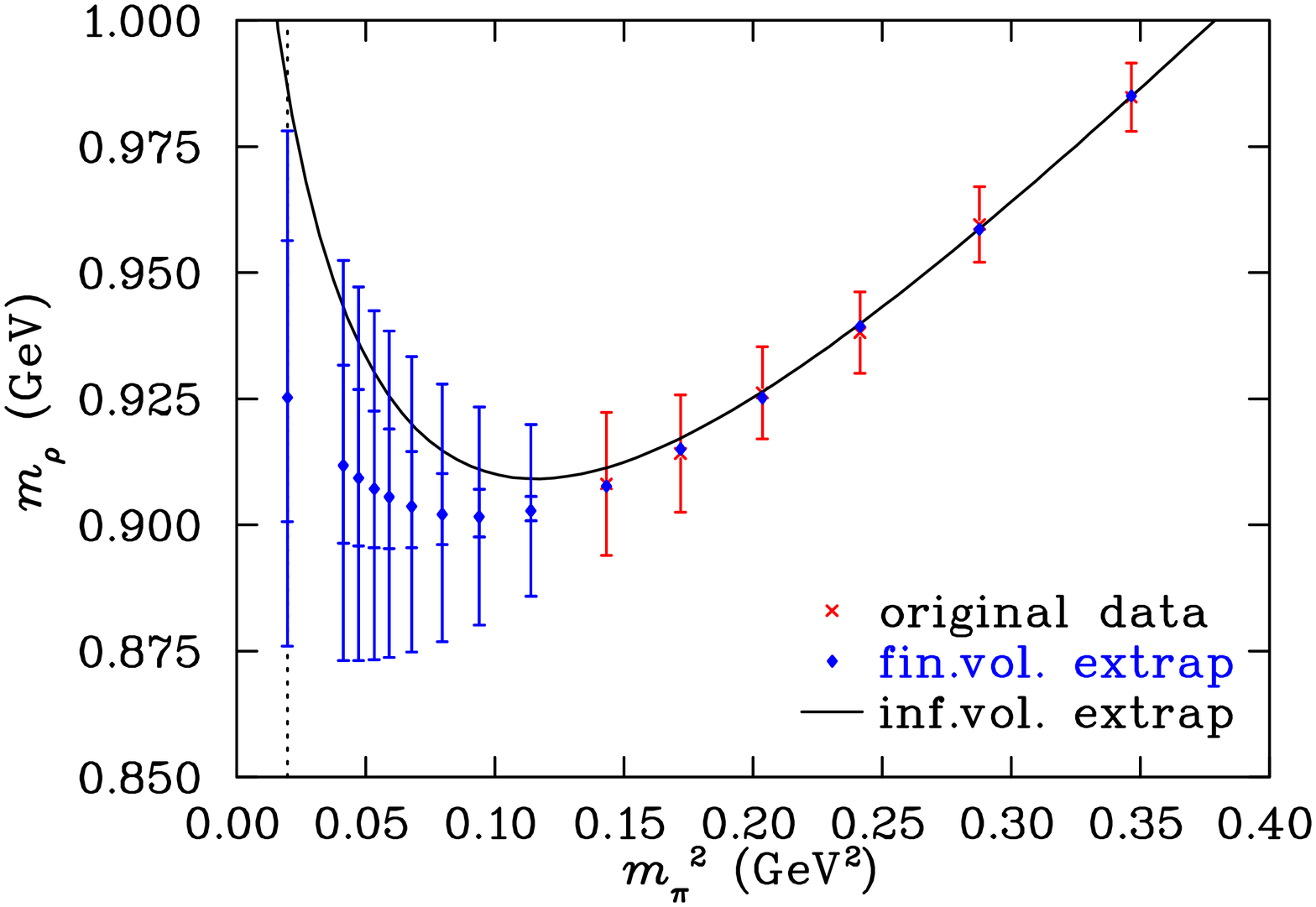}
\vspace{-11pt}
\caption{\footnotesize{   Extrapolation at $\La_{\ro{scale}} = 0.67^{+0.09}_{-0.08}$ GeV based on Kentucky Group data, and using the optimal number of data points, corresponding to $\hat{m}_{\pi,\ro{max}}^2 = 0.35$ GeV$^2$. The inner error bar on the extrapolation points represents purely the systematic error from parameters. The outer error bar represents the systematic and statistical error estimates added in quadrature.}}
\label{fig:initialextrap}
%\end{figure}
%
%final extrapolation
%\begin{figure}[tp]
\vspace{5pt}
\includegraphics[height=0.70\hsize,angle=90]{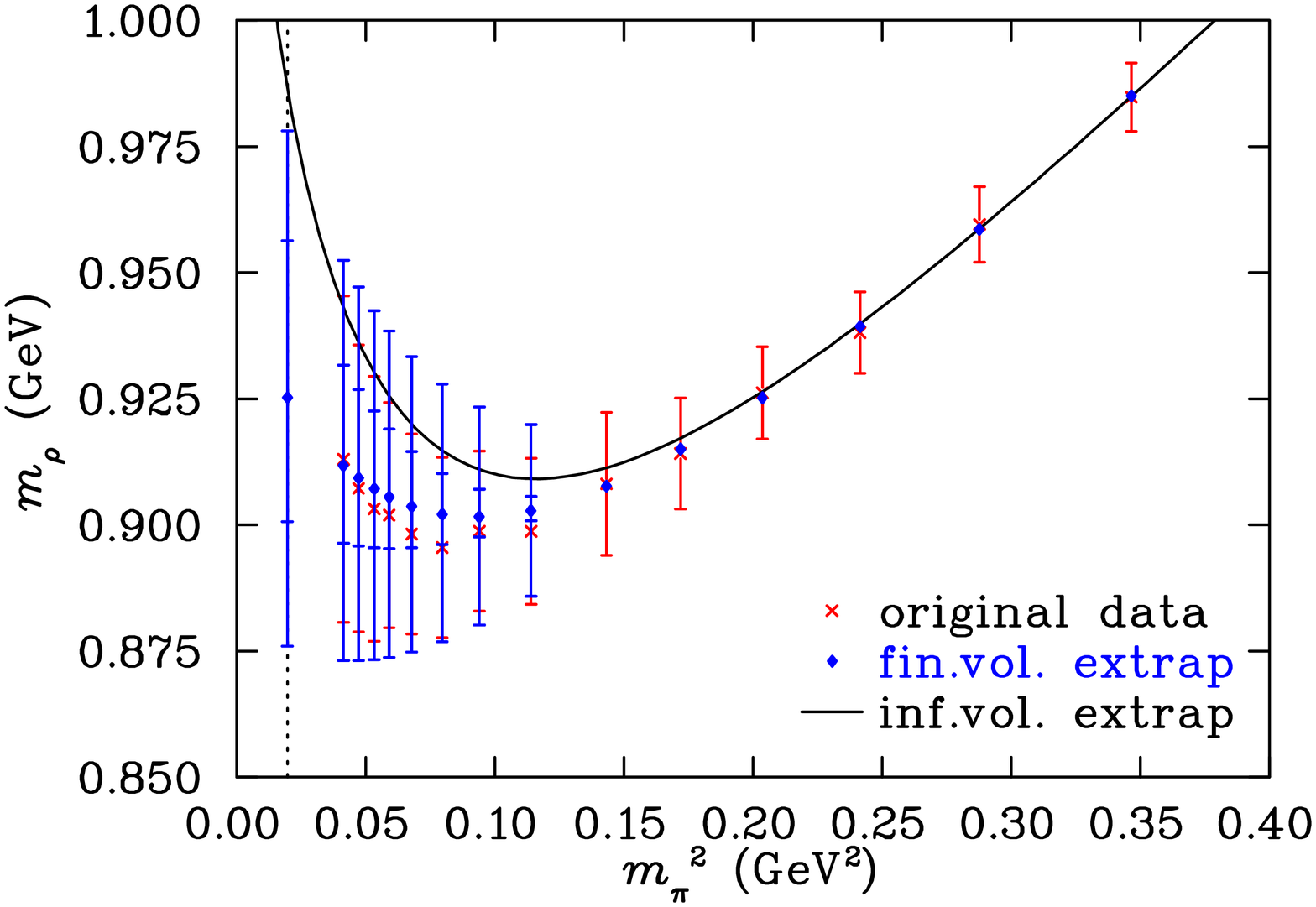}
\vspace{-11pt}
\caption{\footnotesize{   Comparison of chiral extrapolation predictions (blue diamond) with Kentucky Group data (red cross).  Extrapolation is performed at $\La_{\ro{scale}} = 0.67^{+0.09}_{-0.08}$ GeV, and using the optimal number of data points, corresponding to $\hat{m}_{\pi,\ro{max}}^2 = 0.35$ GeV$^2$. The inner error bar on the extrapolation points represents purely the systematic error from parameters. The outer error bar represents the systematic and statistical error estimates added in quadrature.}}
\label{fig:finalextrap}
\end{center}
\end{figure}

The result of the final extrapolation, using the estimate of the optimal 
regularization scale 
$\La_{\ro{scale}} = 0.67^{+0.09}_{-0.08}$ GeV, and using the initial data set
to predict the low-energy data points, is shown in Figure 
\ref{fig:initialextrap}. %The missing original data points 
% are predicted by the extrapolated values. 
The extrapolation to the physical point obtained for this quenched data set is: 
%$m_{\rho,Q}(m_{\pi,\ro{phys}}^2) = 0.915$ 
%($\pm \,0.036$) GeV,
$m_{\rho,Q}^{\ro{ext}}(m_{\pi,\ro{phys}}^2) = 0.925^{+0.053}_{-0.049}$ GeV,  
%($\pm \,0.037$) GeV,
an uncertainty 
of less than $6$\%. 

\begin{table}[tp]
 \caption{\footnotesize{Values of the central, upper and lower regularization scales, in GeV, obtained from the $\chi^2_{dof}$ analysis of $c_0$, $c_2$ and $c_4$, displayed in Figures \ref{fig:Kehfeichisqdofc0} through \ref{fig:Kehfeichisqdofc4}.}}
  \newcommand\T{\rule{0pt}{2.8ex}}
  \newcommand\B{\rule[-1.4ex]{0pt}{0pt}}
  \begin{center}
    \begin{tabular}{llll}
      \hline
      \hline
       \T\B 
      scale (GeV) \qquad & $c_0$ (Fig.\ref{fig:Kehfeichisqdofc0}) $\,\,\,\,\,$& $c_2$ (Fig.\ref{fig:Kehfeichisqdofc2}) $\,\,$& $c_4$ (Fig.\ref{fig:Kehfeichisqdofc4})   \\
      \hline
      %$\La_\ro{central}$  &\T $0.64$ & $0.64$ & $0.64$ \\
      %$\La_\ro{upper}$   &\T $0.76$ & $0.70$ & $0.68$ \\
      %$\La_\ro{lower}$   &\T $0.52$ & $0.58$ & $0.59$ \\
      $\La_\ro{central}$  &\T $0.67$ & $0.67$ & $0.67$ \\
      $\La_\ro{upper}$   &\T $0.78$ & $0.75$ & $0.75$ \\
      $\La_\ro{lower}$   &\T $0.58$ & $0.59$ & $0.60$ \\
      \hline
    \end{tabular}
  \end{center}
\vspace{-6pt}
  \label{table:messcales}
\end{table}

%Figure \ref{fig:finalextrap} shows the same result, now with the missing
% original data points included on the plot for comparison.
%
%It is helpful to remove much of the correlated statistical error at each data 
%point. The uncertainty in each data point is 
%subtracted from the uncertainty of a central data point of, 
%say, $m_\pi^2 = 0.143$ GeV$^2$.
%The result is shown in Figure \ref{fig:finaldeltaextrap}.
%
%Figure \ref{fig:finaldeltaextrap} shows the data plotted with error bars 
%correlated relative to the lightest data point in the original set, 
%$m_\pi^2 = 0.143$ GeV$^2$.
% Note that 
%all of the missing original data points lie within the extrapolations' 
%systematic error bars.
Each extrapolation point  %has 
displays two error bars.
 The inner error bar corresponds to the 
systematic uncertainty %coming from the uncertainty 
in the parameters only, 
 and the outer error bar corresponds
to the systematic and statistical uncertainties 
of each point added in quadrature.
Also, the infinite-volume extrapolation curve is displayed in order to 
illustrate the effect of finite-volume corrections to the loop integrals.

%initial extrapolation
%\vspace{5pt}
%final extrapolation delta
\begin{figure}[tp]
\begin{center}
%\vspace{6mm}
\includegraphics[height=0.70\hsize,angle=90]{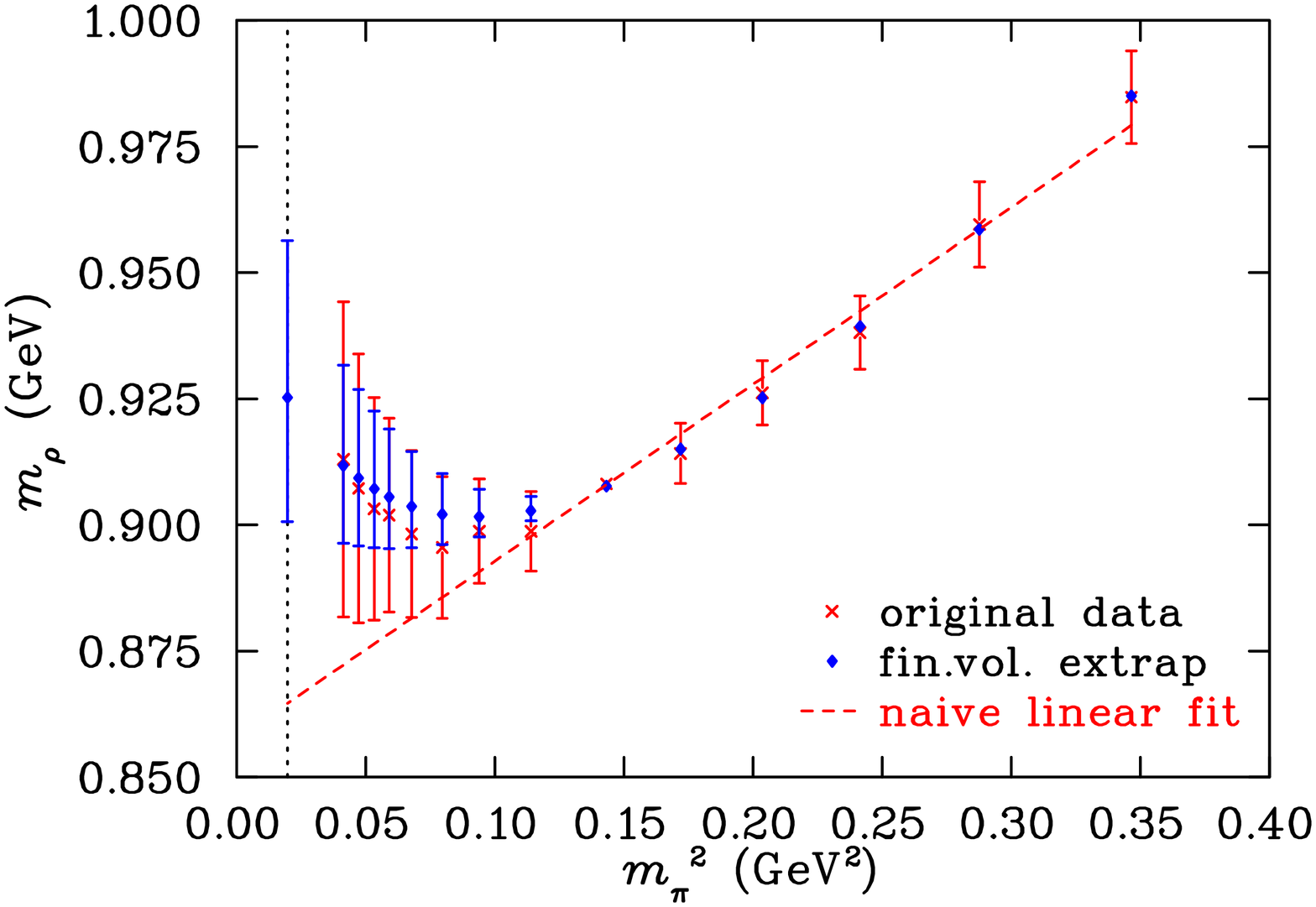}
\vspace{-11pt}
\caption{\footnotesize{   Comparison of chiral extrapolation predictions (blue diamond) with Kentucky Group data (red cross), with errors correlated relative to the point at $m_\pi^2 = 0.143$ GeV$^2$.  Extrapolation is performed at $\La_{\ro{scale}} = 0.67^{+0.09}_{-0.08}$ GeV, and using the optimal number of data points, corresponding to $\hat{m}_{\pi,\ro{max}}^2 = 0.35$ GeV$^2$. The error bar on the extrapolation points represents the systematic error only. A simple linear fit, on the optimal pion mass region, is included for comparison.}}
\label{fig:finaldeltaextrap}
\end{center}
\end{figure}
%\vspace{5pt}

In Figure \ref{fig:finalextrap}, %and 
%\ref{fig:finalextrapvar}, 
the extrapolation predictions are compared 
against the actual simulation results, which were not included in the fit. 
%Figure \ref{fig:finalextrap} displays the result using 
%$\La_{\ro{scale}} = 0.67^{+0.09}_{-0.08}$ GeV, while 
%Figure \ref{fig:finalextrapvar} displays the result using 
%$\La_{\ro{scale}} = 0.64$ GeV,  with the systematic uncertainty 
%calculated by varying $\Lambda$ across all suitable values. 
Both the extrapolations and the simulation results display the 
same non-analytic curvature near the physical point. 
Figure \ref{fig:finaldeltaextrap} shows the data plotted with error bars 
correlated relative to the lightest data point in the original set, 
$m_\pi^2 = 0.143$ GeV$^2$. %, using 
%$\La_{\ro{scale}} = 0.67^{+0.09}_{-0.08}$ GeV. 
To highlight the importance of this application of an extended 
$\chi$EFT, %extended 
%beyond the PCR, 
a simple linear fit is included in Figure \ref{fig:finaldeltaextrap}.  
%which shows the data plotted with error bars 
%correlated relative to the lightest data point in the original set, 
%$m_\pi^2 = 0.143$ GeV$^2$.
 By ignoring low-energy chiral physics, the 
linear fit is statistically incorrect at the physical point.
 %Note also that 
All of the missing original data points are consistent with the extrapolations' 
systematic uncertainties. %, even though Figure \ref{fig:finaldeltaextrap} 
%displays the results using the more restricted estimate of the 
%optimal regularization scale $\La_{\ro{scale}} = 0.67^{+0.09}_{-0.08}$ GeV.
%
%One should note that in Figures \ref{fig:initialextrap} through  
%and \ref{fig:finaldeltaextrap} the extrapolation procedure correctly 
%reproduces the low-energy curvature of the original lattice data.
%In addition, even after %error bar subtraction, 
% statistical correlations are subtracted, the extrapolated points have an 
%error bar almost a third the size of the lattice data points. 
% In order to match this precision at low energies, the time
% required in lattice simulations would increase by an order of magnitude.
After %error bar subtraction, 
 statistical correlations are subtracted, the extrapolated points correspond to
 an 
error bar almost half %a third 
the size of that of the lattice data points. 
 In order to match this precision at low energies, the time
 required in lattice simulations would increase by approximately four times.

%NEW SET c0,c2,c4 vs Lambda:
\begin{figure}[tp]
\centering
\includegraphics[height=0.70\hsize,angle=90]{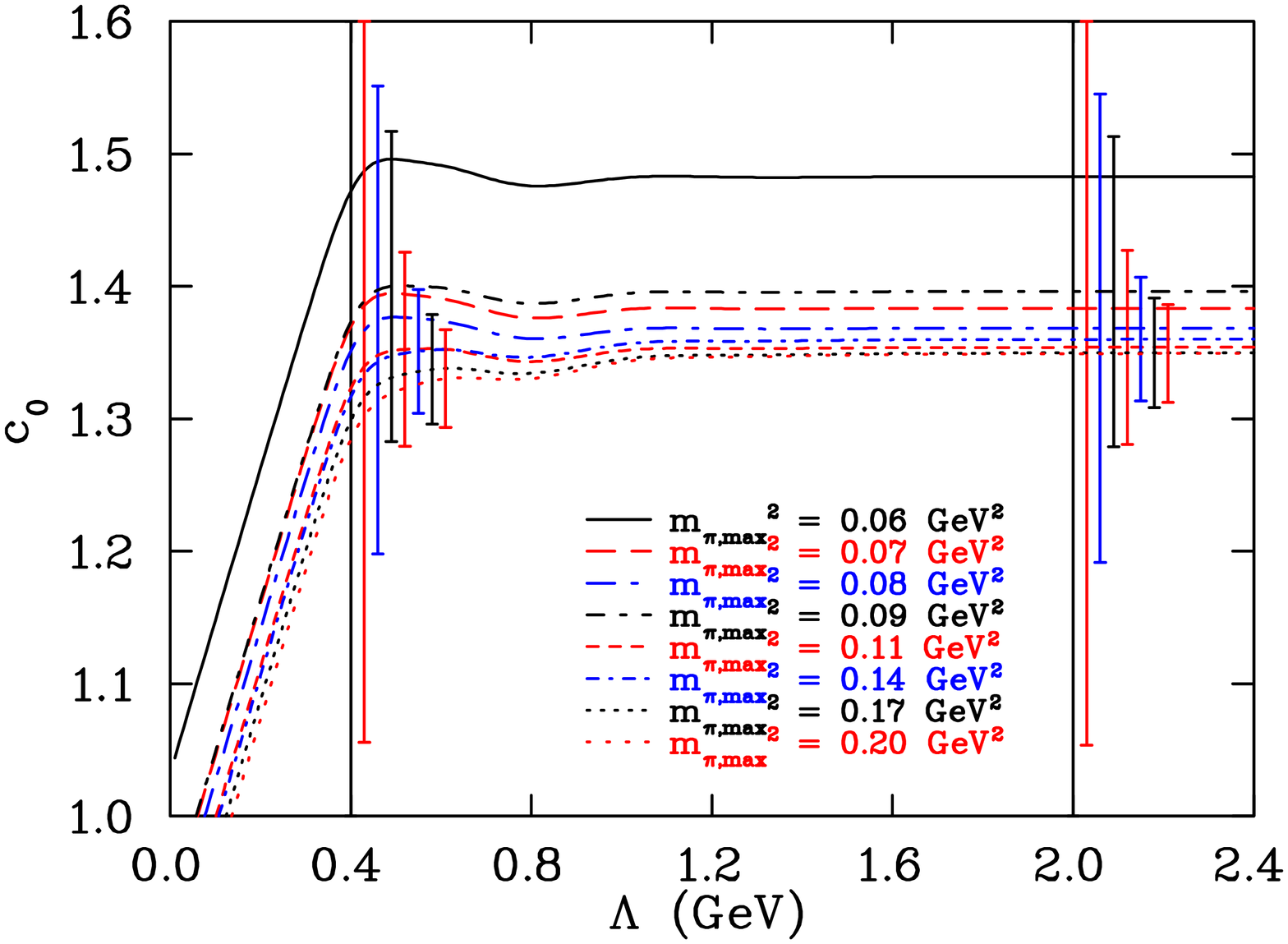}
\vspace{-11pt}
\caption{\footnotesize{Behaviour of $c_0$ vs.\ $\La$ for the initially excluded low-energy data. A triple-dipole regulator is used. A few points are selected to indicate the general size of the statistical error bars.}}
\label{fig:Kehfeic0new}
\vspace{5pt}
\includegraphics[height=0.70\hsize,angle=90]{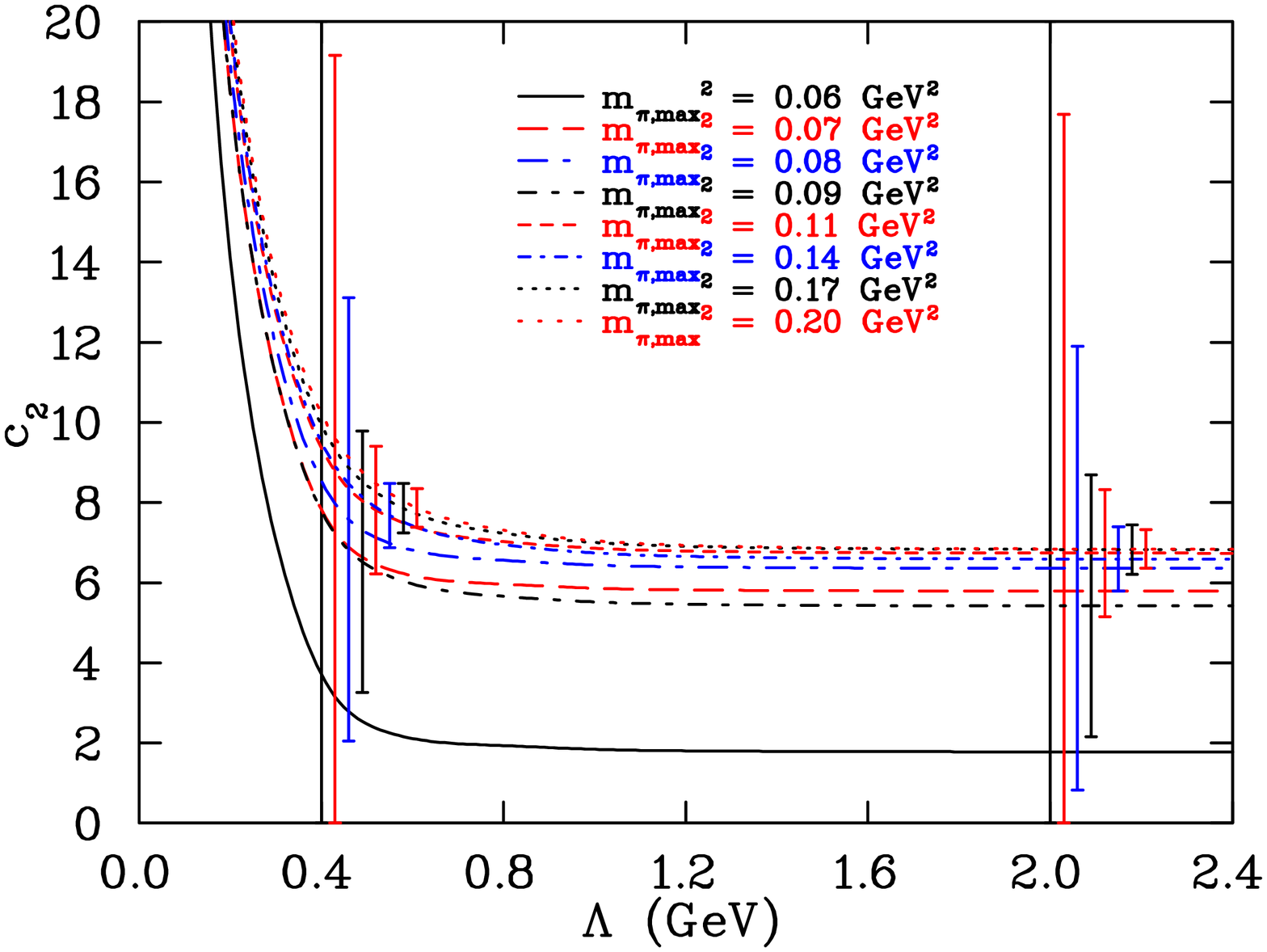}
\vspace{-11pt}
\caption{\footnotesize{Behaviour of $c_2$ vs.\ $\La$ for the initially excluded low-energy data. A triple-dipole regulator is used. A few points are selected to indicate the general size of the statistical error bars.}}
\label{fig:Kehfeic2new}
\vspace{5pt}
\includegraphics[height=0.70\hsize,angle=90]{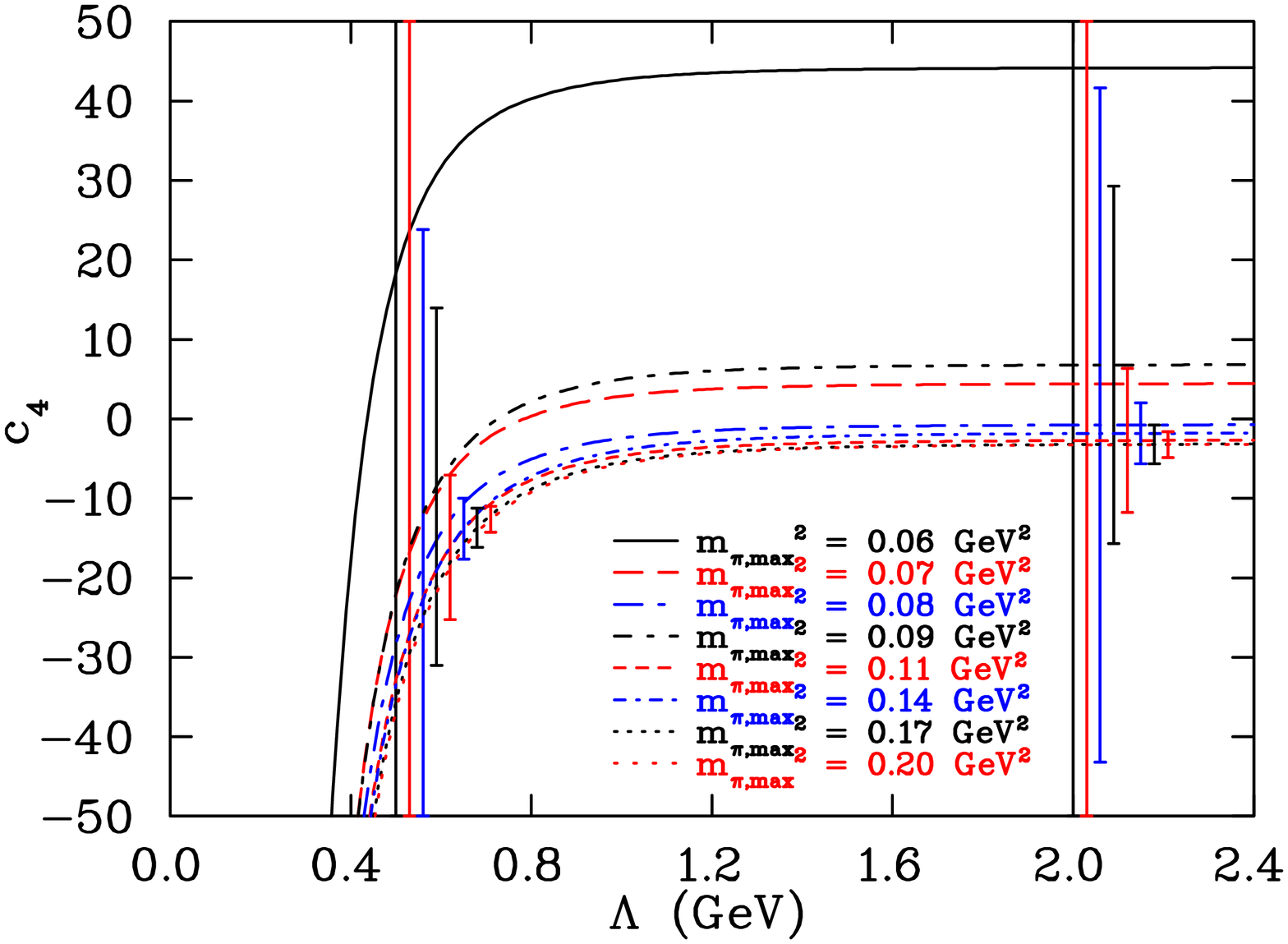}
\vspace{-11pt}
\caption{\footnotesize{Behaviour of $c_4$ vs.\ $\La$ for the initially excluded low-energy data. A triple-dipole regulator is used. A few points are selected to indicate the general size of the statistical error bars.}}
\label{fig:Kehfeic4new}
\end{figure}

\begin{figure}[tp]
\begin{minipage}[t]{0.5\linewidth} % A minipage that covers half the page
\centering
\includegraphics[height=0.9\hsize,angle=90]{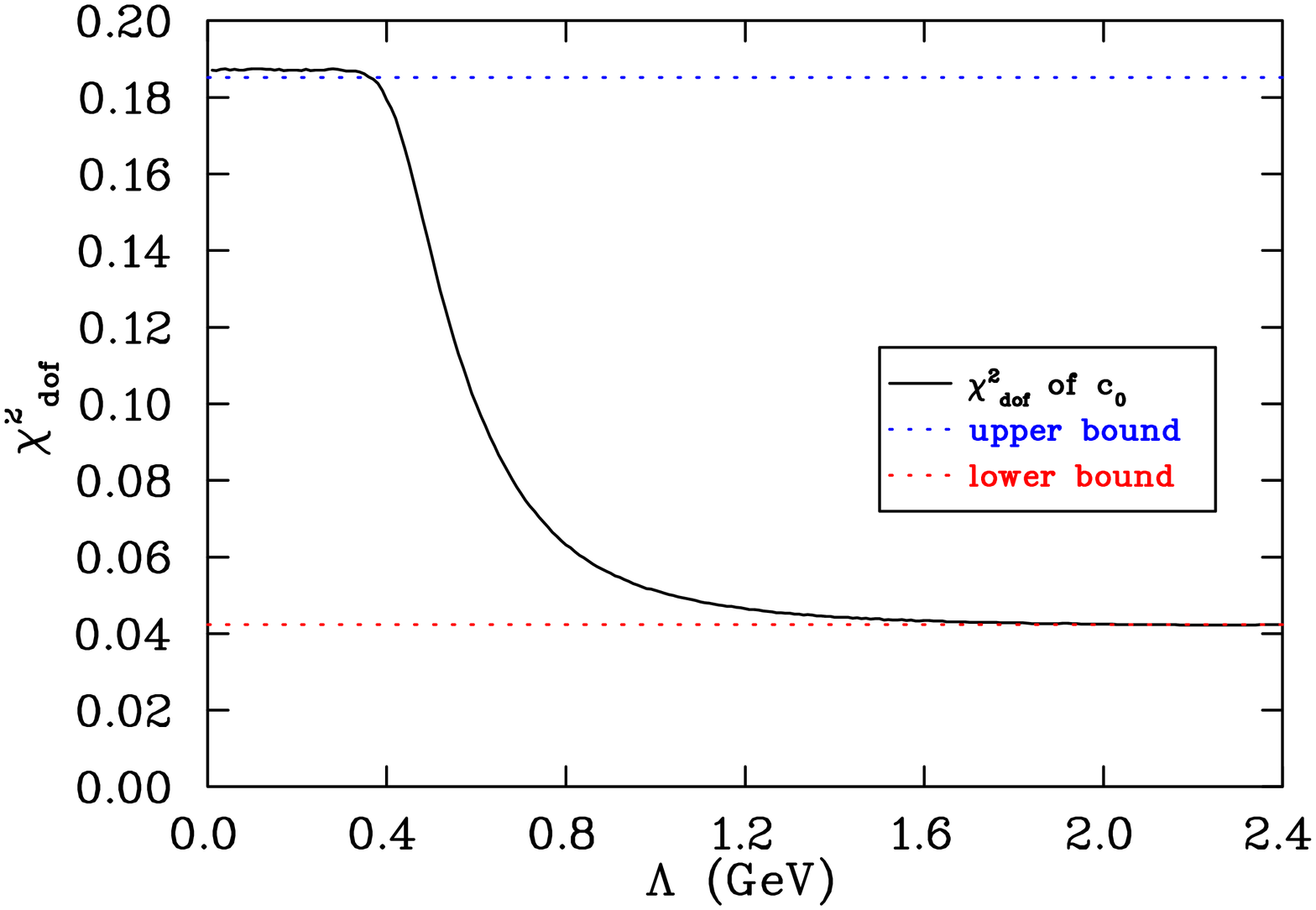}
\vspace{-11pt}
\caption{\footnotesize{$\chi^2_{dof}$, for $c_0$ versus $\La$, corresponding to the renormalization flow curves displayed in Figure \ref{fig:Kehfeic0new}. A lower bound for the regularization scale is found: $\La^{c_0}_{\ro{lower}} = 0.39$ GeV.}}
\label{fig:Kehfeichisqdofc0new}
\vspace{5pt}
\includegraphics[height=0.9\hsize,angle=90]{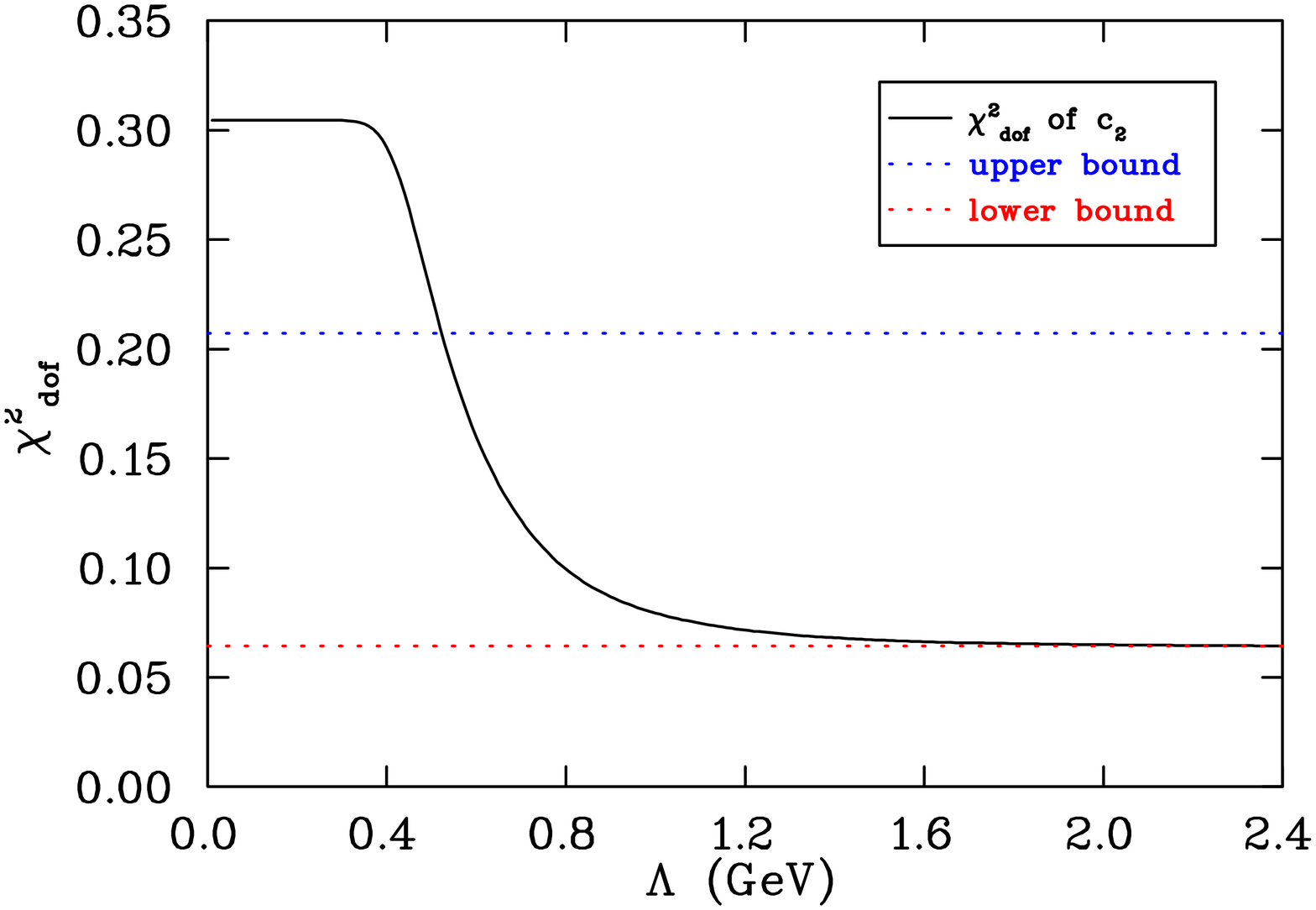}
\vspace{-11pt}
\caption{\footnotesize{$\chi^2_{dof}$, for $c_2$ versus $\La$, corresponding to the renormalization flow curves displayed in Figure \ref{fig:Kehfeic2new}. A lower bound for the regularization scale is found: $\La^{c_2}_{\ro{lower}} = 0.52$ GeV.}}
\label{fig:Kehfeichisqdofc2new}
\vspace{5pt}
\includegraphics[height=0.9\hsize,angle=90]{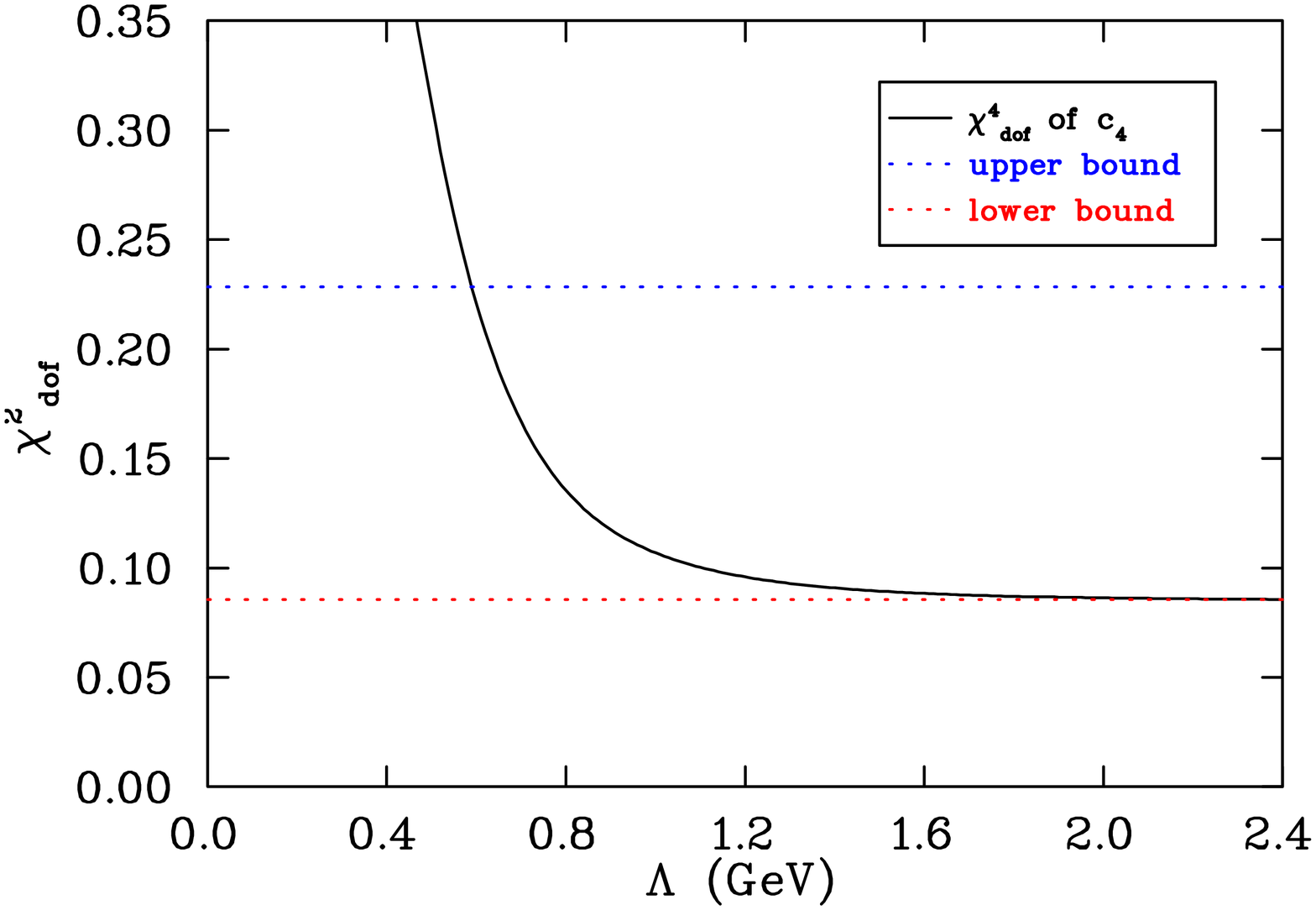}
\vspace{-11pt}
\caption{\footnotesize{$\chi^2_{dof}$, for $c_4$ versus $\La$, corresponding to the renormalization flow curves displayed in Figure \ref{fig:Kehfeic4new}. A lower bound for the regularization scale is found: $\La^{c_4}_{\ro{lower}} = 0.59$ GeV.}}
\label{fig:Kehfeichisqdofc4new}
%\end{figure}
%
\end{minipage}
\hspace{5mm}
\begin{minipage}[t]{0.5\linewidth}
\centering
%
%\begin{figure}[tp]
\includegraphics[height=0.9\hsize,angle=90]{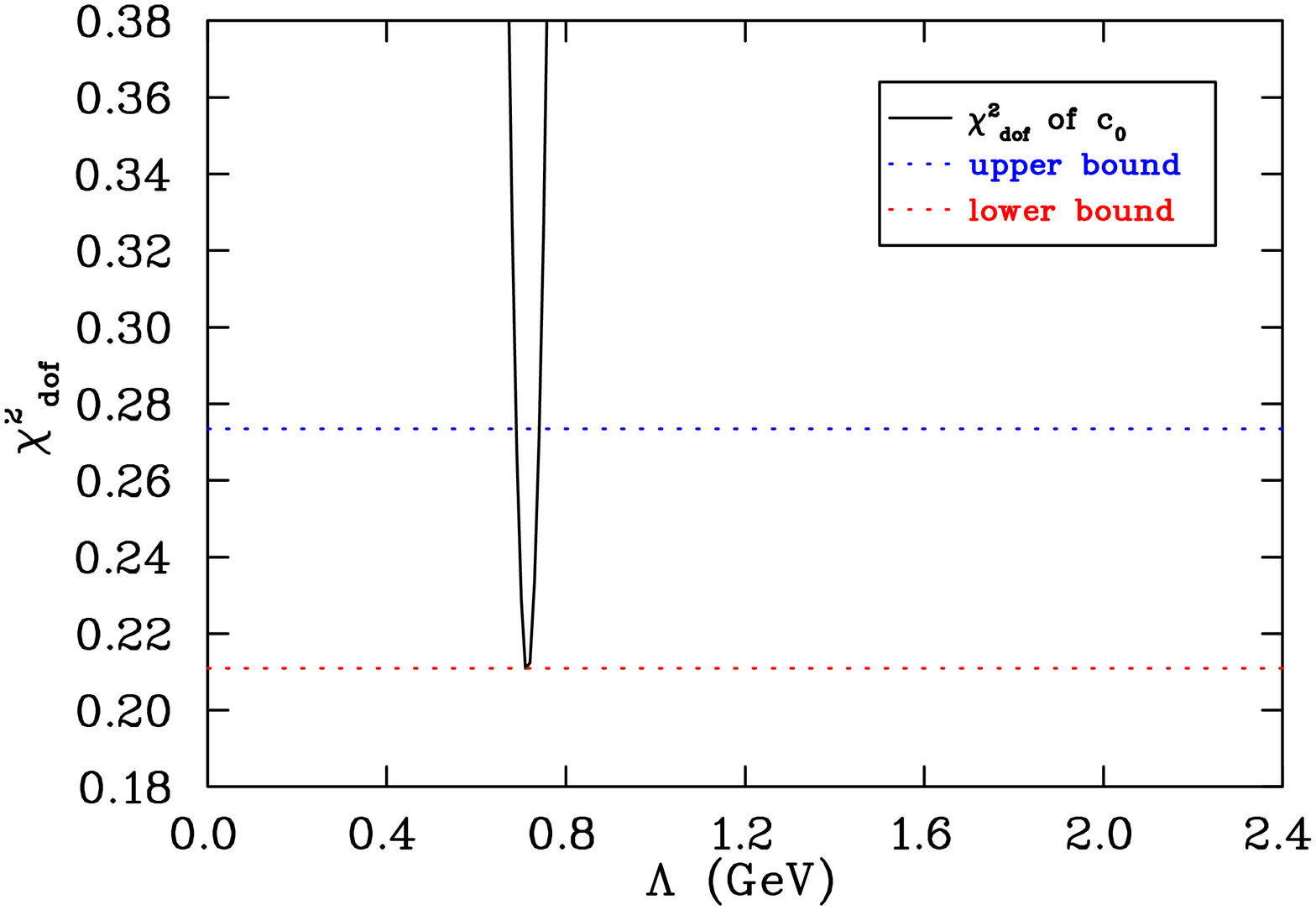}
\vspace{-11pt}
\caption{\footnotesize{$\chi^2_{dof}$, for $c_0$ versus $\La$, corresponding to all available data, including the low-energy set.}}
\label{fig:Kehfeichisqdofc0new17}
\vspace{5pt}
\includegraphics[height=0.9\hsize,angle=90]{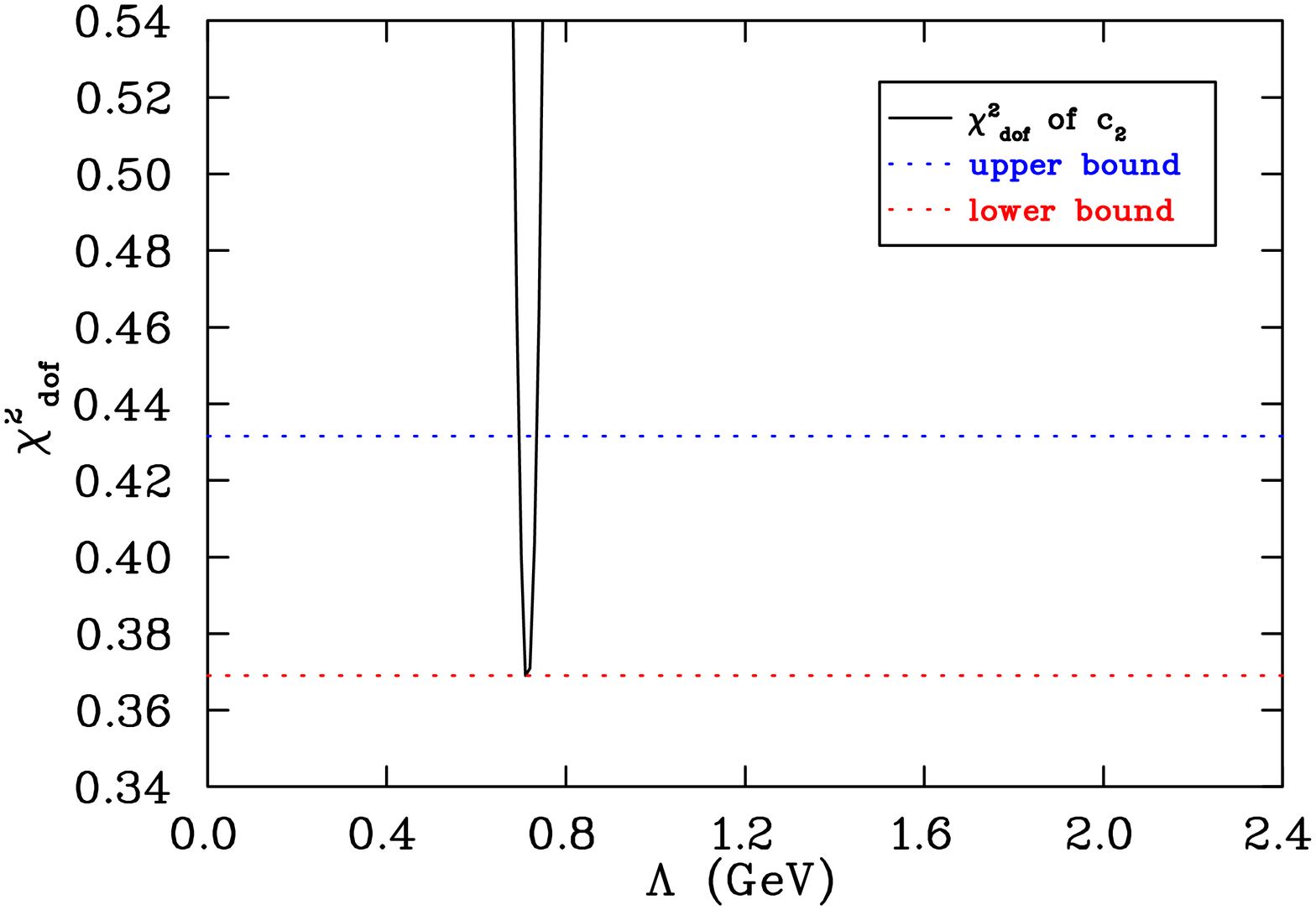}
\vspace{-11pt}
\caption{\footnotesize{$\chi^2_{dof}$, for $c_2$ versus $\La$, corresponding to all available data, including the low-energy set.}}
\label{fig:Kehfeichisqdofc2new17}
\vspace{5pt}
\includegraphics[height=0.9\hsize,angle=90]{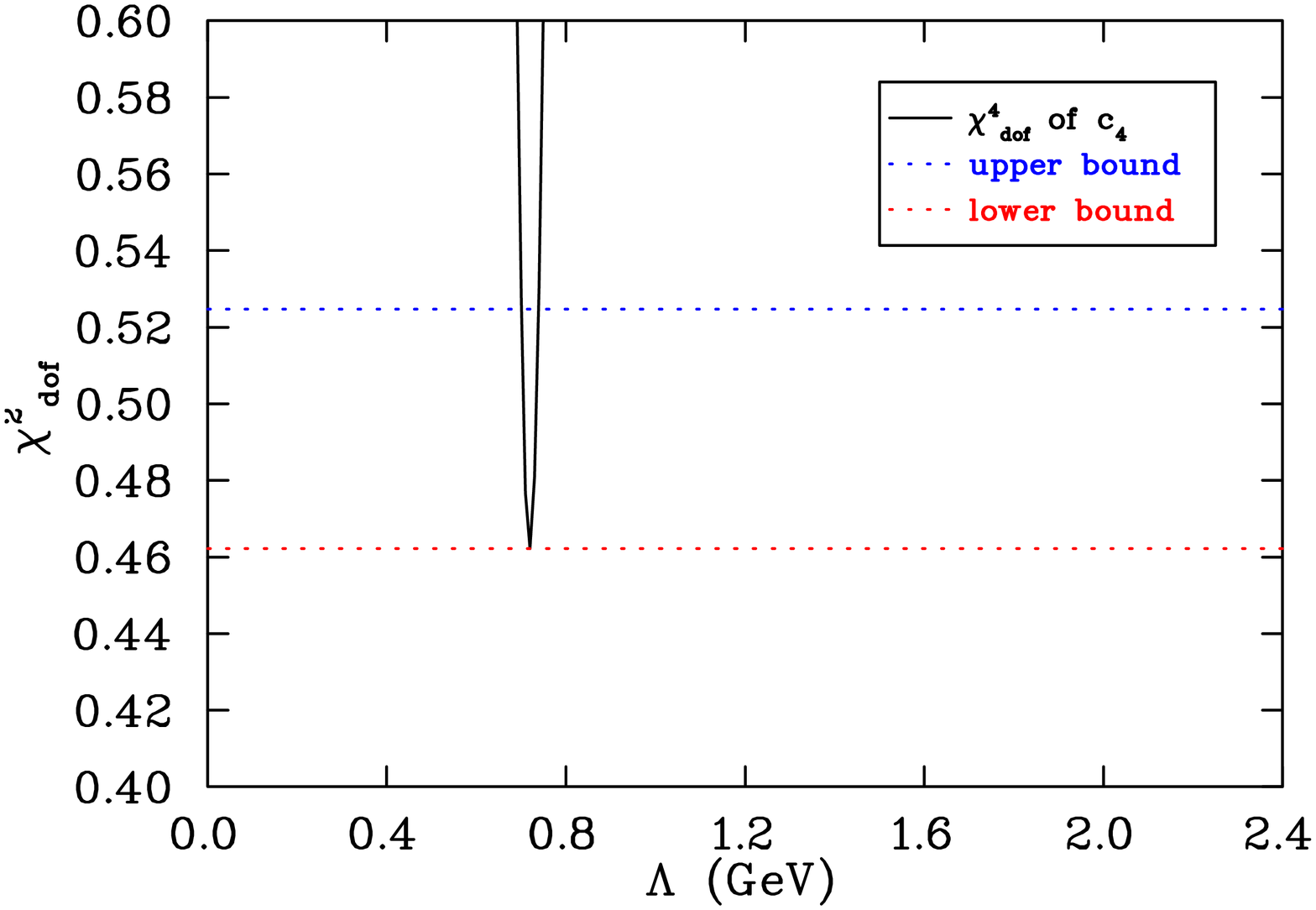}
\vspace{-11pt}
\caption{\footnotesize{$\chi^2_{dof}$, for $c_4$ versus $\La$, corresponding to all available data, including the low-energy set.}}
\label{fig:Kehfeichisqdofc4new17}
\end{minipage}
\end{figure}

In order to check if scheme-independence is recovered using data within the 
PCR, the low-energy data that were initially excluded from analysis 
can now be treated in the same way. That is, renormalization flow curves can 
be constructed as a function of $\La$ 
for sequentially increasing $m_{\pi,\ro{max}}^2$. The results are 
shown in Figures \ref{fig:Kehfeic0new} through \ref{fig:Kehfeic4new}. Clearly, 
the renormalization flow curves for each plot corresponding to $c_0$, $c_2$ 
and $c_4$ are flatter than those of the initial analysis, 
indicating a reduction in the regularization 
scale-dependence due to the use of data closer to the PCR. 
%reduction in sensitivity.
One is not able to extract an optimal regularization scale from these 
plots, as shown in the behaviour of $\chi^2_{dof}$, displayed  
in Figures \ref{fig:Kehfeichisqdofc0new} through \ref{fig:Kehfeichisqdofc4new}. 
However, each $\chi^2_{dof}$ curve provides a lower bound for the regularization 
scale, where FRR breaks down \cite{Hall:2010ai}, 
as discussed in Section \ref{subsect:curves}. 
These lower bounds are: $\La^{c_0}_{\ro{lower}} = 0.39$ GeV, 
$\La^{c_2}_{\ro{lower}} = 0.52$ GeV and $\La^{c_4}_{\ro{lower}} = 0.59$ GeV.  

The statistical error bars of the low-energy coefficients 
 corresponding to a small number of data points in 
Figures \ref{fig:Kehfeic0new} through \ref{fig:Kehfeic4new}  are large, 
and a statistical 
difference among the curves does not appear until 
$m_{\pi,\ro{max}}^2 \approx 0.11$ GeV$^2$. 
Thus the identification of an optimal regularization scale will be aided by  
incorporating data corresponding to even larger values of $m_{\pi,\ro{max}}^2$. 
By considering \emph{all} of the available data, the behaviour of 
$\chi^2_{dof}$, 
 as displayed in Figures \ref{fig:Kehfeichisqdofc0new17} through 
\ref{fig:Kehfeichisqdofc4new17}, resolve precise optimal regularization 
scales: $\La^{c_0}_\ro{central} = 0.72$ GeV, $\La^{c_2}_\ro{central} = 0.71$ GeV 
and $\La^{c_4}_\ro{central} = 0.71$ GeV. 
The systematic errors obtained 
from each $\chi^2_{dof}$ curve seem arbitrarily constrained as a consequence 
of including more data points, 
which extend well outside the chiral regime, and possibly 
outside the applicable region of FRR techniques. This issue is addressed 
in the ensuing section.

%NEW SET graph of chisq/dof for c0,c2,c4

\subsection{Optimal Pion Mass Region and Systematic Uncertainties}
\label{subsect:err2}

In this section,  a robust method for determining an optimal range of 
pion masses is presented. This range
corresponds to an optimal number of simulation results to be used for fitting. 
First, consider 
the extrapolation
of the quenched $\rho$ meson mass, which can now be completed.
The statistical uncertainties in the values of $c_0$, $c_2$, $c_4$ 
are dependent on $m_{\pi,\ro{max}}^2$. As a consequence, the uncertainty in 
the extrapolated $\rho$ meson mass $m_\rho^{\ro{ext}}$ must also be 
dependent on $m_{\pi,\ro{max}}^2$. 
Since the estimate of the statistical uncertainty in an extrapolated point 
will tend to decrease as more data are included in the fit, one might 
na\"{i}vely choose to use the largest $m_{\pi,\ro{max}}^2$ value possible 
in the data set. However, at some large value of $m_{\pi,\ro{max}}^2$, 
FRR $\chi$EFT will not provide a valid model for obtaining a suitable fit. 
At this upper bound of applicability for FRR $\chi$EFT, the uncertainty 
in an extrapolated point is dominated by the systematic error in the 
underlying parameters. %, as shown in Figures \ref{fig:Kehfeimrhovsmpisq} and 
%\ref{fig:Kehfeimrhovsmpisqnew}. 
This is due to a greater scheme-dependence in extrapolations 
using data extending 
 outside the PCR, meaning that the extrapolations
    are more sensitive to changes in the parameters of the loop integrals. 
Thus there is a balance point $m_{\pi,\ro{max}}^2 \!= \hat{m}_{\pi,\ro{max}}^2$,  
where the statistical and systematic uncertainties (added in quadrature)
 in an extrapolation are minimized. 

\begin{figure}[tp]
\centering
\includegraphics[height=0.70\hsize,angle=90]{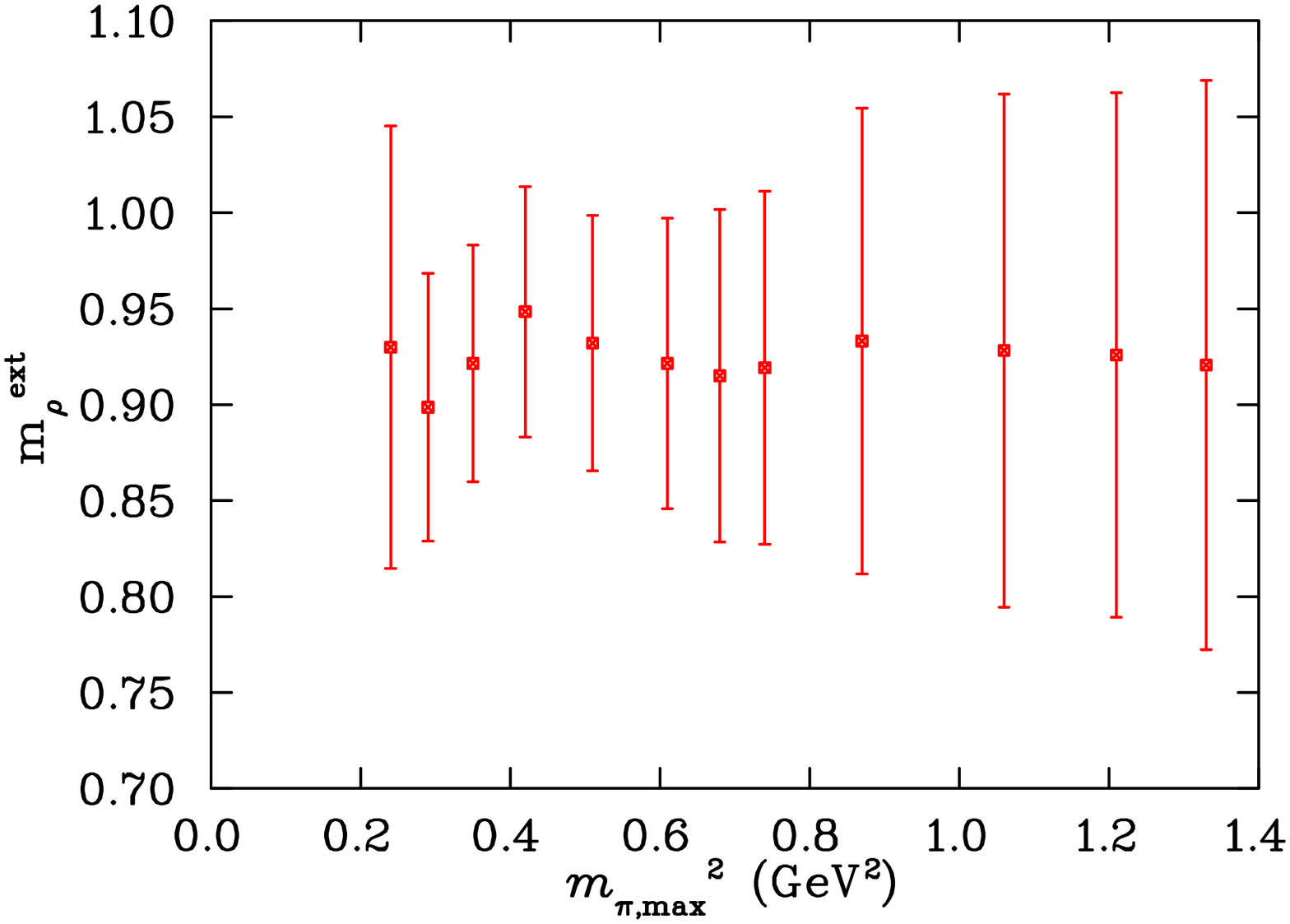}
\vspace{-11pt}
\caption{\footnotesize{  Behaviour of the extrapolation of the quenched $\rho$ meson mass to the physical point $m_{\rho,Q}^{\ro{ext}}(m_{\pi,\ro{phys}}^2)$ vs.\ $m_{\pi,\ro{max}}^2$ using the initial data set, which excludes the lowest mass data points. In each case, $c_0$ is obtained using the scale $\La_{\ro{central}}$ (for a triple-dipole regulator) as obtained from the $\chi^2_{dof}$ analysis. The error bars include the statistical and systematic uncertainties in $c_0$ added in quadrature. The optimal value $\hat{m}_{\pi,\ro{max}}^2 = 0.35$ GeV$^2$.}}
\label{fig:Kehfeimrhovsmpisq}
\vspace{5pt}
\includegraphics[height=0.70\hsize,angle=90]{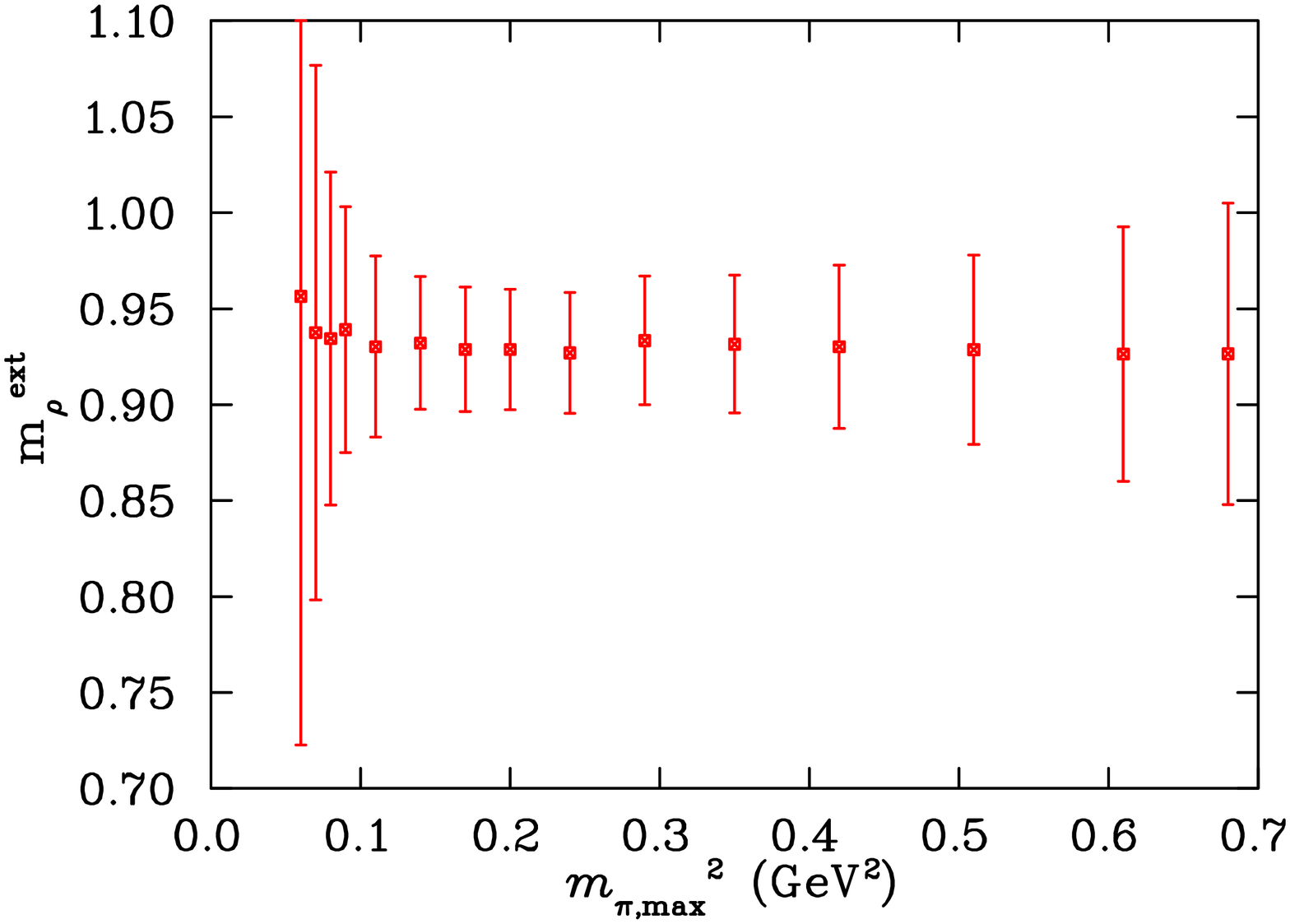}
\vspace{-11pt}
\caption{\footnotesize{  Behaviour of the extrapolation of the quenched $\rho$ meson mass to the physical point $m_{\rho,Q}^{\ro{ext}}(m_{\pi,\ro{phys}}^2)$ vs.\ $m_{\pi,\ro{max}}^2$ using the complete data set, which includes the lowest mass data points. In each case, $c_0$ is obtained using the scale $\La_{\ro{central}}$ (for a triple-dipole regulator) as obtained from the $\chi^2_{dof}$ analysis. The error bars include the statistical and systematic uncertainties in $c_0$ added in quadrature. The optimal value $\hat{m}_{\pi,\ro{max}}^2 = 0.20$ GeV$^2$.}}
\label{fig:Kehfeimrhovsmpisqnew}
\end{figure}

In order to obtain this value $\hat{m}_{\pi,\ro{max}}^2$, consider 
the behaviour of the extrapolation of the $\rho$
 meson mass to the physical point $m_{\rho,Q}^{\ro{ext}}(m_{\pi,\ro{phys}}^2)$, 
as a function of $m_{\pi,\ro{max}}^2$. 
Treating the parameters $g_2$, $g_4$, $M_0^2$, $A_0$ and $\La^\ro{scale}$ as 
independent, their systematic uncertainties from these sources are 
 added in quadrature. In addition, the systematic uncertainty due 
to the choice of the regulator functional form is roughly estimated 
by comparing the results using the double-dipole and the step function.  
These functional forms are the two most different forms of the 
various regulators considered, since the dipole was excluded due to 
the extra non-analytic contributions it introduces.
The results for the initial and 
complete data sets are shown in 
Figures \ref{fig:Kehfeimrhovsmpisq} and \ref{fig:Kehfeimrhovsmpisqnew}, 
respectively. 
%Since the initial data set will be used as a check to extrapolate 
%to the low-energy data points, the low-energy data points 
%This Figure uses the initial data set, which will be used in the final 
%extrapolations in order to check the results of this method with the  
%low-energy data. 
%The plot
  Figure \ref{fig:Kehfeimrhovsmpisq}
indicates an optimal value 
$\hat{m}_{\pi,\ro{max}}^2 = 0.35$ GeV$^2$, which will be used in the final 
extrapolations, in order to check the results of this method with the  
low-energy data. 
By using only the data contained in the optimal pion mass region, 
constrained by $\hat{m}_{\pi,\ro{max}}^2$, an estimate of 
the optimal regularization scale may be calculated with a 
more generous corresponding systematic 
uncertainty. 
The value $\La^\ro{scale} = 0.64$ GeV
is the average of $\La^\ro{scale}_{c_0}$, $\La^\ro{scale}_{c_2}$ and 
$\La^\ro{scale}_{c_4}$ using this method.  The $\chi^2_{dof}$ analysis 
does not provide an upper or lower bound at this value of 
$\hat{m}_{\pi,\ro{max}}^2$.  
These two estimates of the optimal regularization 
scale are consistent with each other. Both shall be used and 
compared in the final analysis. 
Figure \ref{fig:Kehfeimrhovsmpisqnew}
indicates an optimal value 
$\hat{m}_{\pi,\ro{max}}^2 = 0.20$ GeV$^2$ for the complete data set.  
A higher density of data in the low-energy region serves to decrease 
the statistical error estimate of extrapolations to the low-energy region. 
The corresponding value of $\La^\ro{scale}$ is unconstrained in this case, 
since the data lie close to the PCR.

The values of $c_0$, $c_2$ and $c_4$ for both the original data set 
and the complete data set are shown in Table \ref{table:cscomp}, 
with statistical 
error estimate quoted first, and systematic uncertainty due to the parameters 
$g_2$, $g_4$, $M_0^2$, $A_0$, $\La^\ro{scale}$ and the regulator functional
 form quoted second. In the case of the original data set, the value of 
$c_4$ is not well determined, due to the small number of data points used. 
In the case of the complete data set, the results are dominated by 
statistical uncertainty, and this also results in an almost 
unconstrained value of $c_4$. The coefficients of the complete set are less 
well-determined due to the fact that $\hat{m}_{\pi,\ro{max}}^2 = 0.20$ GeV$^2$, 
leaving only low-energy results with large statistical uncertainties 
for fitting.

%mrho vs mpisq

\begin{figure}[tp]
\begin{center}
%\vspace{-5mm}
%\includegraphics[height=0.70\hsize,angle=90]{graphics/mesonmass/Kehfei.Ext6pts.TRIP.0.64.INITIAL.D%ARK.CORR.ps}
%\vspace{-11pt}
%\caption{\footnotesize{   Extrapolation at $\La_{\ro{scale}} = 0.%64$ GeV, varied across the whole range of $\Lambda$ values, based on Kentucky G%roup data, and using the optimal number of data points, corresponding to $\hat{%m}_{\pi,\ro{max}}^2 = 0.35$ GeV$^2$. The inner error bar on the extrapolation p%oints represents purely the systematic error from parameters. The outer error b%ar represents the systematic and statistical error estimate added in quadrature%.}}
%\label{fig:initialextrapvar}
%\end{figure}
%
%final extrapolation
%\begin{figure}[tp]
%\vspace{5pt}
\includegraphics[height=0.70\hsize,angle=90]{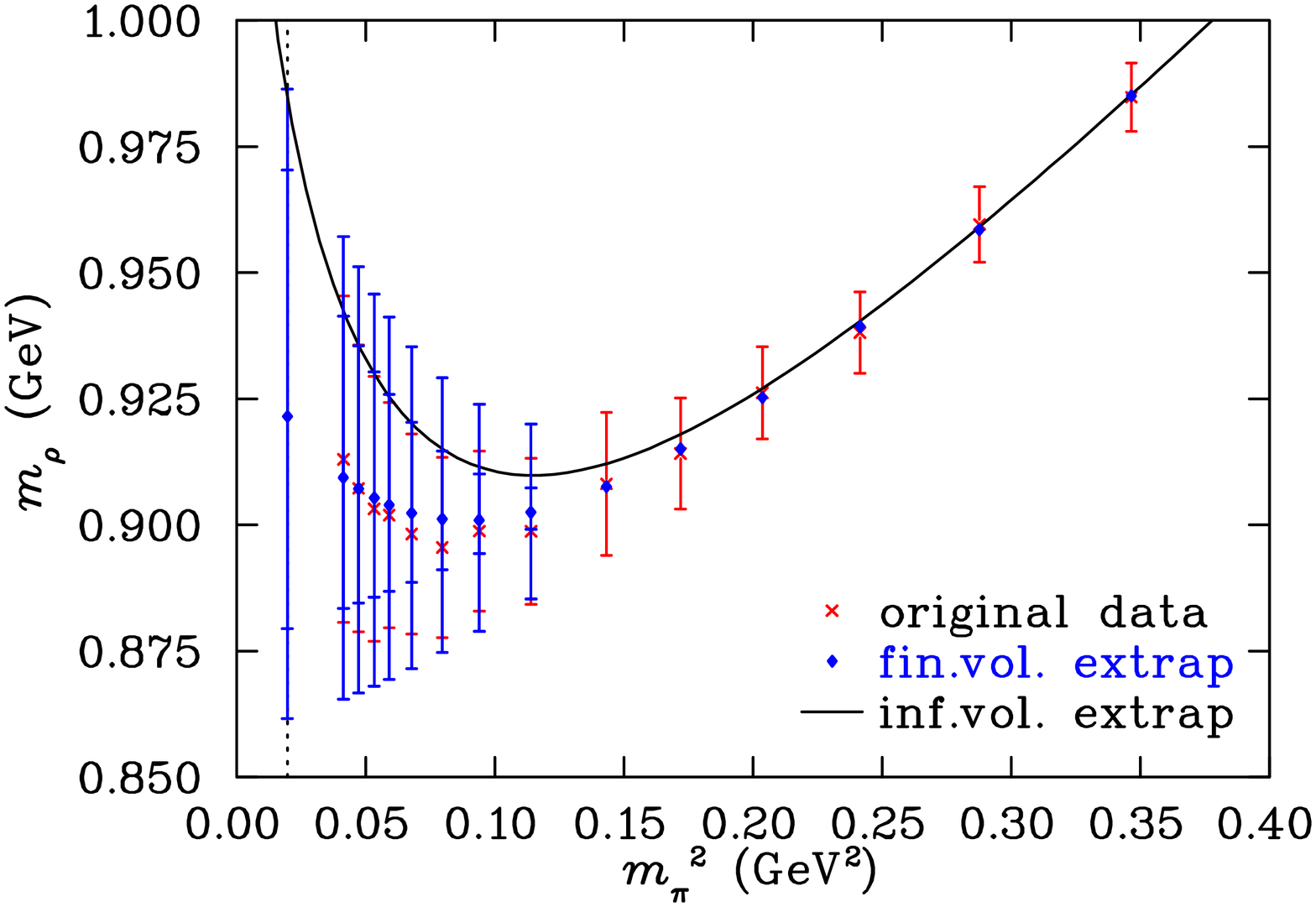}
\vspace{-11pt}
\caption{\footnotesize{   Comparison of chiral extrapolation predictions (blue diamond) with Kentucky Group data (red cross).  Extrapolation is performed at $\La_{\ro{scale}} = 0.64$ GeV, varied across the whole range of $\Lambda$ values, and using the optimal number of data points, corresponding to $\hat{m}_{\pi,\ro{max}}^2 = 0.35$ GeV$^2$. The inner error bar on the extrapolation points represents purely the systematic error from parameters. The outer error bar represents the systematic and statistical error estimates added in quadrature.}}
\label{fig:finalextrapvar}
\end{center}
%\end{figure}
%\vspace{5pt}
%\begin{figure}
\begin{center}
\includegraphics[height=0.70\hsize,angle=90]{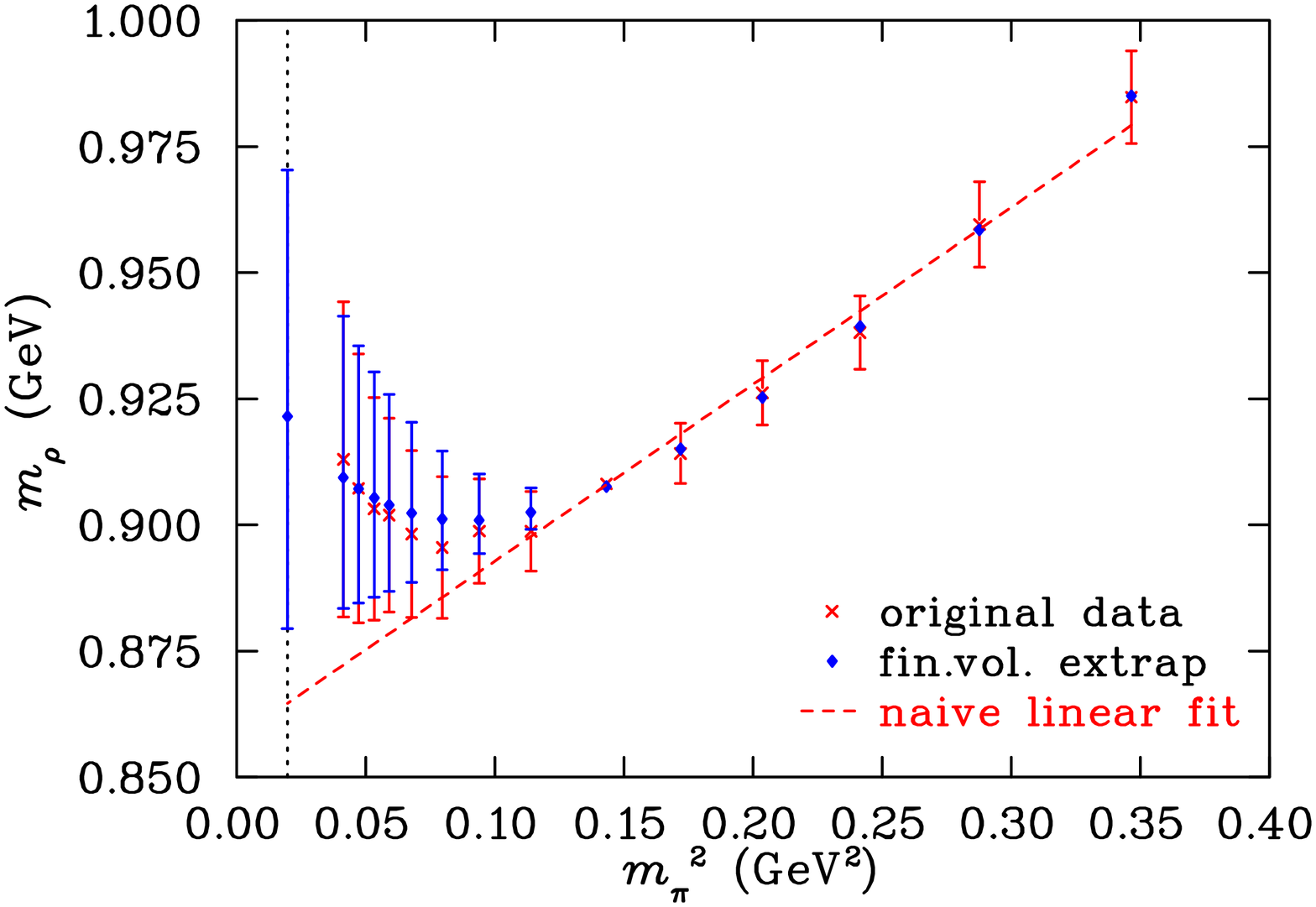}
\vspace{-11pt}
\caption{\footnotesize{   Comparison of chiral extrapolation predictions (blue diamond) with Kentucky Group data (red cross), with errors correlated relative to the point at $m_\pi^2 = 0.143$ GeV$^2$.  Extrapolation is performed at $\La_{\ro{scale}} = 0.64$ GeV, varied across the whole range of $\Lambda$ values, and using the optimal number of data points, corresponding to $\hat{m}_{\pi,\ro{max}}^2 = 0.35$ GeV$^2$. The error bar on the extrapolation points represents the systematic error only. A simple linear fit, on the optimal pion mass region, is included for comparison.}}
\label{fig:finaldeltaextrapvar}
\end{center}
\end{figure}

The result using the estimate of the optimal regularization scale 
$\La^\ro{scale} = 0.64$ GeV, with the systematic uncertainty 
calculated by varying $\Lambda$ across all suitable values, 
and using the initial data set, is shown in Figure \ref{fig:finalextrapvar}.
The extrapolation to the physical point obtained for this quenched data set is: 
%$m_{\rho,Q}(m_{\pi,\ro{phys}}^2) = 0.915$ 
%($\pm \,0.036$) GeV,
$m_{\rho,Q}^{\ro{ext}}(m_{\pi,\ro{phys}}^2) = 0.922^{+0.065}_{-0.060}$ GeV,  
an uncertainty 
of approximately $7$\%. 
Figure \ref{fig:finaldeltaextrapvar} shows the data plotted with error bars 
correlated relative to the lightest data point in the original set, 
$m_\pi^2 = 0.143$ GeV$^2$, using 
$\La_{\ro{scale}} = 0.64$ GeV, and varying $\Lambda$ across its full range 
of values. This naturally increases the estimate of the systematic 
uncertainty of the extrapolations, but also serves to demonstrate 
how closely the results from lattice QCD and $\chi$EFT match.

\begin{table}[tp]
 \caption{\footnotesize{The values of $c_0$, $c_2$ and $c_4$ as obtained from both the original data set and the complete set, which includes the low-energy data. In each case, the coefficients are evaluated using the scale $\La_{\ro{central}}$ (for a triple-dipole regulator) as obtained from the $\chi^2_{dof}$ analysis. The value of $m_{\pi,\ro{max}}^2$ used is that which yields the smallest error bar in adding statistical and systematic uncertainties in quadrature. For the initial data set,  $\hat{m}_{\pi,\ro{max}}^2 = 0.35$ GeV$^2$. For the complete data set, $\hat{m}_{\pi,\ro{max}}^2 = 0.20$ GeV$^2$. The statistical uncertainty is quoted in the first pair of parentheses, and the systematic uncertainty is quoted in the second pair of parentheses. For the original data set, $c_4$ is not well determined, with only a small number of data. For the complete data set, large statistical uncertainties result in an almost unconstrained value of $c_4$. The coefficients of the complete set are less well-determined due to the fact that $\hat{m}_{\pi,\ro{max}}^2 = 0.20$ GeV$^2$, leaving only low-energy results with large statistical uncertainties for fitting.}}
  \newcommand\T{\rule{0pt}{2.8ex}}
  \newcommand\B{\rule[-1.4ex]{0pt}{0pt}}
  \begin{center}
    \begin{tabular}{llll}
      \hline
      \hline
       \T\B 
       \quad & $c_0$(GeV$^2$)  & \quad$c_2$ & \,\,$c_4$(GeV$^{-2}$)   \\
      \hline
      %without estimate of sys error from reg
      %original set  &\T $1.40(3)(39)$ & \quad$6.54(14)(110)$ & \,\,$-2.26(19)(109)$ \\
      %low-energy set   &\T $1.36(3)(109)$ & \quad$6.68(24)(297)$ & \,\,$-3.23(56)(3727)$ \\
%
      %original set  &\T $1.31(5)(17)$ & \quad$7.89(39)(256)$ & \,\,$-16.15(75)(3819)$ \\
      %low-energy set   &\T $1.35(4)(241)$ & \quad$6.84(48)(310)$ & \,\,$-3.28(165)(3607)$ \\
original set  &\T $1.31(5)(17)$ & \quad$7.9(4)(26)$ & \,\,$-16.2(8)(382)$ \\
      complete set   &\T $1.35(4)(241)$ & \quad$6.8(5)(31)$ & \,\,$-3.3(17)(361)$ \\
      \hline
    \end{tabular}
  \end{center}
\vspace{-6pt}
  \label{table:cscomp}
\end{table}

%initial extrapolation

%\section{Analysis and Considerations}
\section{Summary and Specific Issues for the Quenched $\rho$ Meson}

%$\chi$EFT has been shown to be an important tool for 
%the investigation of the chiral properties of hadrons. 
%Finite-range regularization is an instrumental tool for using lattice QCD 
%data beyond the PCR. Using an extrapolation scheme combining 
%chiral effective field theory and quenched lattice QCD data from the Kentucky 
%Group collaboration, an optimal regulator $\La^\ro{scale}$ was obtained.
% Thus the mass of the $\rho$ meson was calculated in the low-energy region, 
%including the physical point.
%The result of the extrapolation correctly predicted the low-energy curvature 
%that was observed when the low-energy lattice data points were examined. 

%The ability of the extrapolation scheme
% to make predictions without any phenomenologically motivated assumptions 
%has been demonstrated.
%The results clearly indicate a successful procedure for using precision 
%lattice QCD data to extrapolate an observable to the low-energy region. 

A technique for isolating an optimal regularization scale
 was investigated in 
QQCD through an examination of the quenched $\rho$ meson mass.
 The result is a successful extrapolation
 based on an extended $\chi$EFT procedure. 
%By using recent precision, 
%quenched lattice results from the
%Kentucky Group, an optimal regulator $\La^\ro{scale}$ was obtained, 
%and an extrapolation performed using quenched 
%lattice QCD data that extends beyond the chiral power-counting regime.
By using %recent precision, 
quenched lattice QCD results that extended beyond the PCR, 
an optimal regularization scale 
was obtained from the renormalization flow of the 
low-energy coefficients $c_0$, $c_2$ and $c_4$.  The optimal scale is found 
to be  $\La^{\ro{scale}} = 0.67^{+0.09}_{-0.08}$ GeV. An optimal value of the 
maximum pion 
mass used for fitting was also calculated, and was found to be 
$\hat{m}_{\pi,\ro{max}}^2 = 0.35$ GeV$^2$. 
By using only the data contained in the optimal pion mass region, 
constrained by $\hat{m}_{\pi,\ro{max}}^2$, a value $\La^\ro{scale} = 0.64$ GeV 
is estimated for the optimal regularization scale, with a wider 
systematic uncertainty corresponding to the entire range of 
values of $\Lambda$. These two estimates of the optimal regularization 
scale are consistent with each other.

%
%Chiral effective field theory was shown to be an important tool for 
%the investigation of the chiral properties of hadrons. 
%Finite-range regularization is an instrumental tool for using lattice QCD 
%data beyond the power-counting regime. Using an extrapolation scheme combining 
%chiral effective field theory and quenched lattice QCD data from the Kentucky 
%Group, an optimal regulator $\La^\ro{scale}$ was obtained.
 The mass of the $\rho$ meson was calculated in the low-energy region. 
At the physical point, the result of the extrapolation, using 
$\La^{\ro{scale}} = 0.67^{+0.09}_{-0.08}$ GeV, is: 
$m_{\rho,Q}^{\ro{ext}}(m_{\pi,\ro{phys}}^2) = 0.925^{+0.053}_{-0.049}$ GeV. 
The result of the extrapolation, using $\La^\ro{scale} = 0.64$ GeV, 
with the systematic uncertainty 
calculated by varying $\Lambda$ across all suitable values,  is: 
$m_{\rho,Q}^{\ro{ext}}(m_{\pi,\ro{phys}}^2) = 0.922^{+0.065}_{-0.060}$ GeV. 
%$m_{\rho,Q}(m_{\pi,\ro{phys}}^2) = 0.915$ 
%($\pm \,0.036$) GeV. 
%$m_{\rho,Q}^{\ro{ext}}(m_{\pi,\ro{phys}}^2) = 0.920$ 
%($\pm \,0.037$) GeV. 
 The extrapolation also correctly predicts the low-energy curvature 
that was observed when the low-energy lattice simulation results were revealed.
%The result of the extrapolation correctly predicted the low-energy curvature 
%that was observed when the low-energy lattice simulation results were examined.
%
%
%The ability of the extrapolation scheme
% to make predictions without any phenomenologically motivated assumptions 
%was demonstrated.
%The results clearly indicate a successful procedure for using  
%lattice QCD data to extrapolate an observable to the low-energy region. 

Since there exists no experimental value for the mass of a particle in 
the quenched approximation, 
this analysis demonstrates the ability of the technique 
 to make predictions without phenomenologically motivated bias. 
The results clearly indicate a successful procedure for using  
lattice QCD data outside the PCR 
to extrapolate an observable to the chiral regime. 

%Develop?

%% file: nucleonmagmom.tex
\chapter{Electromagnetic Properties of the Nucleon}
\label{chpt:nucleonmagmom}

\textit{``[W]e can establish the key to our conclusion: \emph{the datum and the result are logically equivalent.}''}
(Omn\`{e}s, R. 2002. \textit{Quantum Philosophy: Understanding and Interpreting Contemporary Science} p.209) \cite{Omnes}

%Introduce
%how many flavours are we going to consider? JZ: 2 flavour. remember: 
%kaon loops switched off
%Also, isovector? explain.

In this chapter, the focus is turned to the magnetic moment and the 
electric charge radius of the nucleon. 
The magnetic moment is often studied for the %is of interest because of the 
physical significance 
of its anomalous component, obtained from the Pauli form factor $F_2$ 
(defined in Equation (\ref{eqn:matelemem})). 
Since 
 electrically charged pions with non-zero angular momentum 
 %\emph
{dress} the  nucleon, 
they contribute non-trivially to its magnetic moment, 
altering the value from its semi-classical Dirac value. 
Likewise, the electric charge radius, or more precisely, the 
gradient of the Sachs electric form factor $G_E$ in the soft-photon limit, 
provides a phenomenological test of quantum chromodynamics (QCD) 
theory. The leading-order 
low-energy contributions from virtual processes provide non-analytic 
behaviour in the chiral expansion.
 Chiral extrapolations for an infinite-volume box 
agree with experiment at the physical 
point, as will be evident later in this chapter. 
It is of interest in this investigation 
to determine if an optimal regularization scale 
may also be extracted from lattice QCD results  
for these two observables. % in the same way as for the nucleon mass. 
If so, it would provide compelling evidence for the existence 
of an intrinsic scale for the source of the pion cloud of the nucleon.

In lattice QCD, the isovector combination of the nucleon 
%($p - n$) 
is often calculated, as described in Section \ref{subsect:qqcd}. 
%convenient to calculate for computational reasons. 
Feynman diagrams including any photons coupling to 
sea-quark loops cancel in the case of the isovector, and 
the distinction between VQCD and full QCD vanishes. This is fortunate, 
since the calculation of the disconnected loops is computationally 
expensive. 
As a result, preliminary lattice QCD isovector results 
for two-flavor  $\ca{O}(a)$-improved
 Wilson  
quark action from the QCDSF Collaboration are analyzed.

%In the same way as for the nucleon mass, 
The magnetic moment and 
the electric charge radius can each be written %separately 
as  
chiral expansions, ordered in $m_\pi^2$, due to the 
Gell-Mann$-$Oakes$-$Renner Relation from Equation (\ref{eqn:GOR}) 
in Chapter \ref{chpt:chieft}. %The expansions each 
Each expansion comprises a  
 polynomial residual series, and loop integrals that contribute to non-analytic 
chiral behaviour. The diagrams that correspond to the leading-order
 loop integrals 
are shown in Figures \ref{fig:emSEa} through \ref{fig:emSEc}.

\section{Renormalization of the Magnetic Moment}

%\vspace{-5mm}
\subsection{Chiral Expansion of the Magnetic Moment}

\begin{figure}
\centering
\includegraphics[height=105pt]{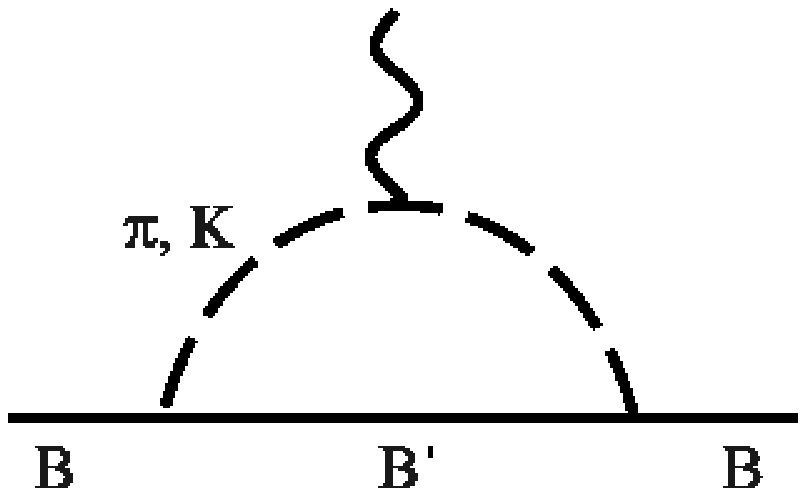}
\vspace{-3mm}
\caption{\footnotesize{The pion/kaon loop contribution (with photon attachment)
 to the magnetic moment 
and the electric charge radius of 
an octet baryon $B$, allowing a transition to a baryon $B'$. %,
% providing the leading non-analytic contribution to
%    the magnetic moment and the electric charge radius. 
All charge conserving transitions are implicit.}}
\label{fig:emSEa}
%\end{figure}
%
%\begin{figure}
\centering
\vspace{6mm}
\includegraphics[height=105pt,angle=0]{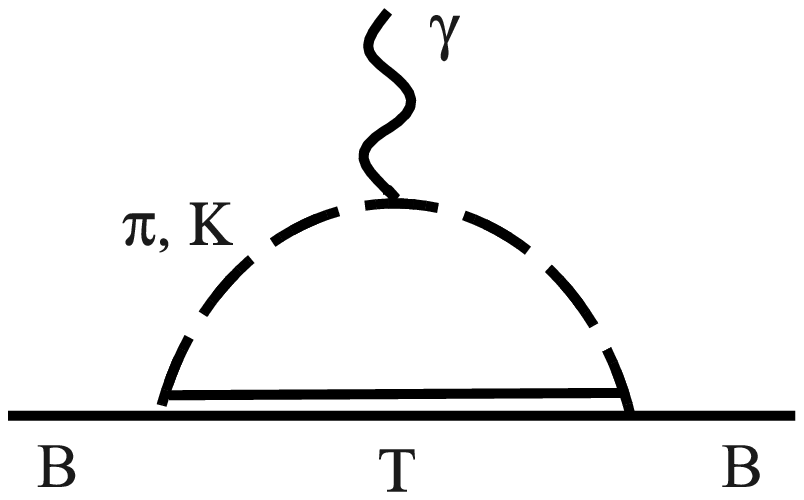}
\vspace{-3mm}
\caption{\footnotesize{The pion/kaon loop contribution 
(with photon attachment) 
to the magnetic moment 
and the electric charge radius 
    of an octet baryon $B$,   
allowing a transition to a nearby and
    strongly-coupled decuplet baryon $T$.}}
\label{fig:emSEb}
%\end{figure}
%
%\begin{figure}
\centering
\vspace{6mm}
\includegraphics[height=105pt,angle=0]{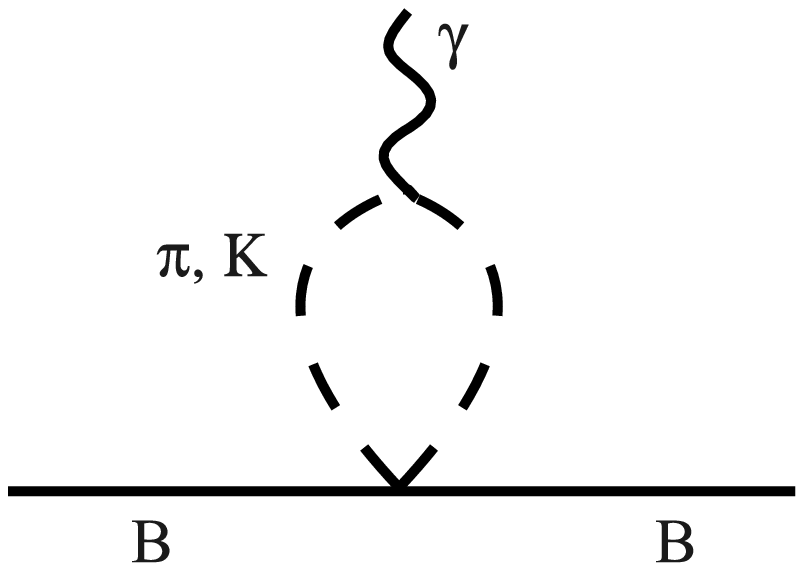}
\vspace{-3mm}
\caption{\footnotesize{The tadpole contribution  
at $\ca{O}(m_q)$ (with photon attachment) to 
the electric 
charge radius of an octet 
baryon $B$.}}
\label{fig:emSEc}
\end{figure}

%WRITE OUT FORMULA
Recalling %from Chapter \ref{chpt:chieft} 
the definition of the 
magnetic moment of the isovector nucleon in Equation (\ref{eqn:nucmagmom}),
 the chiral expansion is as follows:
\eqb
\label{eqn:chiralmag}
\mu^\ro{isov}_n = a_0^\La + a_2^\La\,m_\pi^2 + \ca{T}^\mu_N(m_\pi^2\,;\La) 
+ \ca{T}^\mu_\De
(m_\pi^2\,;\La) + 
 \ca{O}(m_\pi^4)\,,
\eqe
for loop integrals denoted ($\ca{T}$) to differentiate them from the 
self energies. 
In this instance, only two free parameters are chosen, since the 
non-analytic contributions are included only to chiral order 
$\ca{O}(m_\pi^2\,\ro{log}\,m_\pi)$. For a process with 
zero mass-splitting, such as that shown in 
the diagram in Figure \ref{fig:emSEa}, the leading-order 
non-analytic term is proportional to $m_\pi$; a lower chiral order than the 
leading-order term in the nucleon mass expansion. As a result, greater  
chiral curvature is expected, and the automatic renormalization process 
introduced 
in Chapter \ref{chpt:intrinsic} 
will be constructed only to order $\ca{O}(m_\pi^0)$, that is, for the 
chiral 
coefficient $c_0$. The fully renormalized chiral expansion may be written 
to leading non-analytic order $\ca{O}(m_\pi)$ as:
%
%CHECK THIS!!!
\eqb
\label{eqn:mpiexpn}
\mu^{\ro{isov}}_n = c_0 + \chi_N^\mu m_\pi + %a_2^\La\,m_\pi^2 + \f{\chi_\De^\mu}{\De}
%m_\pi^2\log\f{m_\pi}{\mu} +  
% \ca{O}(m_\pi^3)
 \ca{O}(m_\pi^2)\,,
\eqe
where $\mu$ is an implicit mass scale. 
Note also that the diagram in Figure \ref{fig:emSEc} does not 
contribute to the magnetic moment of the nucleon since, in this case,  
the photon couples to spinless pseudo-Goldstone bosons that have no 
orbital angular momentum.

\subsection{Chiral Loop Integrals}

Each loop integral has a solution in the form of a polynomial expansion
analytic in $m_\pi^2$ plus non-analytic terms, of which the leading-order term 
is of particular interest:
\begin{align}
\label{eqn:Sia}
\ca{T}^\mu_N(m_\pi^2\,;\La) 
&=  b_0^{\mu,\La,N} + \chi^\mu_N\,m_\pi + b_2^{\mu,\La,N}\,m_\pi^2 + 
\ca{O}(m_\pi^3)\,,\\
\label{eqn:Sib}
\ca{T}^\mu_\De(m_\pi^2\,;\La)  &= b_0^{\mu,\La,\De} +  b_2^{\mu,\La,\De}\,m_\pi^2 + 
\chi^\mu_\De\,m_\pi^2\,\ro{log}\,m_\pi/\mu
+ \ca{O}(m_\pi^3)\,,
\end{align}
where $\mu$ is a mass scale associated with the logarithm. %, chosen here to be $1$ GeV.

%describe how we transformed the integrals
%

The corresponding loop integrals can %once again 
be 
expressed in a convenient form by taking 
the non-relativistic heavy-baryon limit, and 
performing the pole integration for $k_0$.
%Renormalization is achieved by subtracting the relevant $b_0^\La$ term 
%from the integral, effective absorbing it into the corresponding 
%renormalized coefficient $c_0$. 
The integral corresponding to the diagram in  
Figure \ref{fig:emSEa} takes the form \cite{Wang:2007iw,Wang:2008vb}:
\begin{align}
\tilde{\ca{T}}^\mu_N(m_\pi^2\,;\La) 
&=  \f{-\chi_N^\mu}{\pi^2}\int\!\!\ud^3\! k\f{(\hat{q}\times \vec{k})^2 
u^2(k\,;\La)}{{(k^2 + m_\pi^2)}^2} - b_0^{\mu,\La,N}\\
&= \f{-\chi_N^\mu}{\pi^2}\int\!\!\ud^3\! k\f{k_\perp^2u^2(k\,;\La)}
{{(k^2 + m_\pi^2)}^2} - b_0^{\mu,\La,N}\\
&= \f{-\chi_N^\mu}{3\pi^2}\int\!\!\ud^3\! k\f{k^2u^2(k\,;\La)}
{{(k^2 + m_\pi^2)}^2} - b_0^{\mu,\La,N}\,,
\end{align}
where $\hat{q}$ is the direction of the external momentum introduced 
by an incoming photon.
The argument for this substitution of the perpendicular part $k_\perp$ 
is expounded in Appendix \ref{app:magang}.
The function $u(k\,;\La)$ is the regulator, with associated 
momentum cut-off scale $\La$. In this case, a dipole regulator will be used 
(corresponding to 
a choice of $n=1$ in Equation (\ref{eqn:ndipole}) in Chapter 
\ref{chpt:chieft}). 
%With fewer fit parameters, and non-analytic terms occuring at a lower chiral 
%order than in the case of the nucleon mass, 
Since the working-order $\ca{O}(m_\pi^2\,\ro{log}\,m_\pi)$ of the calculation is 
less than in the case of the nucleon mass analysis, there is a reduced 
possibility of extra scale-dependent non-analytic terms frustrating the 
chiral fit. Thus, ensuring that these scale-dependent 
non-analytic terms %do not occur in 
are removed from the chiral expansion is not so vital, 
and a dipole form is an acceptable choice of regulator.  
%Similarly, 
The integral corresponding to the diagram in 
Figure \ref{fig:emSEb} takes the form:
\begin{align}
\tilde{\ca{T}}^\mu_\De(m_\pi^2\,;\La)  &= 
\f{-\chi_\De^\mu}{\pi^2}\int\!\!\ud^3\! k\f{k_\perp^2 
(2\omega(k) + \Delta)u^2(k\,;\La)}{2\omega^3(k)[\omega(k) + \Delta]^2} 
- b_0^{\mu,\La,\De}\\
&=  \f{-\chi_\De^\mu}{3\pi^2}\int\!\!\ud^3\! k\f{k^2 
(2\omega(k) + \Delta)u^2(k\,;\La)}{2\omega^3(k)[\omega(k) + \Delta]^2} 
- b_0^{\mu,\La,\De}\,,
\end{align}
where $\om(k) = \sqrt{k^2 + m_\pi^2}$ and $\De$ is the mass-splitting.
The chiral coefficients $\chi_N^\mu$ and $\chi_\De^\mu$ are 
constants in terms of the chiral Lagrangian of Equation (\ref{eqn:octdeclag}) 
in Chapter \ref{chpt:chieft}: 
%and the relevant Clebsch-Gordan coefficients, as summarized by Wang 
%\cite{Wang:2008vb}:
%
\begin{align}
\chi_N^{\mu,p} &= -\f{M_N}{8\pi f_\pi^2}(D+F)^2 = -\chi_N^{\mu,n},\\
\chi_\De^{\mu,p} &= -\f{M_N}{8\pi f_\pi^2}\f{2\ca{C}^2}{9} = -\chi_\De^{\mu,n}.
\end{align}
On the finite-volume lattice, each momentum component is quantized in 
units of $2\pi/L$, that
is, $k_i=n_i2\pi/L$ for integers $n_i$.  Finite-volume corrections 
$\de^{\mu,\ro{FVC}}$
 are written as the difference between the finite sum
and the corresponding integral. It is known that the finite-volume 
corrections saturate to a fixed result for large values of regularization scale 
\cite{Hall:2010ai}. As before, this is 
achieved in practice 
by evaluating the finite-volume corrections with fixed regularization 
scale: $\La' = 2.0$ GeV. 
%Following the example set by this article, 
%The value $\La'= 2.0$ GeV is chosen
%to evaluate all finite volume corrections independent of the integral cutoff
 %scale $\La$ in Equations (\ref{eqn:tSia}) and (\ref{eqn:tSib}).
The finite-volume version of Equation (\ref{eqn:chiralmag}) 
 can thus be expressed as:
\begin{align}
\label{eqn:finchiral}
\mu_n &= c_0 + a_2^\La\,m_\pi^2 + (\tilde{\ca{T}}^\mu_N(m_\pi^2\,;\La) + 
\de^{\mu,\ro{FVC}}_{N}(m_\pi^2;\La'))\nn\\ 
&+ (\tilde{\ca{T}}_\De(m_\pi^2\,;\La) + \de^{\mu,\ro{FVC}}_{\De}(m_\pi^2;\La')) + 
 \ca{O}(m_\pi^4)\,.
\end{align}

%Put argument from PhD Work 6 as to how we  simplified it.
%Can also put the more general q-dep result, in spc, in app.

%
%ON THE LATTICE
%Since lattice simulations are necessarily carried out on a discrete
%spacetime, any extrapolations performed should take into account
%finite volume effects.  The low-energy EFT is ideally suited for
%characterising the leading infrared effects associated with the finite
%volume. In order to achieve this, each of the 3-dimensional integrals
%can be transformed to its form on the lattice using a finite-sum of
%discretized momenta, see Allton \textit{et al.}  \cite{Armour:2005mk}
%for instance:
%
%\eqb
%\int\!\!\ud^3 k \ra \f{{(2\pi)}^3}{L_x L_y L_z} \sum_{k_x,k_y,k_z}. 
%\eqe
%

\section{Evidence for an Intrinsic Scale in the Magnetic Moment}

%work with two flavour data from QCDSF
The analysis of the magnetic moment of the nucleon provides an excellent 
check for the identification of an intrinsic scale in the nucleon-pion 
interaction. Using chiral effective field theory ($\chi$EFT), it 
has been demonstrated in Chapter \ref{chpt:nucleonmass} 
that lattice QCD results for the nucleon mass 
have an energy scale embedded within them. 
This property is %intimately related to 
 a consequence of the small size of the %existence of a
 power-counting regime (PCR), where the expansion formulae 
of chiral perturbation theory ($\chi$PT) hold formally. Since a selection 
of lattice QCD results reasonable for fitting an extrapolation invariably 
extend outside the restrictive PCR \cite{Leinweber:2005xz}, 
the validity of a formal scheme 
for extrapolation, and for identifying the leading-order terms in the 
chiral expansion, is %jeopardized.
compromised.
Fortunately, a finite-range regularization (FRR) scheme, in conjunction 
with $\chi$EFT as described in Chapter \ref{chpt:intrinsic}, 
provides a robust method for achieving an extrapolation 
to physical quark masses, and identifying an intrinsic scale embedded 
within lattice QCD results. %This %so-called 
%\emph{extended} 
%extended effective field 
%theory %(E$^2$FT) 
%

Recall that the method 
proceeds by analyzing the behaviour of the renormalization 
of one or more low-energy coefficients of the chiral expansion as a
function of the FRR scale. Ideally, that is, with lattice QCD 
results constrained entirely within the PCR, the renormalized 
coefficients should be independent of regularization scale. However, 
in practice, a scale-dependence is observed; %curvature; 
particularly for data sets including %involving 
data points corresponding to large quark masses. 
%By analyzing a variety of subsets of the lattice QCD results %for varying 
%quark masses, 
By truncating the lattice QCD results at different values of $m_{\pi,\ro{max}}^2$, 
an optimal FRR scale can be identified. 
%points, each corresponding 
%to a 
This optimal scale is the value of $\La$ at which the low-energy coefficient 
under analysis 
is least sensitive to
 the truncation of the lattice data.
%Due to consistent results among various the analysis of the magnetic 
%moment and the nucleon mass \cite{Hall:2010ai}, 
If the optimal scale is consistent among the analyses of magnetic moment 
and the nucleon mass in Chapter \ref{chpt:nucleonmass}, 
it provides evidence for an intrinsic scale in the nucleon.

The preliminary QCDSF results for the magnetic moment at a variety of 
 $m_\pi^2$ values are displayed in Figure \ref{fig:magdata}. 
The experimental value is also marked. 
The set of data 
is listed in 
Appendix 
\ref{chpt:appendix4}, Table \ref{table:magdata}.
 The lattice sizes of each 
data point vary from $1.43$ to $3.04$ fm using $N_f = 2$ and 
 $\ca{O}(a)$-improved
 Wilson quark action.
A simple linear fit is included in this plot, which does not take into account 
the chiral loop integrals, nor the finite-volume corrections to the data.
Therefore, it is not surprising that the 
 linear fit fails to reach the experimental value of the 
magnetic moment at the physical pion mass.
Since the lattice QCD results extend outside the PCR, the result of 
an extrapolation that includes the chiral loop integrals 
will be scale-dependent.  However, the scale-dependence
 may be ameliorated using the procedure, which obtains both an optimal 
regularization 
scale and an estimate of its systematic uncertainty, constrained 
by the lattice results. %\cite{Hall:2010ai}.
%can be handled by obtaining an optimal regulator in the extrapolations, 
%as described in Ref. \cite{Hall:2010ai}..

\begin{figure}[tp]
\centering
\includegraphics[height=0.7\hsize,angle=90]{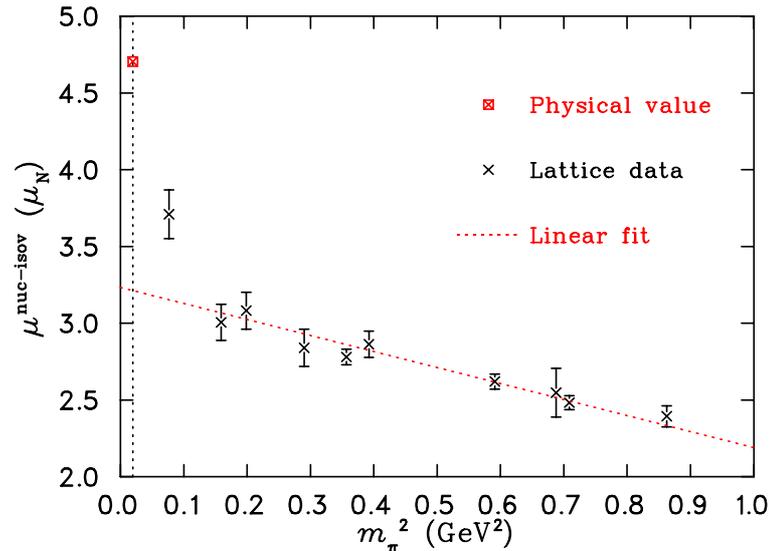}
\vspace{-11pt}
\caption{\footnotesize{Preliminary lattice QCD data for $\mu_n^\ro{isov}$ 
from QCDSF,  with the physical value from experiment as marked.}}
\label{fig:magdata}
\end{figure}

\subsection{Renormalization Flow Analysis}
\label{subsect:renormflowmom}

\begin{figure}[tp]
\centering
\includegraphics[height=0.7\hsize,angle=90]{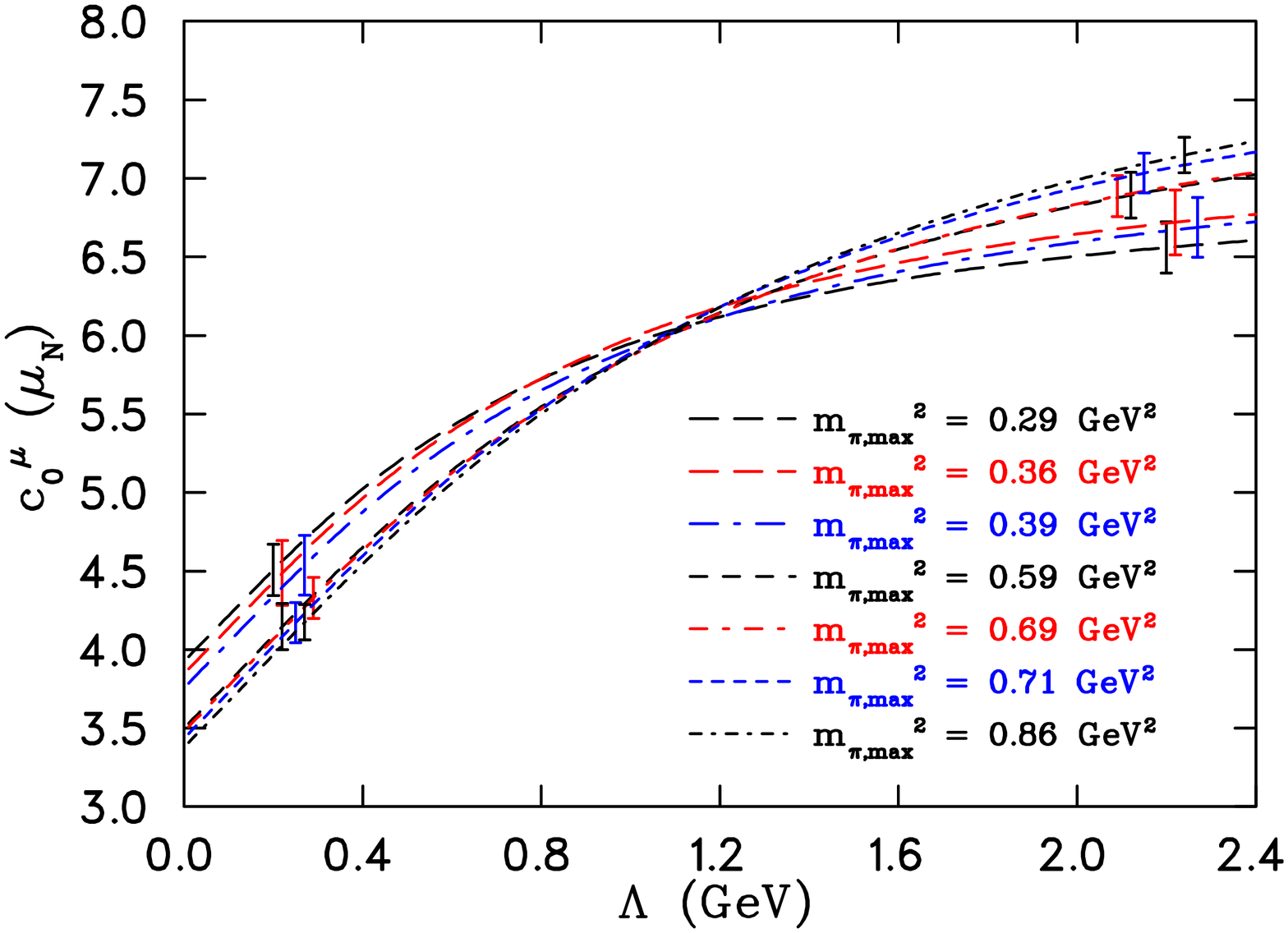}
\vspace{-11pt}
\caption{\footnotesize{ The renormalization flow of
  $c_0$ for $\mu_n^\ro{isov}$ obtained using a dipole regulator, based on 
 lattice QCD data from QCDSF.}}
\label{fig:c0}
%\end{figure}
\vspace{6mm}
%\begin{figure}[tp]
\includegraphics[height=0.7\hsize,angle=90]{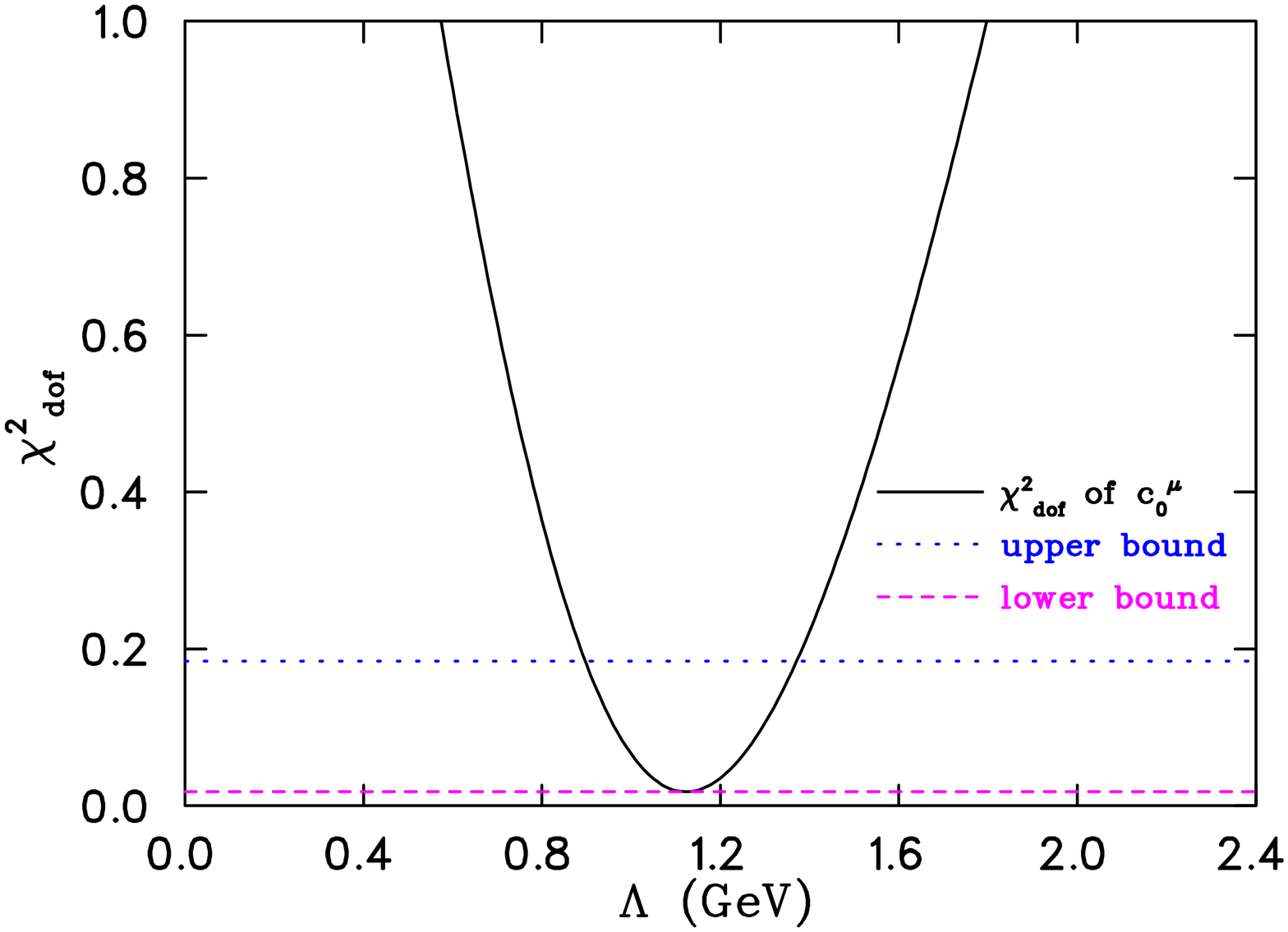}
\vspace{-11pt}
\caption{\footnotesize{ $\chi^2_{dof}$ for 
 the renormalization flow of
  $c_0$ for $\mu_n^\ro{isov}$ obtained using a dipole regulator, based on 
 lattice QCD data from QCDSF.}}
\label{fig:c0chisqdof}
\end{figure}

In order to obtain the optimal regularization scale, 
the low-energy coefficient $c_0$ 
from Equation (\ref{eqn:finchiral}) will be calculated across a range 
of values of regularization scale $\La$. 
Thus the renormalization flow can be constructed. 
Multiple renormalization flow curves may be obtained by constraining the 
fit window by a maximum, $m_{\pi,\ro{max}}^2$, and sequentially adding  
data points to extend further outside the PCR. 
The renormalization flow curves for a dipole regulator 
are plotted on the same set of axes 
in Figure \ref{fig:c0}. 
Note that each data
point plotted has an associated error bar, but for the sake of clarity 
only a few points are selected to indicate the general size of the 
statistical error bars. 
As more data are included in the fit, a greater degree of scale-dependence 
is observed. There is a reasonably well-defined value of $\La$ 
at which the renormalization of $c_0$ is least sensitive to the truncation 
of the data: $\La^{\ro{scale}} \approx 1.1$ GeV. 
This indicates %that there exists an 
the optimal regularization scale 
embedded 
within the lattice QCD results themselves.

\subsection{Analysis of Systematic Uncertainties}
\label{subsect:statmom}

The optimal regularization scale for a dipole form 
can be more precisely extracted from Figure \ref{fig:c0} 
using the chi-square-style analysis. 
Such an analysis will also provide 
a measure of the systematic uncertainty in the optimal regularization scale.
By plotting $\chi^2_{dof}$ against the regularization scale $\La$, 
where $dof$ equals 
the number of curves $n$ minus one for the fit parameter 
$c_0$, a measure of the spread of the 
renormalization 
flow curves can be calculated, and the intersection point obtained.
%
%A plot of $\chi^2_{dof}$ is constructed at each $\La$ for 
% $c_0$ (with uncertainty $\de c_0$).
% 
%Figure \ref{fig:c0chisqdof} shows the $\chi^2_{dof}$ plot 
%% for a dipole regulator are shown in 
%corresponding to the renormalization flow of Figure \ref{fig:c0}.
The $\chi^2_{dof}$ plot % for a dipole regulator are shown in 
corresponding to Figure \ref{fig:c0} is shown in 
Figure \ref{fig:c0chisqdof}. 
The optimal regularization scale  $\La^\ro{scale}$ 
is taken to be the central value $\La^\ro{central}$ of the plot, and 
the upper and lower bounds obey 
 the condition $\chi^2_{dof} < \chi^2_{dof, min} + 1/(dof)$. 
Thus the optimal regularization scale for a dipole regulator  
is: %$\La^{\ro{scale}} = 1.125 - 0.215 +0.195$
$\La^{\ro{scale}} = 1.13^{+0.22}_{-0.20}$ %(\scriptsize{$\stackrel{+0.22}{-0.20}$}) GeV.
% \scriptsize{$\Big\{\begin{matrix}+0.22\\-0.20 \end{matrix}$}} \normalsize GeV.
%$\Big\{\begin{matrix}+0.22\\-0.20 \end{matrix}$  
 GeV.
This value is consistent with the optimal regularization scale obtained 
for the nucleon mass using a dipole form, based on 
lattice QCD results in Chapter \ref{chpt:nucleonmass}. Recall that 
the mean value for the optimal regularization scale from the nucleon mass 
analysis is: $\bar{\La}^\ro{scale}_\ro{dip} \approx 1.3$ GeV. 
This provides evidence that %,
%for a given functional form, 
the optimal 
regularization scale 
is associated with an intrinsic scale characterizing the size of 
the nucleon, as probed by the pion.

\subsection{Chiral Extrapolation Results}
\label{subsect:chiextrapmom}

\begin{figure}[tp]
\centering
\includegraphics[height=0.7\hsize,angle=90]{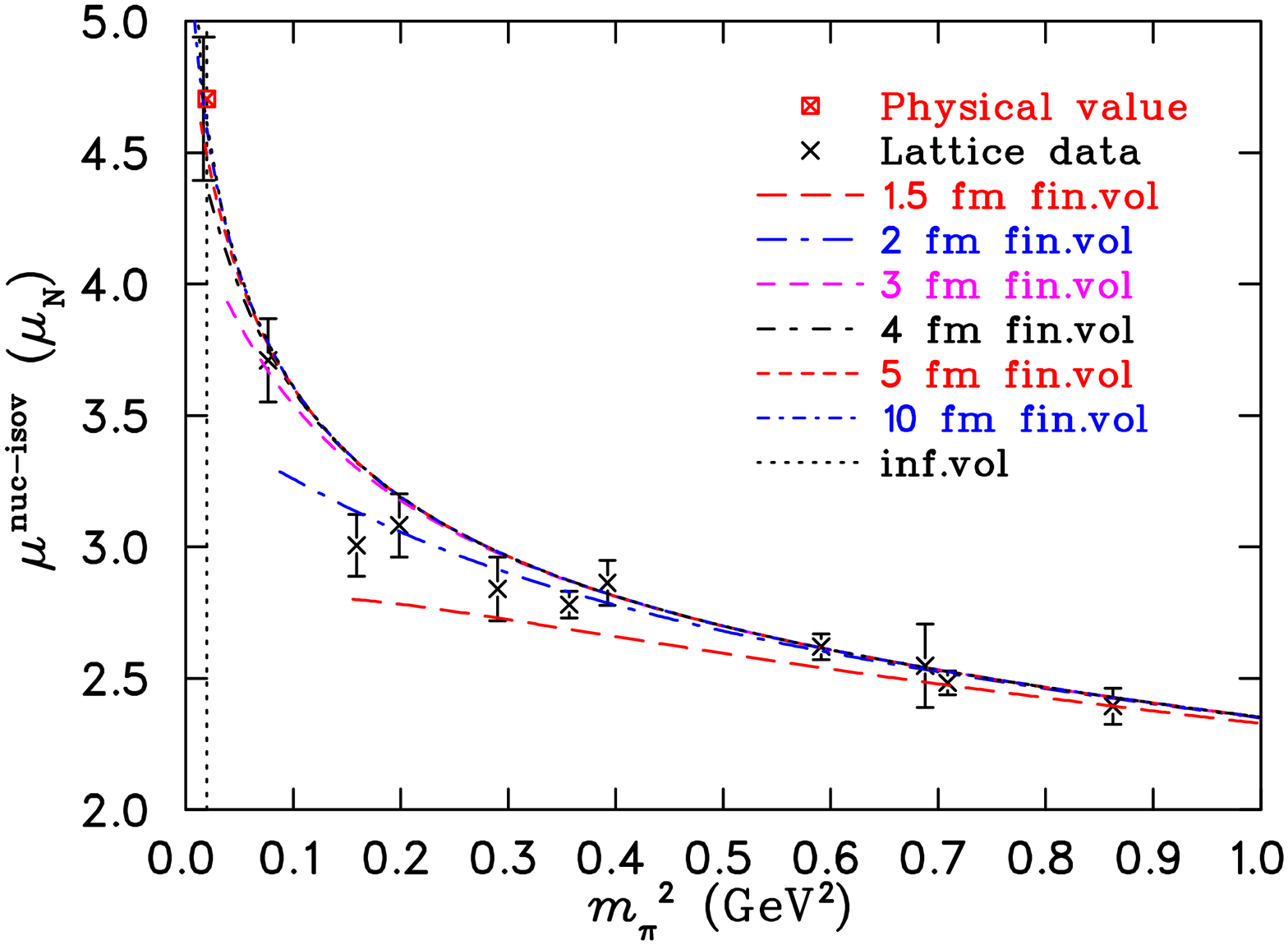}
\vspace{-11pt}
\caption{\footnotesize{ Extrapolations of $\mu_n^\ro{isov}$ 
at different finite volumes and infinite volume, using a dipole regulator, 
based on lattice QCD data from QCDSF, lattice sizes: $1.43-3.04$ fm. The provisional constraint $m_\pi L > 3$ 
is used. The physical value from experiment is marked. 
An estimate in the uncertainty in the extrapolation due to $\La^{\ro{scale}}$ 
 has been calculated from Figure \ref{fig:c0chisqdof}, and is indicated at the physical value of $m_\pi^2$. The curve corresponding to a lattice size of $10$ fm is almost indistinguishable from the infinite-volume curve.}}
\label{fig:extraps}
\vspace{12mm}
\includegraphics[height=0.7\hsize,angle=90]{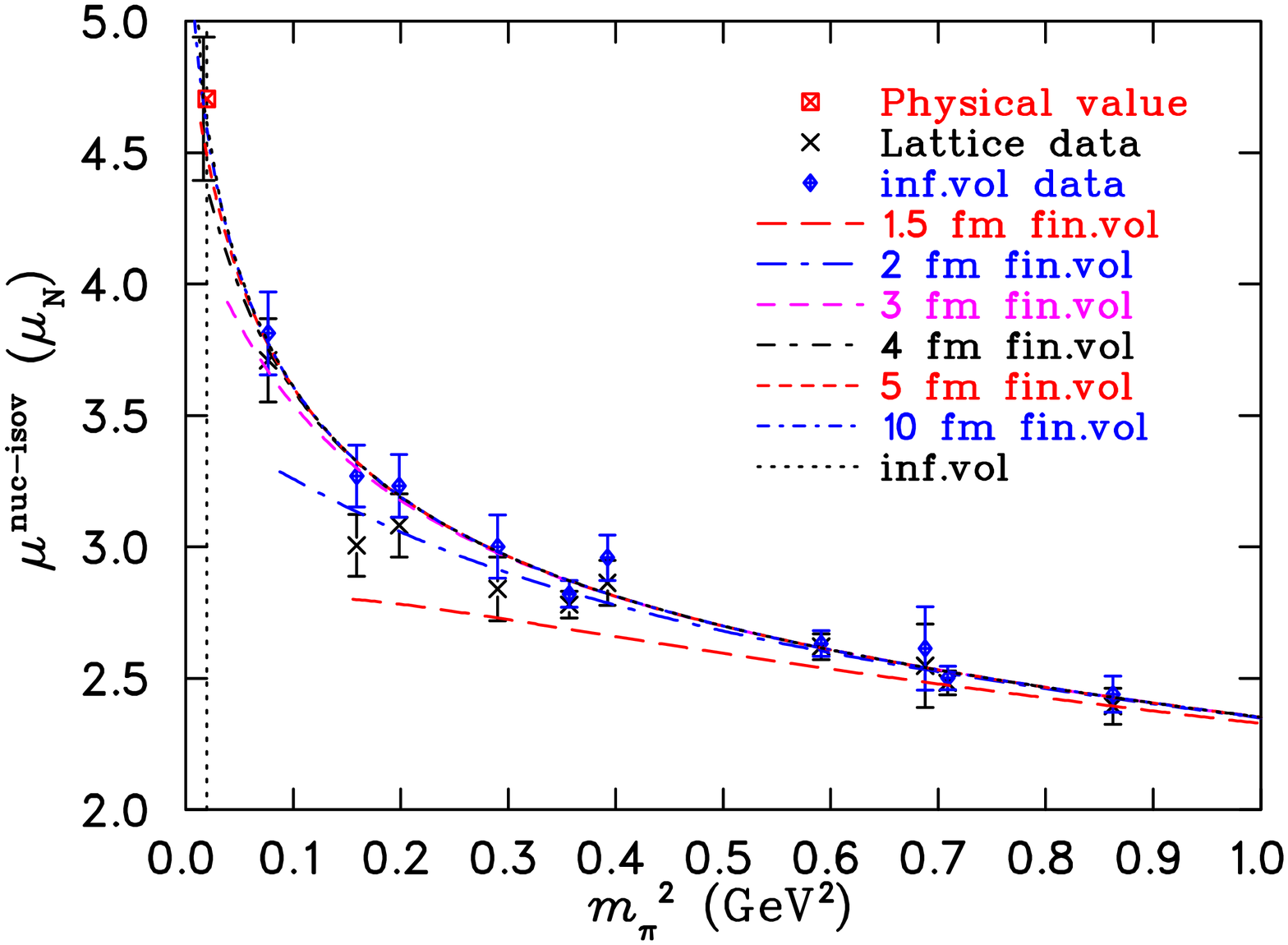}
\vspace{-11pt}
\caption{\footnotesize{ Extrapolations of $\mu_n^\ro{isov}$ 
at different finite volumes and infinite volume, using a dipole regulator, 
based on lattice QCD data from QCDSF, lattice sizes: $1.43-3.04$ fm. The provisional constraint $m_\pi L > 3$ 
is used. The infinite-volume corrected data points are shown. 
The physical value from experiment is marked. An estimate in the uncertainty in the extrapolation due to $\La^{\ro{scale}}$ 
 has been calculated from Figure \ref{fig:c0chisqdof}, and is indicated at the physical value of $m_\pi^2$. The curve corresponding to a lattice size of $10$ fm is almost indistinguishable from the infinite-volume curve.}}
\label{fig:extrapsinf}
\end{figure}

Using the optimal regularization scale, 
extrapolations or interpolations can be made to any quark mass. 
 Consider the behaviour of the magnetic moment as a function of the quark mass 
as shown in Figure \ref{fig:extraps} (in physical units). 
Here, the finite-volume expansion of Equation (\ref{eqn:finchiral}) 
is constrained 
by the lattice results from several different volumes. 
Extrapolation curves 
are then plotted for infinite volume and a variety of finite volumes at 
which current lattice QCD results are produced. 
For each curve, only the values for which $m_\pi L > 3$ are plotted, 
 provisionally, to avoid undesired effects of the $\epsilon$-regime.
The infinite-volume extrapolation to the physical point 
is within $2\%$ of the experimentally 
derived value: $\mu_n^{\ro{isov}} = 4.6798 \mu_N$.
The finite-volume extrapolations are useful for estimating the 
result of a lattice QCD calculation at certain box sizes. This can  
provide a benchmark for estimating the outcome of a lattice QCD simulation 
at larger and untested box sizes.
Note that even a relatively standard 
$3$ fm lattice box length will differ significantly from the experimental 
value at the physical point.
Since the data points in Figure \ref{fig:extraps} are at differing finite 
volumes, the infinite-volume corrected data are also displayed 
 in Figure \ref{fig:extrapsinf}.

%WRITE OUT FORMULA

%integrals @end of Work 4
%integral orders @chiPTmodnew.f90, and @centre of Work 5

%\subsection{Finite Volume Considerations for Form Factors on the Lattice}
\section{Finite-Volume Considerations for the Electric Charge Radius}
%
%@centre-end of Work 6

%MOTIVATION want to do extraps @ any mpisq, @ any L val.
%Extrapolations of lattice QCD simulation 
%results are a useful tool in examining the 
%chiral properties of observables. 
Reliable extrapolations take into account finite-volume effects, 
as well as leading-order chiral loop corrections. 
In many cases, calculating the finite-volume corrections to loop integrals 
poses 
no essential problems. %, and is merely limited to computational resources. 
Examples of $\chi$EFT analyses accounting for 
finite-volume effects can be found in References 
\cite{Beane:2004tw,Hall:2010ai}. %add more

However, the treatment of the electric 
charge radius is more challenging. Once form factors have been extracted from 
the lattice simulation, they are typically converted directly 
into charge radii, %ready 
%for analysis. 
The essential difficulty lies in the definition of the 
charge `radius' at finite volume. In order to define the radius, a 
derivative in the momentum transfer $Q^2 = \vec{q}^2-q_0^2$ (at $Q^2 = 0$) 
must be applied to the electric 
form factor. %Thus, one assumes the existence of a 
%well-defined `small $Q^2$ expansion' for the form factor. 
This approach breaks down on the lattice, where only discrete momentum values 
are allowed. 

 In this chapter, a method is outlined 
for handling finite-volume corrections to a given lattice simulation result. 
%set of electric 
%charge radius lattice data. 
It will be discovered that the finite-volume 
corrections to the loop integrals must be applied before the conversion 
from form factor to charge radius. By applying the finite-volume 
corrections directly to the electric form factor, and ensuring that 
the procedure preserves the electric charge normalization, %the dipole 
%Ansatz 
an extrapolation in $Q^2$ 
may be used to construct an infinite-volume charge radius. 
The infinite-volume charge radius can be %is allowed to be 
defined as normal. %in the usual way. 
A finite-volume charge radius may also be defined, as long as an allowed value 
of $Q^2$ is used in the conversion from infinite to finite volume.

\begin{figure}
\centering
\includegraphics[height=0.7\hsize, angle = 90]{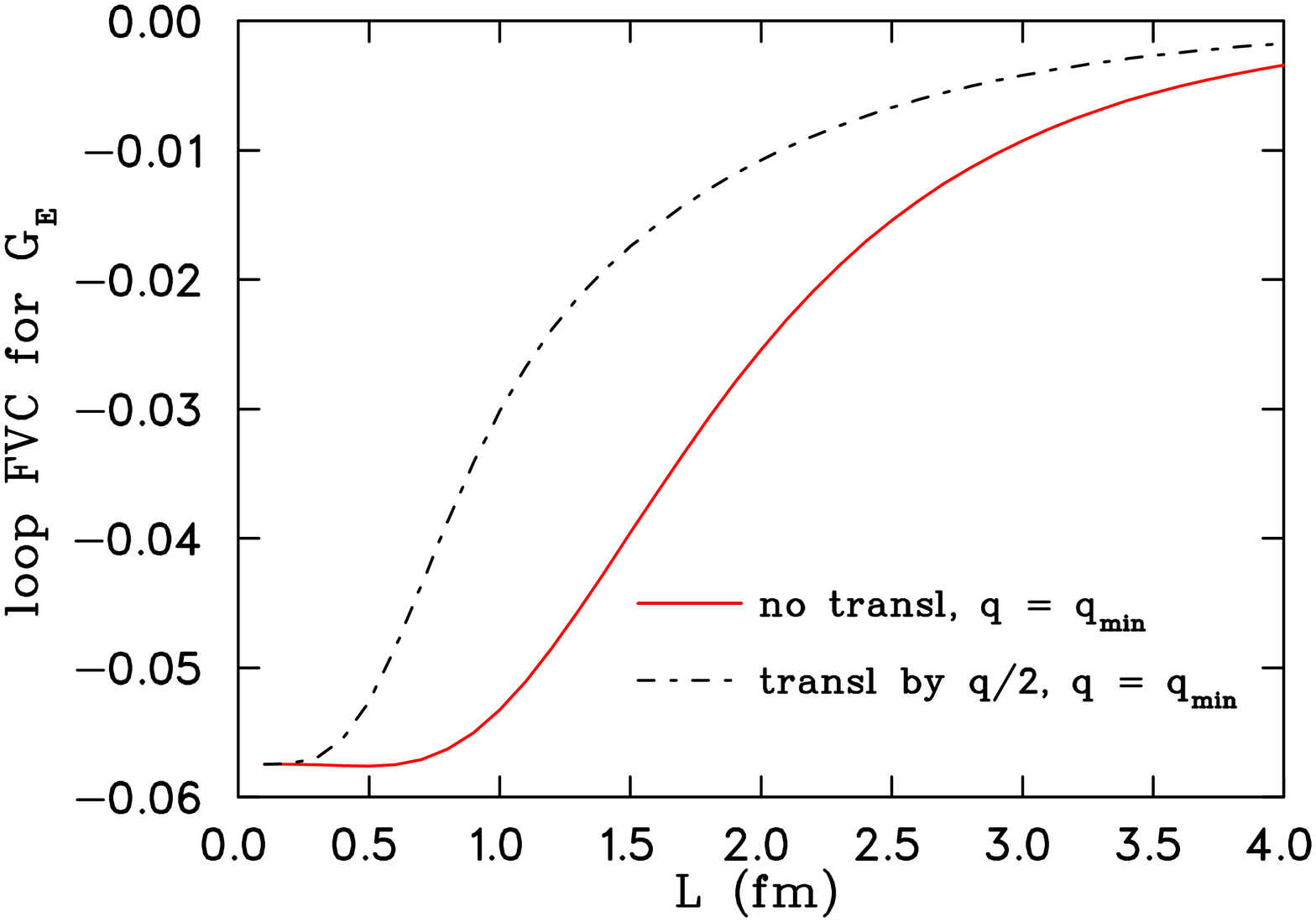}
\caption{\footnotesize{Finite-volume correction  for the loop integral contributing to $G_E$, with $q = q_{\ro{min}}$. The choice of $q/2 = q_{\ro{min}}/2$ is not an allowed value on the lattice. The momentum translated and untranslated behaviour of the finite-volume correction are inconsistent with each other.}}
\label{fig:loopfvcQsq}
\vspace{12mm}
\includegraphics[height=0.7\hsize, angle = 90]{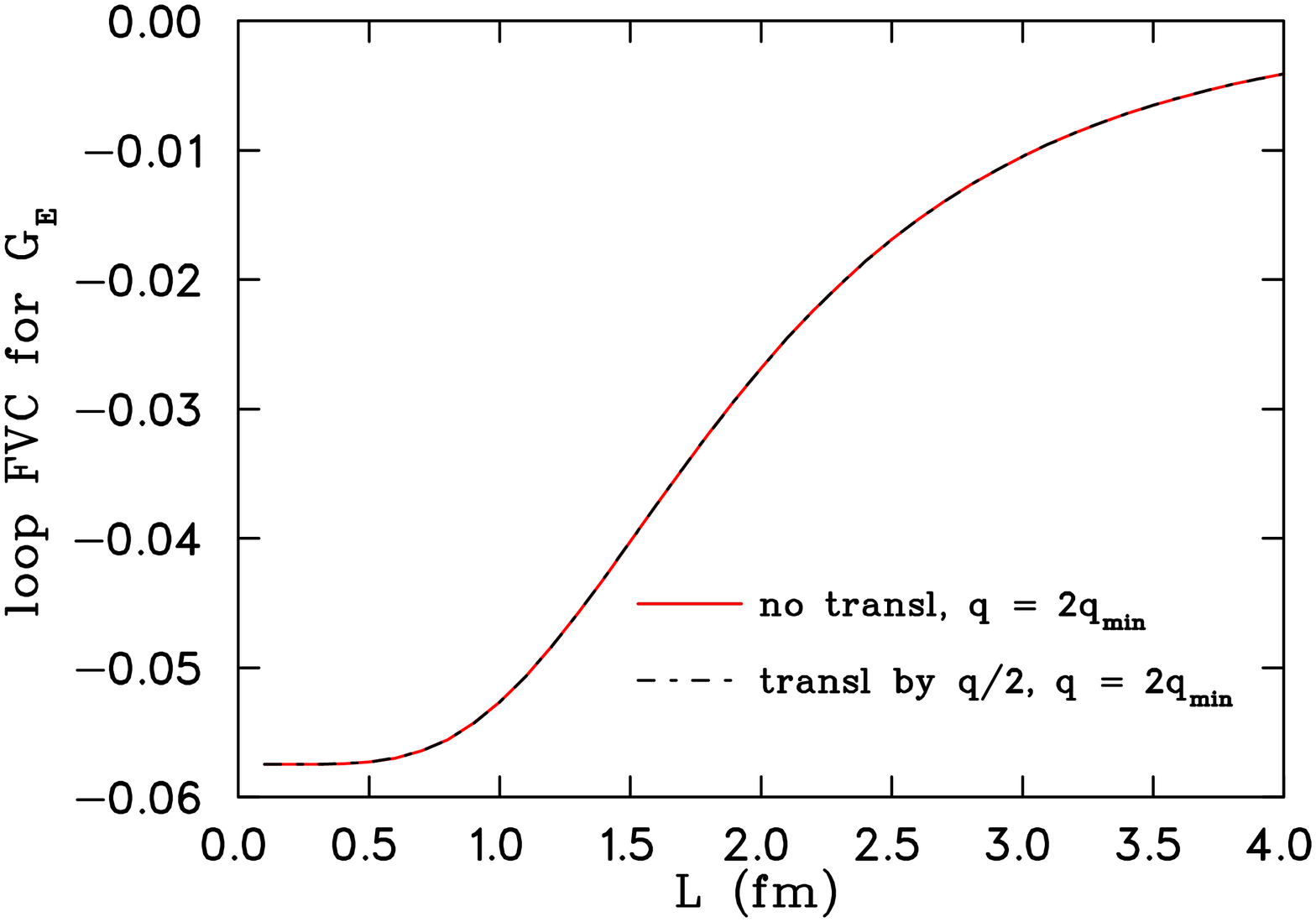}
\caption{\footnotesize{Finite-volume correction for the loop integral contributing to $G_E$, with $q = 2q_{\ro{min}}$. The choice of $q/2 = q_{\ro{min}}$ is an allowed value on the lattice. Therefore, the momentum translated and untranslated behaviour of the finite-volume correction is identical.}}
\label{fig:loopfvc4Qsq}
\end{figure}

%The problem one faces 
The first challenge involves the definition of the electric charge radius
 in terms of this derivative in Equation (\ref{eqn:raddefn}). Since only certain, 
discrete values of momenta are allowed on the lattice, the derivative may 
only be constructed from these allowed momenta when calculating finite-volume 
corrections.
This crucial observation becomes apparent when a comparison is made between 
the loop integrals evaluated at allowed, and unallowed, values of momentum 
transfer, respectively. The comparison is shown in Figures 
\ref{fig:loopfvcQsq} and \ref{fig:loopfvc4Qsq}, for momentum 
$q=(2\pi\, n/L)$, on the lattice. Here, $n$ is an integer. 
%\ref{fig:loopfvcQsq} and \ref{fig:loopfvc4Qsq}. 
%
%re: plots
Note that, if there is a momentum-translation in the loop integrals, 
$k\rightarrow k-q/2$, 
the choice of $q = q_{\ro{min}} = (2\pi/L)$ (for box length $L$), means 
that $q/2$ is  %\emph
 {no longer} 
 an allowed value on the lattice, and these finite-volume corrections will be 
inconsistent with the 
untranslated result. 
Under such a momentum-translation, 
 external momenta of $q/2$ flow through the loop integral, and 
one should choose at least a value of $q = 2q_{\ro{min}}$ to define a 
consistent discrete derivative for use in the definition of the charge radius 
in Equation (\ref{eqn:raddefn}).
However, choosing a momentum transfer of $q = q_{\ro{min}}$ 
 for a moderate lattice size of $3$ fm leads to a relatively large value: 
% Q^2 = q^2 - q_0^2. q_0 = sqrt(MN^2 - 700 MeV^2) - MN = 0.373 MeV
%Q^2 = 699.86 = 700, to 2 figs.
 $Q^2 \approx 700$ MeV$^2$. 
In defining the charge radius, 
the necessary extrapolation to $Q^2 = 0$ will be made more reliable 
by choosing 
a value of $Q^2$ to be as small as possible. 
This situation differs from the infinite-volume calculation of loop integrals, 
where true momentum-translation invariance is restored, and a continuous 
derivative may be used as normal.

\subsection{Chiral Loop Integrals}

Though loop integrals in the continuum
limit are invariant under momentum translations 
$k\rightarrow k+c\,q$, $c\in\mathbb{Z}$ (for internal loop momentum $k$),
a finite-volume loop sum must not include any values of $q$ less than 
$q_{\ro{min}}=(2\pi/L)$. Therefore, 
to obtain a suitable charge radius 
one chooses a definition of the loop integrals %such that finite volume converges 
%correctly to infinite volume as $L\rightarrow \infty$ for $Q^2 = (2\pi/L)^2$:
such that no factors of $\vec{q}/2$ appear. In fact, as long as no fractions 
of $\vec{q}$ appear in the integrand, the finite-volume version will converge  
correctly to the infinite-volume version as the box length is taken to 
infinity, %$L\rightarrow \infty$, 
for $q  = (2\pi\, n/L)$, $n\in \mathbb{Z}$:
\begin{align}
\label{eqn:SiL}
\ca{T}^E_{N}(Q^2) &=  
\frac{\chi^E_{N}}{3\pi}\!\int\!\!\mathrm{d}^3 k 
\frac{(k^2-\vk\cdot\vq)\,u(\vk\,;\La)\,u(\vk-\vq\,;\La)}
{\omega_{\vk}\omega_{\vk-\vq}(\omega_{\vk}
+\omega_{\vk-\vq})},\\
\ca{T}^E_{\De}(Q^2) &=  
\frac{\chi^E_{\De}}{3\pi}\!\int\!\!\mathrm{d}^3 k 
\frac{(k^2-\vk\cdot\vq)\,u(\vk\,;\La)\,u(\vk-\vq\,;\La)}
{(\omega_{\vk} + \Delta)(\omega_{\vk-\vq} + \Delta)(\omega_{\vk}
+\omega_{\vk-\vq})},\\
\ca{T}^E_{\ro{tad}}(Q^2) &=  \frac{\chi^E_{t}}{\pi}\!\int\!\!\mathrm{d}^3 
k \frac{u^2(\vk\,;\La)}
{\omega_{\vk}+\omega_{\vk-\vq}},%,\\
%\Sigma &= \Sigma_{\ro{loop}} + \Sigma_{\ro{tad}}.
\label{eqn:SiT}
\end{align}
where $\om_{\vec{k}} = \sqrt{{\vec{k}}^2 - m^2_\pi}$, %$m_\phi$ 
%is the pion or kaon mass, 
and $\De$ is the mass-splitting. 
The chiral coefficients $\chi_N^E$ and $\chi_\De^E$ and $\chi_t^E$ 
 are summarized by Wang 
\cite{Wang:2008vb}:
\begin{align}
\chi_N^{E,p} &= \f{5}{16\pi^2f_\pi^2}(D+F)^2 = -\chi_N^{E,n},\\
\chi_\De^{E,p} &= -\f{5}{16\pi^2f_\pi^2}\f{4\ca{C}^2}{9} = -\chi_\De^{E,n},\\
\chi_t^{E,p} &= -\f{1}{16\pi^2f_\pi^2} = -\chi_t^{E,n}.
\end{align}
%
%Note that the Feynman Diagrams in Figures \ref{fig:emSEa} and \ref{fig:emSEb} 
%are both incorporated into the general expression for $\ca{T}_{}$ in 
%Equation (\ref{eqn:SiL}).
%The finite-volume correction is defined through the convention:
%
%\eqb
% \delta_L[\ca{T}^E] = \chi\left[\f{{(2\pi)}^3}{L_x L_y L_z}\sum_{k_x,k_y,k_z} -% 
%\int\!\mathrm{d}^3k\right]\ca{I}\,,
%\eqe

%for integrand $\ca{I}$. 
The integrals which contribute to the electric charge radius, denoted ($T^E$),
 %which occur in the electric charge radius' chiral expansion 
%in Equation (\ref{eqn:chiralexp}) 
are exactly analogous to the integrals
 ($\ca{T}^E$) defined in Equations 
(\ref{eqn:SiL}) through (\ref{eqn:SiT}), that 
correspond to the electric form factor $G_E$. To obtain the integrals that 
contribute to the charge radius, 
one simply takes the derivative with respect to momentum transfer $Q^2$ at 
vanishingly small values of $Q^2$. This is allowed in the infinite-volume 
limit:
\eqb
T^E = \lim_{Q^2\rightarrow 0}-6\f{\cd \ca{T}^E(Q^2)}{\cd Q^2}.
\label{eqn:siSi}
\eqe 
Note that the ensuing procedure for calculating the finite-volume corrected 
electric charge radius uses only the infinite-volume versions of the 
chiral loop integrals. 
Fitting methods need only be applied at infinite volume. %, since the 
%residual coefficients ($a_i$) are expected to be volume-independent quantities.
Thus, the external momentum derivative in Equation (\ref{eqn:siSi}) need not be 
discretized, but may remain a continuous derivative.

To achieve a chiral extrapolation, 
it is convenient to subtract the coefficients 
$b_0^\La$ from the respective loop integrals that contribute to the 
electric charge radius:
\begin{align}
\tilde{T}_N^E &= T_N^E - b_0^{\La,N},\\
\tilde{T}_\De^E &= T_\De^E - b_0^{\La,\De},\\
\tilde{T}_\ro{tad}^E &= T_\ro{tad}^E - b_0^{\La,t}.
\end{align}
This %effectively shuffles 
removes the regularization scale-dependence from the lowest-order fit parameter 
%into higher-order terms 
of the chiral expansion. This technique provides 
an advantage in easily extracting the low-energy coefficient $c_0$ 
from the chiral expansion, described in Section \ref{subsect:elecexpn}.

%
%Now describe what `tilde' means. This will only be used for the \si 
%integrals for the charge radius, though.

%explain why we do fvc's on G_E. 
%Go through the procedure.
%For observables such as the nucleon mass, the peculiarity of the finite-volume %corrections as 
%discussed below does not feature, and the calculation of the finite-volume corrections occur in a 
%more straightforward manner [\cite{Hall:2010ai}]. 
%The problem lies in the definition of a `radius' on the lattice. 
%One needs to find a suitable and consistent definition of this radius.
%In this discussion, the radius is defined in terms of the dipole 
%Ansatz, evaluated at a value of $Q^2$ chosen to be the smallest 
%value of momentum available on the lattice, $Q^2_{\ro{min}}$.

%In the case of the electric charge radius, 
As emphasized already, 
Figures \ref{fig:loopfvcQsq} and \ref{fig:loopfvc4Qsq} 
show that the finite-volume corrections to the loop integrals cannot be applied 
directly to the charge radius itself. %Doing so results 
%in unphysical behaviour such as enhancement of the radius 
%as the box size becomes small. This is due to the inapplicability 
%of a small $Q^2$ expansion on the lattice. 
The momentum discretization ruins 
the circular symmetry in $q$ except at the values coinciding with lattice 
momentum values $(2\pi\, n/L)$, $n\in \mathbb{Z}$. 
The finite-volume corrections should be applied to the electric form 
factor $G_E(Q^2)$ instead.
A momentum convention in the loop integral is chosen such that $q$ may be 
chosen to be $q_{\ro{min}} = (2\pi\, n/L)$. %, thus avoiding the issue
% as illustrated in Figures 2 and 3. 
%\ref{fig:loopfvcQsq} and \ref{fig:loopfvc4Qsq}.
The procedure for achieving the correct finite-volume corrections %, at each 
%value of $m_\pi^2$,
 is outlined below.

First, the lattice finite-volume charge radius $\rad_E^L$ must be converted 
into a finite-volume 
form factor $G_E^L(Q^2)$, using $q = q_{\ro{min}} = (2\pi/L)$. 
This is achieved 
through use of an extrapolation in $Q^2$. As an example, a dipole Ansatz 
yields the following formula:
\eqb
G_E^L(Q^2_{\ro{min}}) = 
{\left(1 + \frac{Q^2_{\ro{min}}\rad_E^L}{12}\right)}^{-2}\,,
\eqe
where $Q^2_{\ro{min}} = \vec{q}^2_{\ro{min}} - (E_N - M_N)^2$. 
In many cases, this simply reverses the steps used to convert lattice 
results to charge radii. 
In this investigation, the electric form factor was fortunately obtained 
directly from the preliminary lattice QCD data from QCDSF.  
The next step is to transform the finite-volume form factor  
$G_E^L(Q^2_{\ro{min}})$ to an infinite-volume form factor 
$G_E^\infty(Q^2_{\ro{min}})$, so that the infinite-volume charge radius can be 
calculated. %This is so that the charge radius 
%can be calculated at any box size of our choosing. For example, 
%once the infinite volume charge radius has been obtained 
This is achieved by subtracting
the electric charge symmetry-preserving finite-volume correction, defined by:
\eqb
\label{eqn:fvc}
 \Delta_L(Q^2_{\ro{min}},0) = 
%\delta_L\left[\ca{T}(Q^2_{\ro{min}},\La') - \ca{T}(0,\La')\right]. 
\delta_L\left[\ca{T}^E(Q^2_{\ro{min}}) - \ca{T}^E(0)\right]. 
\eqe
The
second term of Equation (\ref{eqn:fvc}) ensures that both infinite- and 
finite-volume form 
factors are correctly normalized, % to $1$ when $Q^2 = 0$, 
that is,  
$G_E^{L,\infty}(0) = 1$. %, 
%as expected. 
Thus, the infinite-volume electric form factor can be calculated 
using the equation:
\eqb
G_E^\infty(Q^2_{\ro{min}}) = G_E^L(Q^2_{\ro{min}}) - \Delta_L(Q^2_{\ro{min}},0).
%\delta_L\left[\ca{T}(Q^2_{\ro{min}},\La') - 
%\ca{T}(0,\La')\right].
\eqe

\subsection{Chiral Expansion of the Electric Charge Radius}
\label{subsect:elecexpn}

The infinite-volume charge radius $\rad_E^\infty$ can be recovered 
from the form factor by using the extrapolation in $Q^2$. Once the 
infinite-volume charge radius has been obtained, 
a chiral extrapolation can be performed if needed. The chiral loop integrals 
corresponding to the charge radius are those defined by Equation 
(\ref{eqn:siSi}). Using the dipole Ansatz:
%dipole Ansatz again:
%
\eqb
\rad_E^\infty = \frac{12}{Q^2_{\ro{min}}}
\left(\sqrt{\frac{1}{G_E^\infty(Q^2_{\ro{min}})}}
-1\right)\,.%\,[\mathrm{fm}^2].
\eqe
This infinite-volume radius, %is suitable for use with an extrapolation. 
 calculated at multiple values of $m_\pi^2$,
 can be used for fitting and obtaining 
coefficients from the chiral expansion: %(renormalized to order $\ca{O}(m_\pi^0)$:
 %:
%
%\eqb
%\rad_E^\infty = \{a_0 + a_2m_\pi^2 \} + {}(m_\pi^2,\La)
%+ \mathcal{O}(m_\pi^3).%,
%\eqe
%\begin{align}
%\eqb
%\label{eqn:chiralexp}
%\rad_E^\infty %&
%= \{a_0^\La + a_2^\La m_\pi^2\} + T^E_{N} + + T^E_\De %\nn\\
%&
%+ T^E_{\ro{tad}} + \ca{O}(m_\pi^3),
%\eqe
%\end{align}
%
%\begin{align}
\eqb
\label{eqn:chiralexprenorm}
\rad_E^\infty %&
= \{c_0^{(\mu)} + a_2^\La m_\pi^2\} + \tilde{T}^E_{N}(m_\pi^2\,;\La) + 
\tilde{T}^E_\De(m_\pi^2\,;\La) %\nn\\
%&
+ \tilde{T}^E_{\ro{tad}}(m_\pi^2\,;\La) + \ca{O}(m_\pi^4),
\eqe
%\end{align}
where the expansion has been renormalized %to order $\mathcal{O}(1)$ 
in anticipation of the analysis of the 
renormalization flow of the coefficient $c_0$.
%The loop integrals $\tilde{\sigma}$ are obtained from the aforementioned 
%expressions by multiplying by a normalization of $-6$ 
% and taking a derivative with respect 
%to $Q^2$, which now can be treated as a continuous derivative.% The `$\sim$' 
%denotes the subtraction of the constant term of the loop integrals 
%in the particular FRR scheme.
%
This expansion contains an analytic polynomial in $m_\pi^2$ plus 
the leading-order chiral loop integrals, 
from which non-analytic behaviour arises.

By evaluating the loop integrals, the fully renormalized 
chiral expansion can be written in terms of a polynomial in $m_\pi^2$ 
and non-analytic terms. To leading non-analytic order $\ca{O}(\ro{log}\,m_\pi)$:
%
%CHECK THIS!!!
\eqb
\label{eqn:logexpn}
\rad_E^\infty = c_0^{(\mu)} + (\chi_N^E + 
\chi_t^E)\log\f{m_\pi}{\mu} + \ca{O}(m_\pi^2)\,.
%c_2m_\pi^2 +\f{\chi_D^E}{\De^2}m_\pi^2\log\f{m_\pi}{\mu}+ \ca{O}(m_\pi^3)\,.
\eqe
Since %the loop integrals contribute a logarithm to 
the chiral 
expansion of Equation (\ref{eqn:logexpn}) contains a logarithm, 
the value of $c_0$ can only be extracted relative to some mass 
scale $\mu$, 
which is chosen to be $1$ GeV.

Finally, the finite-volume charge radius can be evaluated 
%at any  value of $m_\pi^2$ and 
by adding the finite-volume correction to the form factor at 
any box length $\tilde{L}$, and corresponding momentum transfer on the 
lattice, 
${\tilde{Q}}^2_{\ro{min}}$: %converting the charge radius to a 
%form factor, at infinite volume. %, using the dipole Ansatz. 
\eqb
G_E^{\tilde{L}}({\tilde{Q}}^2_{\ro{min}}) = G_E^\infty({\tilde{Q}}^2_{\ro{min}}) 
+ \Delta_{\tilde{L}}({\tilde{Q}}^2_{\ro{min}},0).
%\delta_{\tilde{L}}\left[\ca{T}({\tilde{Q}}^2_{\ro{min}},\La') - \ca{T}(0,\La')\right],
\eqe
%
%The minimum non-zero momentum transfer on the lattice, 
%${\tilde{Q}}^2_{\ro{min}}$, corresponds to the chosen box length $\tilde{L}$.
%
%\eqb
%G_E^\infty ({Q'}^2_{\ro{min}}) = {\left(1 + 
%\frac{{Q'}^2_{\ro{min}}\rad_E^\infty}{12}\right)}^{-2}\,.
%\eqe
%The finite-volume corrections are then added back in order to obtain the 
%finite-volume form factor,
%
%
The finite-volume charge radii are obtained from the chosen extrapolation  
formula at box size $\tilde{L}$. 
%and finally, the dipole Ansatz used to convert the 
% form factor 
%into a  charge radius, at finite volume, with box length $\tilde{L}$:
%
%\eqb
%\rad_E^{\tilde{L}} = \frac{12}{{\tilde{Q}}^2_{\ro{min}}}
%\left(\sqrt{\frac{1}{G_E^{\tilde{L}}({\tilde{Q}}^2_{\ro{min}})}}-1\right)\,[\mathrm{fm}^2]\,.
%\eqe
%
An electric charge radius may be calculated at any desired value of  
%$m_\pi^2$ and 
box length, based on lattice QCD simulation results. %In doing so, 
%the electric charge radius may be extrapolated to the physical point or 
%to the chiral limit, and
Thus,  
the finite-volume behaviour of the charge radius 
may be analyzed. %for any lattice size $L$, yielding theoretical predictions 
%for lattice QCD results at that volume.
 \\

\section{Evidence for an Intrinsic Scale in the  Electric Charge Radius}

%\section{Finite Volume Corrections in the Chirqal Limit}
%Put derivation of Beane's Master Formula in Appendix \ref{app:fvc}

The 
 preliminary 
QCDSF results for the electric charge radius of the nucleon  
are displayed, %on the same plot as the 
%along-side 
 with the experimental value marked, in Figure 
\ref{fig:elecdata}. The set of data 
is also listed in 
Appendix 
\ref{chpt:appendix4}, Table \ref{table:elecdata}.
%at a variety of 
% $m_\pi^2$ values are displayed in Figure \ref{fig:data}.
 The lattice sizes of each 
data point vary from $1.92$ to $3.25$ fm using $N_f = 2$ and  
$\ca{O}(a)$-improved
 Wilson quark action.  
%\cite{}[pending QCDSF].
A simple linear fit is included in this plot, which does not take into account 
the non-analytic behaviour of the 
chiral loop integrals, nor the finite-volume corrections to the data.
Just as for the case of the magnetic moment, the 
 linear fit does not reach the experimental value of the 
electric charge radius at the physical pion mass.
Since the lattice QCD results extend outside the PCR, the result of 
an extrapolation will be scale-dependent. However, this scale-dependence 
can be handled by obtaining an optimal regularization scale using the 
aforementioned procedure. 
%in the extrapolations, 
%as described in Chapter \ref{chpt:intrinsic}

\subsection{Renormalization Flow Analysis}
\label{subsect:renormflowelec}

\begin{figure}[tp]
\centering
\includegraphics[height=0.7\hsize,angle=90]{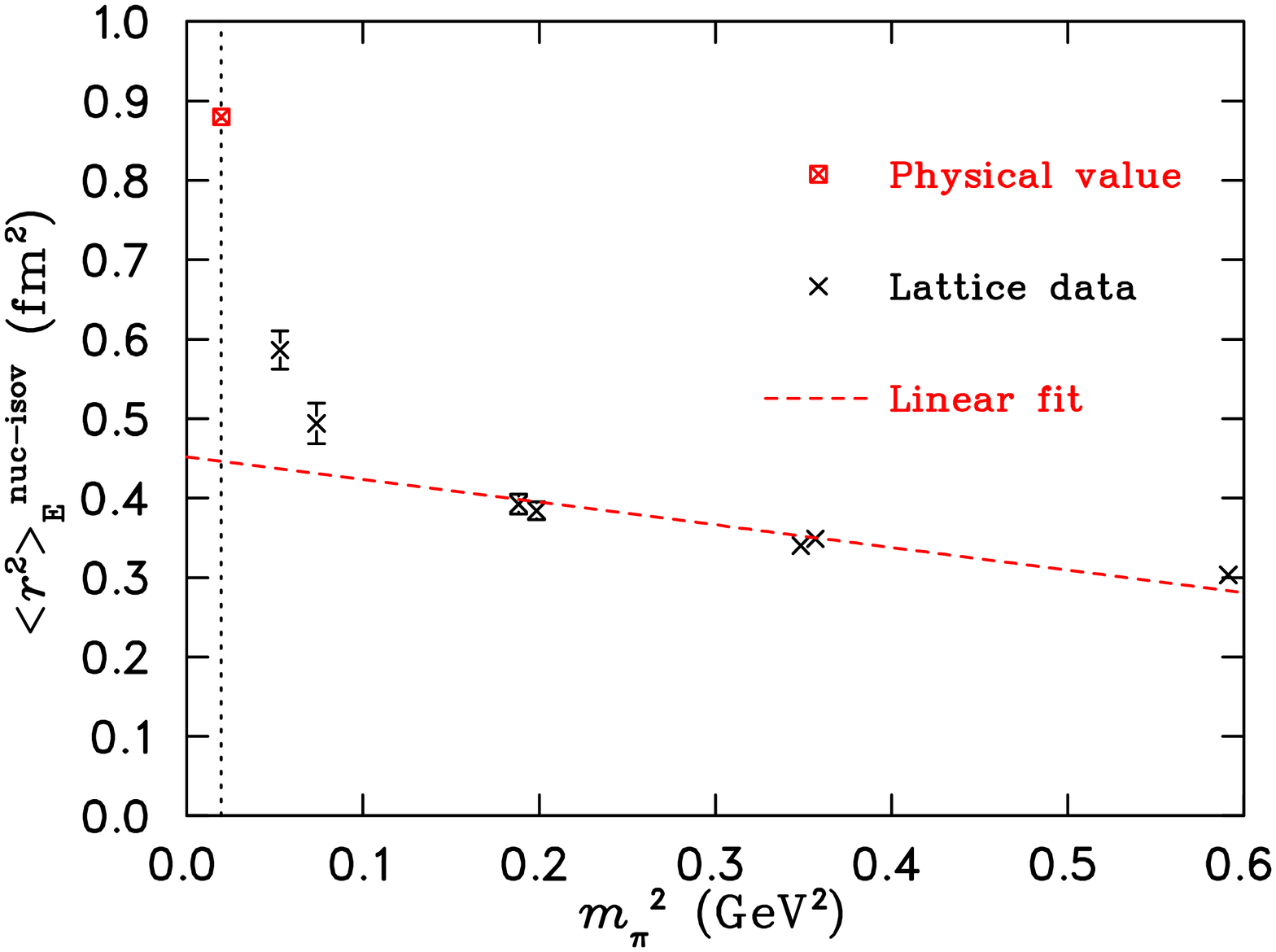}
\vspace{-11pt}
\caption{\footnotesize{Preliminary lattice QCD data for $\rad_E^{\ro{isov}}$ 
from QCDSF, with physical value from experiment as marked.}}
\label{fig:elecdata}
%\end{figure}
%
\vspace{12mm}
%\begin{figure}[tp]
\centering
\includegraphics[height=0.7\hsize,angle=90]{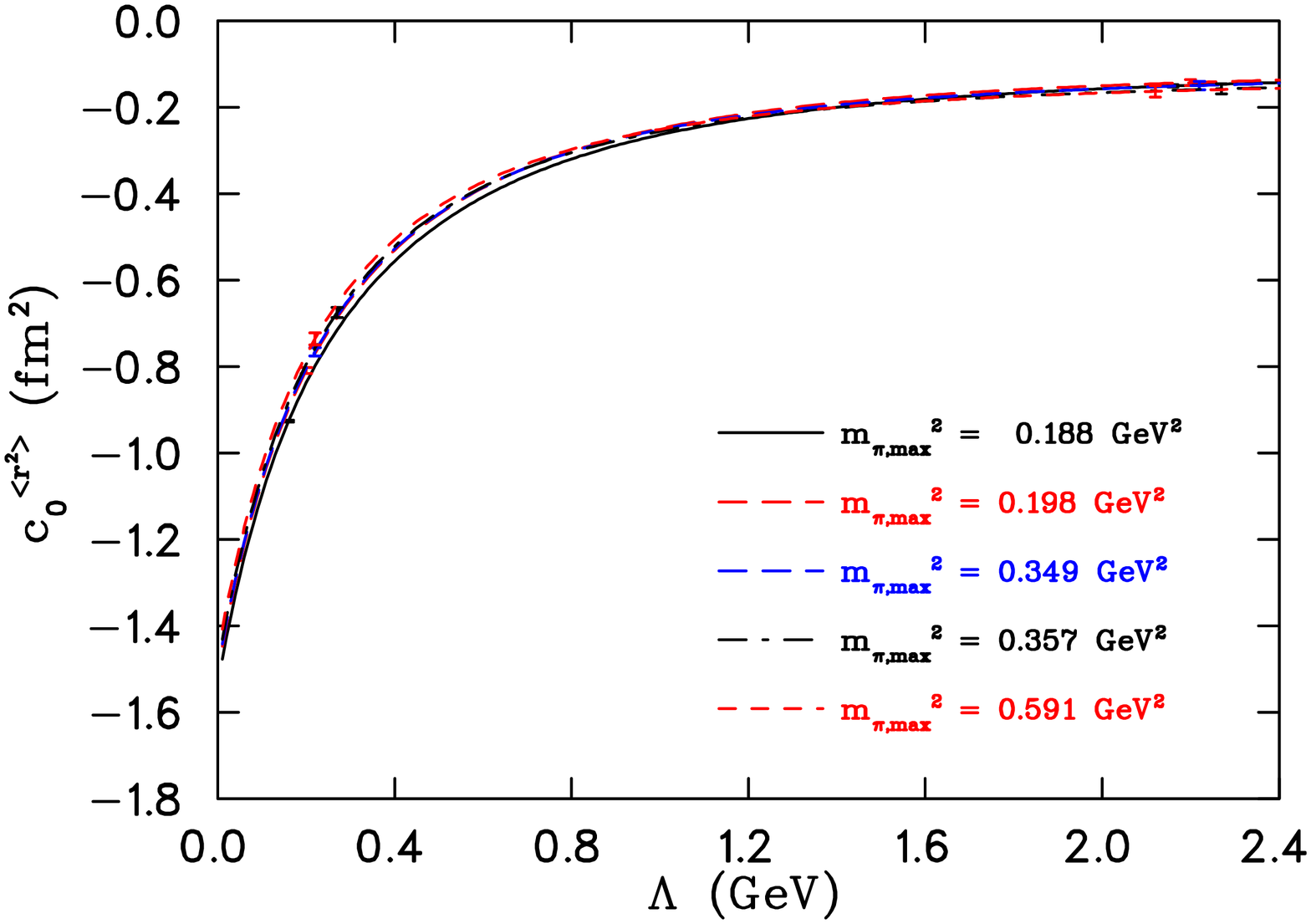}
\vspace{-11pt}
\caption{\footnotesize{ The renormalization flow of
  $c_0$ for $\rad_E^{\ro{isov}}$ obtained using a dipole regulator, based on preliminary 
 lattice QCD data from QCDSF. $c_0$ is calculated relative to the energy scale 
$\mu = 1$ GeV.}}
\label{fig:c0elec}
\end{figure}

In order to obtain an optimal regularization scale, 
the low-energy coefficient $c_0^{(\mu)}$ 
from Equation (\ref{eqn:chiralexprenorm}) will be calculated across a range 
of regularization scale values. 
%Thus the renormalization flow can be constructed. 
Multiple renormalization flow curves may be obtained by constraining the 
fit window by a maximum, $m_{\pi,\ro{max}}^2$, and sequentially adding  
data points to extend further outside the PCR. 
The renormalization flow curves for a dipole regulator 
are plotted on the same set of axes 
in Figure \ref{fig:c0elec}. 
Note that each data
point plotted has an associated error bar, but for the sake of clarity 
only a few points are selected to indicate the general size of the 
statistical error bars. 
%As more data are included in the fit, a greater degree of scale-dependence 
%is observed. %Note that there is a reasonably well-defined regularization scale 
%value 
%at which the renormalization of $c_0$ is insensitive to the truncation 
%of the data. This indicates that there exists an optimal regulator embedded 
%in the lattice QCD results themselves.
Note that, unlike the analysis of the nucleon mass  
and the magnetic moment, 
there is no distinct intersection point 
in the renormalization flow curves. In addition, the regularization 
scale-dependence of 
the coefficient $c_0^{(\mu)}$ is very slight, 
as long as the regularization scale 
is not too small, as discussed in Section \ref{sec:lowerbound}. 
%(A discussion of the best estimate for a lower bound on 
%the regularization scale can be found in Ref. \cite{Hall:2010ai}.)
 This lack of scale-dependence is a natural consequence of the logarithm 
in the chiral expansion of Equation (\ref{eqn:logexpn}), 
which is slowly-varying with respect to the 
regularization scale.
%UP TO HERE!!! 4/2/11

\subsection{Analysis of Systematic Uncertainties}
\label{subsect:statelec}

\begin{figure}[tp]
\centering
\includegraphics[height=0.7\hsize,angle=90]{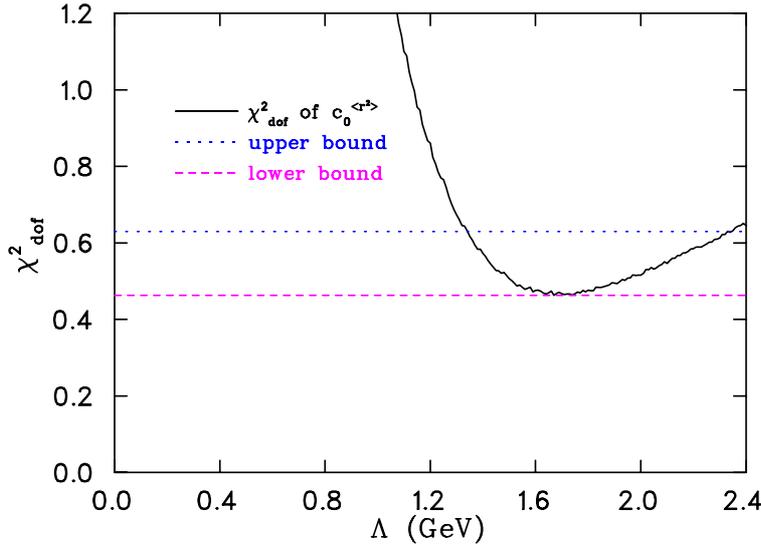}
\vspace{-11pt}
\caption{\footnotesize{ $\chi^2_{dof}$ for 
 the renormalization flow of
  $c_0$ for $\rad_E^{\ro{isov}}$ obtained using a dipole regulator, based on preliminary 
 lattice QCD data from QCDSF. $c_0$ is calculated relative to the energy scale 
$\mu = 1$ GeV.}}
\label{fig:c0chisqdofelec}
\end{figure}

An optimal regularization scale for a dipole form 
can nevertheless be extracted from Figure \ref{fig:c0elec} 
using the chi-square-style analysis. The analysis also provides 
a measure of the systematic uncertainty in the optimal scale.
By plotting $\chi^2_{dof}$ against the regularization scale 
$\La$, where $dof$ equals 
the number of curves $n$ minus one, a measure of the spread of the 
renormalization 
flow curves can be calculated, and the intersection point obtained. 
The function $\chi^2_{dof}$ is constructed in the same way as 
Equations (\ref{eqn:chisqwm}) and (\ref{eqn:chisq}). 
The $\chi^2_{dof}$ plot % for a dipole regulator are shown in 
corresponding to Figure \ref{fig:c0elec} is shown in 
Figure \ref{fig:c0chisqdofelec}.
%The optimal regularization scale $\La^\ro{scale}$ 
%is taken to be the central value $\La^\ro{central}$ of the plot, and 
%the upper and lower bounds obey 
% the condition $\chi^2_{dof} < \chi^2_{dof, min} + 1/(dof)$. 
Thus the optimal dipole regularization scale for a dipole regulator 
is: %$\La_{\ro{scale}} = 1.125 - 0.215 +0.195$
$\La^{\ro{scale}} = 1.67^{+0.66}_{-0.33}$ %(\scriptsize{$\stackrel{+0.22}{-0.20}$}) GeV.
% \scriptsize{$\Big\{\begin{matrix}+0.22\\-0.20 \end{matrix}$}} \normalsize GeV.
%$\Big\{\begin{matrix}+0.22\\-0.20 \end{matrix}$  
 GeV. 
This value, though larger than optimal dipole regularization scale 
values obtained from the previous analyses of the nucleon mass 
and the magnetic moment, is nevertheless 
consistent, with one-standard-deviation agreement. %with the optimal dipole regularization scale 
%values obtained 
%for the nucleon mass  %\cite{Hall:2010ai}, 
%using lattice QCD data from 
%JLQCD \cite{Ohki:2008ff}, PACS-CS \cite{Aoki:2008sm} and 
%CP-PACS\cite{AliKhan:2001tx}, 
%and the magnetic moment. %, %\cite{}, 
%using lattice QCD data from QCDSF \cite{}[pending QCDSF]. 
Thus, strong evidence is found that %,
%for a given functional form, 
the optimal 
regularization scale 
indicates the existence of an intrinsic scale, which characterizes  
the nucleon-pion interaction.
%

%

%\vspace{12mm}

\subsection{Chiral Extrapolation Results}
\label{subsect:chiextrapelec}

\begin{figure}[tp]
\centering
\includegraphics[height=0.7\hsize,angle=90]{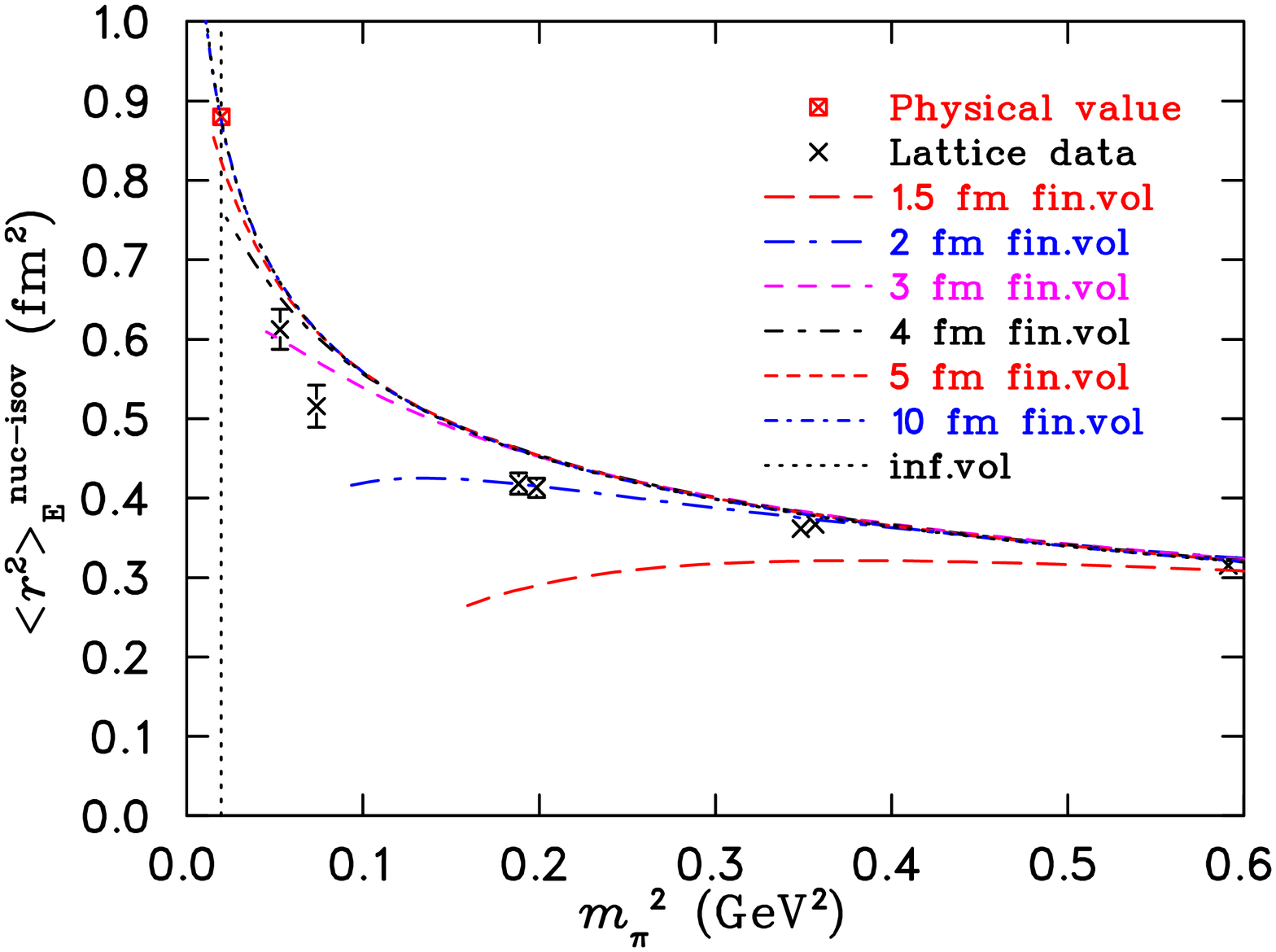}
\vspace{-11pt}
\caption{\footnotesize{ Extrapolations of $\rad_E^{\ro{isov}}$ 
at different finite volumes and infinite volume, using a dipole regulator, 
based on preliminary lattice QCD data from QCDSF, lattice sizes: $1.92-3.25$ fm. The provisional constraint $m_\pi L > 3$ 
is used. The physical value from experiment is marked. The curve corresponding to a lattice size of $10$ fm is almost indistinguishable from the infinite-volume curve. 
%An estimate in the uncertainty in the extrapolation due to $\La^{\ro{scale}}$ 
% has been calculated from Figure \ref{fig:c0chisqdof}.
}}
\label{fig:extrapselec}
\vspace{12mm}
\includegraphics[height=0.7\hsize,angle=90]{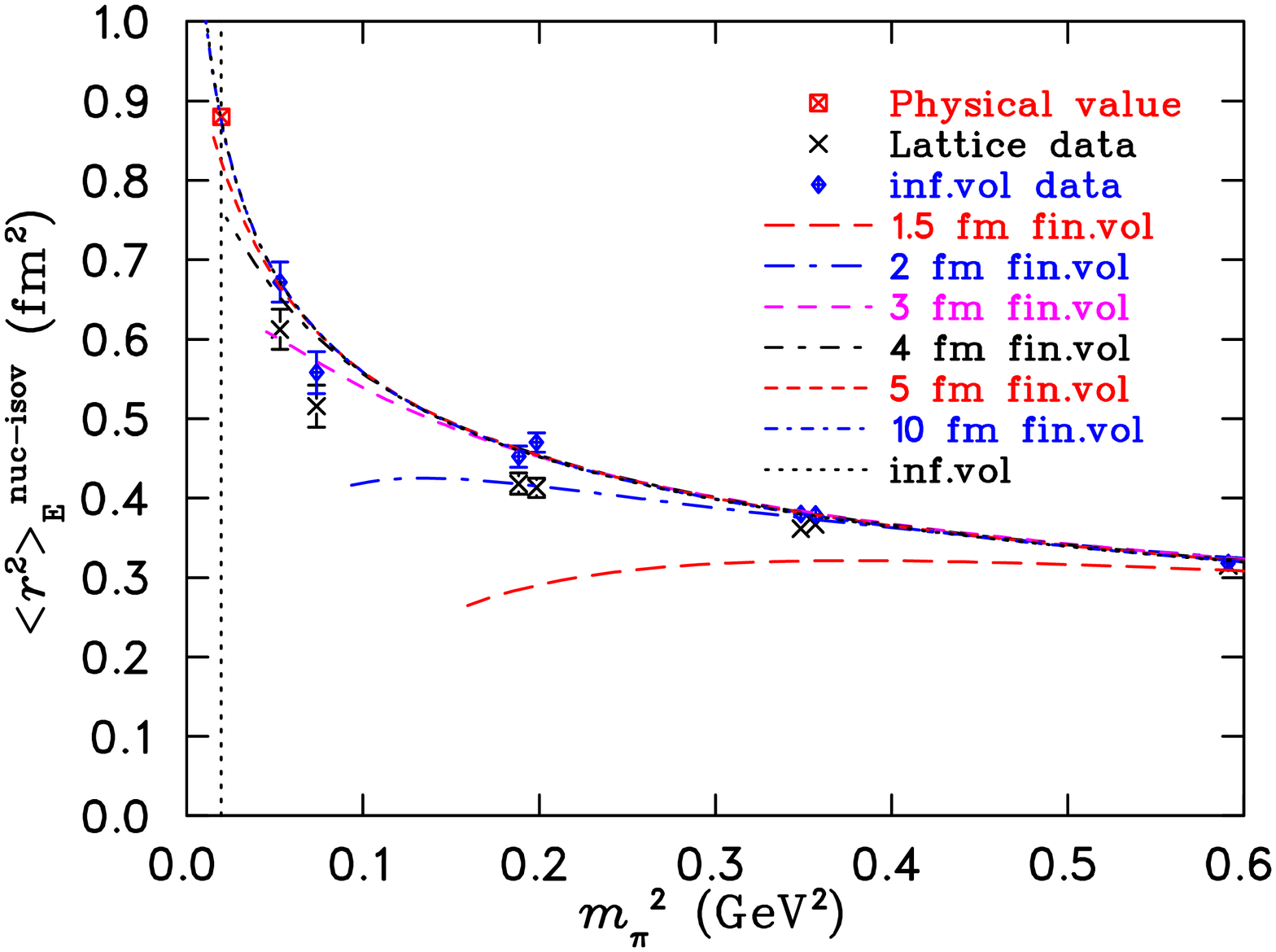}
\vspace{-11pt}
\caption{\footnotesize{ Extrapolations of $\rad_E^{\ro{isov}}$ 
at different finite volumes and infinite volume, using a dipole regulator, 
based on preliminary lattice QCD data from QCDSF, lattice sizes: $1.92-3.25$ fm. The provisional constraint $m_\pi L > 3$ 
is used. The infinite-volume corrected data points are shown. 
The physical value from experiment is marked. The curve corresponding to a lattice size of $10$ fm is almost indistinguishable from the infinite-volume curve. %An estimate in the uncertainty in the extrapolation due to $\La^{\ro{scale}}$ 
% has been calculated from Figure \ref{fig:c0chisqdof}.
}}
\label{fig:extrapsinfelec}
\end{figure}

Using the optimal regularization scale, 
a reliable chiral extrapolation can be performed, 
with the systematic uncertainty in the optimal regularization scale 
taken into account. 
 Consider the behaviour of the electric charge radius 
as a function of the quark mass 
as shown in Figure \ref{fig:extrapselec} (in physical units). 
Extrapolation curves 
are plotted for infinite-volume, and a variety of finite-volumes at 
which current lattice QCD results are produced. 
For each curve, only the values for which $m_\pi L > 3$ %true!(checked) 
are plotted, 
 provisionally, to avoid undesired effects of the $\epsilon$-regime.
%The infinite-volume extrapolation is merely $0.5\%$ different from 
%the experimentally 
%derived value: $\langle r^2\rangle_E^\ro{isov} = 0.88$ fm$^2$. 
The infinite-volume extrapolation to the physical point 
differs from the experimentally 
derived value: $\langle r^2\rangle_E^\ro{isov} = 0.88$ fm$^2$, by merely 
$0.5\%$.
The finite-volume extrapolations are also useful for estimating the 
result of a lattice QCD calculation at certain box sizes. This can also 
provide a benchmark for estimating the outcome of a lattice QCD simulation 
at larger and untested box sizes.
Note that the result of an extrapolation to the 
physical point, using an optimistic  
$4$ fm lattice box length, will differ significantly from the experimental 
value.
Since the data points in Figure \ref{fig:extrapselec} are at differing finite 
volumes, the infinite-volume corrected data points are displayed 
  in Figure \ref{fig:extrapsinfelec}.

\begin{figure}[tp]
\centering
\includegraphics[height=0.7\hsize,angle=90]{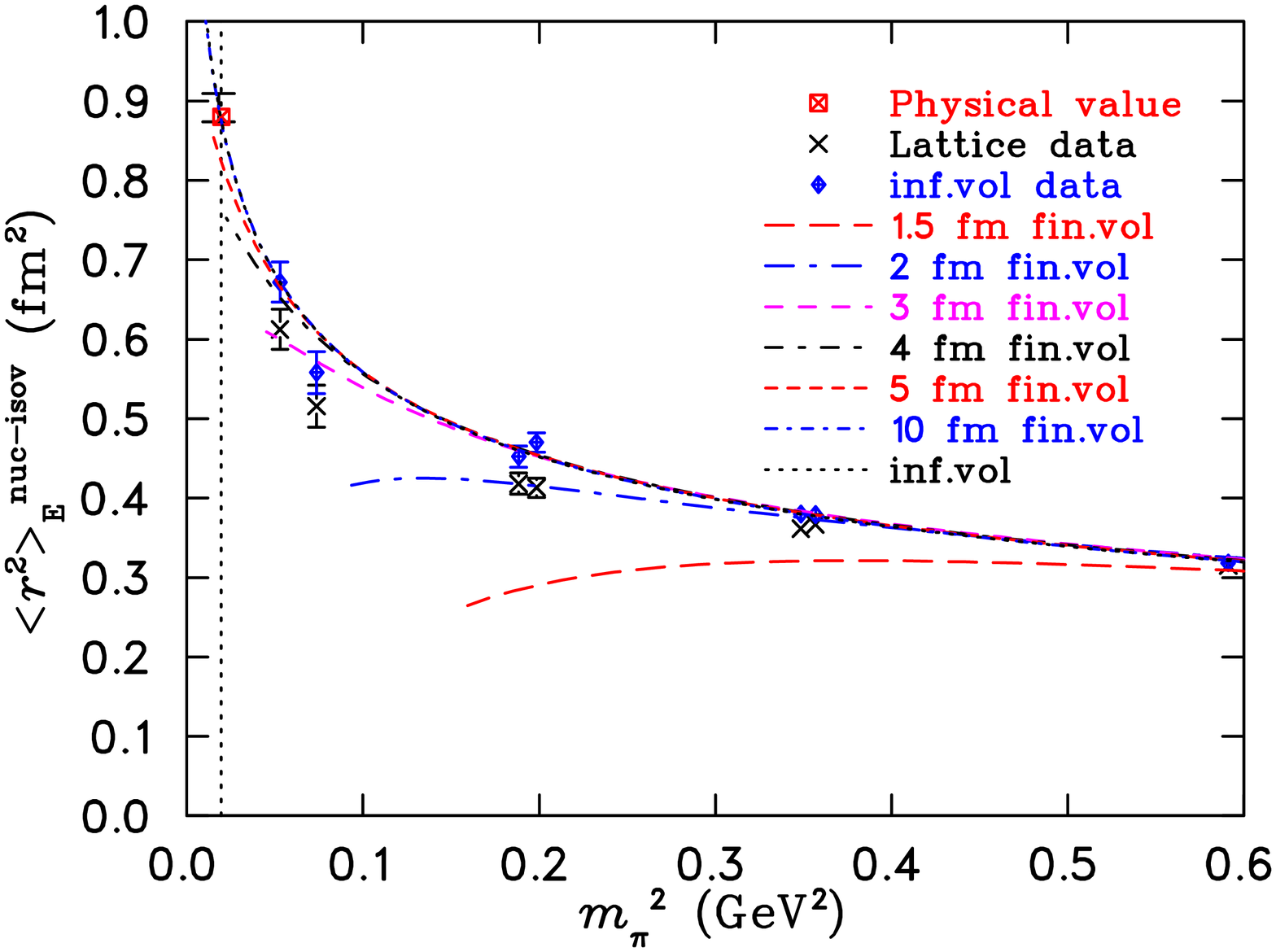}
\vspace{-11pt}
\caption{\footnotesize{ Extrapolations of $\rad_E^{\ro{isov}}$ 
at different finite volumes and infinite volume, using a dipole regulator, 
based on preliminary lattice QCD data from QCDSF, lattice sizes: $1.92-3.25$ fm. The provisional constraint $m_\pi L > 3$ 
is used. The infinite-volume corrected data points are shown. 
The physical value from experiment is marked. An estimate in the uncertainty in the extrapolation, due to $\La^{\ro{scale}}$, 
 has been calculated from Figure \ref{fig:c0chisqdofelec}, and is indicated at the physical value of $m_\pi^2$. The curve corresponding to a lattice size of $10$ fm is almost indistinguishable from the infinite-volume curve.}}
\label{fig:extrapsinferrelec}
\vspace{12mm}
\includegraphics[height=0.7\hsize,angle=90]{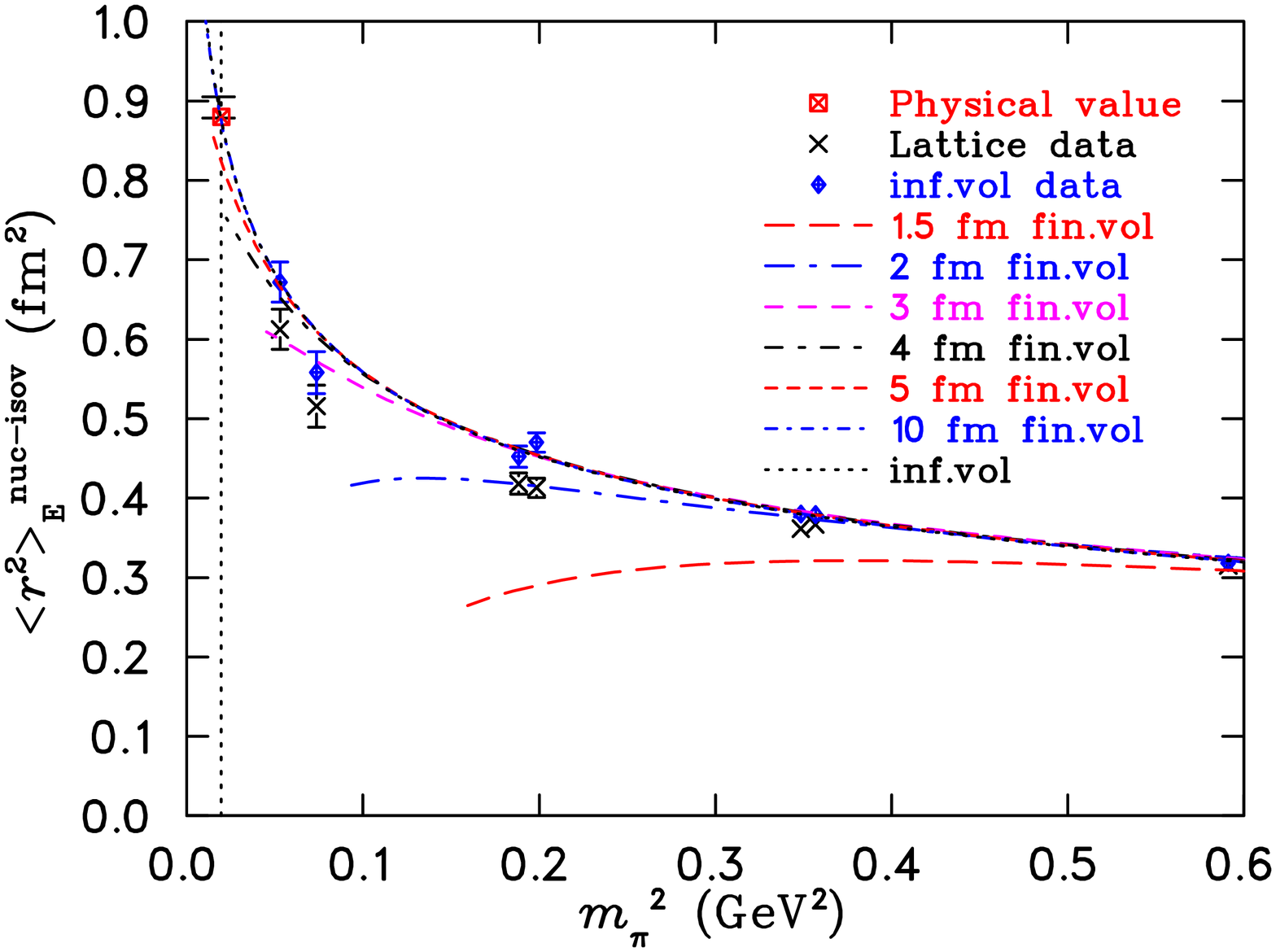}
\vspace{-11pt}
\caption{\footnotesize{ Extrapolations of $\rad_E^{\ro{isov}}$ 
at different finite volumes and infinite volume, using a dipole regulator, 
based on preliminary lattice QCD data from QCDSF, lattice sizes: $1.92-3.25$ fm. The provisional constraint $m_\pi L > 3$ 
is used. The physical value from experiment is marked. 
An estimate of the statistical uncertainty in the extrapolation is marked at the physical value of $m_\pi^2$. The curve corresponding to a lattice size of $10$ fm is almost indistinguishable from the infinite-volume curve.}}
\label{fig:extrapsstatelec}
\end{figure}

To highlight the insensitivity of the extrapolation to the regularization 
scale $\La^{\ro{scale}}$, an estimate of the systematic uncertainty in the 
extrapolation to the physical point solely due to $\La^{\ro{scale}}$ is displayed 
in Figure \ref{fig:extrapsinferrelec}. The size of the error bar at the physical point is comparable 
to that due to statistical uncertainty, as shown in Figure 
\ref{fig:extrapsstatelec}.
This indicates that, in the case of the 
electric charge radius, the identification of an intrinsic scale is 
borderline, due to the dominance of the logarithm in the chiral expansion, 
and its slowly varying property in the large $m_\pi$ regime. 
Therefore, chiral extrapolations of the electric charge radius are more robust, 
in the sense that the scale-dependence in the result is suppressed, and the 
identification of an intrinsic scale is not so vital as in the case of 
the nucleon mass or magnetic moment.

%\section{Electromagnetic Summary and Specific Issues}
\section{Summary and Specific Issues for the Electromagnetic Properties of the Nucleon}

%-echo abstract
%It was discovered that finite-volume corrections are ill-defined on the 
%lattice. The use of continuous derivatives in constructing the electric 
%charge radius 
%leads to inconsistent results in the finite-volume corrections. 
% It was discovered that finite-volume corrections must be applied to 
%the form factor rather than to the charge radius directly. 
%Therefore, a procedure was developed to apply finite-volume corrections 
%to the electric form factor by using a discrete derivative involving momenta 
%available on the lattice. The resultant finite-volume corrected form factor 
%may then be converted into a charge radius using an extrapolation in $Q^2$. 
%A new procedure was obtained for calculating the finite volume corrections  
%of the electric charge radii of octet baryons, using effective 
%field theory. The procedure allows extrapolations to be made at 
%infinite volume, or at any chosen finite volume or quark mass.

It was discovered that finite-volume corrections for charge radii %, 
%defined by the derivative of $G_E(Q^2)$ at $Q^2 = 0$, 
are ill-defined on the 
lattice. The use of continuous derivatives in constructing the electric 
charge radius 
leads to inconsistent results for %in a finite volume. 
the finite-volume corrections. 
 It was discovered that the finite-volume corrections must be applied to 
the electric form factors rather than to the charge radii directly. 
Therefore, a procedure was developed to apply finite-volume corrections 
to the electric form factor, % by using a discrete derivative, 
strictly 
involving momenta 
available on the lattice. The resultant finite-volume corrected form factor 
may then be converted into a charge radius using an extrapolation in 
 momentum transfer $Q^2$.

The technique for obtaining an optimal regularization scale 
from lattice QCD data  
has been investigated in 
the context of the magnetic moment and the electric charge radius 
of the isovector nucleon. By using 
recent, preliminary lattice QCD results from QCDSF, an optimal 
regularization scale 
for a dipole regulator was obtained. This was achieved, in each case, 
by analyzing the 
renormalization flow of the 
low-energy coefficient $c_0$ of the relevant chiral expansion with respect 
to the scale $\La$, 
whilst extending the data step-wise beyond the PCR.  
A regularization scale was discovered, for both the magnetic moment 
and the electric charge radius, for which the 
renormalization of each $c_0$ is least sensitive to the truncation of 
the lattice QCD data. 
The values of the optimal regularization scale 
were consistent with each other, as well as with the results from the 
nucleon mass analysis. 
Thus an intrinsic scale has been uncovered, which characterizes 
the size of the nucleon, as probed by the pion.

Using the value of the intrinsic scale, the extrapolation of
 the magnetic moment and the electric charge radius 
to the physical pion mass and the infinite-volume limit 
%lattice box size 
is consistent with experiment. The finite-volume 
extrapolations provide a benchmark for estimating the outcome of a 
lattice QCD simulation at realistic or currently %unattainable 
optimistic lattice box sizes.

The results clearly demonstrate a successful procedure for using  
lattice QCD data to extrapolate an observable to the low-energy region 
of QCD. 

%% file: conclusion.tex
\chapter{Conclusion}
\label{chpt:conclusion}

\textit{``Recall that in our theoretical construction those probabilities appeared simply as a logical, or linguistic, tool. It is only at this stage that they finally acquire the empirical significance they were lacking, and that chance enters the theoretical framework.''}
(Omn\`{e}s, R. 2002. \textit{Quantum Philosophy: Understanding and Interpreting Contemporary Science} p.209) \cite{Omnes}

\section{Evaluation and Summary Analysis}

Chiral effective field theory ($\chi$EFT) 
offers unique insights into the low-energy behaviour of hadrons. 
By using $\chi$EFT in conjunction with lattice quantum chromodynamics 
(lattice QCD) results, a deeper 
understanding of the underlying chiral interactions may be derived. 
%ascertained. 
In particular, the mathematical behaviour of the chiral expansion of an 
observable, within a power-counting scheme (PCR), was investigated. 
This led to the development of a method for identifying the PCR, 
where the renormalization of the low-energy coefficients of the chiral 
expansion are independent of the regularization scale. 
Novel methods for identifying a preferred renormalization scheme 
allowed the extrapolation of an observable to the chiral regime, 
and to infinite-volume lattice box sizes, without introducing a 
regularization scale in an \textit{ad hoc} fashion.

In this thesis, a procedure was established whereby an 
optimal regularization scale could be obtained from lattice QCD data. 
By constructing some ideal pseudodata using a known functional form, 
and based on actual lattice simulation results, 
 the behaviour of the low-energy coefficients, with respect to the 
regularization scale, indicated an optimal value for the scale. 
By considering pseudodata sets that extended increasingly beyond the PCR, 
there was a value of regularization scale at which the renormalization 
was least sensitive to this extension. This optimal scale is the value at 
which the correct values of the low-energy coefficients are recovered.  

Actual lattice simulation results 
for the nucleon mass, magnetic moment and electric charge radius were also 
analyzed using the same procedure. In each case, the analysis led to 
a consistent value of optimal regularization scale. In cases where multiple 
low-energy coefficients were analyzed, the optimal scale realized from each 
matched exactly: a non-trivial result. 

The analysis of lattice simulation results 
for the mass of the quenched $\rho$ meson was used to test the robustness  
of the method. A reliable technique for determining an  
optimal regularization scale, and performing infinite-volume and chiral 
extrapolations, was established. 

Comparing the optimal scales obtained from the 
nucleon mass, magnetic moment and electric charge radius analyses, 
a consistent optimal regularization scale was found. 
This indicates the existence of an intrinsic energy scale that 
characterizes the nucleon-pion interaction: %the division between 
%the pion cloud and the core of the nucleon. 
the size of the nucleon as probed by the pion. 

%nucleon mass
In the analysis of the nucleon mass, as described in Chapter 
\ref{chpt:nucleonmass}, 
it was demonstrated that a preferred regularization scheme 
exists only for data sets
extending outside the PCR. 
However, it is not always possible to identify this scale. 
The scale-dependence of an observable can be weakened by working 
to a higher chiral order. 
%Such a preferred 
%regularization scheme 
% is associated with an intrinsic scale %\emph{intrinsic scale}
% for the size of the pion dressings of the nucleon.
 The aforementioned procedure was
 used to calculate the nucleon mass at the physical point,
 the low-energy coefficients $c_0$ and $c_2$, and their associated statistical 
 and systematic errors.  Several different
functional forms of regulator were considered, and lattice QCD data from
 JLQCD, PACS-CS %\cite{Aoki:2008sm}
 and CP-PACS were used in the analyses. % \cite{AliKhan:2001tx}.
By working to chiral order $\ca{O}(m_\pi^3)$, 
 an optimal cut-off scale $\La^\ro{scale}$
for each set of lattice QCD data was obtained, and 
 an estimate of the systematic error in the choice of 
renormalization scheme was calculated, using a chi-square-style analysis. 
 A mean value for the optimal regularization scale of 
$\bar{\La}^\ro{scale}_\ro{dip} \approx 1.3$ GeV
 was obtained for the dipole, $\bar{\La}^\ro{scale}_\ro{doub} \approx 1.0$ GeV 
for the double-dipole and $\bar{\La}^\ro{scale}_\ro{trip} \approx 0.9$ GeV
 for the triple-dipole.
An analysis of the lowest suitable value for a regularization scale 
 allowed the identification of a breakdown region of finite-range 
regularization (FRR). 
The existence of a breakdown region %merely 
indicates that the ultraviolet regularization scale 
is low enough to remove or suppress the low-energy chiral behaviour being 
analyzed.

%rho
The robustness of the procedure for determining an optimal regularization 
scale and performing chiral extrapolations was tested in 
Chapter \ref{chpt:mesonmass}. 
In order to establish the predictive power of the %chiral extrapolation scheme 
%and the identification of an optimal regularization scale, 
procedure, 
the quenched $\rho$ meson mass was considered. 
 Because an experimental value of this observable does not exist, 
its calculation served to demonstrate the 
ability of the procedure 
 to make predictions without prior bias. 
Using lattice 
simulation results from the Kentucky Group, the procedure was tested, and  
 the interesting low-energy simulation results were \emph{predicted correctly}. 
By restricting the procedure to use only higher energy simulation data 
points, the low-energy coefficients $c_0$, $c_2$ and $c_4$ were considered 
and an optimal regularization scale was identified: 
%$\La^{\ro{scale}}_\rho = 0.64^{+0.08}_{-0.07}$ GeV
$\La^{\ro{scale}}_{\rho,\ro{trip}} = 0.67^{+0.09}_{-0.08}$ GeV. 
 An optimal value of the 
maximum pion 
mass used for fitting was also calculated, and was found to be 
$\hat{m}_{\pi,\ro{max}}^2 = 0.35$ GeV$^2$. 
By using only the data contained in the optimal pion mass region, 
constrained by $\hat{m}_{\pi,\ro{max}}^2$, a value 
$\La^\ro{scale}_{\rho,\ro{trip}} = 0.64$ GeV 
is estimated for the optimal regularization scale, with a wider 
systematic uncertainty corresponding to the entire range of 
suitable values of $\Lambda$. These two estimates of the optimal regularization 
scale are consistent with each other.
 
Upon revealing the omitted low-energy data, the extrapolations were  
compared to the simulation results at each value of pion mass. 
The correct chiral curvature was reproduced by the extrapolations, indicating 
the non-analytic chiral behaviour of the loop integrals. 
The results of extrapolations using $\chi$EFT, and 
the results of lattice QCD simulations were demonstrated to be consistent. 
The extrapolation to the physical point obtained for this quenched data set, 
using 
$\La^{\ro{scale}}_{\rho,\ro{trip}} = 0.67^{+0.09}_{-0.08}$ GeV, is: 
$m_{\rho,Q}^{\ro{ext}}(m_{\pi,\ro{phys}}^2) = 0.925^{+0.053}_{-0.049}$ GeV, 
an uncertainty 
of less than $6$\%.  
The result of the extrapolation, using 
$\La^\ro{scale}_{\rho,\ro{trip}} = 0.64$ GeV, 
with the systematic uncertainty 
calculated by varying $\Lambda$ across all suitable values,  is: 
$m_{\rho,Q}^{\ro{ext}}(m_{\pi,\ro{phys}}^2) = 0.922^{+0.065}_{-0.060}$ GeV, 
an uncertainty 
of only $7$\%. 
%$m_{\rho,Q}(m_{\pi,\ro{phys}}^2) = 0.915$ 
%($\pm \,0.036$) GeV, with an uncertainty of approximately $4$\%.

%e-m 
In the case of the electromagnetic properties of the nucleon, 
preliminary results from QCDSF were used.  
 The magnetic moment of the isovector nucleon was analyzed for a dipole 
regulator. A well-defined optimal regularization scale was obtained: 
$\La^{\ro{scale}}_{\mu,\ro{dip}} = 1.13^{+0.22}_{-0.20}$ GeV, for chiral order 
$\ca{O}(m_\pi^2\,\ro{log}\,m_\pi)$, 
and a successful extrapolation to the physical pion mass and infinite-volume 
was achieved, and compared to the experimental value. 
The infinite-volume extrapolation to the physical point 
was within $2\%$ of the experimentally 
derived value.

When considering charge radii, there are 
subtleties in performing finite-volume corrections. %Because of the definition 
%of the radius as the slope of the form factor for small momentum transfer 
%$Q^2$,  
 In defining the charge radius, 
the finite-volume corrections must be applied before 
an extrapolation to $Q^2 = 0$ is taken. Thus the finite-volume corrections 
must be applied to the form factors directly. 
Using this method, the electric charge radius of the isovector nucleon 
was analyzed for a dipole regulator. Assuming the regularization scale 
is not within the breakdown region of FRR, the scale-dependence of the 
low-energy coefficient $c_0^\mu$ (up to some scale $\mu$ of the chiral 
logarithm) is weak. The leading-order non-analytic behaviour of the 
logarithm in the chiral expansion is slowly varying with respect 
to the regularization scale. Nevertheless, an optimal regularization scale 
was obtained: $\La^{\ro{scale}}_{E,\ro{dip}} = 1.67^{+0.66}_{-0.33}$ GeV,  
working to chiral order $\ca{O}(m_\pi^2\,\ro{log}\,m_\pi)$. 
 A successful extrapolation to the physical pion mass and infinite-volume 
 was achieved, and compared to the experimental value. 
 The infinite-volume extrapolation was merely $0.5\%$ different from
 the experimentally 
 derived value. 
%The infinite-volume extrapolation differs from the experimentally 
%derived value: $\langle r^2\rangle_E^\ro{isov} = 0.88$ fm$^2$, by merely 
%$0.5\%$.

\begin{figure}[tp]
\centering
\includegraphics[height=0.7\hsize,angle=90]{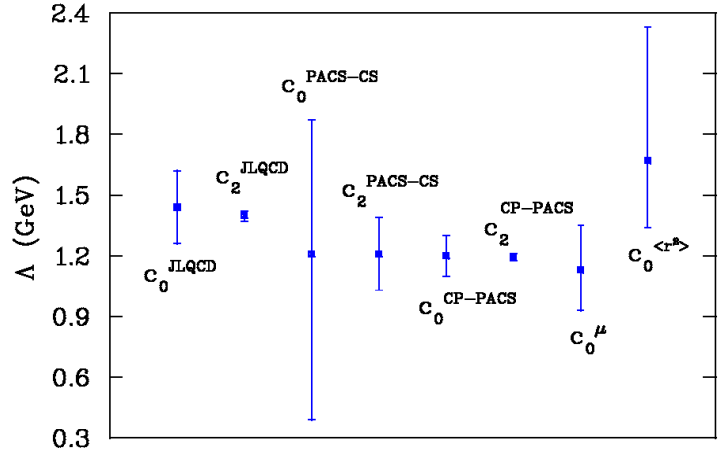}
\vspace{-11pt}
\caption{\footnotesize{Collated values for the intrinsic scale obtained from each data set for the nucleon mass, magnetic moment and electric charge radius, by analyzing a variety of low-energy coefficients. Each point, with its associated  systematic error bar, is labelled by the low-energy coefficient analyzed. The results from the analyses of the nucleon mass are further denoted by the collaboration whose lattice results are used. $c_0^\mu$ and $c_0^{\langle r^2\rangle}$ denote the intrinsic scale obtained from the analysis of the low-energy coefficient $\,c_0\,$ corresponding to the magnetic moment, and the electric charge radius expansions, respectively. A dipole regulator is used. }}
\label{fig:lambdaplottot}
\end{figure}

Figure \ref{fig:lambdaplottot} collates the values of the intrinsic scale 
for a dipole regulator obtained 
from each of the three sets of lattice results from the nucleon mass 
analysis, the magnetic moment analysis and the electric charge radius 
analysis.  
 In summary, a method for determining the existence of a well defined intrinsic 
scale has been discovered.  
It has also been illustrated how its value can be determined from lattice
 QCD results. 

\section{Future Studies and Further Developments}
\label{sect:fut}

The research presented in this thesis encourages several avenues for 
further investigation. In the heavy-baryon formulation of chiral 
perturbation theory ($\chi$PT), 
presented first for the renormalization of the mass of the nucleon in 
Chapter \ref{chpt:intrinsic}, 
the finite-volume corrections to the %order $\ca{O}(m_q)$ 
tadpole 
contribution are not evaluated. This is due to a technical subtlety 
associated with the large $m_\pi$ behaviour of the finite-volume correction, 
%Don't introduce anything new in the conclusion!
%which converges to zero, as expected, using the 'master formula' 
%as described by Beane \cite{Beane:2004tw} (see Appendix \ref{app:master}). 
%Yet the formula is badly behaved 
due to the $m_\pi^2$ coefficient occurring in Equations (\ref{eqn:tad}) through 
(\ref{eqn:tadexpnsi}). %At chiral order $\ca{O}(m_\pi^2)$, the 
The tadpole 
finite-volume corrections diverge as $m_\pi^4$. 
Since it is known that the finite-volume corrections 
must converge \cite{Beane:2004tw}, and lattice QCD simulations do not exhibit 
any divergence associated with large $m_\pi$ on a finite volume, 
 higher-order terms, for example, those occurring at order $\ca{O}(m_\pi^4)$, 
must act to reduce the estimated value of the finite-volume correction. 
%Look at previous versions of nuc-article...

In the analysis of the renormalization flow of the nucleon mass, 
 it was discovered that the 
scale dependence was weakened by working to a sufficiently 
high chiral order. 
It was also found, however, that the residual scale-dependence 
persisted as a significant component of the systematic uncertainty.
 For efficient propagation of this uncertainty, an interesting future direction
 would be to consider Bayesian methods of marginalization over the 
scale-dependence \cite{Schindler:2008fh}.

%
%covariance chisq for estimation of sys error in choice of lambda
%%%
%An estimation of the systematic uncertainty in the optimal regularization 
%scale is provided by a chi-square-style analysis. The sets of pion masses 
%analyzed, which comprise the degrees of freedom, are treated as independent 
%data sets. Realistically, these data sets are correlated, since a new 
%set is formed simply by including an additional higher mass data point 
%to the existing set. The result is an underestimate of the systematic 
%uncertainty in the optimal regularization scale, since the assumption 
%that each data set is independent constrains the possible fits to the data 
%for each regularization scale. A more accurate estimate of the systematic 
%error could be calculated from a covariance chi-square analysis, 
%which builds the correlations associated with adding a new data point 
%into the analysis.

More generally, this research provides a strong basis for the investigation 
of baryon resonances by analysing lattice QCD simulations. Resonances 
of the nucleon, such as the Roper Resonance, are not well understood 
in terms of effective field theory. The structure and behaviour of the 
resonances lend themselves to a fruitful future area of research. 
Indeed, it is not possible to link the finite-volume results of lattice 
QCD to experiment without understanding their relation to the multi-particle 
states that dress the resonances. FRR $\chi$EFT is particularly 
well-suited to exploring this important area of research. 

%\newpage 
\section{Codetta}
%\textit{``All kinds of motion, every combination,\\
%They came at last into such disposition \\
%As now establishes the sum of things.''}
%\textit{``Nature's work \\
%Is done by means of particles unseen.''} %p29
%\textit{``These particles, incapable themselves \\
%Of separate status, are held together, \\
%Are indissoluble by any force.''} \\ %p37
%(Titus Lucretius Carus. \textit{De Rerum Natura} p.37, 
%translation by Rolfe Humphries.)

\textit{``[T]he collective efforts of numerous physicists have revealed 
some of nature's best-kept secrets. And once revealed, these explanatory gems 
have opened vistas on a world we thought we knew, but whose splendor we had not 
even come close to imagining.''} (Greene, B. 1999. \textit{The Elegant Universe} p.386) \cite{Greene}

The dynamics of quantum chromodynamics 
provide a rich framework for the investigation of 
the properties of hadrons. In particular, low-energy effective 
field theory allows one to glean insights into the physical behaviour 
of subatomic particles and the structure of matter. By incorporating 
the fundamental symmetries of quantum chromodynamics 
into the action, chiral perturbation theory 
provides 
a robust method for the calculation of hadronic observables within 
the power-counting regime. 
In this thesis, finite-range regularized chiral effective field theory
 was used to develop a procedure 
for performing calculations beyond the power-counting regime, 
and handling any subsequent 
finite-range regularization scale-dependence.
Using chiral effective field theory in conjunction with the non-perturbative
 approach 
of lattice quantum chromodynamics, 
chiral extrapolations; finite-volume effects; 
the consequences of dynamical chiral 
symmetry breaking on subatomic behaviour; the importance of strangeness; 
vacuum polarizations; and many other phenomena yield fruitful understanding 
into the inner workings of the universe.

\textbf{Concluding Statement}
%\section{Concluding Statement}

Chiral effective field theory allows the identification of an 
intrinsic energy scale in the nucleon-pion interaction from lattice 
simulation results. An optimal finite-range regularization scale, 
obtained from analyzing the renormalization flow of the low-energy coefficients 
of the chiral expansion, allows successful extrapolations to be made 
to the chiral regime and to the infinite-volume limit. There is strong evidence 
 to suggest 
that  the optimal scale characterizes the intrinsic energy scale of the 
interaction between the pion and the nucleon.

The datum, the results and the rigorous theory integrate to form a strong 
argument. Chiral effective field theory extended beyond the power 
counting-regime allows the identification of an intrinsic energy scale, 
and leads to a robust method for chiral and infinite-volume extrapolations.
This is the original contribution of this thesis. 

\textit{``We have thus achieved the point where the theory may finally be compared with experience, and the road leading from formalism to concrete reality is at last complete.''}
(Omn\`{e}s, R. 2002. \textit{Quantum Philosophy: Understanding and Interpreting Contemporary Science} p.209) \cite{Omnes}

%% file: appendix2.tex
%\chapter{Appendix 2: Conventions}
\vspace{-5mm}
\chapter{Conventions}
\label{chpt:appendix2}

\vspace{-10mm}
\section{Dirac and Pauli Spin Matrices}
\label{app:spin}

The Pauli matrices are usually chosen as such:
\begin{align}
\tau^1 &= \begin{pmatrix}0&1\cr 1&0\end{pmatrix} \\
\tau^2 &= \begin{pmatrix}0&-i\cr i&0\end{pmatrix} \\
\tau^3 &= \begin{pmatrix}1&0\cr 0&-1\end{pmatrix} 
\label{eqn:paulispinmatrices}
\end{align}
There are several conventions for the definition of the Dirac matrices
 (such as Weyl/Chiral
or the Majorana Representation). Here is the Dirac Representation:
\begin{align}
\gamma^0 &= \begin{pmatrix}\mathbb{I}& 0\cr 0& -\mathbb{I}\end{pmatrix} \\
\gamma^i &= \begin{pmatrix}0&\sigma^i\cr -\sigma^i&0\end{pmatrix} \\
\gamma_5 &= \begin{pmatrix}0&\mathbb{I}\cr \mathbb{I}&0\end{pmatrix} \quad = \quad i \gamma^0 \gamma^1 \gamma^2 \gamma^3 \,. 
\label{eqn:diracrepresentation}
\end{align}
All representations of these matrices satisfy the requirement of Clifford Algebra due to the conditions imposed in the derivation of the Dirac Equation \cite{P&S}.
\begin{align}
\left \{\gamma^\mu, \gamma^\nu \right \} &= 2 g^{\mu\nu}\,,\\
\left \{\gamma^\mu, \gamma_5 \right \} &= 0 \,.
\end{align}

\section{$\SU(3)$ Gell-Mann Matrices}
\label{app:struc}

The generators of the Lie Group $\SU(3)$ satisfy
the commutator relations:
\eqb
[\la^a, \la^b] = i f^{abc}\la^c\,.
\eqe
This, combined with the relevant Jacobi Indentities
for the generators,
defines the structure constants \cite{P&S}:
\eqb
f^{ade}f^{bcd} + f^{bde}f^{cad}+ f^{cde}f^{abd}\,.
\eqe

\section{Spinor Fields}
\label{app:spinors}

The equal-time canonical anti-commutation relations for Dirac spinor fields 
are:
\begin{align}
\{\psi(x), \bar{\psi}(y) \}_{x_{0} = y_{0}} &= \hbar \delta^{3}(\vec{x} - \vec{y})\,,  \\
\{\psi(x), \psi(y) \}_{x_{0} = y_{0}} &= 0 \,.
\end{align}
The fields take the form \cite{P&S}:
\begin{equation}
\psi(x) = \int\!\!\frac{\ud^3p}{(2\pi)^3}\frac{1}{\sqrt{2\omega_{\vec{p}}}} \sum_{s} \bigg( a_{\vec{p}}^{s} u^{s}(p) e^{-ip\cdot x} + b_{\vec{p}}^{s \dagger} v^{s} (p) e^{ip \cdot x} \bigg)\,,
\end{equation}
and the canonical anti-commutation relations expressed in terms of the
 Pauli-Jordan function:
\begin{equation}
\{ \psi(x), \bar{\psi}(y)\} = (i\slashed{\partial}_{x} + m)i\Delta(x-y\,;m)\,.
\end{equation}

The Grassmann algebra is defined by the anticommutation rule between 
Grassmann variables $\p$ and a commutation rule with non-Grassmann 
numbers $c$:
%
%\begin{align}
\eqb
\{\p_i, \p_j\} = 0 = %&= 0\,,%\\
[\p_i, c\,] \,. %&= 0\,.
\eqe
%\end{align}
%
For Berezin integration over fermion spinor fields $\psi$ and $\bar{\psi}$, 
the follow rules are adopted:
\begin{align}
&\bullet\quad\int \!\! \ud \psi_{i} \psi_{j} = \int \!\! \ud \bar{\psi}_{i} 
\bar{\psi}_{j}  =  c\,\delta_{ij}\,, \\
&\bullet\quad\int\!\!\ud\p_i \f{\cd f}{\cd \p_j} = 0\,,
\label{eqn:Berez}
\end{align}
where the non-Grassmann 
constant $c$ is chosen, by convention, to be equal to $1$ and  
the function $f$ is defined on the Grassmann algebra. 
As a consequence of Equation (\ref{eqn:Berez}), the Berezin integral 
over unity vanishes:
\eqb
\int\!\! \ud \psi = \int\!\! \ud \bar{\psi}  =  0\,. 
\eqe

%
%\begin{align}
%\int\!\!\ud\p_i (c_1f_1(\p_j) + c_2f_2(\p_j)) &= 
%c_1\int\!\!\ud\p_i f_1(\p_j) + c_2\int\!\!\ud\p_i f_2(\p_j) 
%\mbox{(linearity)}\\
%
%\end{align}
%

\section{Meson and Baryon Field Definitions}
\label{app:fields}

The $\SU(3)$ mixed-symmetric meson octet fields $\pi(x) = \pi^a(x) \la^a$
 can be encoded in a traceless $3\times 3$
matrix of the form:
\eqb
\pi(x) = \sqrt{2}\begin{pmatrix}
\f{1}{\sqrt{2}}\pi^{0} + \f{1}{\sqrt{6}}\eta & \pi^{+} & K^{+} \cr
\pi^{-} & -\f{1}{\sqrt{2}}\pi^0 + \f{1}{\sqrt{6}}\eta & K^{0} \cr
K^{-} & \bar{K}^{0} & \f{-2}{\sqrt{6}}\eta 
\end{pmatrix}\,,
\eqe
In $\SU(2)$ the pions form the triplet representation 
$(\pi^{-},\pi^{0},\pi^{+})$ which can be written by summing
over the Pauli spin matrices in Appendix (\ref{app:spin}):
\eqb
\pi(x) = \tau^a \pi^a(x) =\begin{pmatrix}
\pi^{0} & \sqrt{2}\pi^{+} \cr
\sqrt{2} & -\pi^{0}
\end{pmatrix}\,,
\eqe
 Using the convention for Clebsch-Gordan coefficients from 
Wang \emph{et al.} \cite{Wang:2008vb}, the 
the mixed-symmetric baryon octet matrix has the form:
\eqb
B(x) = \begin{pmatrix}
\f{1}{\sqrt{2}}\Si^0 + \f{1}{\sqrt{6}}\La & \Si^{+} & p \cr
\Si^{-} & -\f{1}{\sqrt{2}}\Si^0 + \f{1}{\sqrt{6}}\La & n \cr
\Xi^{-} & \Xi^{0} & \f{-2}{\sqrt{6}}\La 
\end{pmatrix}\,,
\eqe
The maximally symmetric decuplet tensor (suppressing
Lorentz indices) has elements defined by:
\begin{align}
T_{111}&=\De^{++},\, T_{112}=\f{1}{\sqrt{3}}\De^{+},\,
T_{122}=\f{1}{\sqrt{3}}\De^{0},\, T_{222}=\De^{-},\nn\\
T_{113}&=\f{1}{\sqrt{3}}\Si^{*,+},\quad 
T_{123}=\f{1}{\sqrt{6}}\Si^{*,0},\quad
T_{223}=\f{1}{\sqrt{3}}\Si^{*,-},\nn\\
T_{133}&=\f{1}{\sqrt{3}}\Xi^{*,0},\quad 
T_{233}=\f{1}{\sqrt{3}}\Xi^{*,-},\quad
T_{333}= \Om^{-}.
\end{align}

%\section{Higher Order Lagrangians from Chiral Perturbation Theory}
%\label{app:lag}

%The second order nucleon-pion lagrangian, including electromagnetic 
%contributions, takes the following form of seven terms:
%
%\eqb
%\cL_{\pi N} = \bar{\Psi}\left\{\Tr[\ca{M}_+}]\right\}\Psi
%\eqe

%% file: appendix3.tex
\vspace{-5mm}
\chapter{Integration Techniques}
\label{chpt:appendix3}

\vspace{-10mm}
\section{Magnetic Quantities}
\label{app:magmom}

%1. loop and sum simplification, getting rid of k_perp @start of Work 6 &@ 
%(^v^)centre of Work 6<--- but I don't use the electric one!
%2. q-dep result in spc @startish of Work 5
%3. computation simplifcation for the sum @startish of Work 6

\subsection{Angular Components of Magnetic Moment Loop Integrals}
\label{app:magang}

In anticipation of applying finite-volume corrections to chiral loop integrals 
by comparing them to their respective summations on the lattice, 
the time-component of the $\ud^4 k$ integral is evaluated using 
Cauchy's Integral Formula, and a $\ud^3 k$ integral remains for analysis, 
as in Chapters \ref{chpt:intrinsic} through \ref{chpt:mesonmass}.

When calculating the magnetic moment in the heavy-baryon limit, 
without explicitly specifying a regularization scheme, 
the one-loop integral (corresponding 
to Figure \ref{fig:emSEa}) 
takes the following form: 
\eqb
\label{eqn:loopang}
\ca{T}^\mu_N = -\f{\chi^\mu_N}{2\pi^2}
\int\!\!\ud^3 k \f{(\hat{q}\times\vk)^2}{{(k^2 + m_\pi^2)}^2}\,.
\eqe
It is useful to be able to simplify the angular part of the integral, 
formed by the cross product of external momentum direction $\hat{q}$ 
with the loop momentum $\vk$, into a numerical coefficient. 
In order for this to be valid in calculating finite-volume corrections, 
the simplification must hold in both integral and sum forms 
of the loop diagram. Evaluating the angular part of Equation 
(\ref{eqn:loopang}) yields:
\begin{align}
\ca{T}^\mu_N &= -\f{\chi^\mu_N}{2\pi^2}
\int_{0}^{2\pi}\!\!\!\ud\varphi \int_0^\infty\!\!\!\ud k \int_{-1}^{+1}\!\!\ud x
\f{k^4 (1-x^2)}{{(k^2 + m_\pi^2)}^2}\\
&= -\f{\chi^\mu_N}{\pi} \int_0^\infty\!\!\ud k \int_{-1}^{1}\ud x\!\!
\f{k^4 (1-x^2)}{{(k^2 + m_\pi^2)}^2}\\
&= -\f{4\chi^\mu_N}{3\pi}\int_0^\infty\!\!\ud k\f{k^4}{{(k^2+m_\pi^2)}^2}\,.
\end{align}
Now, this one-dimensional integral can be transformed into a three-dimensional 
integral simply by adding in a na\"{i}ve solid angle component, using the 
identity: 
$\f{1}{4\pi}\int\!\!\ud\Omega = 1$:
\begin{align}
%\f{1}{4\pi}\int\!\!\ud\Omega &=& 1,\\
\ca{T}^\mu_N &= -\f{4\chi^\mu_N}{3\pi}\f{1}{4\pi}\int\!\!\ud\Omega\int_0^\infty
\!\!\!\ud k\f{k^4}{{(k^2+m_\pi^2)}^2}\\
\label{eqn:loopfac}
&= -\f{\chi^\mu_N}{3\pi^2}\int\!\!\ud^3\! k \f{k^2}{{(k^2+m_\pi^2)}^2}\,.
\end{align}
Comparing Equations (\ref{eqn:loopang}) and (\ref{eqn:loopfac}) shows 
that the objective has been achieved for the integral case. 
For finite volume sums, the result may not hold in general, and so must 
be checked independently. Define the following sum for box length $L$:
\eqb
\ca{T}^\mu_{N,L} = -\f{\chi^\mu_N}{2\pi^2}\left(\f{2\pi}{L}\right)^3 
\sum_{\vk}\f{(\hat{q}\times\vk)^2}{{(k^2+m_\pi^2)}^2}.
\eqe
Because $k^2 \equiv \vk^2$ is symmetrical in directions $\hat{k}_{x,y,z}$, 
it follows that:
\begin{align}
\vk^2 = k_x^2 + k_y^2 + k_z^2 &= 3 \,k_i^2\quad (i=x,y,z),\\
\mbox{and since}\quad (\hat{q}\times\vk) &= k_\perp^2,\\
\mbox{it follows}\quad k_\perp^2 &= 2 \,k_i^2.
\end{align}
Thus:
\begin{align}
\ca{T}^\mu_{N,L} &= -\f{\chi^\mu_N}{\pi^2}\left(\f{2\pi}{L}\right)^3 
\sum_{\vk}\f{k_i^2}{{(k^2+m_\pi^2)}^2}\\
&= -\f{\chi^\mu_N}{3\pi^2}\left(\f{2\pi}{L}\right)^3 
\sum_{\vk}\f{k^2}{{(k^2+m_\pi^2)}^2},
\end{align}
which is the finite-volume equivalent of Equation (\ref{eqn:loopfac}). 

\subsection{Combinatorial Simplification}
\label{app:comb}

The calculation of the three-dimensional finite sum can be made more efficient 
computationally, by transforming it to a one-dimensional sum in terms 
of the new variable $n^2 = k^2 (2\pi/L)^2$. It does, however, require 
calculation of the number of configurations of the squares of 
 $k_x$, $k_y$ and $k_z$ 
to obtain each value of $n^2$, denoted $C^{(3)}(n^2)$.
Thus, for an integrand $\ca{I}(\vk)$:
%need to know the number of ways of combining any values of i,j,k to get 
% a certain number n^2 = k^2(2\pi/L)^2
%
\eqb
\left(\f{2\pi}{L}\right)^3\,\, \sum_{\vk}^{k_{\ro{max}}}\ca{I}(\vk) = 
\left(\f{2\pi}{L}\right)^3\,\, \sum_{n^2}^{n^2_{\ro{max}}}C^{(3)}\ca{I}(n^2)\,,
\eqe
 where $n^2_{\ro{max}} = k^2_{\ro{max}} (2\pi/L)^2$.

\subsection{Sachs Magnetic Form Factors at Finite $Q^2$}
\label{app:Qsq}

Consider calculations involving the leading-order pion loop contributions 
 to 
the magnetic form factor $G_M(Q^2)$ at finite $Q^2$, 
(allowing non-zero mass splitting $\De$).
 The following integral %, with FRR function $u(k\,;\La)$ 
can be made more easily calculable using spherical polar 
coordinates (using $\om(\vk) = \sqrt{k^2 + m_\pi^2}$):
\begin{align}
\ca{T}^\mu_N(Q^2) &= -\f{\chi^\mu_N}{\pi^2}
\int\!\!\ud^3\! k \Big[ \nn\\
&\f{k_y^2\,[\om(\vk) + \om(\vk+\vq) + \De]
%\,u(\vk\,;\La)u(\vk+\vq\,;\La)}
}
{\om(\vk)[\om(\vk)+\De]\,\om(\vk+\vq)
[\om(\vk+\vq)+\De]\,[\om(\vk) + \om(\vk+\vq)]}\Big]\\
\label{eqn:spcmag}
&= -\f{\chi^\mu_N}{\pi^2}\int_0^{2\pi}\!\!\!\ud\varphi\int_0^{\pi}\!\!\!\ud\theta
\int_0^\infty\!\!\!\ud k \Big[\nn\\
&
\f{k_y^2\,k^2\sin\theta\,[\om(k) + \om(k+q) + \De]
%\,u(k\,;\La)u(k+q\,;\La)}
}
{\om(k)[\om(k)+\De]\,\om(k+q)
[\om(k+q)+\De]\,[\om(k) + \om(k+q)]}\Big]\,.\qquad\quad
\end{align}
The integral can be further altered to remove the infinite integral 
under the change of variables $k \rightarrow k/(1-k)$. For arbitrary function 
$f$:
%
%\eqb
%\int_0^\infty\!\!\!\ud x\,\,f(x) = 
%\int_0^1\!\!\ud x'\,\,\f{f(x'/(1-x'))}{(1-x')^2}\,.
%\eqe
%
\eqb
\int_0^\infty\!\!\!\ud k\,\,f(k) = 
\int_0^1\!\!\ud k\,\,\f{f(k/(1-k))}{(1-k)^2}\,.
\eqe

Thus, defining (for convenience) $p(k) \equiv k/(1-k)$, 
Equation (\ref{eqn:spcmag}) becomes:
\begin{align}
\ca{T}^\mu_N(Q^2) &=&\nn\\
& -\f{\chi^\mu_N}{\pi^2}\int_0^{2\pi}\!\!\!\ud\varphi\int_{-1}^{1}\!\!
\!\!\ud x%(\cos\theta)
\int_0^{1}\!\!\!\ud k \Big[\nn\\
& \f{k_y^2\,p^2\,[\om(p) + \om(p+q) + \De]
%\,u(p\,;\La)u(p+q\,;\La)}
}
{\om(p)[\om(p)+\De]\,\om(p+q)
[\om(p+q)+\De]\,[\om(p) + \om(p+q)]\,(1-k)^2}\Big]\,.\qquad\quad
\end{align}
%

%\section{Electric}
%\label{app:cr}

\section{Electric Charge Radius Integral Expansions}
\label{app:intexp}
%inf vol only for charge rad

%@ centre of Work 6 and onwards...

For the infinite-volume electric charge radius, the chiral loop integrals 
must be calculated for use with the chiral expansion of 
Equation %(\ref{eqn:chiralexp}) and 
(\ref{eqn:chiralexprenorm}). 
%This requires a $Q^2$ derivative, which 
%in this case is performed 
%analytically, and the limit of  
%vanishing $Q^2$ taken. % (see Equation \ref{eqn:siSi}). 
Each loop integrand is expanded out for small $Q^2$, and the derivative 
 in the limit of vanishing $Q^2$ is extracted. 
Using the notation of Chapter 
\ref{chpt:nucleonmagmom}, and a dipole regulator, the one-loop contribution 
takes the following form: 
\begin{align}
T^E_{N} &= \lim_{Q^2\rightarrow 0}-6\f{\cd \ca{T}^E_{N}(Q^2)}{\cd Q^2}\nn\\
&= \f{6\chi_N^E}{5\pi}\int\!\!\ud^3 k\f{\cd}{\cd q^2} \left[
\f{(\vk + \vq/2)\cdot(\vk-\vq/2)\,u_{\ro{dip}}(\vk+\vq/2\,;\La)\,
u_{\ro{dip}}(\vk-\vq/2\,;\La)}
{\om(\vk+\vq/2)\om(\vk-\vq/2)[\om(\vk+\vq/2) + \om(\vk-\vq/2)]}\right]
\Bigg|_{q^2 = 0}\\
&= \f{6\chi_N^E}{5\pi}\int\!\!\ud^3 k \Big(\La^8\{-\om^2(\vk)(\vk^2 + \La^2)
(13\vk^4 + 2m_\pi^2\La^2 + 5\vk^2(2m_\pi^2 + \La^2)) \nn\\
&+ \vk^4
(21\vk^4 + 16m_\pi^4 + 5\La^4 + 2\vk^2(16m_\pi^2 + 5\La^2))\cos^2\theta\}\Big)
\Big(16\om^7(\vk)(\vk^2 + \La^2)^6\Big)^{-1}\,.
\end{align}
If a mass-splitting is included:
\begin{align}
T^E_\De &= \f{6\chi_\De^E}{5\pi}\!\int\!\!\ud^3 k\f{\cd}{\cd q^2}\!\left[
\f{(\vk + \vq/2)\cdot(\vk-\vq/2)\,u_{\ro{dip}}(\vk+\vq/2\,;\La)\,
u_{\ro{dip}}(\vk-\vq/2\,;\La)}
{(\om(\vk+\vq/2) + \De)(\om(\vk-\vq/2)+ \De)
[\om(\vk+\vq/2) + \om(\vk-\vq/2)]}\right]
\Bigg|_{q^2 = 0}\\
&= \f{6\chi_\De^E}{5\pi}\int\!\!\ud^3 k \Big(\La^8\{-\om(\vk)(\vk^2+\La^2)
[13\vk^6 + 2m_\pi^2(m_\pi^2 + \De(2\om(\vk)+\De))\La^2 \nn\\
&+ 
\vk^4(23m_\pi^2+24\om(\vk)\De+11\De^2+5\La^2)\nn\\
&+\vk^2(10m_\pi^4+\De(8\om(\vk)+3\De)\La^2 + m_\pi^2(20\om(\vk)\De
+10\De^2+7\La^2) )]\nn\\
&+ \vk^4[21\vk^6+16m_\pi^6 + 16m_\pi^4\De(2\om(\vk)+\De)+5m_\pi^2\La^4
+\De(4\om(\vk)+\De)\La^4\nn\\
&+\vk^4(53m_\pi^2+36\om(\vk)\De + 17\De^2
+ 10\La^2)\nn\\
&+\vk^2(48m_\pi^4+\La^2(8\om(\vk)\De + 2\De^2+5\La^2) +2m_\pi^2
(32\om(\vk)\De + 16\De^2 + 5\La^2) )]\cos^2\theta\}\Big)\nn\\
& \times\Big(16\om^5(\vk)(\om(\vk)+\De)^4(\vk^2 + \La^2)^6\Big)^{-1}\,.
\end{align}
Similarly, the tadpole contribution takes the following form:
\begin{align}
T_{\ro{tad}}^E &= \f{6\chi_t^E}{\pi}\int\!\!\ud^3 k\f{\cd}{\cd q^2} \left[
\f{u^2_{\ro{dip}}(\vk\,;\La)}{\om(\vk+\vq/2) + \om(\vk-\vq/2)}
\right]\Bigg|_{q^2 = 0}\\
&= \f{6\chi_t^E}{\pi}\int\!\!\ud^3 k \f{\vk^2\cos^2\theta - \om^2(\vk)}
{16\om^5(\vk)}u^2_{\ro{dip}}(\vk\,;\La).
\end{align}

%\section{Finite-Volume Corrections using the 'Master Formula'}
%\label{app:master}

\section{Finite Volume Corrections to Tadpole Amplitudes}
\label{app:tadfvc}

\begin{figure}
\centering
\includegraphics[height=0.7\hsize,angle=90]{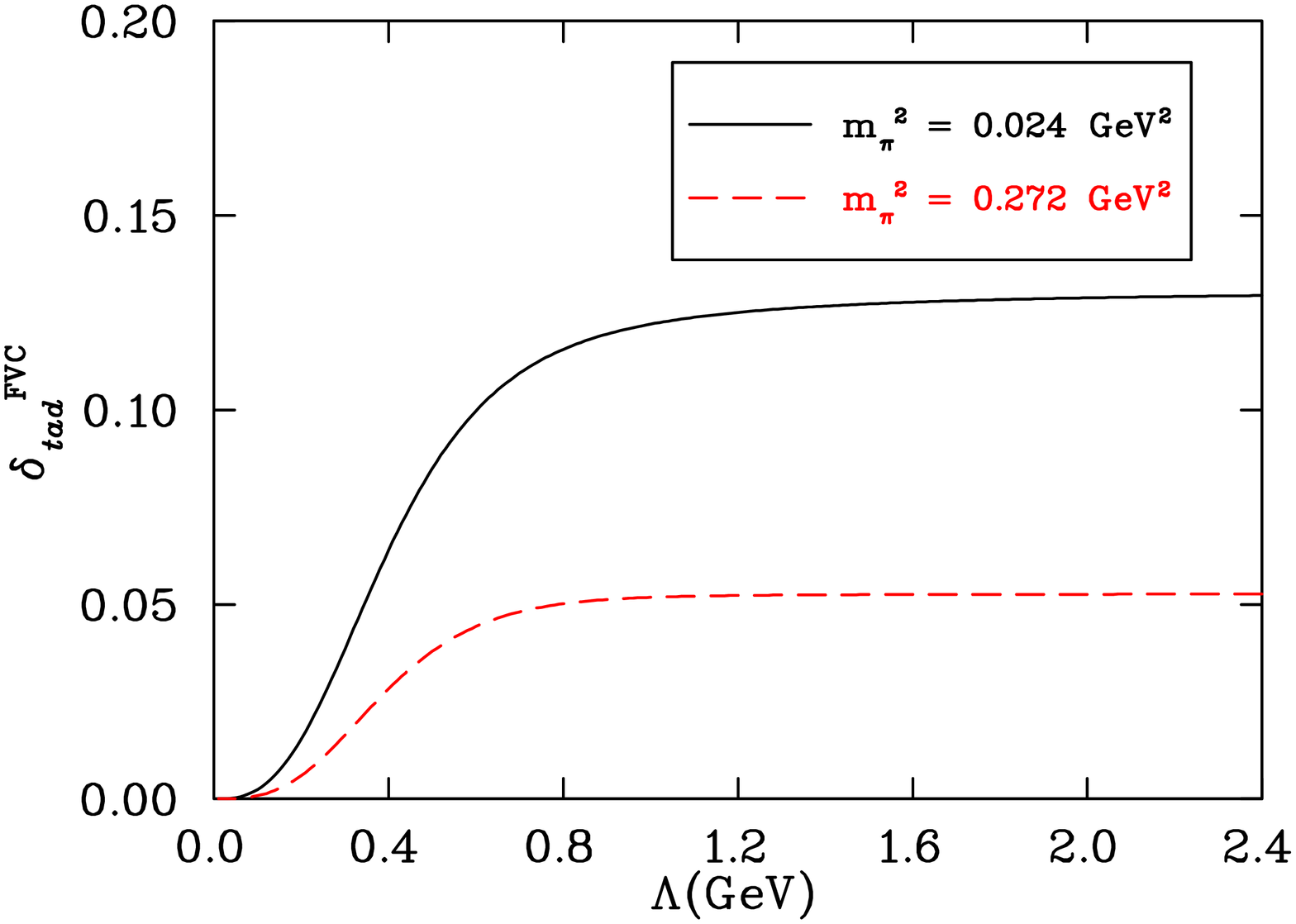}
\vspace{-12pt}
\caption{ Behaviour of the finite-volume
    corrections $\de_{tad}^\ro{FVC}$ vs.\ $\La$ on a $2.9$ fm box using a dipole regulator. Results for two different values of $m_\pi^2$ are shown.}
\label{fig:fvctad}
\vspace{6mm}
\includegraphics[height=0.7\hsize,angle=90]{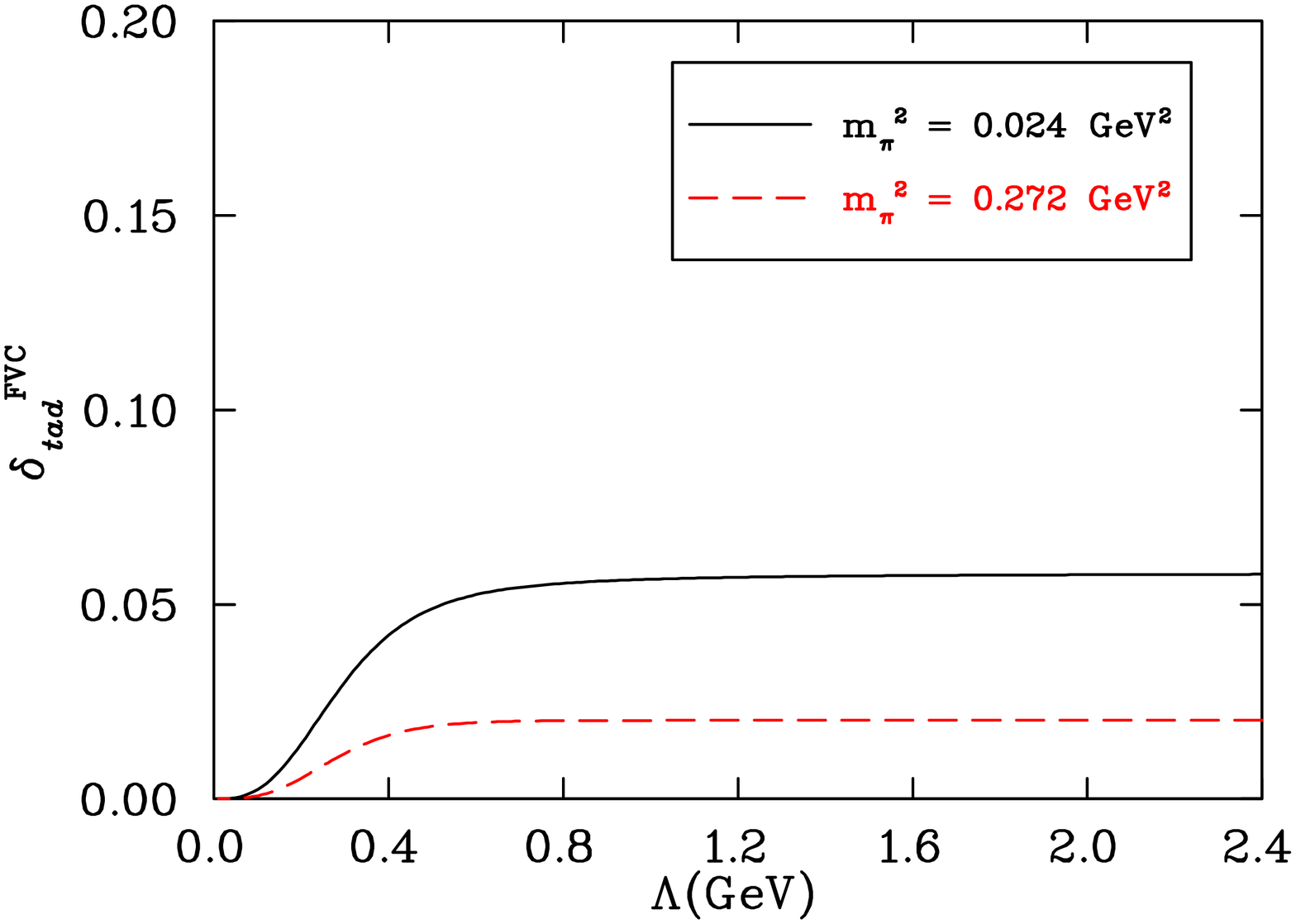}
\vspace{-12pt}
\caption{ Behaviour of finite-volume
    corrections $\de_{tad}^\ro{FVC}$ vs.\ $\La$ on a $4.0$ fm box using a dipole regulator. Results for two different values of $m_\pi^2$ are shown.}
\label{fig:fvctad4fm}
\end{figure}

Finite-volume corrections should vanish as $m_\pi^2$ becomes large,
as observed in lattice quantum chromodynamics (lattice QCD) 
simpulations. This has also been observed, 
in turn, for each
of the finite-volume corrections involved in the extrapolation of the
nucleon mass to fourth-order. However, the tadpole finite-volume correction,
$\de_{tad}^{\ro{FVC}}$, is different in that it is multiplied by a factor of 
$m_\pi^2$, 
as evident in Equation (\ref{eqn:tadunmod}). 
 The product  
$c_2m_\pi^2\de_{tad}^{\ro{FVC}}$ is not convergent for large $m_\pi$. Figures 
\ref{fig:fvctad} and \ref{fig:fvctad4fm} show the behaviour of the 
tadpole 
finite-volume correction for a $2.9$ fm box and a $4.0$ fm box, respectively. 

The finite-volume estimate of $c_2$, denoted $c_2^V$, is not in general 
the same value as the infinite-volume $c_2$.
Thus the finite-volume correction of the tadpole cannot be written as
simply the difference between the finite volume sum and the infinite
volume integral, but must distinguish between $c_2^V$ and  $c_2$:

\eqb
c_2m_\pi^2\de_{tad}^\ro{FVC} = c_2 m_\pi^2
\left( \f{c_2^V}{c_2} \Si_{tad}^V - \Si_{tad} \right)\,.
\eqe

%Consider the loop integrals of 
%Eqs. (\ref{eqn:NN}) - (\ref{eqn:tadexpnsi}) but without making the
%subtrations of the $b_i$ terms.
Since $c_2$  is by definition the coefficient of the $m_\pi^2$
term in the nucleon mass expansion,
the renormalization of the residual coefficient $a_2$ by the contributions 
from the integrals $\Si_{N}$, $\Si_\De$ and $\Si_{tad}$, defined in Equations 
(\ref{eqn:NNunmod}) through (\ref{eqn:tadexpnSiunmod}), 
can be written as follows: 
\begin{align}
c_2 m_\pi^2 &= (a_2 + b_2^{N} + b_2^{\De}
 + c_2 b_2^{t})m_\pi^2\,,\\
\Rightarrow c_2 &= \f{a_2^\La + b_2^{N} + b_2^{\De}}{1-b_2^{t}}\,.
\label{eqn:nonlinremormiv}
\end{align}
%

%as well as the coefficient of the tadpole integral,
%the following relation holds for $c_2$:
%

%
An analogous relation exists for the finite volume $c_2^V$:
\eqb
 c_2^V = \f{a_2 + b_2^{V,N} + b_2^{V,\De}}{1-b_2^{V,t}}\,.
\label{eqn:nonlinremormfv}
\eqe
%
%All $b_2^V$ coefficients are easily obtained from the loop integrals
%themselves.  
%
By simultaneously solving for $a_2$ and $c_2^V$, 
the ratio $c_2^V/c_2$ can be calculated in principle,
and the tadpole finite volume corrections are tractable.
It should be noted however, that this does not resolve the problem of divergent
 behaviour for large $m_\pi$. % Therefore, it has been decided
 % that this finite
% volume correction in the present analysis should not be included.

%\section{The Analytic Expression for Finite Volume Corrections}
%\label{app:fvc}
%re: Beane

%% file: appendix4.tex
%\vspace{-5mm}
\chapter{Lattice Simulation Results}
\label{chpt:appendix4}

%\vspace{-10mm}
%\section{The Nucleon Mass}
%\label{app:nucmass}

%\subsection{JLQCD Results}
%\label{app:JLQCD}

\begin{table}[tp]
 \caption{\footnotesize{JLQCD \cite{Ohki:2008ff} 
lattice QCD simulation results for
    the nucleon mass $M_N$ %and the associated statistical error $\delta M_N$   
 at various pion mass squared values $m_\pi^2$. 
The lattice spacing is $0.118$ fm and the spatial lattice length is $1.90$ fm. %Note that data points 
%with $m_\pi L < 4$ are not used in the analysis.
}}
  \newcommand\T{\rule{0pt}{2.8ex}}
  \newcommand\B{\rule[-1.4ex]{0pt}{0pt}}
  \begin{center}
    \begin{tabular}{lll}
      \hline
      \hline
       \T\B 
      $m_\pi^2$(GeV$^2$) &  $M_N$(GeV)  & $m_\pi L$  \\
      \hline
% $0.567$ & $1.62$ & $0.006$ & $7.25$ \\
%$0.386$ & $1.46$ & $0.006$ & $5.98$ \\
%$0.273$ & $1.35$ & $0.006$ & $5.03$ \\
%$0.191$ & $1.25$ & $0.006$ & $4.20$ \\
%$0.135$ & $1.16$ & $0.008$ & $3.54$ \\
%$0.084$ & $1.11$ & $0.010$ & $2.78$ \\
%
$0.567$ & $1.615$($6$)  & $7.25$ \\
$0.386$ & $1.456$($6$)  & $5.98$ \\
$0.273$ & $1.350$($6$)  & $5.03$ \\
$0.191$ & $1.255$($6$)  & $4.20$ \\
$0.135$ & $1.164$($8$)  & $3.54$ \\
$0.084$ & $1.111$($10$) & $2.78$ \\
      \hline
    \end{tabular}
  \end{center}
\vspace{-6pt}
  \label{table:JLQCDdata}
\end{table}

%\subsection{PACS-CS Results}
%\label{app:PACS-CS}

\begin{table}[tp]
 \caption{\footnotesize{PACS-CS \cite{Aoki:2008sm} 
lattice QCD simulation results for
    the nucleon mass $M_N$ %and the associated statistical error $\delta M_N$  
 at various pion mass squared values $m_\pi^2$. 
The lattice spacing is $0.0907$ fm and the 
spatial lattice length is $2.90$ fm. %Note that data points 
%with $m_\pi L < 4$ are not used in the analysis.
}}
  \newcommand\T{\rule{0pt}{2.8ex}}
  \newcommand\B{\rule[-1.4ex]{0pt}{0pt}}
  \begin{center}
    \begin{tabular}{lll}
      \hline
      \hline
       \T\B 
      $m_\pi^2$(GeV$^2$) &  $M_N$(GeV) & $m_\pi L$  \\
      \hline
%$0.492$ & $1.58$ & $0.005$  & $10.32$ \\
%$0.325$ & $1.41$ & $0.012$  & $8.38$ \\
%$0.169$ & $1.21$ & $0.012$  & $6.05$ \\
%$0.087$ & $1.09$ & $0.019$  & $4.35$ \\
%$0.024$ & $0.93$ & $0.078$  & $2.29$ \\
%
$0.492$ & $1.583$($5$)   & $10.32$ \\
$0.325$ & $1.411$($12$)  & $8.38$ \\
$0.169$ & $1.215$($12$)  & $6.05$ \\
$0.087$ & $1.093$($19$)  & $4.35$ \\
$0.024$ & $0.932$($78$)  & $2.29$ \\
      \hline
    \end{tabular}
  \end{center}
\vspace{-6pt}
  \label{table:PACS-CSdata}
\end{table}

%\subsection{CP-PACS Results}
%\label{app:CP-PACS}

\begin{table}[tp]
 \caption{\footnotesize{CP-PACS \cite{AliKhan:2001tx} 
lattice QCD simulation results for
    the nucleon mass $M_N$, %the associated statistical error $\delta M_N$, 
 the lattice spacing $a$ and the 
spatial lattice length $L$    
 at various pion mass squared values $m_\pi^2$.
%The lattice spacing is $0.0907$ fm. %Note that data points 
%with $m_\pi L < 4$ are not used in the analysis.
}}
  \newcommand\T{\rule{0pt}{2.8ex}}
  \newcommand\B{\rule[-1.4ex]{0pt}{0pt}}
  \begin{center}
    \begin{tabular}{lllll}
      \hline
      \hline
       \T\B 
      $m_\pi^2$(GeV$^2$) &  $M_N$(GeV)  & $a$(fm) & $L$(fm) & $m_\pi L$ \\
      \hline
%$0.940$ & $1.81$ & $0.015$ & $0.102$ & $2.45$ & $12.03$ \\
%$0.913$ & $1.80$ & $0.004$ & $0.130$ & $3.12$ & $15.11$ \\
%$0.704$ & $1.65$ & $0.009$ & $0.099$ & $2.38$ & $10.10$ \\
%$0.689$ & $1.64$ & $0.005$ & $0.123$ & $2.95$ & $12.42$ \\
%$0.539$ & $1.52$ & $0.009$ & $0.095$ & $2.28$ & $8.49$ \\
%$0.502$ & $1.50$ & $0.006$ & $0.118$ & $2.83$ & $10.17$ \\
%$0.353$ & $1.35$ & $0.012$ & $0.092$ & $2.21$ & $6.65$ \\
%$0.272$ & $1.27$ & $0.007$ & $0.111$ & $2.66$ & $7.04$ \\
%
$0.940$ & $1.809$($15$) & $0.102$ & $2.45$ & $12.03$ \\
$0.913$ & $1.798$($4$) & $0.130$ & $3.12$ & $15.11$ \\
$0.704$ & $1.652$($9$) & $0.099$ & $2.38$ & $10.10$ \\
$0.689$ & $1.643$($5$) & $0.123$ & $2.95$ & $12.42$ \\
$0.539$ & $1.519$($9$) & $0.095$ & $2.28$ & $8.49$ \\
$0.502$ & $1.497$($6$) & $0.118$ & $2.83$ & $10.17$ \\
$0.353$ & $1.348$($12$) & $0.092$ & $2.21$ & $6.65$ \\
$0.272$ & $1.275$($7$) & $0.111$ & $2.66$ & $7.04$ \\
      \hline
    \end{tabular}
  \end{center}
\vspace{-6pt}
  \label{table:CP-PACSdata}
\end{table}

\begin{table}[tp]
 \caption{\footnotesize{Quenched lattice QCD data for
    the $\rho$ meson mass $m_\rho$
 at various pion mass squared values $m_\pi^2$. 
The lattice size is 
 $20^3 \times 32$, with a lattice spacing of $0.153$ fm. %Note that data points 
%with $m_\pi L < 4$ are not used in the analysis.
}}
  \newcommand\T{\rule{0pt}{2.8ex}}
  \newcommand\B{\rule[-1.4ex]{0pt}{0pt}}
  \begin{center}
    \begin{tabular}{lll}
      \hline
      \hline
       \T\B 
      $m_\pi^2$(GeV$^2$) &  $m_\rho$(GeV)  & $m_\pi L$  \\
      \hline
$3.150$ & $2.001(1)$ & $27.53$ \\
$2.187$ & $1.700(2)$ & $22.94$ \\
$1.742$ & $1.548(2)$ & $20.47$ \\
%The top three are not used really
$1.329$ & $1.399(2)$ & $17.88$ \\
$1.212$ & $1.354(2)$ & $17.08$ \\
$1.062$ & $1.294(2)$ & $15.98$ \\
$0.867$ & $1.214(3)$ & $14.44$ \\
$0.743$ & $1.162(4)$ & $13.37$ \\
$0.676$ & $1.133(4)$ & $12.75$ \\
$0.610$ & $1.103(5)$ & $12.12$ \\
$0.515$ & $1.060(5)$ & $11.13$ \\
$0.422$ & $1.016(6)$ & $10.07$ \\
$0.347$ & $0.985(7)$ & $9.13$ \\
$0.288$ & $0.960(8)$ & $8.32$ \\
$0.241$ & $0.938(8)$ & $7.62$ \\
$0.204$ & $0.926(9)$ & $7.00$ \\
$0.172$ & $0.914(11)$ & $6.43$ \\
$0.143$ & $0.908(14)$ & $5.87$ \\
$0.114$ & $0.899(15)$ & $5.24$ \\
$0.094$ & $0.899(16)$ & $4.75$ \\
$0.080$ & $0.896(18)$ & $4.38$ \\
$0.068$ & $0.898(20)$ & $4.04$ \\
$0.059$ & $0.902(22)$ & $3.77$ \\
$0.053$ & $0.903(26)$ & $3.58$ \\
$0.047$ & $0.907(28)$ & $3.37$ \\
$0.041$ & $0.913(32)$ & $3.15$ \\
      \hline
    \end{tabular}
  \end{center}
\vspace{-6pt}
  \label{table:rhodata}
\end{table}

\begin{table}[tp]
 \caption{\footnotesize{Preliminary  
lattice QCD simulation results from QCDSF for
    the isovector nucleon magnetic moment $\mu_n^v$,  
 the lattice spacing $a$ and the 
spatial lattice length $L$    
 at various pion mass squared values $m_\pi^2$.
%The lattice spacing is $0.0907$ fm. %Note that data points 
%with $m_\pi L < 4$ are not used in the analysis.
}}
  \newcommand\T{\rule{0pt}{2.8ex}}
  \newcommand\B{\rule[-1.4ex]{0pt}{0pt}}
  \begin{center}
    \begin{tabular}{lllll}
      \hline
      \hline
       \T\B 
      $m_\pi^2$(GeV$^2$) &  $\mu_n^v$($\mu_N$) & $a$(fm) & $L$(fm) & $m_\pi L$ \\
      \hline
%$0.863$ & $2.39$ & $0.069$ & $0.089$ & $1.43$ & $6.73$ \\
%$0.709$ & $2.48$ & $0.045$ & $0.073$ & $1.76$ & $7.50$ \\
%$0.688$ & $2.55$ & $0.159$ & $0.091$ & $1.45$ & $6.11$ \\
%$0.591$ & $2.62$ & $0.049$ & $0.084$ & $2.01$ & $7.85$ \\
%$0.392$ & $2.86$ & $0.086$ & $0.070$ & $1.67$ & $5.30$ \\
%$0.357$ & $2.78$ & $0.051$ & $0.084$ & $2.03$ & $6.13$ \\
%$0.290$ & $2.84$ & $0.121$ & $0.070$ & $1.67$ & $4.57$ \\
%$0.198$ & $3.08$ & $0.120$ & $0.081$ & $1.96$ & $4.42$ \\
%$0.159$ & $3.01$ & $0.118$ & $0.077$ & $1.84$ & $3.72$ \\
%$0.077$ & $3.71$ & $0.158$ & $0.076$ & $3.04$ & $4.26$ \\
%
%
%2 d.p.
%$0.863$ & $2.39$($7$) $0.089$ & $1.43$ & $6.73$ \\
%$0.709$ & $2.48$($4$) $0.073$ & $1.76$ & $7.50$ \\
%$0.688$ & $2.55$($16$) $0.091$ & $1.45$ & $6.11$ \\
%$0.591$ & $2.62$($5$) $0.084$ & $2.01$ & $7.85$ \\
%$0.392$ & $2.86$($9$) $0.070$ & $1.67$ & $5.30$ \\
%$0.357$ & $2.78$($5$) $0.084$ & $2.03$ & $6.13$ \\
%$0.290$ & $2.84$($12$) $0.070$ & $1.67$ & $4.57$ \\
%$0.198$ & $3.08$($12$) $0.081$ & $1.96$ & $4.42$ \\
%$0.159$ & $3.01$($12$) $0.077$ & $1.84$ & $3.72$ \\
%$0.077$ & $3.71$($16$) $0.076$ & $3.04$ & $4.26$ \\
%
%Or: 3 d.p:
$0.863$ & $2.394$($69$) & $0.089$ & $1.43$ & $6.73$ \\
$0.709$ & $2.483$($45$) & $0.073$ & $1.76$ & $7.50$ \\
$0.688$ & $2.548$($159$) & $0.091$ & $1.45$ & $6.11$ \\
$0.591$ & $2.621$($49$) & $0.084$ & $2.01$ & $7.85$ \\
$0.392$ & $2.863$($86$) & $0.070$ & $1.67$ & $5.30$ \\
$0.357$ & $2.781$($51$) & $0.084$ & $2.03$ & $6.13$ \\
$0.290$ & $2.840$($121$) & $0.070$ & $1.67$ & $4.57$ \\
$0.198$ & $3.082$($120$) & $0.081$ & $1.96$ & $4.42$ \\
$0.159$ & $3.006$($118$) & $0.077$ & $1.84$ & $3.72$ \\
$0.077$ & $3.711$($158$) & $0.076$ & $3.04$ & $4.26$ \\
      \hline
    \end{tabular}
  \end{center}
\vspace{-6pt}
  \label{table:magdata}
\end{table}

\begin{table}[tp]
 \caption{\footnotesize{Preliminary  
lattice QCD simulation results from QCDSF for
    the isovector nucleon electric charge radius $\rad_E$, 
 the lattice spacing $a$ and the 
spatial lattice length $L$    
 at various pion mass squared values $m_\pi^2$.
%The lattice spacing is $0.0907$ fm. %Note that data points 
%with $m_\pi L < 4$ are not used in the analysis.
}}
  \newcommand\T{\rule{0pt}{2.8ex}}
  \newcommand\B{\rule[-1.4ex]{0pt}{0pt}}
  \begin{center}
    \begin{tabular}{lllll}
      \hline
      \hline
       \T\B 
      $m_\pi^2$(GeV$^2$) &  $\rad_E$(fm$^2$) & $a$(fm) & $L$(fm) & $m_\pi L$ \\
      \hline
%$0.591$ & $0.30$ & $0.008$ & $0.084$ & $2.01$ & $7.85$ \\
%$0.357$ & $0.35$ & $0.006$ & $0.084$ & $2.03$ & $6.13$ \\
%$0.349$ & $0.34$ & $0.005$ & $0.080$ & $1.92$ & $5.75$ \\
%$0.198$ & $0.38$ & $0.011$ & $0.081$ & $1.96$ & $4.42$ \\
%$0.188$ & $0.39$ & $0.012$ & $0.068$ & $2.19$ & $4.81$ \\
%$0.074$ & $0.49$ & $0.025$ & $0.076$ & $3.04$ & $4.18$ \\
%$0.053$ & $0.59$ & $0.024$ & $0.068$ & $3.25$ & $3.79$ \\
%
$0.591$ & $0.303$($8$) & $0.084$ & $2.01$ & $7.85$ \\
$0.357$ & $0.349$($6$) & $0.084$ & $2.03$ & $6.13$ \\
$0.349$ & $0.340$($5$) & $0.080$ & $1.92$ & $5.75$ \\
$0.198$ & $0.384$($11$) & $0.081$ & $1.96$ & $4.42$ \\
$0.188$ & $0.392$($12$) & $0.068$ & $2.19$ & $4.81$ \\
$0.074$ & $0.494$($25$) & $0.076$ & $3.04$ & $4.18$ \\
$0.053$ & $0.586$($24$) & $0.068$ & $3.25$ & $3.79$ \\
      \hline
    \end{tabular}
  \end{center}
\vspace{-6pt}
  \label{table:elecdata}
\end{table}